\documentclass[a4paper,12pt]{article}
\pdfoutput=1
\usepackage{epsfig,amsmath,amsfonts,amsthm}
\usepackage[figuresright]{rotating}
\usepackage[left=28mm,right=20mm,top=30mm,bottom=20mm]{geometry}
\usepackage{subfig}
\usepackage{graphicx}
\graphicspath{{RoMnewfigures/}}
\usepackage{rotating}
\usepackage{booktabs}
\usepackage{moreverb}

\DeclareGraphicsExtensions{.pdf,.png,.jpg}
\usepackage{natbib}
\numberwithin{equation}{section}

\def\var{\hbox{Var}}

\providecommand{\keywords}[1]{\textbf{{Keywords}} #1}
\usepackage[nokeyprefix]{refstyle}
\usepackage{varioref}
\usepackage{xr}
\usepackage{hyperref}

\begin{document}

\title{Simulation study of estimating between-study variance and overall effect in meta-analyses of log-response-ratio for lognormal data}

\author{Ilyas Bakbergenuly, David C. Hoaglin,  and Elena Kulinskaya}

\date{\today}

\maketitle



\abstract{Methods for random-effects meta-analysis require an estimate of the between-study variance, $\tau^2$. The performance of estimators of $\tau^2$ (measured by bias and coverage) affects their usefulness in assessing heterogeneity of study-level effects, and also the performance of related estimators of the overall effect. For the effect measure log-response-ratio (LRR, also known as the logarithm of the ratio of means, RoM), we review four point estimators of $\tau^2$ (the popular methods of DerSimonian-Laird (DL), restricted maximum likelihood, and Mandel and Paule (MP), and the less-familiar method of Jackson), four interval estimators for $\tau^2$ (profile likelihood, Q-profile, Biggerstaff and Jackson, and Jackson), five point estimators of the overall effect (the four related to the point estimators of $\tau^2$ and an estimator whose weights use only study-level sample sizes), and seven interval estimators for the overall effect (four based on the point estimators for $\tau^2$, the Hartung-Knapp-Sidik-Jonkman (HKSJ) interval, a modification of HKSJ that uses the MP estimator of $\tau^2$ instead of the DL estimator, and an interval based on the sample-size-weighted estimator).  We obtain empirical evidence from extensive simulations of data from lognormal distributions.}


\section{Introduction}\label{sec:Intro}
Meta-analysis is a statistical methodology for combining estimated effects from several studies in order to assess their heterogeneity and obtain an overall estimate. In this paper we focus on the log-response-ratio (LRR, also known as the logarithm of the ratio of means, RoM) as the effect measure. In ecology almost half of all meta-analyses use this outcome measure (\cite{KorichevaG-2014, Nakagawa2012}).

{  The LRR was originally introduced by \cite{hedges1999meta} and  rediscovered later as RoM by  \cite{Friedrich2008} assuming underlying normality of the raw data.}   However, the LRR is not defined for negative values of the study means, and  \cite{lajeunesse2015bias} modeled the data by lognormal distributions. We explore the meta-analysis of LRR under the lognormal distribution in this report. Our results under normality constitute a companion report.

If the studies can be assumed to have the same true effect, a meta-analysis can use a fixed-effect (FE) model (common-effect model) to combine the estimates. Otherwise, the studies' true effects can depart from homogeneity in a variety of ways. Most commonly, a random-effects (RE) model regards those effects as a sample from a distribution and summarizes their heterogeneity via its variance, usually denoted by $\tau^2$.
The between-studies variance, $\tau^2$, has a key role in estimates of the mean of the distribution of random effects; but it is also important as a quantitative indication of heterogeneity \citep{higgins2009re}.
In studying estimation for meta-analysis of LRR, we focus first on $\tau^2$ and then proceed to the overall effect.

\cite{veroniki2015methods} provide a comprehensive overview and recommendations  on methods of estimating $\tau^2$ and its uncertainty. Their review, however, has two important limitations. First, the authors study only \lq\lq methods that can be applied for any type of outcome data.'' However, as we show elsewhere, the performance of the methods varies widely among effect measures. Second, any review of the topic, such as \cite{veroniki2015methods}, currently can draw on only limited empirical information on the comparative performance of the methods. We address both issues for the effect measure LRR.

\cite{veroniki2015methods} (Appendix Table 1) cite no previous simulation studies on the comparative performance of estimates of $\tau^2$ for LRR.

Several studies have considered  the quality of estimation of LRR itself. \cite{Friedrich2008} report extensive simulations for LRR under normality, but they use only the DerSimonian-Laird (DL) method to estimate $\tau^2$ and do not report on its quality. \cite{lajeunesse2015bias} discusses bias correction for LRR and its variance, and provides some simulation results for lognormal distributions, but only under the fixed-effect model. \cite{Doncaster2017} provide some limited simulation results for accuracy of estimation of the heterogeneity variance $\tau^2$, the overall LRR, and its variance, using the DL and restricted maximum-likelihood (REML) methods to estimate $\tau^2$ under normality.  To assess bias of the estimators of LRR, they use mean absolute error, which is not a measure of bias; it is the linear counterpart of mean squared error.

To address this gap in information on methods of estimating the heterogeneity variance for LRR, we use simulation to study four methods recommended by \cite{veroniki2015methods}. These are the well-established methods of \cite{dersimonian1986meta}, restricted maximum likelihood, and \cite{mandel1970interlaboratory} (MP), and the less-familiar method of \cite{jackson2013confidence}. We also study coverage of confidence intervals for $\tau^2$ achieved by four methods: the Q-profile method of \cite{viechtbauer2007confidence}, the methods of \cite{biggerstaff2008exact} and \cite{jackson2013confidence}, and the profile-likelihood-based interval.

For each estimator of $\tau^2$,  we also study bias of the corresponding inverse-variance-weighted estimator of the overall effect. However, it is well known that these inverse-variance-weighted estimators have unacceptable bias for some other effect measures, as \cite{BHK2018SMD} and \cite{Hamman2018} show for the standardized mean difference. Therefore, we added an estimator (SSW) whose weights depend only on the sample sizes of the Treatment and Control arms. We study the coverage of the confidence intervals associated with the inverse-variance-weighted estimators, and also the HKSJ interval (\cite{hartung2001refined, sidik2002simple}), a modification of the HKSJ interval that uses the MP estimator of $\tau^2$ instead of the DL estimator, and an interval centered at SSW that uses the MP estimator of $\tau^2$ in estimating its variance and bases its half-width on a $t$ distribution.

\section{Study-level estimation of log-response-ratio}
We assume that each of the $K$ studies in the meta-analysis consists of two arms, Treatment and Control, with sample sizes $n_{iT}$ and $n_{iC}$. The total sample size in Study $i$ is $n_i=n_{iT}+n_{iC}$. The subject-level data in each arm are assumed to be lognormally distributed with means $\mu_{iT}$ and $\mu_{iC}$ and variances $\sigma_{iT}^2$ and $\sigma_{iC}^2$. The sample means are $\bar{X}_{ij}$, and the sample variances are $s^2_{ij}$, for $i = 1, \ldots, K$ and $j = C$ or $T$.

The response ratio is usually meta-analyzed on a log scale, where the effect measure is
$\lambda_{i}=\log(\mu_{iT}/\mu_{iC})$, estimated by
$\hat\lambda_{i}=\log(\bar{X}_{iT}/\bar{X}_{iC})$, and the population and sample means are assumed to be positive.
The within-study variance estimate of $\hat{\lambda}_{i}$, obtained by the delta method, is \citep{hedges1999meta}
\begin{equation}\label{eq:ROM_var}
v_i^2=\frac{s_{iT}^2}{n_{iT}\bar{X}_{iT}^2}+\frac{s_{iC}^2}{n_{iC}\bar{X}_{iC}^2}= \frac{\hat{V}_{iT}^2}{n_{iT}}+ \frac{\hat{V}_{iC}^2}{n_{iC}},
\end{equation}
where $\hat{V}_{ij}$ is the sample coefficient of variation (CV).

The log transformation introduces bias (as discussed by \cite{bakbergenuly2016inference}): the expected value of $\hat{\lambda}_i$ is not equal to $\lambda_i$ . To eliminate this bias in small samples, \cite{lajeunesse2015bias} proposed two bias-corrected modifications, and he recommended
\begin{equation} \label{eq:lambda_hat_delta}
\hat\lambda^{\Delta}_i=\hat\lambda_i+\frac{1}{2}\left [ \frac{s_{iT}^2}{n_{iT}\bar{X}_{iT}^2}-\frac{s_{iC}^2}{n_{iC}\bar{X}_{iC}^2}\right ],
\end{equation}
and estimated its variance by
\begin{equation} \label{eq:var_lambda_hat_delta}
\widehat{\var}(\hat\lambda^{\Delta}_i)=v_i^2+\frac{1}{2}\left [ \frac{s_{iT}^4}{n_{iT}^2\bar{X}_{iT}^4}-\frac{s_{iC}^4}{n_{iC}^2\bar{X}_{iC}^4}\right ].
\end{equation}

Because $\hat{\lambda}$ is not defined for negative values of the study means, {and dropping negative findings would introduce a bias,}  \cite{lajeunesse2015bias} modeled the data by lognormal distributions. In principle, log-normal distributions often make sense for non-negative data. This choice would eliminate the restricted-range bias, but not the transformation bias of LRR.  Of course, the choice of model should be based on the properties of the data and not on the perceived ease of statistical modeling.

Even though sample means and variances are unbiased estimators of the population means and variances  for lognormal distributions, they are very inefficient, especially as far as variance estimation is concerned \cite[Section 14.4.1, p. 220--222]{J-K-B-1994}. If the data are assumed to come from lognormal distributions, a much more straightforward approach would be to log-transform the individual observations, which would reduce the problem to meta-analysis of mean difference. This would provide much better inference. However, when individual-level data are not available, meta-analyses must work with the sample means and variances.

We provide simulations from lognormal distributions in Section~\ref{sec:simsect}. Simulations from normal distributions are in a separate arXiv report. 

\section{Standard random-effects model} \label{sec:StdREM}
The standard random-effects model assumes that within- and between-study variabilities are accounted for by approximately normal distributions of within- and between-study effects.  For a generic measure of effect,
\begin{equation}\label{eq:standardREM}
\hat{\theta}_{i} \sim N(\theta_{i}, {\sigma}_{i}^2) \quad \text{and} \quad \theta_{i} \sim N(\theta,\tau^2),
\end{equation}
resulting in the marginal distribution $\hat{\theta}_{i} \sim N(\theta, \sigma_{i}^2+\tau^2)$. $\hat{\theta}_{i}$ is the estimate of the effect in Study $i$, and its within-study variance is $\sigma_{i}^2$, estimated by $\hat{\sigma}_{i}^2$, $i=1,\ldots, K$.  The between-study variance, $\tau^{2}$, is estimated by $\hat{\tau}^2$. The overall effect, $\theta$, is customarily estimated by the weighted mean
\begin{equation}\label{eq:WAverChapter6}
\hat{\theta}_{\mathit{RE}}=\frac{\sum\limits_{i=1}^{K}\hat{w}_{i}(\hat{\tau}^2)\hat{\theta}_{i}}{\sum\limits_{i=1}^{K}\hat{w}_{i}(\hat{\tau}^2)},
\end{equation}
where the $\hat{w}_{i}(\hat{\tau}^2)=(\hat{\sigma}_{i}^2+\hat{\tau}^2)^{-1}$ are inverse-variance weights. The FE estimate $\hat{\theta}$  uses weights $\hat{w}_{i} = \hat{w}_{i}(0)$.

If $w_i = 1/\var(\hat{\theta}_i)$, the variance of the weighted mean of the $\hat{\theta}_i$ is $1/ \sum w_{i}$. Thus, many authors estimate the variance of $\hat{\theta}_{\mathit{RE}}$ by $\left[\sum_{i=1}^{K}\hat{w}_{i}(\hat{\tau}^2)\right]^{-1}$.  In practice, however, this estimate may not be satisfactory (\cite{sidik2006robust, li1994bias,rukhin2009weighted}).

\section{Methods of estimating between-study variance}\label{sec:EstVar}

In this section we briefly list the point and interval estimators of the between-studies variance ($\tau^2$) used in our study.

\subsection{Point estimators}

The most popular, but rather biased, estimator of $\tau^2$ is the method-of-moments estimator of \cite{dersimonian1986meta} (DL), denoted by $\hat{\tau}_{\mathit{DL}}^2$.

Assuming  that the $\hat{\theta}_{i}$ are distributed as $N(\theta, \hat{\sigma}_{i}^2+\tau^2)$, the restricted-maximum-likelihood (REML) estimator $\hat{\tau}_{\mathit{REML}}^2$ maximizes the restricted (or residual) log-likelihood function $l_R(\theta,\tau^2)$.    REML is superior to DL because of its balance between unbiasedness and efficiency \citep{viechtbauer2005bias}.

The Mandel-Paule (MP) estimator (\cite{mandel1970interlaboratory}), $\hat{\tau}_{\mathit{MP}}^2$, is another moment-based estimator of the between-study variance. It is estimated iteratively. It is known to be superior to DL (\cite{veroniki2015methods}), but no simulations for LRR have been performed so far.

\cite{dersimonian2007random} generalized DL, replacing the weights $\hat{w}_i$ by arbitrary fixed positive constants, $a_i$. As an option when there is little a priori knowledge about the extent of heterogeneity, but some is anticipated, \cite{jackson2013confidence} proposed the estimator of $\tau^2$  with $a_i = 1/\hat{\sigma}_i$. We refer to this method as J.

\subsection{Interval estimators}

The $95\%$ profile-likelihood (PL) confidence interval for $\tau^2$ consists of the values that are not rejected by the likelihood-ratio test with $\tau^2$ as the null hypothesis.   This interval is usually used with $\hat{\tau}_{REML}^2$.

Similarly, the Q-profile (QP) confidence interval for $\tau^2$ consists of the values that are not rejected by the usual test for heterogeneity based on Cochran's $Q$ (\cite{cochran1954combination}).  The distribution of $Q$ is assumed (incorrectly) to be the chi-squared distribution with $K-1$ degrees of freedom.

For a generic effect measure, \cite{biggerstaff2008exact}  derived the exact distribution of a $Q$ statistic with constant weights $a_i$. That distribution yielded a generalized Q-profile confidence interval. We refer to this interval with $a_i=1/\hat{\sigma}_{i}^2$ as the BJ confidence interval.

\cite{jackson2013confidence} proposed another generalized Q-profile confidence interval (J) for $\tau^2$. The approach is the same as for the BJ interval, but with $a_i = 1/\hat{\sigma}_{i}$.

\section{Methods of estimating overall effect} \label{sec:EstLRR}
Most of the point estimators of the overall effect have corresponding interval estimators, but some do not. Therefore, we describe point estimators and interval estimators in separate sections.

\subsection{Point estimators}
A random-effects method that estimates $\theta$ by a weighted mean with inverse-variance weights, as in Equation~(\ref{eq:WAverChapter6}), is determined by the particular $\hat{\tau}^2$ that it uses in $\hat{w}_{i}(\hat{\tau}^2)$. Because the study-level effects and their variances are related (as in Equation~ (\ref{eq:ROM_var}) for LRR), all inverse-variance-weighted estimators of $\hat{\lambda}$ may have considerable bias. For completeness, we studied DL, REML, MP, and J.

To reduce this bias in estimating $\lambda$, our experience with the bias of inverse-variance-weighted estimators for standardized mean difference (\cite{BHK2018SMD}) led us to include a point estimator whose weights depend only on the studies' sample sizes (\cite{hedges1985statistical, hunter1990methods}). For this estimator (SSW), $w_{i} = \tilde{n}_i = n_{iT}n_{iC}/(n_{iT} + n_{iC})$; that is, $w_i$ substitutes $1$ for the estimated CVs in Equation~(\ref{eq:ROM_var}); $\tilde{n}_i$ is the effective sample size in Study $i$. The estimator of the variance of SSW is
\begin{equation}\label{eq:varianceOfSSW}
\widehat{\var}(\hat{\theta}_{\mathit{SSW}})= \frac{\sum \tilde{n}_i^2 (v_i^2 + \hat{\tau}^2)} {(\sum \tilde{n}_i)^2},
\end{equation}
in which $v_i^2$ comes from Equation (\ref{eq:ROM_var}) and $\hat{\tau}^2 = \hat{\tau}_{\mathit{MP}}^2$.

We also study the behavior of the bias-corrected estimator $\hat{\lambda}^{\Delta}$, Equation~(\ref{eq:lambda_hat_delta}), in lognormal data.

\subsection{Interval estimators}
The point estimators DL, REML, MP, and J  have companion interval estimators of $\theta$. The customary approach estimates the variance of $\hat{\theta}_{\mathit{RE}}$ by $\left[\sum_{i=1}^{K}\hat{w}_{i}(\hat{\tau}^2)\right]^{-1}$ and bases the half-width of the interval on the normal distribution. These intervals are usually too narrow.

\cite{hartung2001refined} and, independently, \cite{sidik2002simple} developed an improved estimator for the variance of $\hat{\theta}_{\mathit{RE}}$. The Hartung-Knapp-Sidik-Jonkman (HKSJ) confidence interval uses this  estimator together with critical values from the $t$ distribution on $K - 1$ degrees of freedom. A potential weakness is that the HKSJ interval uses $\hat{\theta}_{\mathit{DL}}$ as its midpoint, so it will have any bias that is present in $\hat{\theta}_{\mathit{DL}}$. We studied a modification of the HKSJ confidence interval that uses $\hat{\tau}^2_{\mathit{MP}}$ and $\hat{\theta}_{\mathit{MP}}$; we refer to this interval as the HKSJ(MP) confidence interval.

The interval estimator corresponding to SSW (SSW MP) uses the SSW point estimator as its center, and its  half-width equals the estimated standard deviation of SSW under the random-effects model times the critical value from the $t$ distribution on $K - 1$ degrees of freedom.

\section{Simulation study}\label{sec:simsect}
As mentioned in Section~\ref{sec:Intro}, a few studies have used simulation to examine estimators of the overall effect for LRR, but no studies have examined estimators of $\tau^2$.

The range of values of RR may be rather wide. { The empirical study by \cite{Senior-2016} reports values of RR up to $3.72$,  though the second largest value is $1.46$. }
The simulations by \cite{Friedrich2008} used values up to $1.56$ (LRR = 0.445). \cite{lajeunesse2015bias} used means between $0$ and $8$ in both arms and small sample sizes, starting from $n_T+n_C=4$. Our simulation study for LRR uses an interval of $0\leq \lambda \leq 2$ (or $0\leq \hbox{ RR }\leq 7.39$) as  realistic for a range of applications. Unfortunately,  no information is available on the accompanying range of $\tau^2$ values. In their simulations for SMD, \cite{Hamman2018} consider the range from $0$ to $2.5$ as typical for ecology.

\subsection{Design of the simulations}\label{sec:LRRsim}
Our simulation study assesses the performance of four methods for point estimation of the between-studies variance, $\tau^2$ (DL, REML, J, and MP) and four methods of interval estimation of $\tau^2$ (the Q-profile interval, the generalized Q-profile intervals of \cite{biggerstaff2008exact} and \cite{jackson2013confidence}, and the profile-likelihood confidence interval based on REML).

We study bias of the inverse-variance-weighted estimator of the overall effect corresponding to each of the estimators of $\tau^2$ (DL, REML, J, and MP), as well as bias of SSW, whose weights depend only on the sample sizes of the Treatment and Control arms.

We also study coverage of the confidence intervals associated with those inverse-variance-weighted estimators, and also the HKSJ interval (\cite{hartung2001refined, sidik2002simple}), a modification of the HKSJ interval that uses the MP estimator of $\tau^2$ instead of the DL estimator, and an interval centered at SSW that uses the MP estimator of $\tau^2$ in estimating its variance and uses critical values from a $t$ distribution.

Two basic distributions may serve as the source of the data in the Treatment and Control arms: the lognormal distribution (the subject of the present report) and the normal distribution (the subject of a separate report).
We generate $\lambda_i$ from $N(\lambda, \tau^2)$ and set $\mu_{iT} = \exp(\lambda_i)\mu_{iC}$. Then we generate $n_{ij}$ independent observations from the lognormal distributions with means $\mu_{ij}$ and variances $\sigma^2_{ij}$. We obtain the sample means $\bar X_{ij}$  and the sample variances $s^2_{ij}$   and calculate the sample LRR $\hat{\lambda}_{i}=\log(\bar{X}_{iT}/\bar{X}_{iC})$ and their variances $\hat v_{i}^2$ as in  Equation~(\ref{eq:ROM_var}). We also calculate the bias-corrected estimate, $\hat{\lambda}_{i}^{\Delta}$, Equation~(\ref{eq:lambda_hat_delta}), and its variance, Equation~(\ref{eq:var_lambda_hat_delta}) \citep{lajeunesse2015bias}.

For the overall value of LRR, we chose $\lambda=(0,0.2,0.5,1,2)$ (corresponding to $0 \leq \hbox{ RR } \leq 7.39$), as realistic for a range of applications.

When the data are lognormal, proximity to zero does not affect data generation or inferences. Therefore, as the mean of the Control arm we take $\mu_{iC}=1$.

All simulations use the same numbers of studies, small ($K = 5, \;10, \;30$) and large ($K =  50, \;100, \;125$) and,  for each combination of parameters, the same vector of total sample sizes $n = (n_{1},\ldots, n_{K})$ and equal numbers of observations in the Control and Treatment arms.

We study only meta-analyses in which the study size is the same in all $K$ studies.. The  study sizes, $n_i$, start from $4$, because some studies in ecology have such small sample sizes, and they extend to $1000$.  By using the same patterns of sample sizes for each combination of the other parameters, we avoid the additional variability in the results that would arise from choosing sample sizes at random (e.g., uniformly between 100 and 250).

In summary, we vary four parameters: the overall true LRR ($\lambda$), the between-studies variance ($\tau^2$), the number of studies ($K$), and the total sample size ($n$).  We set $\sigma^2_C=\sigma^2_T=1$. Table~\ref{tab:altdataLRR} lists the configurations.

We use a total of $10,000$ repetitions for each combination of parameters. Thus, the simulation standard error for estimated coverage of $\tau^2$ or $\lambda$ at the $95\%$ confidence level is roughly $\sqrt{0.95 \times 0.05 / 10,000} = 0.00218$.

The simulations were programmed in R version 3.3.2 using the University of East Anglia 140-computer-node High Performance Computing (HPC) Cluster, providing a total of 2560 CPU cores, including parallel processing and large memory resources. For each configuration, we divided the 10,000 replications into 10 parallel sets of 1000 replications.


\begin{table}[ht]
	\caption{\label{tab:altdataLRR} \emph{Configurations of parameters in the simulations for LRR.}}
	\begin{footnotesize}
		\begin{center}
			\begin{tabular}
				{|l|l|l|}
				\hline
				Parameter&Equal study sizes&  Full results in\\
				&&Appendix\\
				\hline
				$K$ (number of studies: small/large)& (5, 10, 30) \& (50, 100, 125)&A \& B - small $n$ \\
				$n$ (total study size: small/large) &(4, 10, 20, 40) \& (100, 250, 640, 1000)&\\
				$\sigma_{T}^2$ \& $\sigma_{C}^2$ (within-study variances)& 1 \& 1&C \& D - large $n$\\
				$\lambda$ (overall value of the LRR) &0, 0.2, 0.5, 1, 2&\\
				$\tau^{2}$ (variance of random effect)& 0(0.1)1 &  \\
				Lognormal distribution&&\\
				$\mu_C$ (mean in Control arm)&1&\\
				\hline
				estimation of $\tau^{2}$   & &A\;\&\;C  \\
				estimation of $\lambda$ &&   B\;\&\;D\\
				
				\hline
				
			\end{tabular}
		\end{center}
	\end{footnotesize}
\end{table}

\subsection{Results}
\textbf{Bias and coverage in estimation of $\tau^2$ (Appendices A1--A4 and C1--C4)}
\\[8mm]
{\bf Bias.}
When $n$ is very small (Figures~\ref{BiasTauRoM0ln_smallN_small_K}--\ref{BiasTauRoM2ln_smallN_small_K}), all four estimators of $\tau^2$ have substantial positive bias, increasing linearly with $\tau^2$ (when $\lambda = 0$ and $n = 4$, the intercept is around $0.4$, and the slope is around $0.9$). This pattern persists for $\lambda \leq 1$; but when $\lambda = 2$, the slope is essentially 0. As $n$ increases to 40, the intercept and slope decrease; but the trace for DL begins to diverge from the others, followed by the trace for J, and increasingly as $\lambda$ increases. $K$ has little effect. MP and REML have similar, reasonably small, bias when $n \geq 40$ (Figures~\ref{BiasTauRoM0ln_largeN_small_K}--\ref{BiasTauRoM2ln_largeN_small_K}). When $n \geq 100$, the traces for DL and J bend toward increasingly negative bias as $\tau^2$ increases; their bias becomes worse as $n$ increases and slightly worse as $K$ increases (for example, when $\lambda = 0$, $n = 1000$, and $K \geq 50$, the bias of DL is  $-0.28$ at $\tau^2 = 1$). The bias correction for $\hat{\lambda}_i$ does not reduce the bias (Appendices A2, A4, C2, and C4).
\\[8mm]
{\bf Coverage.}
When $n < 40$ and $K = 5$, the coverage of all four intervals for $\tau^2$ is below the nominal 95\%, especially when $n = 4$ and $\tau^2 < 0.4$; increasing $K$ to 10 and 30 reduces coverage substantially and makes this pattern worse (Figure~\ref{CovTauRoM0ln_smallN_small_K}), and increasing $\lambda$ has little effect (Figures~\ref{CovTauRoM02ln_smallN_small_K}--\ref{CovTauRoM2ln_smallN_small_K}). Increasing $K$ to 50 and beyond reduces coverage further, even to 0 when $n = 4$ and $\tau^2 = 0$ (Figure~\ref{CovTauRoM0ln_smallN_large_K}). When $n \geq 40$ and $K = 5$ or 10, BJ and J generally provide nominal or slightly higher coverage, and QP and PL are slightly lower. Situations with $K \geq 30$ are often quite challenging; BJ has low coverage from $K\geq 30$ (Figure~\ref{CovTauRoM1ln_largeN_small_K}), and for larger $n$ and $K$, J coverage deteriorates similarly to BJ, while QP and PL provide good coverage (Figure~\ref{BiasTauRoM0ln_largeN_large_K}). The bias correction does not improve coverage.
\\[8mm]

\noindent\textbf{Bias and coverage in estimation of $\lambda$ (Appendices B1--B4  and  D1--D4)}
\\[8mm]
{\bf Bias.}
All five estimators of $\lambda$ have bias that shows little dependence on $K$. When $\lambda = 0$ and $\tau^2 = 0$, they all have essentially no  bias. When $\tau^2 > 0$, the bias is very roughly linear in $\tau^2$, with negative slope but a non-negative intercept for the IV-weighted estimators and a negative intercept for SSW. The intercept for the IV-weighted estimators is positive for $n \geq 10$, so their bias is positive for smaller $\tau^2$ and negative for larger $\tau^2$; but the traces flatten as $n$ increases, and by $n = 40$ their bias is positive for $0.1 \leq \tau^2 \leq 1$. The trace for SSW flattens similarly; and when $n = 40$, its bias has smaller magnitude than the IV-weighted estimators when $0.1 \leq \tau^2 \leq 0.5$ and larger magnitude when $0.6 \leq \tau^2 \leq 1$. When $\lambda > 0$, the biases of all five estimators at $\tau^2 = 0$ and the intercepts (i.e., biases) at $\tau^2 =0.1$ increase; for a given $\lambda$ both the intercepts and the slopes decrease as $n$ increases. As a result, when $\lambda \geq 0,5$, the bias of SSW usually has smaller magnitude than the IV-weighted estimators. In relative terms, when $n < 40$, the biases are substantial: as much as 10\% of $\lambda$ in some cases. Here SSW has the least bias, about 10\% for $\lambda \geq 1$ and $n=10$, declining to 5\% for $\lambda \geq 1$ and $n=20$ (Figure~\ref{BiasThetaRoM1ln_smallN_small_K}). The bias correction for $\hat{\lambda}_i$ reduces the bias  (Figure~\ref{BiasThetaRoM1lnCor_smallN_small_K}) and should be used.
\\[8mm]
{\bf Coverage.}
$t$-intervals centered at SSW provide the best coverage of $\lambda$, and that coverage is satisfactory when $n \geq 20$ and $K\leq 30$. Those intervals may have coverage greater than 97\% (primarily when $\tau^2 = 0$ and $K = 5$ or 10 and in a few cases where $\tau^2 = 0.1$, $K = 5$, $n = 20$ or 40, and $\lambda \leq 0.5$) or coverage less than 93\% (mainly when $\tau^2$ is small, $K \geq 50$, $n= 20$ or 40, and $\lambda \geq 0.5$). All other methods have inferior coverage and are not recommended. Coverage of the intervals centered at SSW is better when the bias correction is used for $\hat{\lambda}_i$; then it is good when $n \geq 10$ (Figure~\ref{CovThetaRoM2lnCor_smallN_small_K}). When $n$ is small, $K \geq 50$, and $\lambda=0$, coverage of the standard methods improves somewhat, whereas coverage of SSW MP becomes less than 93\% when $K > 50$, especially for large $\tau^2$ (Figure~\ref{CovThetaRoM0ln_smallN_large_K}). When the bias correction is used, coverage of SSW MP is the best, and it is good overall for small $\lambda$, but it is much below 95\% for $n=4$, $\lambda\geq 0.5$ and small $\tau^2$, where it worsens for larger $K$ (Figure~\ref{CovThetaRoM1ln_smallN_large_K}). For large $K$ and large $n$, coverage of SSW MP is still the best, especially at $\tau^2=0$, and the bias correction still produces better results (Figure~\ref{CovThetaRoM1lnCor_largeN_large_K}).
\\[8mm]

\section{Discussion}

The results of our simulations provide a rather disappointing picture of the current state of meta-analysis of LRR. For such effect measures as LRR and SMD, also popular in ecology, the relation between the studies' estimated effects and their estimated variances has several undesirable results: dependence of the performance of all inverse-variance-based methods on the effect sizes, biased estimation of overall effects, and below-nominal coverage of their confidence intervals, especially for small sample sizes. Our simulations show this clearly.

We show that, for a lognormal underlying distribution, the between-studies variance $\tau^2$ cannot be estimated reliably for sample sizes less than $100$.

Arguably, the main purpose of a meta-analysis is to provide point and interval estimates of an overall effect. For general use, the estimate of overall effect should be unbiased, and the confidence interval should have nominal coverage.

Usually, after estimating the between-study variance $\tau^2$, an inverse-variance-weighted approach is used to estimate the overall effect (and, often, its variance). The origin of the IV approach lies in the fact that, for known variances, and given unbiased estimates of the within-study effects, it provides a uniformly minimum-variance unbiased estimate (UMVUE) of $\theta$. However, in practice, the within-study variances are unknown, and using estimates for them leads to bias in the IV estimate of the overall effect and below-nominal coverage of the confidence interval. Thus, the IV approach is misguided; for most measures of effect, it cannot avoid these shortcomings.

{ The gaps in evidence include the possibility that the variances in the two arms may differ, which is rarely, if ever, reflected in simulations. Due to sheer volume of our simulations, we did not attempt to fill this gap. However, we do not expect  the performance of the IV methods to improve under more challenging scenarios.}

A pragmatic solution to unbiased estimation of $\theta$ uses weights that do not involve estimated variances (for example, weights proportional to the studies' sample sizes $n_i$). Our point estimator SSW uses weights proportional to an effective sample size,  $\tilde{n}_i = n_{iC}n_{iT}/n_i$. Then, the estimate of the overall effect is $\hat{\lambda}_{\mathit{SSW}} = \sum \tilde{n}_i\hat{\lambda}_i/\sum \tilde{n}_i$, and the estimate of its variance comes from Equation~(\ref{eq:varianceOfSSW}). Finally, the t-based confidence interval for $\lambda$ is centered at $\hat{\lambda}_{\mathit{SSW}}$.

SSW, combined with the bias-corrected estimator of $\lambda_i$, works reasonably well for sample sizes as low as $10$ in interval estimation of $\lambda$, and for $n \geq 40$ in point estimation. We recommend this method for further use in applications.


\section {Methods of estimation of $\tau^2$ and $\lambda$ used in simulations}
\subsection*{Point estimators of $\tau^2$}
\begin{itemize}
	\item DL - method of \cite{dersimonian1986meta}
	\item J - method of \cite{jackson2013confidence}
	\item MP - method of \cite{mandel1970interlaboratory}
	\item REML - restricted maximum-likelihood method
\end{itemize}

\subsection*{Interval estimators of $\tau^2$}
\begin{itemize}
	\item BJ - method of \cite{biggerstaff2008exact}
	\item J - method of \cite{jackson2013confidence}
	\item PL - profile-likelihood confidence interval based on $\hat{\tau}_{REML}^2$
	\item QP - Q-profile confidence interval of \cite{viechtbauer2007confidence}
\end{itemize}

\subsection*{Point estimators of $\lambda$}
Inverse-variance-weighted methods with $\tau^2$ estimated by:
\begin{itemize}
	\item DL
	\item J
	\item MP
	\item REML
\end{itemize}
and
\begin{itemize}
	\item SSW - weighted mean with weights that depend only on studies' sample sizes
\end{itemize}

\subsection*{Interval estimators of $\lambda$}
Inverse-variance-weighted methods using normal quantiles, with  $\tau^2$ estimated by:
\begin{itemize}
	\item DL
	\item J
	\item MP
	\item REML
\end{itemize}
Inverse-variance-weighted methods with modified variance of $\hat{\lambda}$ and t-quantiles as in  \cite{hartung2001refined} and \cite{sidik2002simple}
\begin{itemize}
	\item HKSJ (DL) -  $\tau^2$ estimated by DL
	\item HKSJ (MP) -  $\tau^2$ estimated by MP
\end{itemize}
and
\begin{itemize}
	\item SSW MP - SSW point estimator of $\lambda$ with estimated variance given by Equation~(\ref{eq:varianceOfSSW}) and t-quantiles
\end{itemize}

\section*{Funding}
The work by E. Kulinskaya was supported by the Economic and Social Research Council [grant number ES/L011859/1].

\clearpage
\bibliographystyle{plainnat}
\bibliography{XBib_17Apr19,MD_SMD_Bib_13Apr19}%

\clearpage
\renewcommand{\thesection}{A1.\arabic{section}}
\setcounter{section}{0}
\setcounter{figure}{0}
\text{\LARGE{\bf{Appendices}}}
\section*{A: Plots of bias and coverage of estimators of $\tau^2$, small $n$}
\begin{itemize}
	\item A1. Lognormal model, usual estimator of $\lambda_i$, $K=5,10,30$
	\item A2. Lognormal model, bias-corrected estimator of $\lambda_i$, $K=5,10,30$
	\item A3. Lognormal model, usual estimator of $\lambda_i$, $K=50,100,125$
	\item A4. Lognormal model, bias-corrected estimator of $\lambda_i$, $K=50,100,125$
\end{itemize}
\clearpage
\renewcommand{\thefigure}{A1.1.\arabic{figure}}
\section*{A1. Lognormal model, usual estimator of $\lambda_i$, $n= 4, 10, 20, 40$, $K=5,10,30$}
\subsection*{A1.1 Bias of point estimators of $\tau^2$}
Each figure corresponds to a value of $\lambda \;(= 0, 0.2, 0.5, 1, 2)$, a set of values of $n$ (= 4, 10, 20, 40), and a set of values of $K$ (= 5, 10, 30).\\
Each panel corresponds to a value of $n$ and a value of $K$ and has $\tau^2 = 0.0(0.1)1.0$ on the horizontal axis.\\
The point estimators of $\tau^2$ are
\begin{itemize}
	\item DL (DerSimonian-Laird)
	\item REML (restricted maximum likelihood)
	\item MP (Mandel-Paule)
	\item J (Jackson)
\end{itemize}

\clearpage

\begin{figure}[t]
	\includegraphics[scale=0.33]{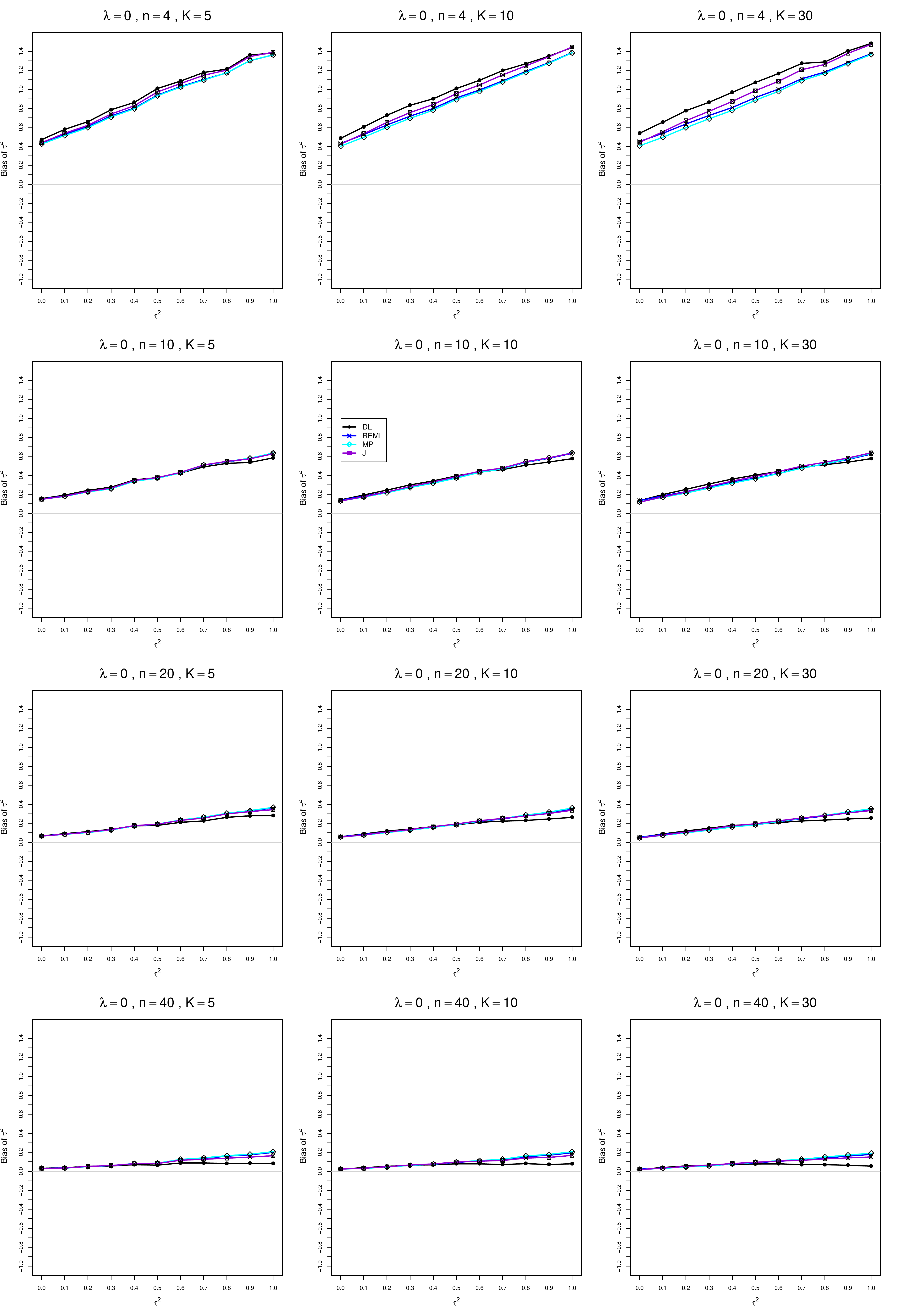}
	\caption{Bias of estimators of between-studies variance $\tau^2$ for $\lambda=0$, $n = 4, \;10, \;20, \;40$, and $K = 5, \;10, \;30$. Usual estimate of $\lambda_i$
		\label{BiasTauRoM0ln_smallN_small_K}}
\end{figure}

\begin{figure}[t]
	\includegraphics[scale=0.33]{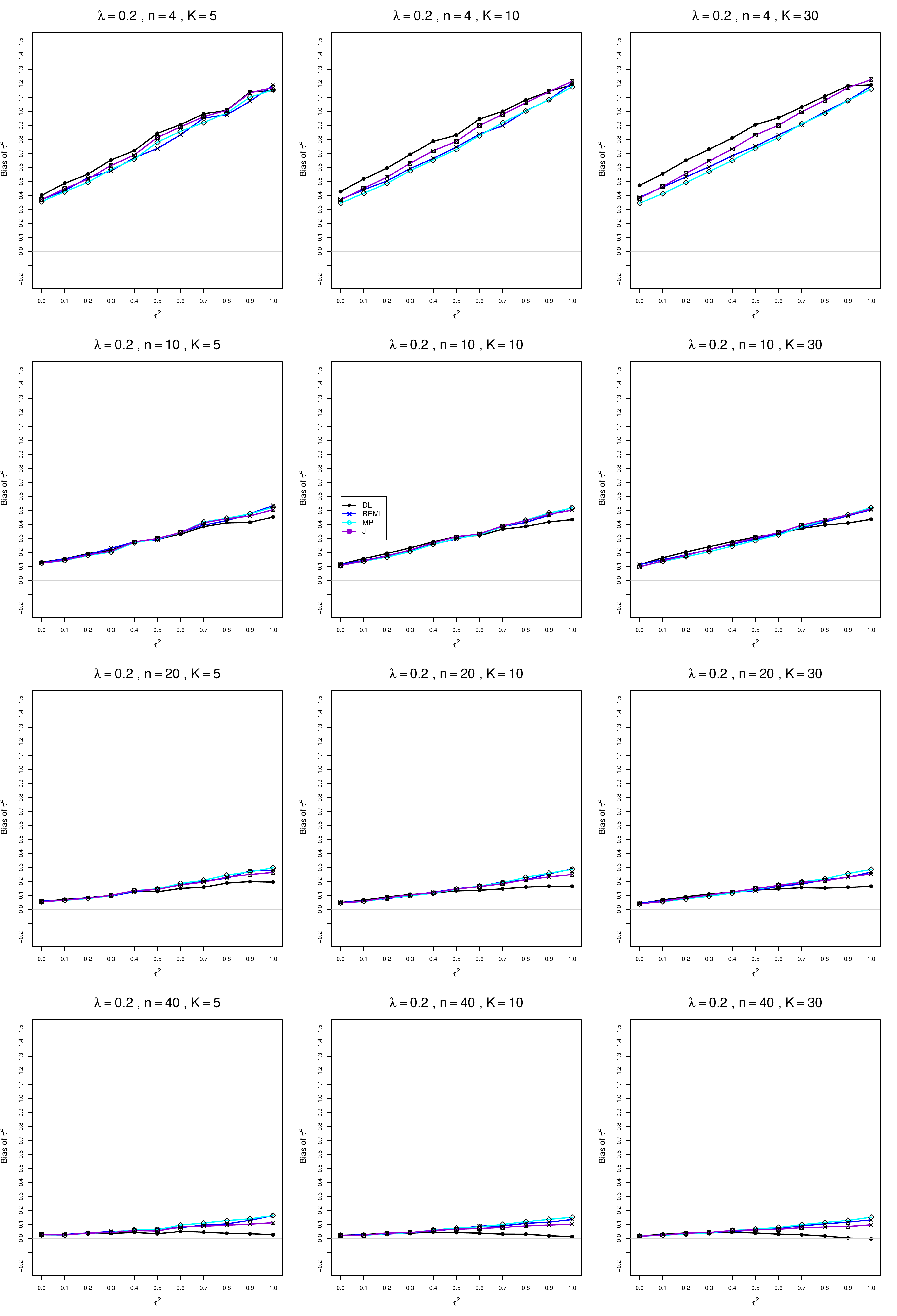}
	\caption{Bias of estimators of between-studies variance $\tau^2$ for $\lambda=0.2$, $n = 4, \;10, \;20, \;40$, and $K = 5, \;10, \;30$. Usual estimate of $\lambda_i$
		\label{BiasTauRoM02ln_smallN_small_K}}
\end{figure}

\begin{figure}[t]
	\includegraphics[scale=0.33]{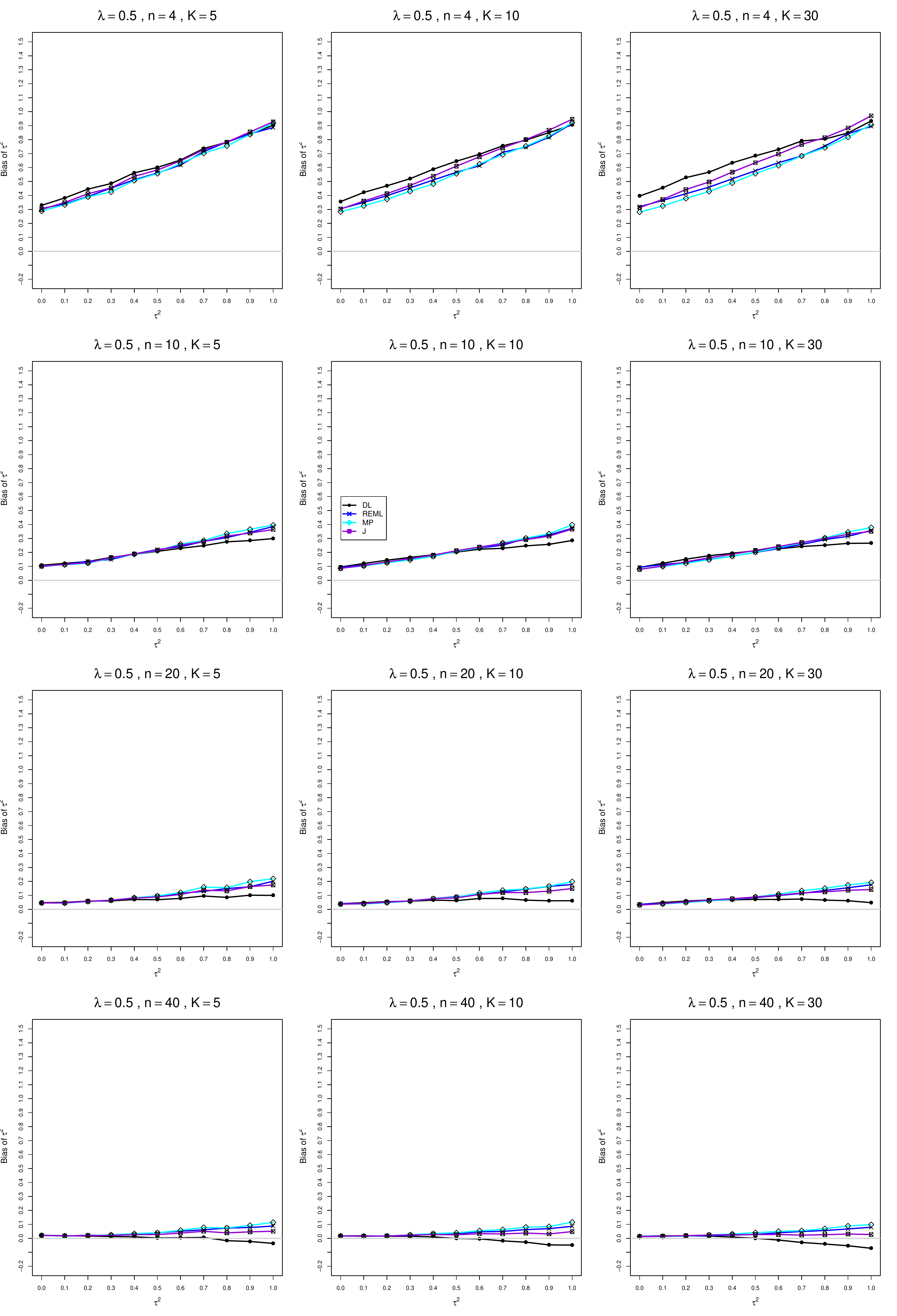}
	\caption{Bias of estimators of between-studies variance $\tau^2$ for $\lambda=0.5$, $n = 4, \;10, \;20, \;40$, and $K = 5, \;10, \;30$. Usual estimate of $\lambda_i$
		\label{BiasTauRoM05ln_smallN_small_K}}
\end{figure}

\begin{figure}[t]
	\includegraphics[scale=0.33]{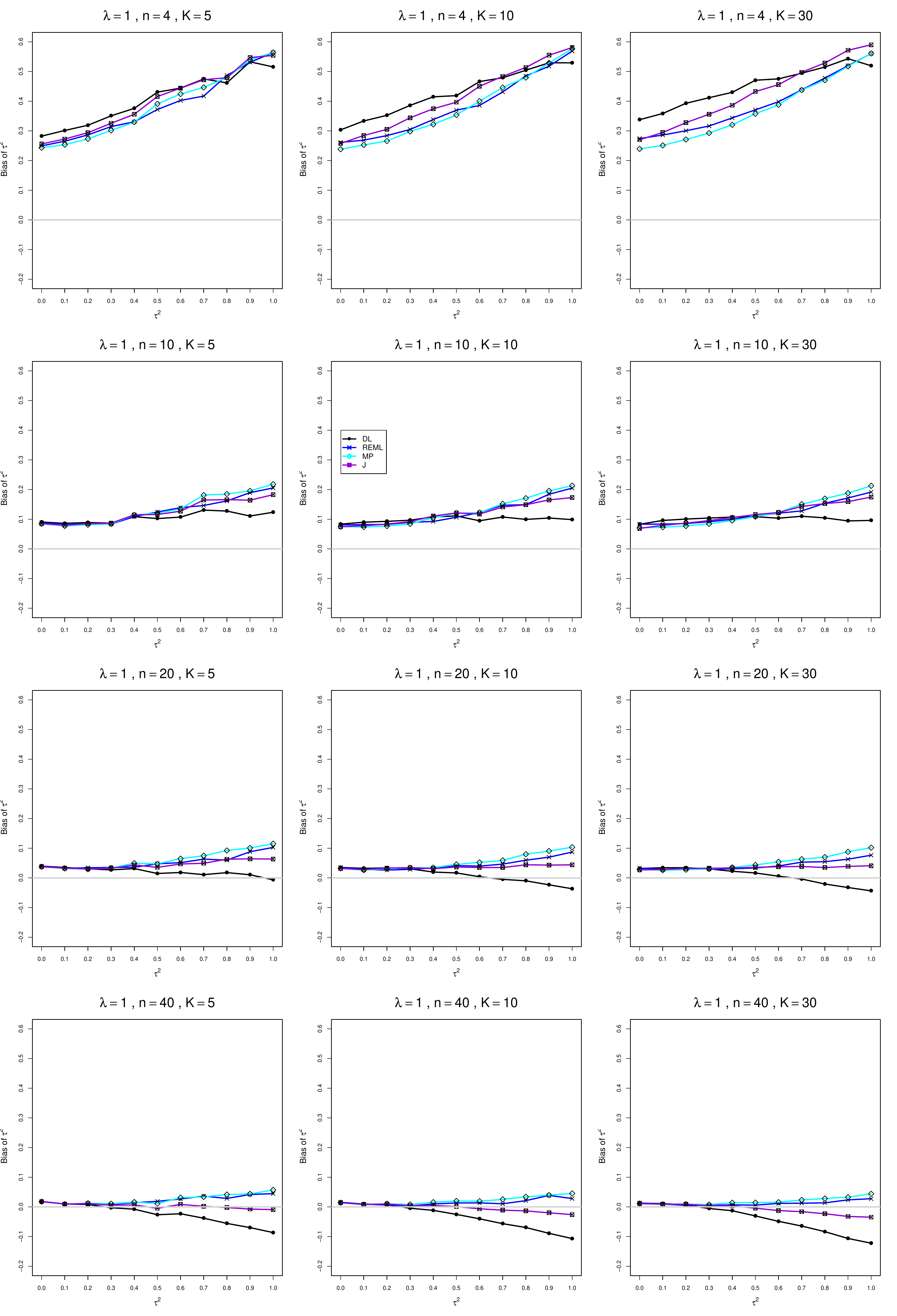}
	\caption{Bias of estimators of between-studies variance $\tau^2$ for $\lambda=1$, $n = 4, \;10, \;20, \;40$, and $K = 5, \;10, \;30$. Usual estimate of $\lambda_i$
		\label{BiasTauRoM1ln_smallN_small_K}}
\end{figure}

\begin{figure}[t]
	\includegraphics[scale=0.33]{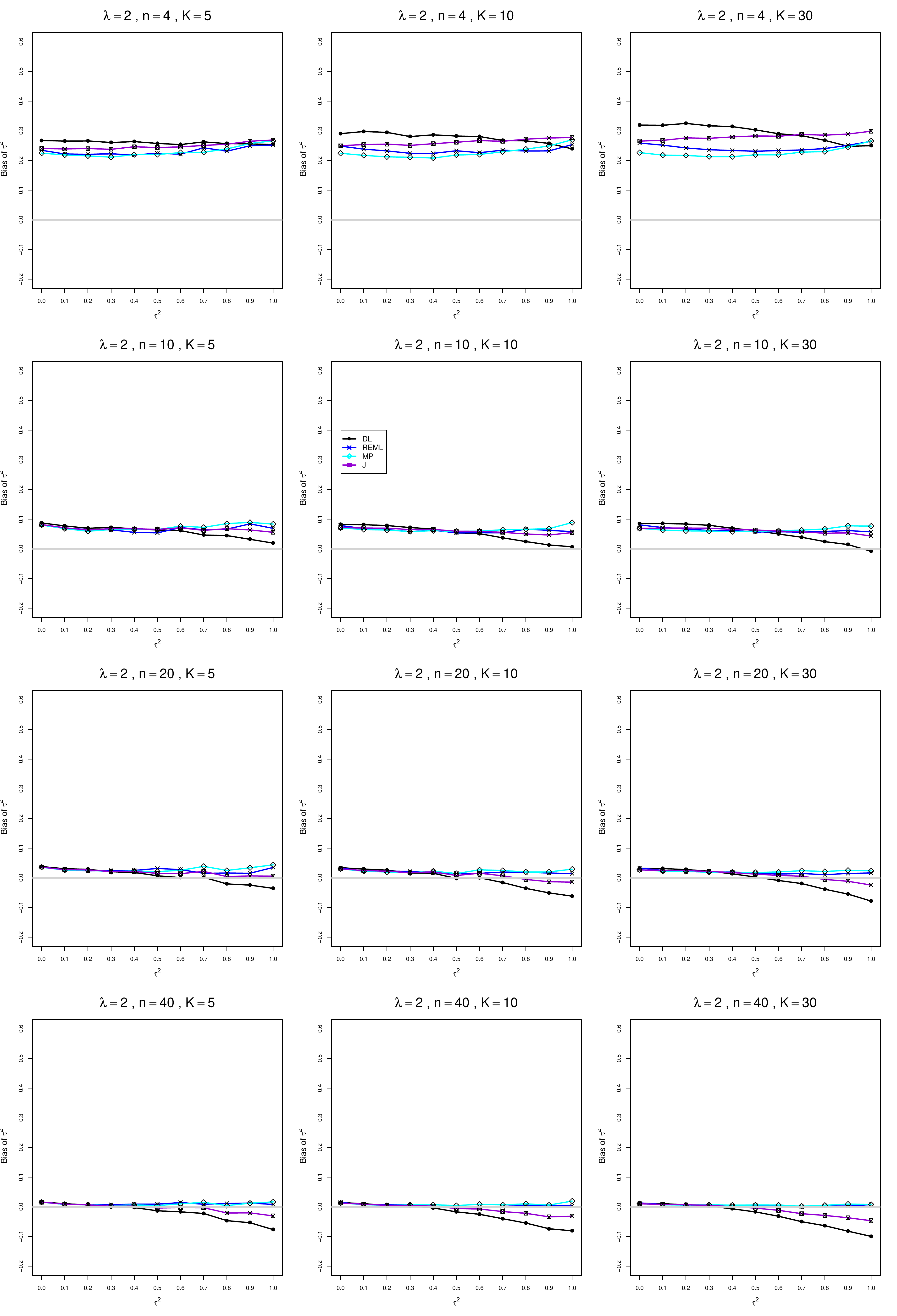}
	\caption{Bias of estimators of between-studies variance $\tau^2$ for $\lambda=2$, $n = 4, \;10, \;20, \;40$, and $K = 5, \;10, \;30$. Usual estimate of $\lambda_i$
		\label{BiasTauRoM2ln_smallN_small_K}}
\end{figure}

\renewcommand{\thefigure}{A1.2.\arabic{figure}}
\setcounter{figure}{0}
\clearpage
\subsection*{A1.2 Coverage of interval estimators of $\tau^2$}
Each figure corresponds to a value of $\lambda \;(= 0, 0.2, 0.5, 1, 2)$, a set of values of $n$ (= 4, 10, 20, 40), and a set of values of $K$ (= 5, 10, 30).\\
Each panel corresponds to a value of $n$ and a value of $K$ and has $\tau^2 = 0.0(0.1)1.0$ on the horizontal axis.\\
The interval estimators of $\tau^2$ are
\begin{itemize}
	\item QP (Q-profile confidence interval)
	\item BJ (Biggerstaff and Jackson interval)
	\item PL (Profile-likelihood interval)
	\item J (Jackson interval)
\end{itemize}

\begin{figure}[t]
	\includegraphics[scale=0.35]{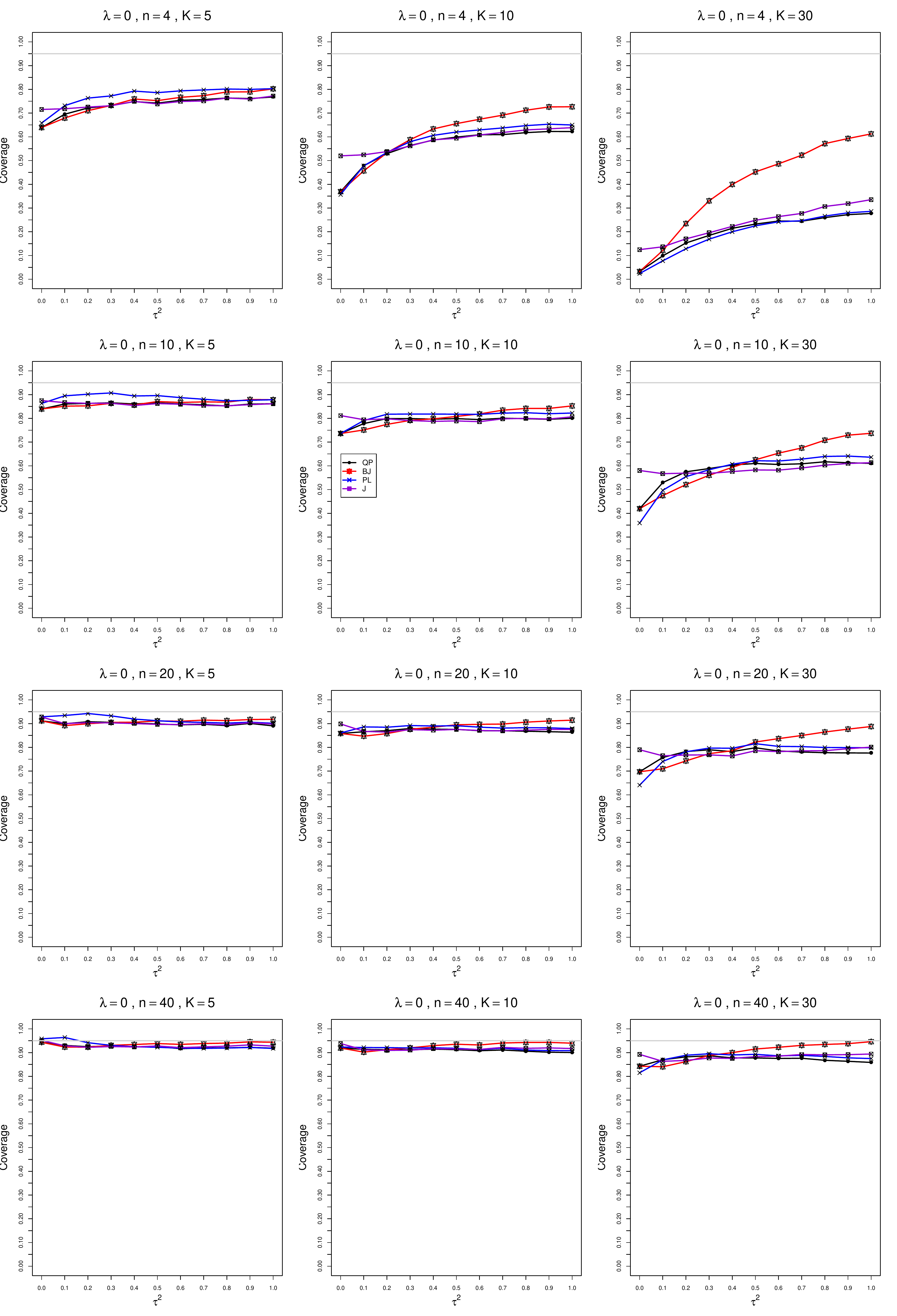}
	\caption{Coverage of 95\% confidence intervals for the between-studies variance $\tau^2$ when $\lambda=0$, $n = 4, \;10, \;20, \;40$, and $K = 5, \;10, \;30$. Usual estimate of $\lambda_i$ \label{CovTauRoM0ln_smallN_small_K}}
\end{figure}
\begin{figure}[t]
	\includegraphics[scale=0.35]{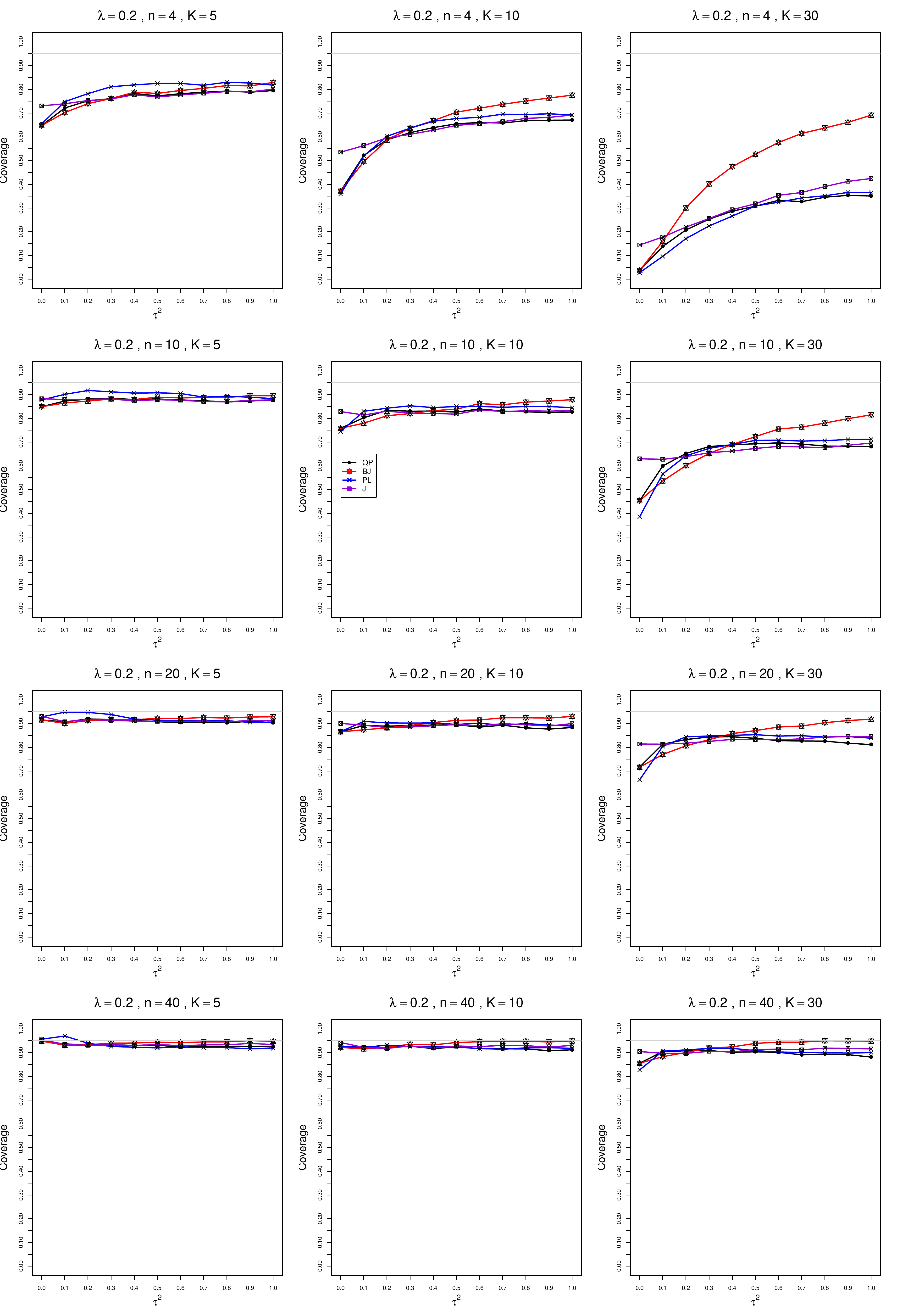}
	\caption{Coverage of 95\% confidence intervals for the between-studies variance $\tau^2$ when $\lambda=0.2$, $n = 4, \;10, \;20, \;40$, and $K = 5, \;10, \;30$. Usual estimate of $\lambda_i$ \label{CovTauRoM02ln_smallN_small_K}}
\end{figure}

\begin{figure}[t]
	\includegraphics[scale=0.35]{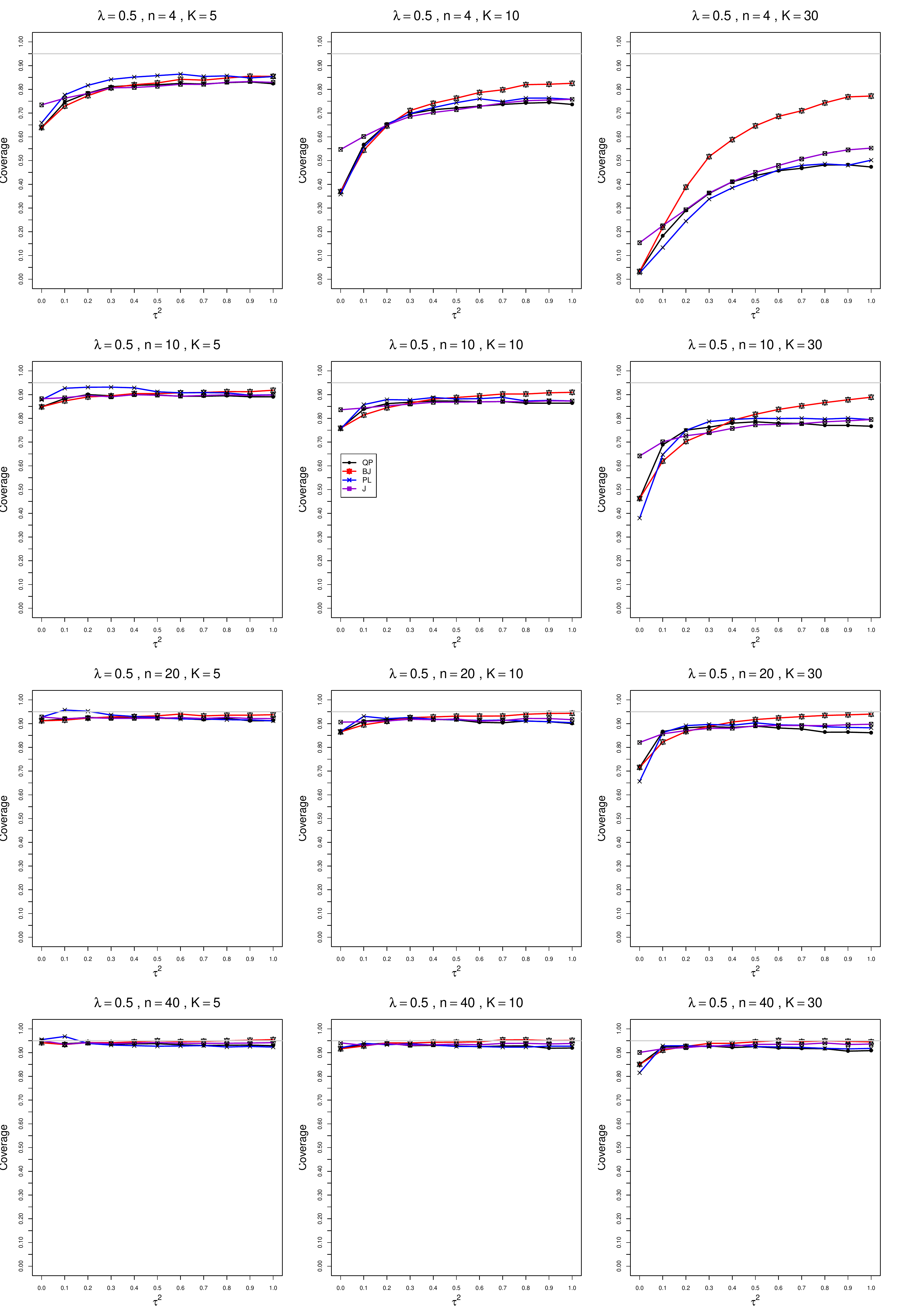}
	\caption{Coverage of 95\% confidence intervals for the between-studies variance $\tau^2$ when $\lambda=0.5$, $n = 4, \;10, \;20, \;40$, and $K = 5, \;10, \;30$. Usual estimate of $\lambda_i$ 		\label{CovTauRoM05ln_smallN_small_K}}
\end{figure}

\begin{figure}[t]
	\includegraphics[scale=0.35]{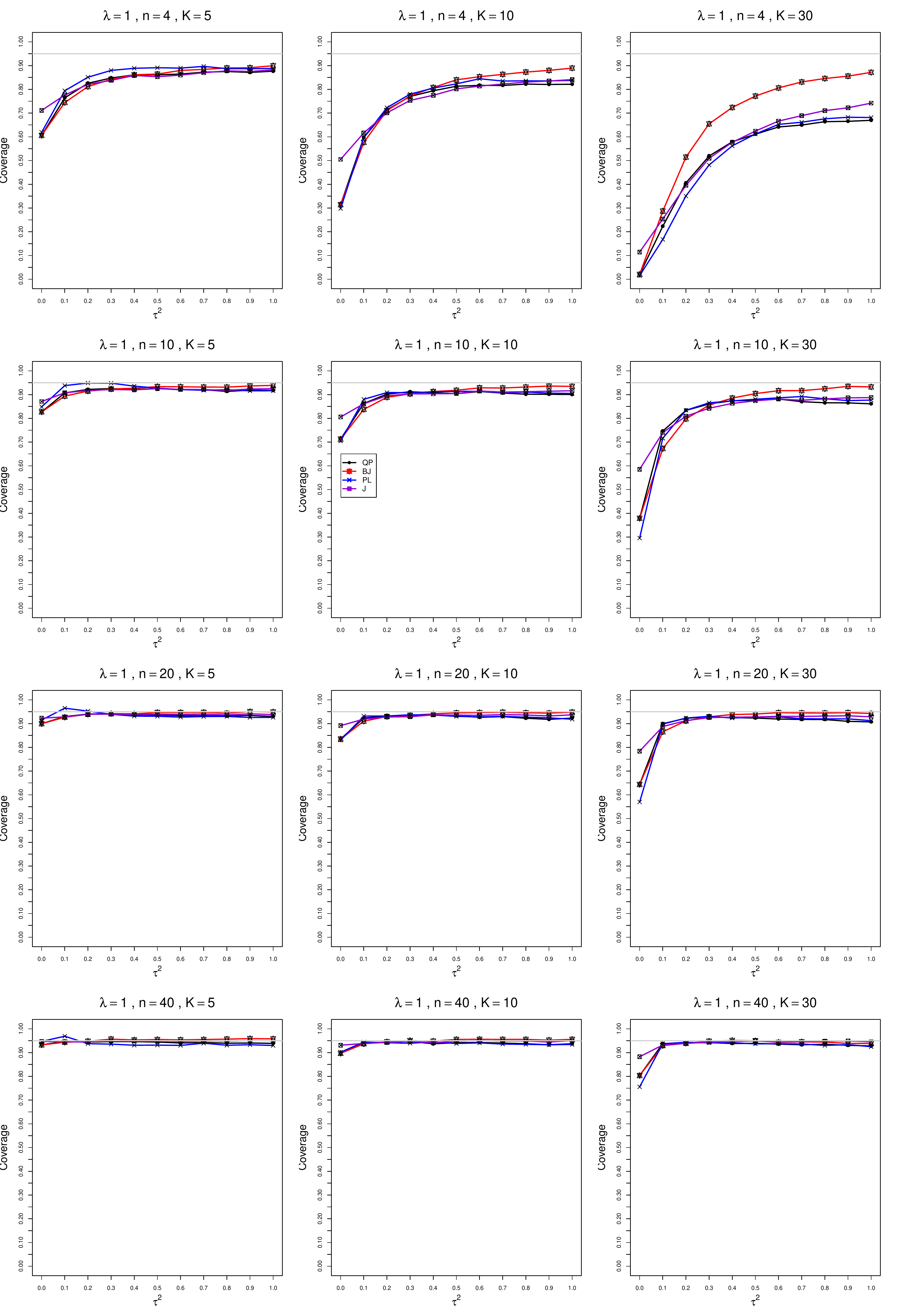}
	\caption{Coverage of 95\% confidence intervals for the between-studies variance $\tau^2$ when $\lambda=1$, $n = 4, \;10, \;20, \;40$, and $K = 5, \;10, \;30$. Usual estimate of $\lambda_i$ 		\label{CovTauRoM1ln_smallN_small_K}}
\end{figure}
\begin{figure}[t]
	\includegraphics[scale=0.35]{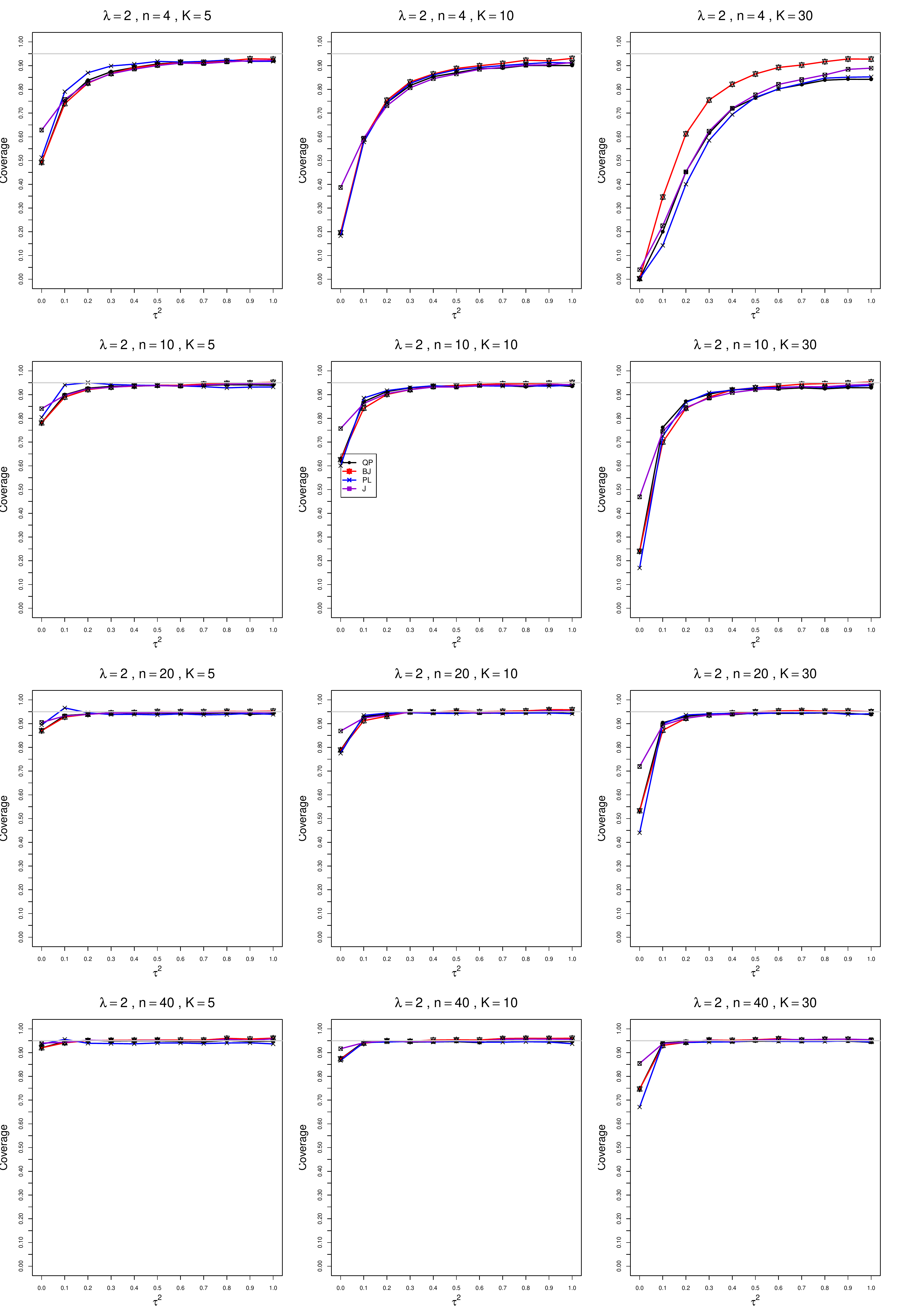}
	\caption{Coverage of 95\% confidence intervals for the between-studies variance $\tau^2$ when $\lambda=2$, $n = 4, \;10, \;20, \;40$, and $K = 5, \;10, \;30$. Usual estimate of $\lambda_i$ 		\label{CovTauRoM2ln_smallN_small_K}}
\end{figure}

\clearpage
\renewcommand{\thefigure}{A2.1.\arabic{figure}}
\setcounter{figure}{0}
\section*{A2. Lognormal model, bias-corrected estimator of $\lambda_i$, $n= 4, 10, 20, 40$, $K=5,10,30$}
\subsection*{A2.1 Bias of point estimators of $\tau^2$}
Each figure corresponds to a value of $\lambda \;(= 0, 0.2, 0.5, 1, 2)$, a set of values of $n$ (= 4, 10, 20, 40), and a set of values of $K$ (= 5, 10, 30).\\
Each panel corresponds to a value of $n$ and a value of $K$ and has $\tau^2 = 0.0(0.1)1.0$ on the horizontal axis.\\
The point estimators of $\tau^2$ are
\begin{itemize}
	\item DL (DerSimonian-Laird)
	\item REML (restricted maximum likelihood)
	\item MP (Mandel-Paule)
	\item J (Jackson)
\end{itemize}

\clearpage

\begin{figure}[t]
	\includegraphics[scale=0.33]{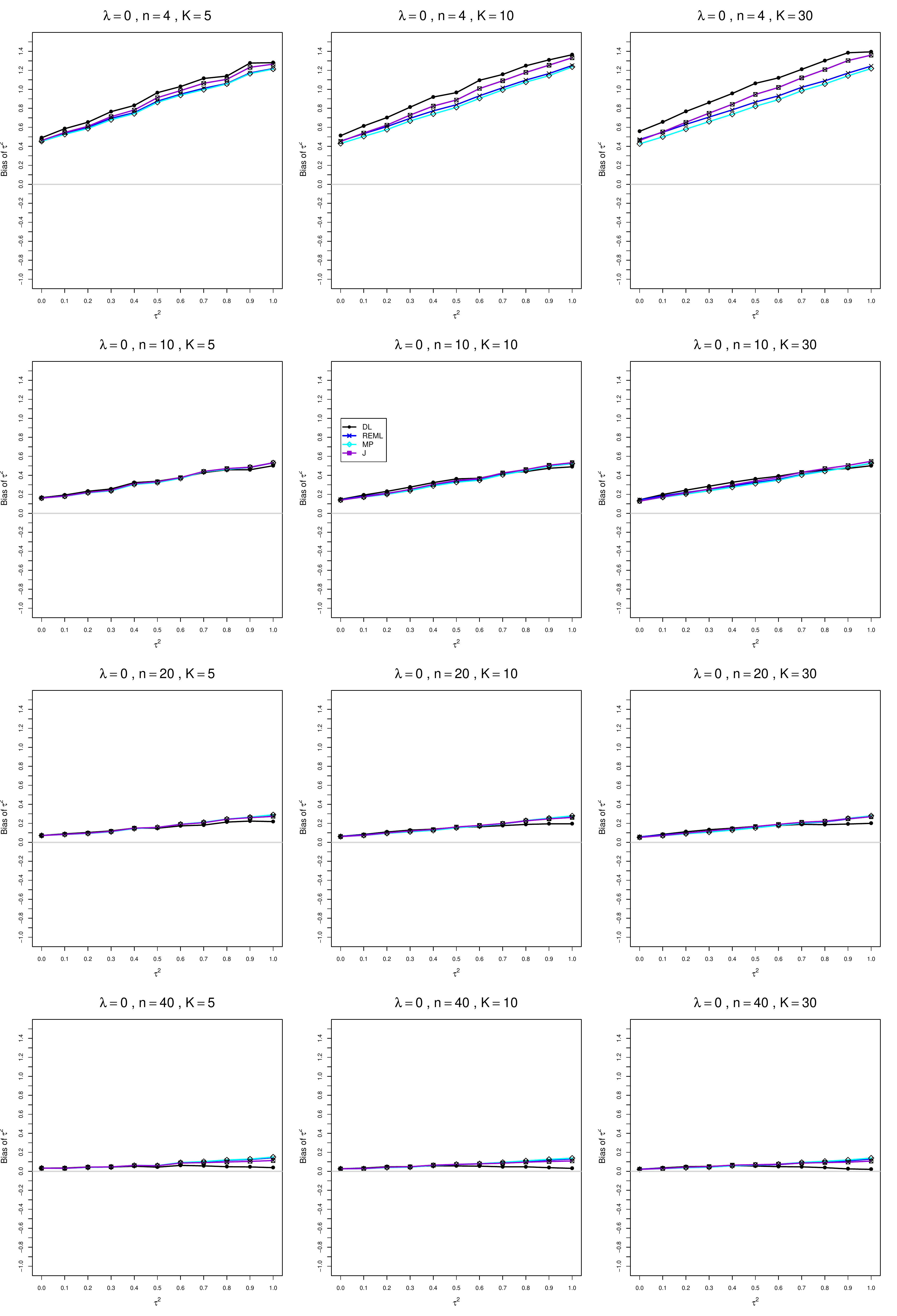}
	\caption{Bias of estimators of between-studies variance $\tau^2$ for $\lambda=0$, $n = 4, \;10, \;20, \;40$, and $K = 5, \;10, \;30$. Bias-corrected estimate of $\lambda_i$
		\label{BiasTauRoM0lnCor_smallN_small_K}}
\end{figure}

\begin{figure}[t]
	\includegraphics[scale=0.33]{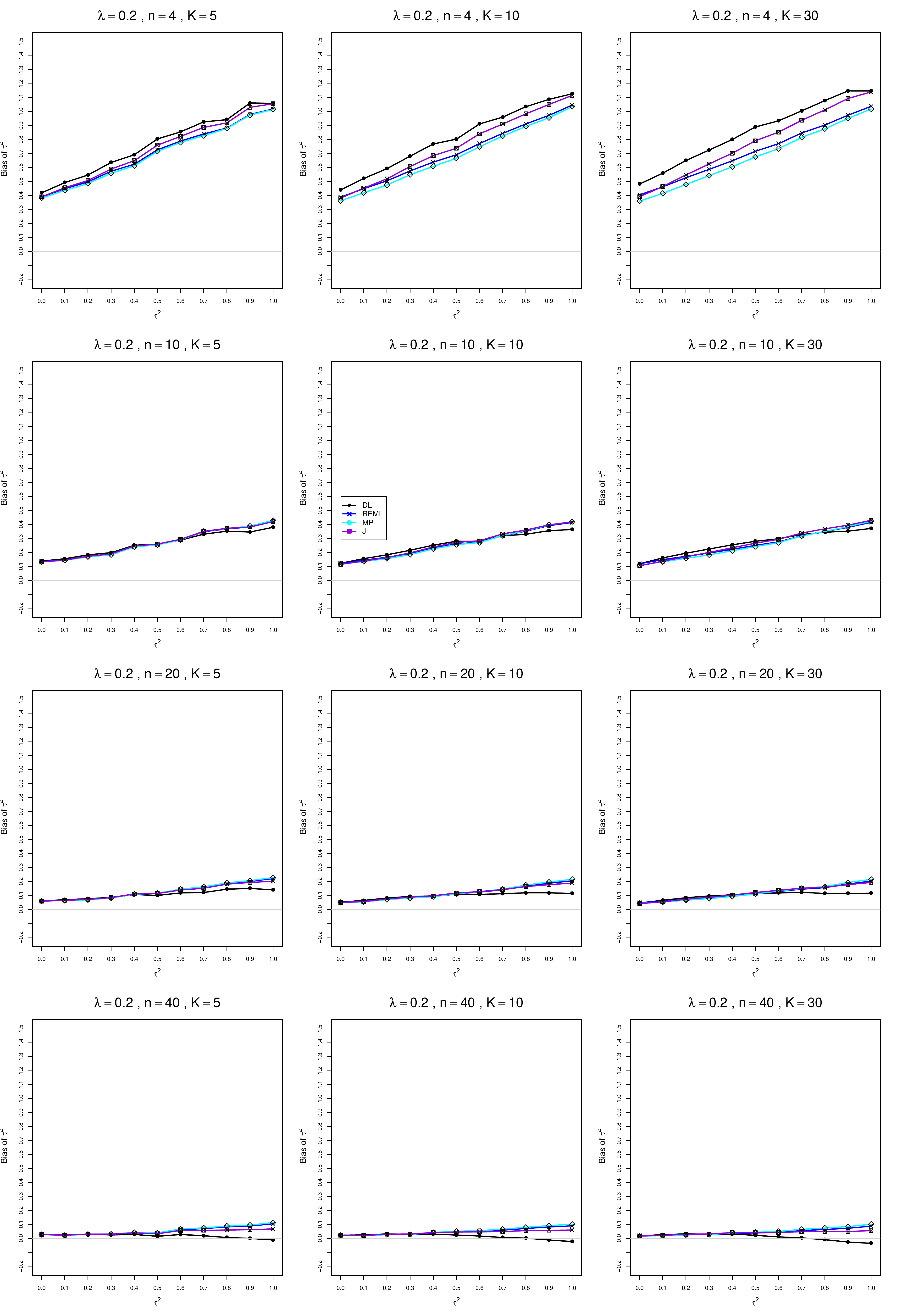}
	\caption{Bias of estimators of between-studies variance $\tau^2$ for $\lambda=0.2$, $n = 4, \;10, \;20, \;40$, and $K = 5, \;10, \;30$. Bias-corrected estimate of $\lambda_i$
		\label{BiasTauRoM02lnCor_smallN_small_K}}
\end{figure}

\begin{figure}[t]
	\includegraphics[scale=0.33]{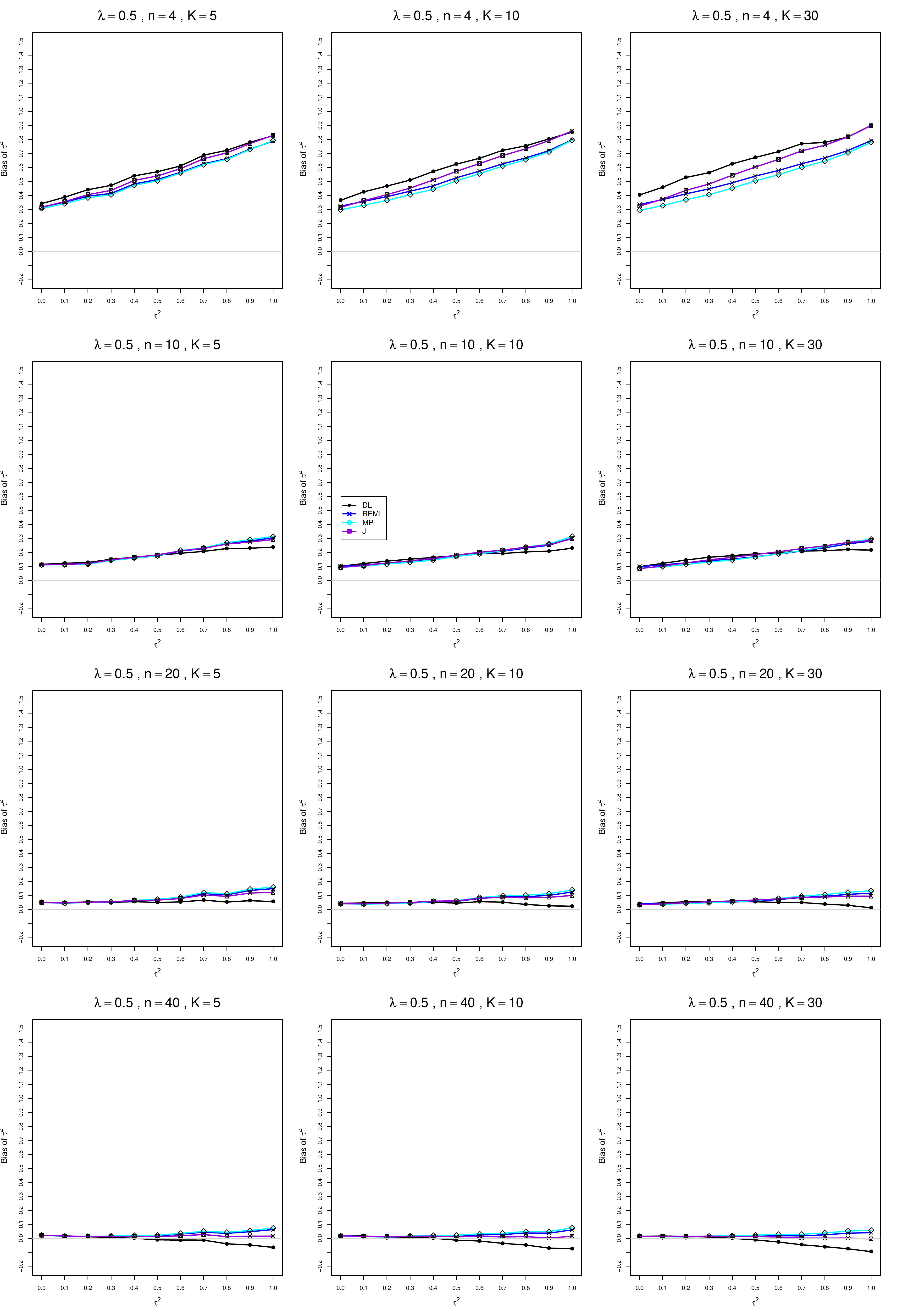}
	\caption{Bias of estimators of between-studies variance $\tau^2$ for $\lambda=0.5$, $n = 4, \;10, \;20, \;40$, and $K = 5, \;10, \;30$. Bias-corrected estimate of $\lambda_i$
		\label{BiasTauRoM05lnCor_smallN_small_K}}
\end{figure}

\begin{figure}[t]
	\includegraphics[scale=0.33]{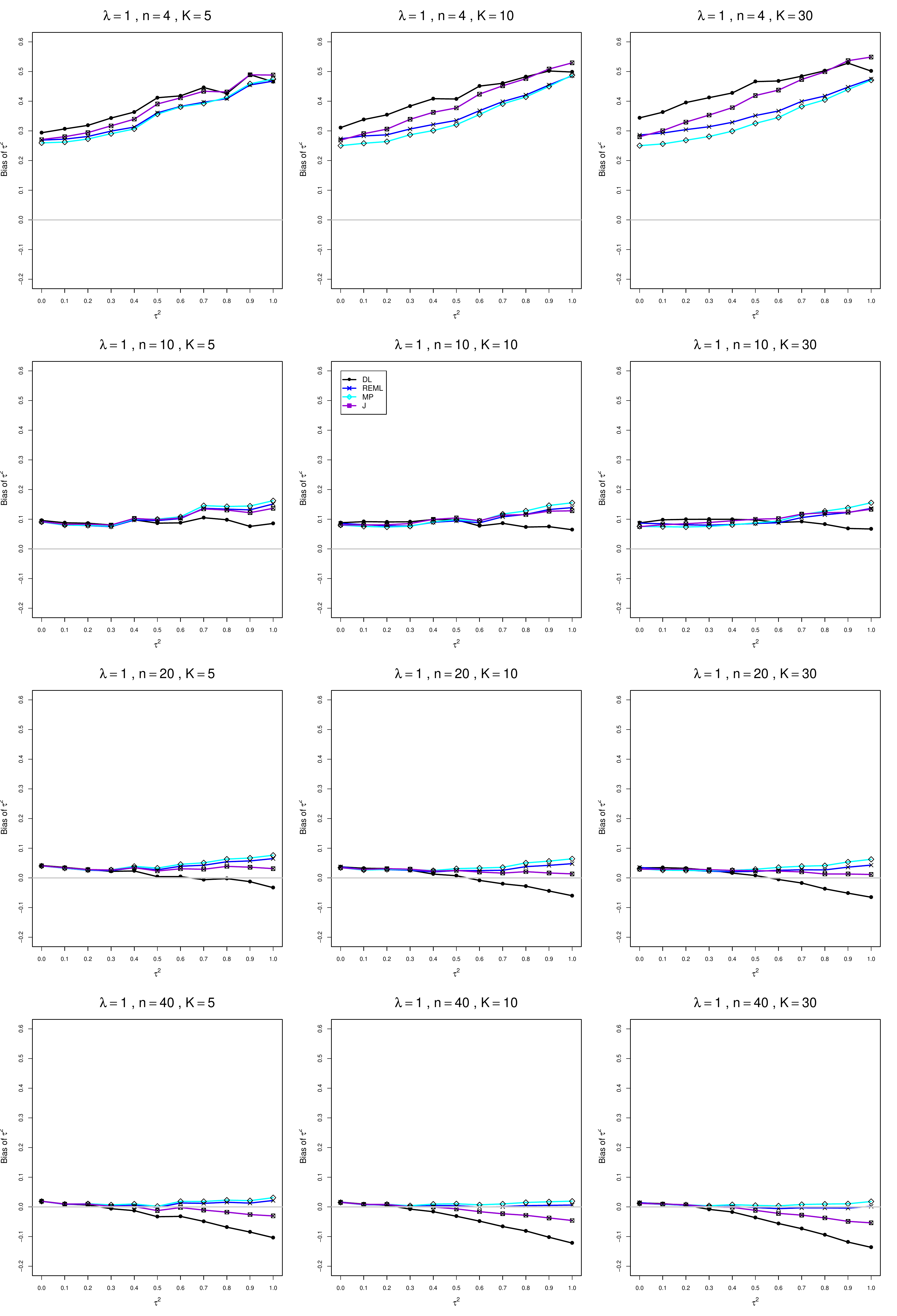}
	\caption{Bias of estimators of between-studies variance $\tau^2$ for $\lambda=1$, $n = 4, \;10, \;20, \;40$, and $K = 5, \;10, \;30$. Bias-corrected estimate of $\lambda_i$
		\label{BiasTauRoM1lnCor_smallN_small_K}}
\end{figure}

\begin{figure}[t]
	\includegraphics[scale=0.33]{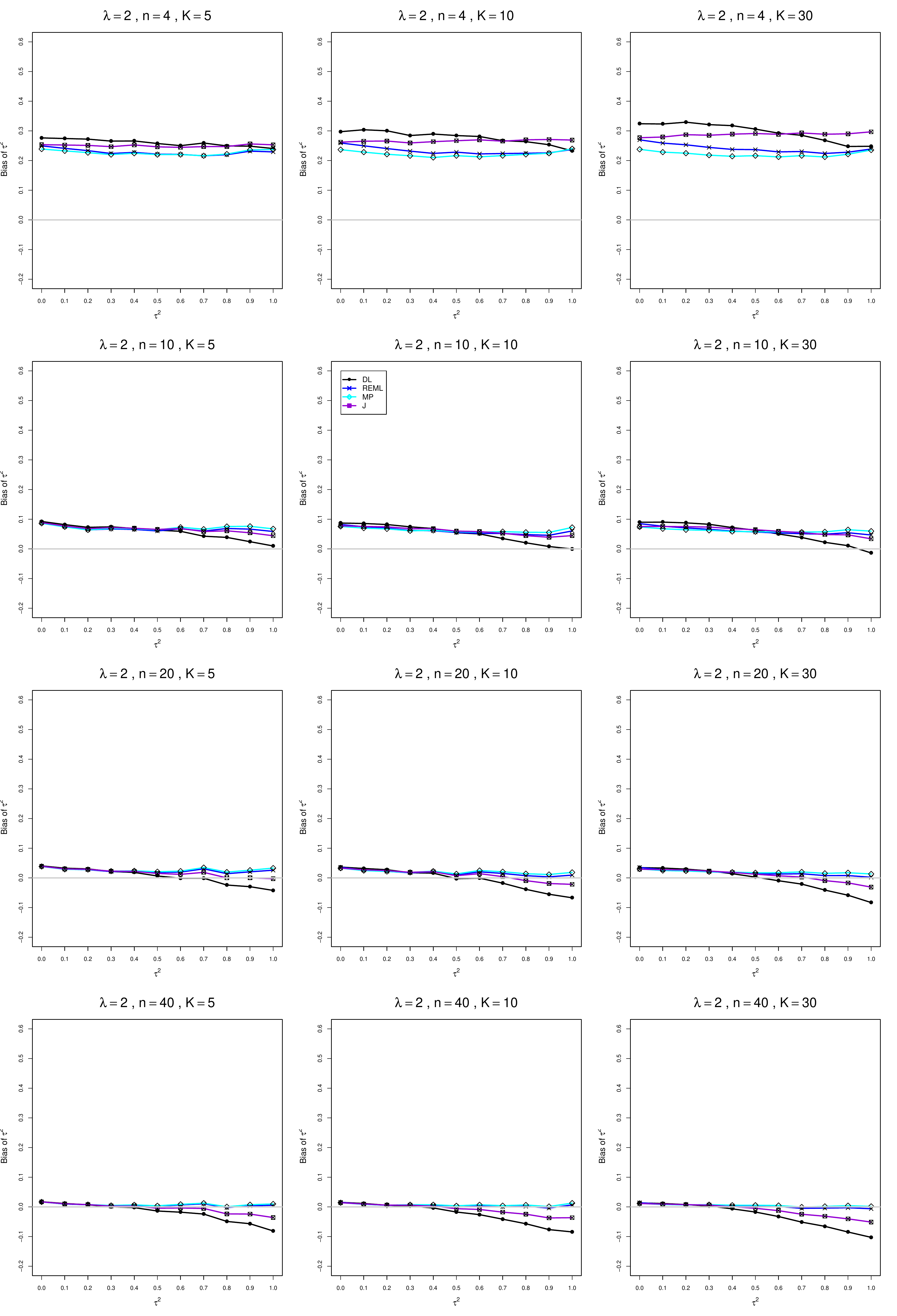}
	\caption{Bias of estimators of between-studies variance $\tau^2$ for $\lambda=2$, $n = 4, \;10, \;20, \;40$, and $K = 5, \;10, \;30$. Bias-corrected estimate of $\lambda_i$
		\label{BiasTauRoM2lnCor_smallN_small_K}}
\end{figure}

\renewcommand{\thefigure}{A2.2.\arabic{figure}}
\setcounter{figure}{0}
\clearpage
\subsection*{A2.2 Coverage of interval estimators of $\tau^2$}
Each figure corresponds to a value of $\lambda \;(= 0, 0.2, 0.5, 1, 2)$, a set of values of $n$ (= 4, 10, 20, 40), and a set of values of $K$ (= 5, 10, 30).\\
Each panel corresponds to a value of $n$ and a value of $K$ and has $\tau^2 = 0.0(0.1)1.0$ on the horizontal axis.\\
The interval estimators of $\tau^2$ are
\begin{itemize}
	\item QP (Q-profile confidence interval)
	\item BJ (Biggerstaff and Jackson interval )
	\item PL (Profile-likelihood interval)
	\item J (Jackson interval)
\end{itemize}
\clearpage

\begin{figure}[t]
	\includegraphics[scale=0.35]{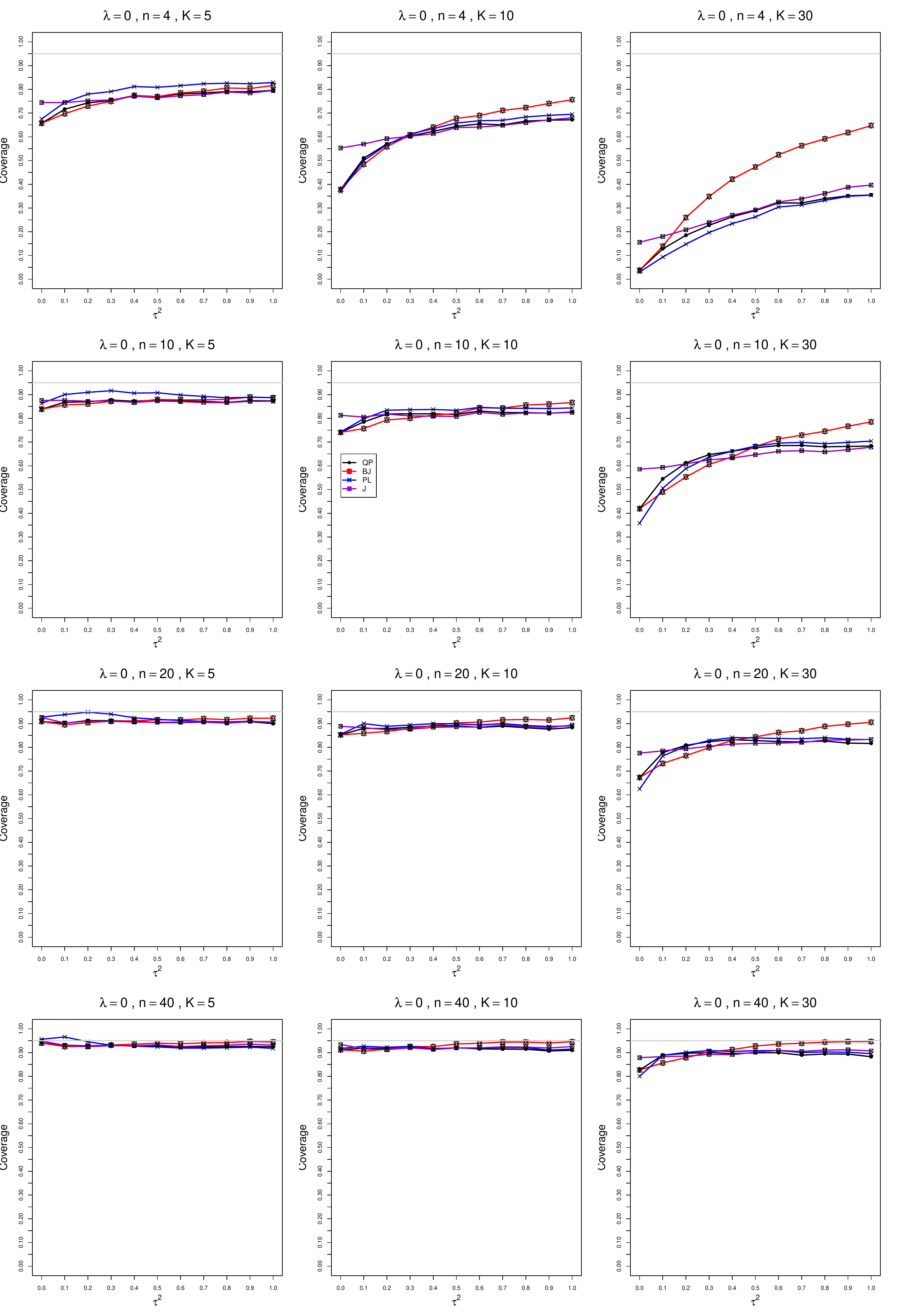}
	\caption{Coverage of 95\% confidence intervals for the between-studies variance $\tau^2$ when $\lambda=0$, $n = 4, \;10, \;20, \;40$, and $K = 5, \;10, \;30$. Bias-corrected estimate of $\lambda_i$
		\label{CovTauRoM0lnCor_smallN_small_K}}
\end{figure}
\begin{figure}[t]
	\includegraphics[scale=0.35]{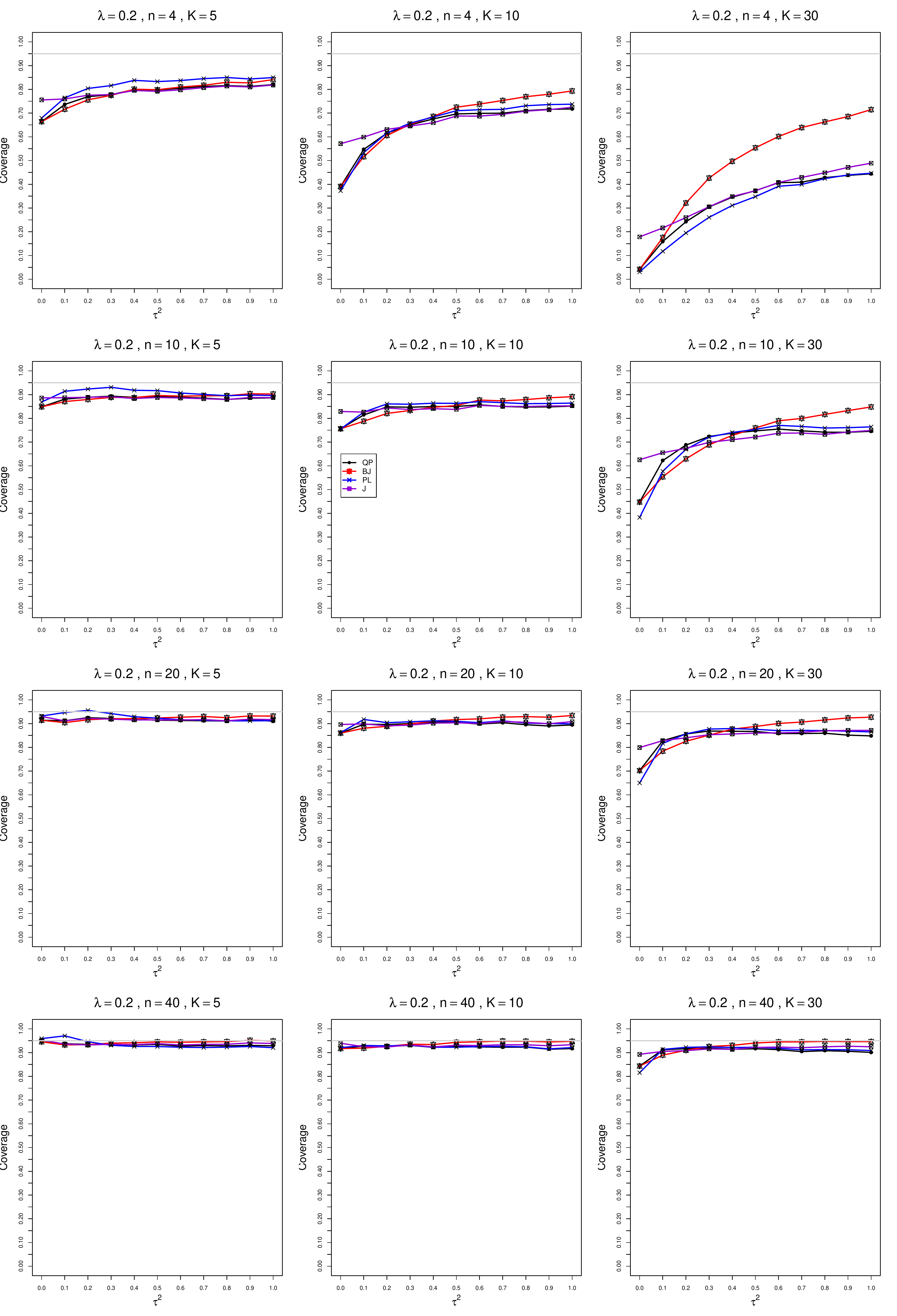}
	\caption{Coverage of 95\% confidence intervals for the between-studies variance $\tau^2$ when $\lambda=0.2$, $n = 4, \;10, \;20, \;40$, and $K = 5, \;10, \;30$. Bias-corrected estimate of $\lambda_i$ 		\label{CovTauRoM02lnCor_smallN_small_K}}
\end{figure}

\begin{figure}[t]
	\includegraphics[scale=0.35]{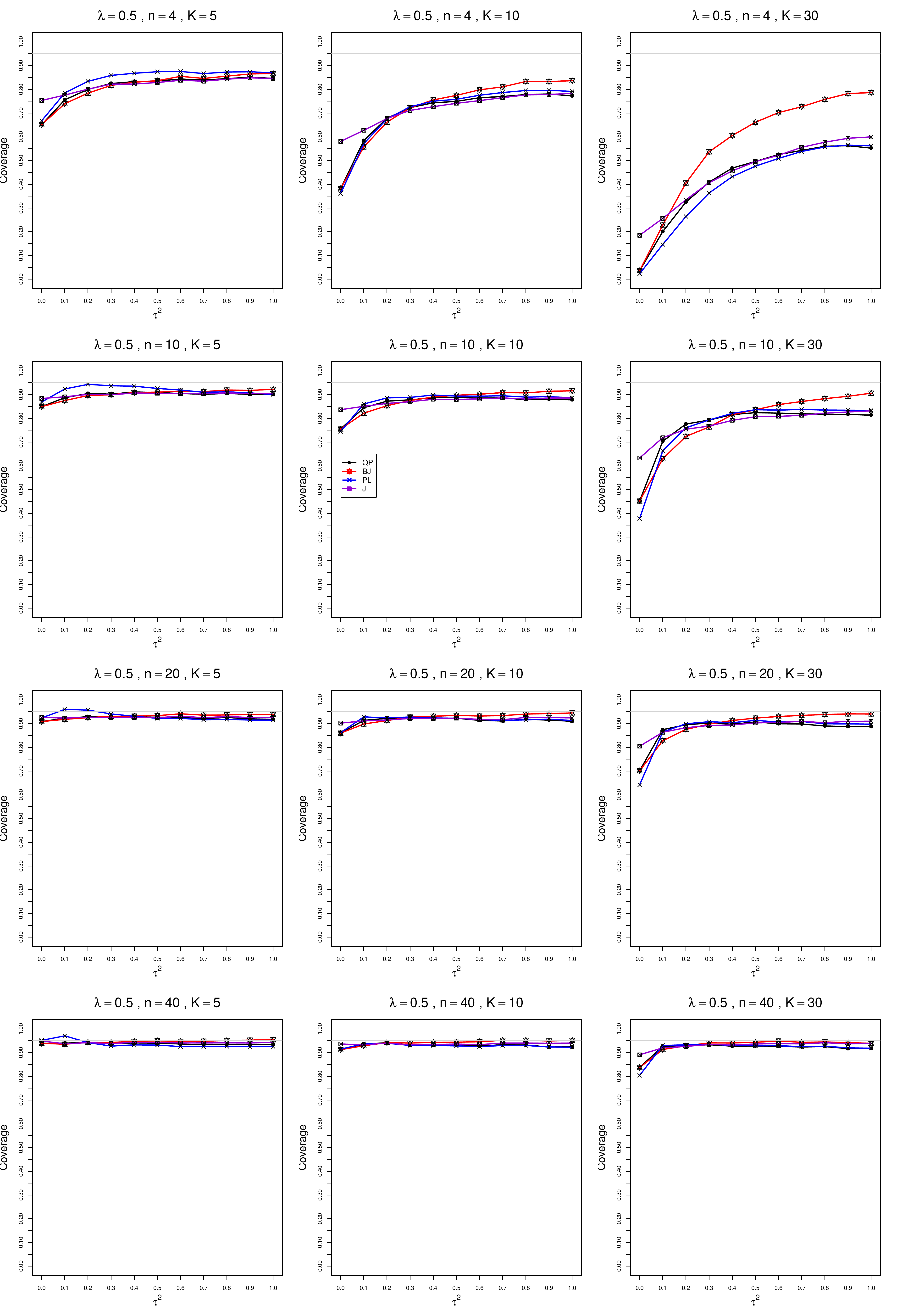}
	\caption{Coverage of 95\% confidence intervals for the between-studies variance $\tau^2$ when $\lambda=0.5$, $n = 4, \;10, \;20, \;40$, and $K = 5, \;10, \;30$. Bias-corrected estimate of $\lambda_i$ 		\label{CovTauRoM05lnCor_smallN_small_K}}
\end{figure}

\begin{figure}[t]
	\includegraphics[scale=0.35]{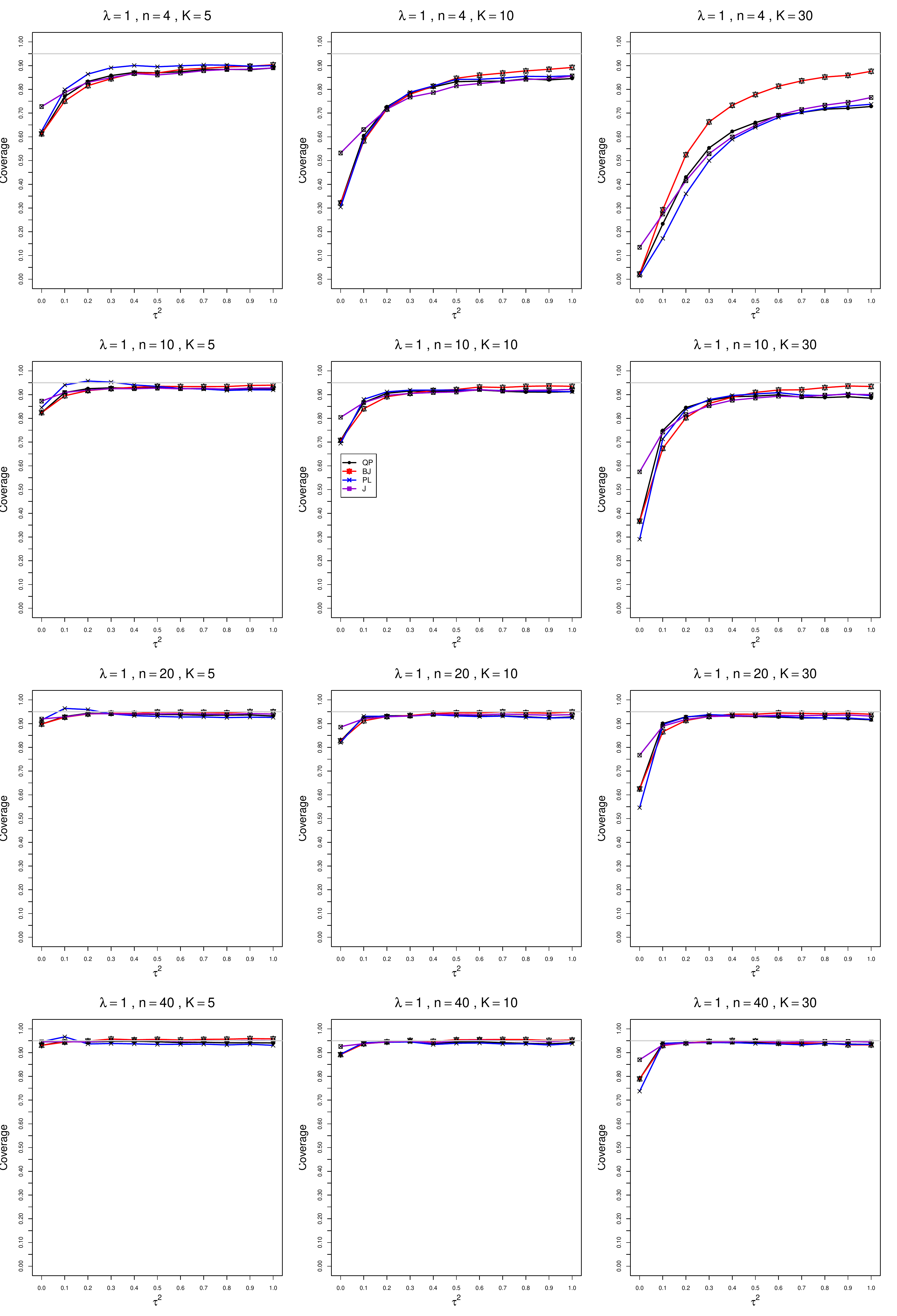}
	\caption{Coverage of 95\% confidence intervals for the between-studies variance $\tau^2$ when $\lambda=1$, $n = 4, \;10, \;20, \;40$, and $K = 5, \;10, \;30$. Bias-corrected estimate of $\lambda_i$ 		\label{CovTauRoM1lnCor_smallN_small_K}}
\end{figure}
\begin{figure}[t]
	\includegraphics[scale=0.35]{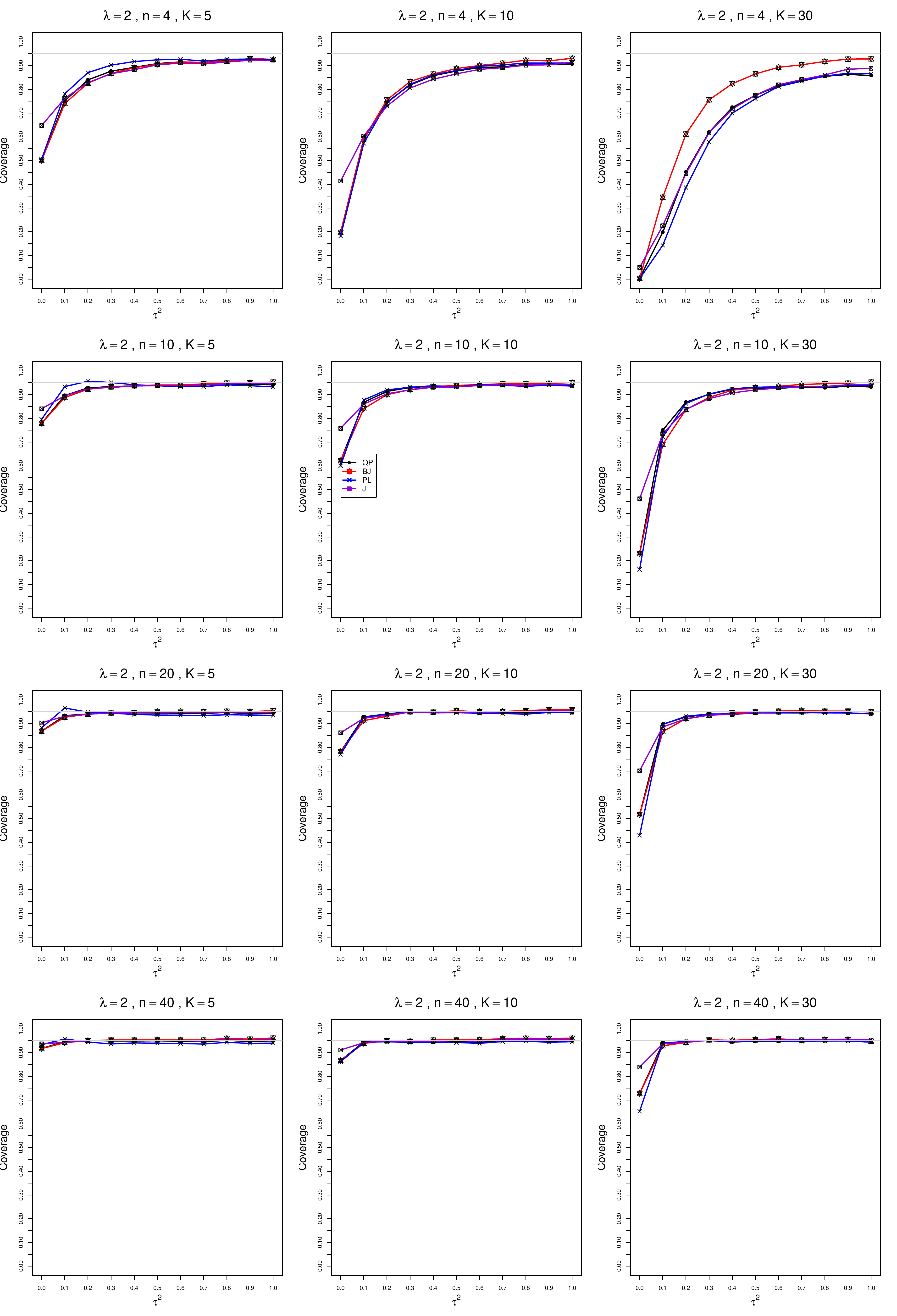}
	\caption{Coverage of 95\% confidence intervals for the between-studies variance $\tau^2$ when $\lambda=2$, $n = 4, \;10, \;20, \;40$, and $K = 5, \;10, \;30$. Bias-corrected estimate of $\lambda_i$
		\label{CovTauRoM2lnCor_smallN_small_K}}
\end{figure}

\clearpage
\renewcommand{\thefigure}{A3.1.\arabic{figure}}
\renewcommand{\thesection}{A3.\arabic{section}}
\section*{A3. Lognormal model, usual estimator of $\lambda_i$, $n= 4, 10, 20, 40$, $K=50,100,125$}
\subsection*{A3.1 Bias of point estimators of $\tau^2$}
Each figure corresponds to a value of $\lambda \;(= 0, 0.2, 0.5, 1, 2)$, a set of values of $n$ (= 4, 10, 20, 40), and a set of values of $K$ (= 50, 100, 125).\\
Each panel corresponds to a value of $n$ and a value of $K$ and has $\tau^2 = 0.0(0.1)1.0$ on the horizontal axis.\\
The point estimators of $\tau^2$ are
\begin{itemize}
	\item DL (DerSimonian-Laird)
	\item REML (restricted maximum likelihood)
	\item MP (Mandel-Paule)
	\item J (Jackson)
\end{itemize}

\clearpage

\begin{figure}[t]
	\includegraphics[scale=0.33]{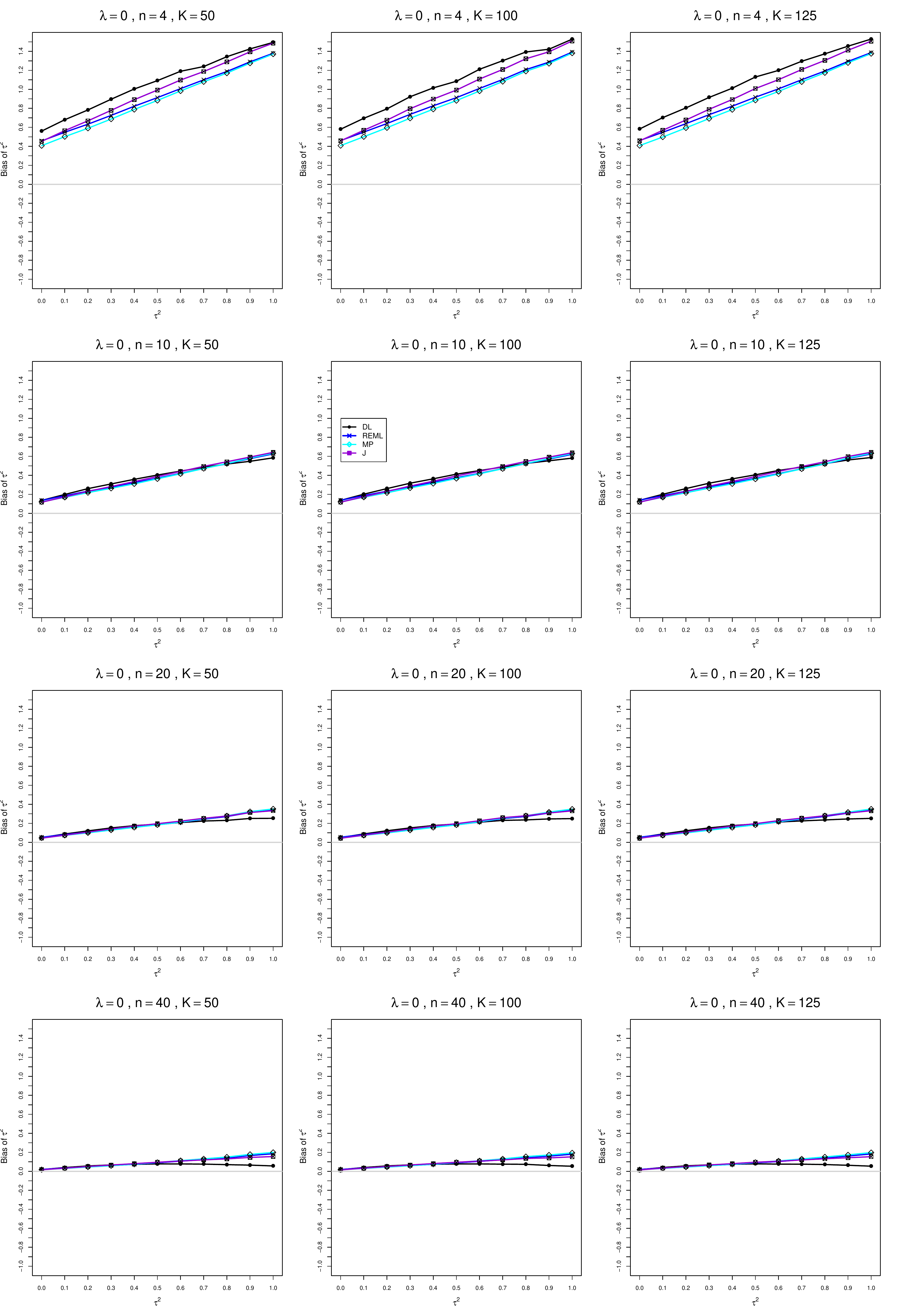}
	\caption{Bias of estimators of between-studies variance $\tau^2$ for $\lambda=0$, $n = 4, \;10, \;20, \;40$, and $K = 50, \;100, \;125$. Usual estimate of $\lambda_i$
		\label{BiasTauRoM0ln_smallN_large_K}}
\end{figure}

\begin{figure}[t]
	\includegraphics[scale=0.33]{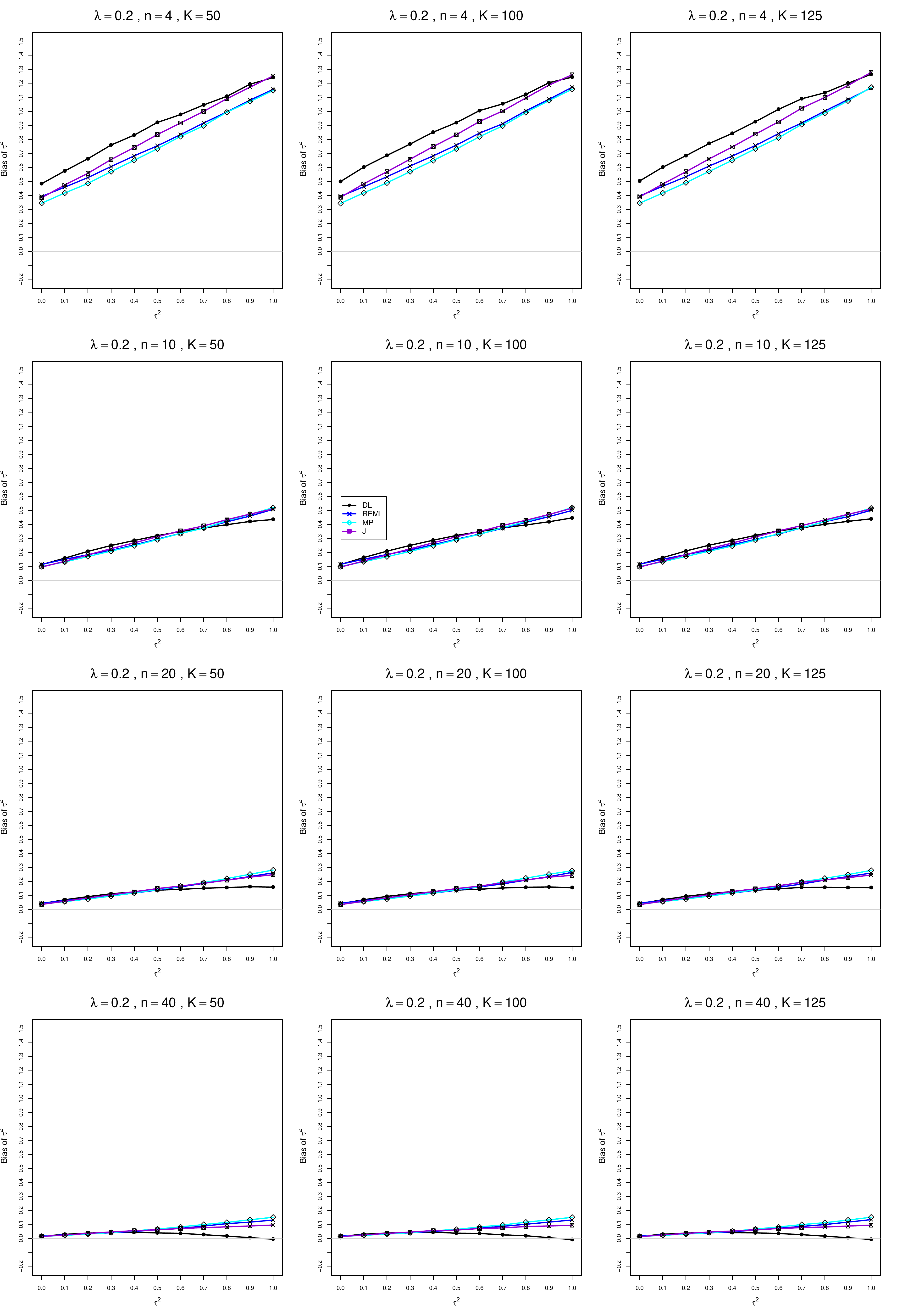}
	\caption{Bias of estimators of between-studies variance $\tau^2$ for $\lambda=0.2$, $n = 4, \;10, \;20, \;40$, and $K = 50, \;100, \;125$. Usual estimate of $\lambda_i$
		\label{BiasTauRoM02ln_smallN_large_K}}
\end{figure}

\begin{figure}[t]
	\includegraphics[scale=0.33]{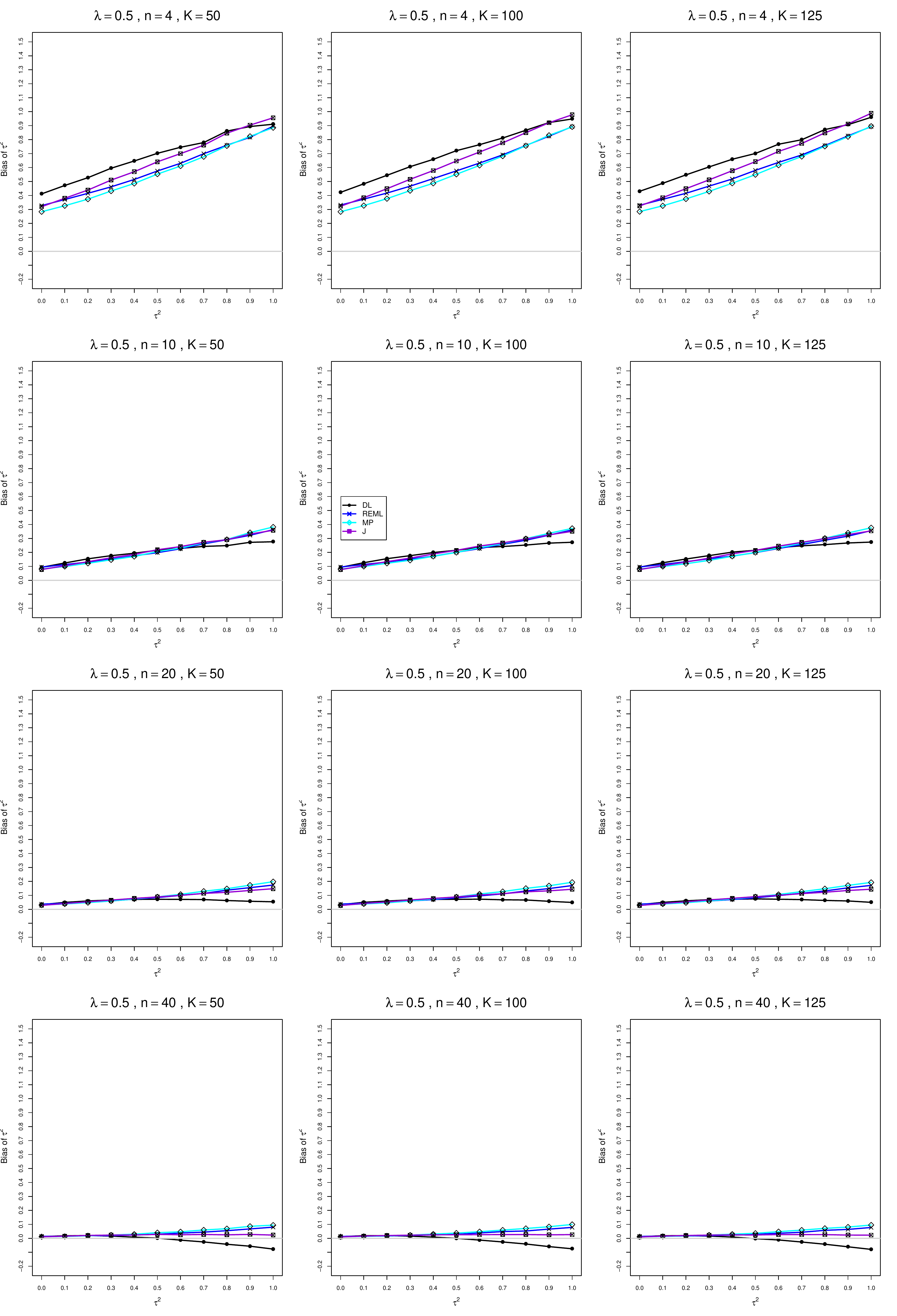}
	\caption{Bias of estimators of between-studies variance $\tau^2$ for $\lambda=0.5$, $n = 4, \;10, \;20, \;40$, and $K = 50, \;100, \;125$. Usual estimate of $\lambda_i$
		\label{BiasTauRoM05ln_smallN_large_K}}
\end{figure}

\begin{figure}[t]
	\includegraphics[scale=0.33]{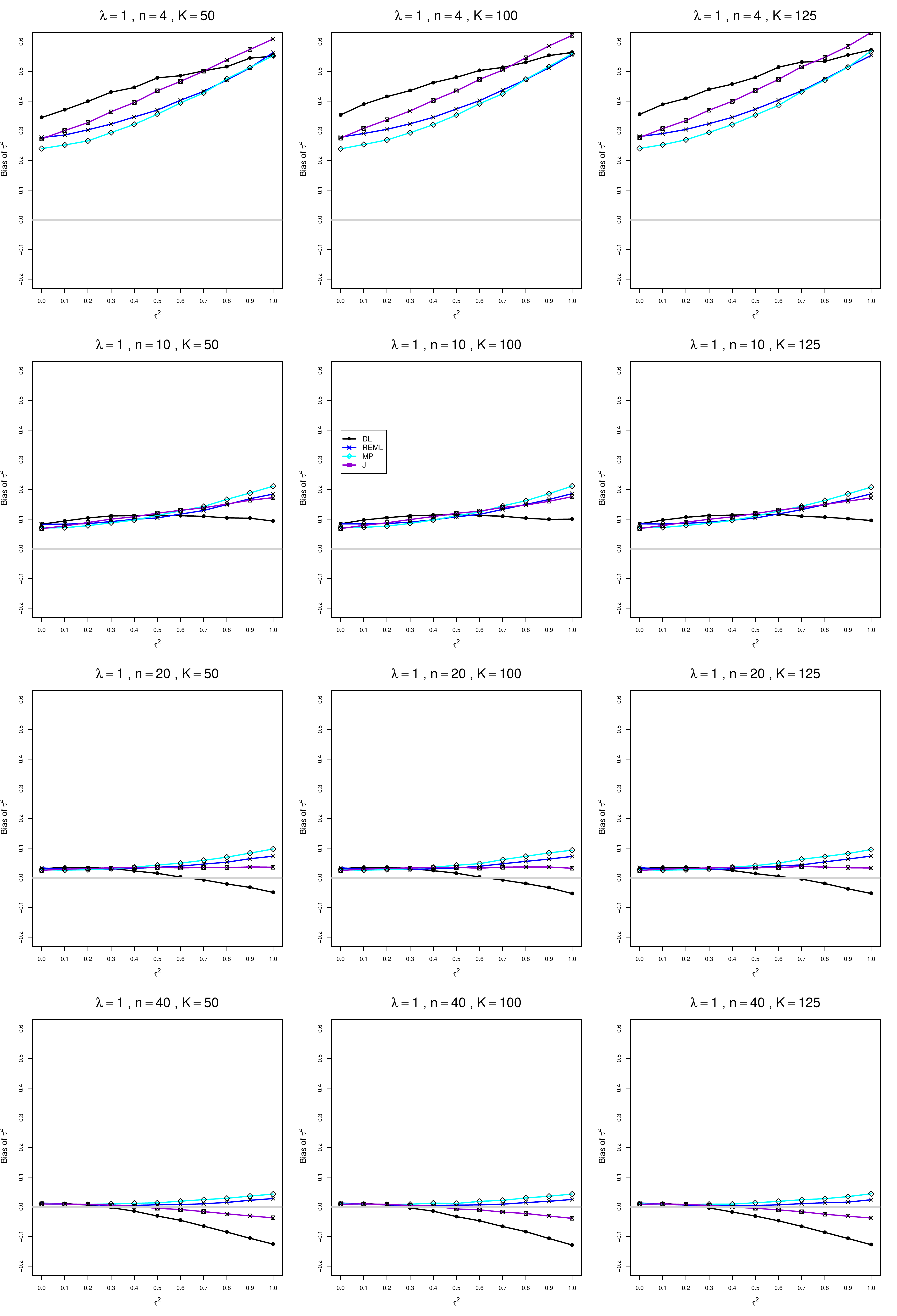}
	\caption{Bias of estimators of between-studies variance $\tau^2$ for $\lambda=1$, $n = 4, \;10, \;20, \;40$, and $K = 50, \;100, \;125$. Usual estimate of $\lambda_i$
		\label{BiasTauRoM1ln_smallN_large_K}}
\end{figure}

\begin{figure}[t]
	\includegraphics[scale=0.33]{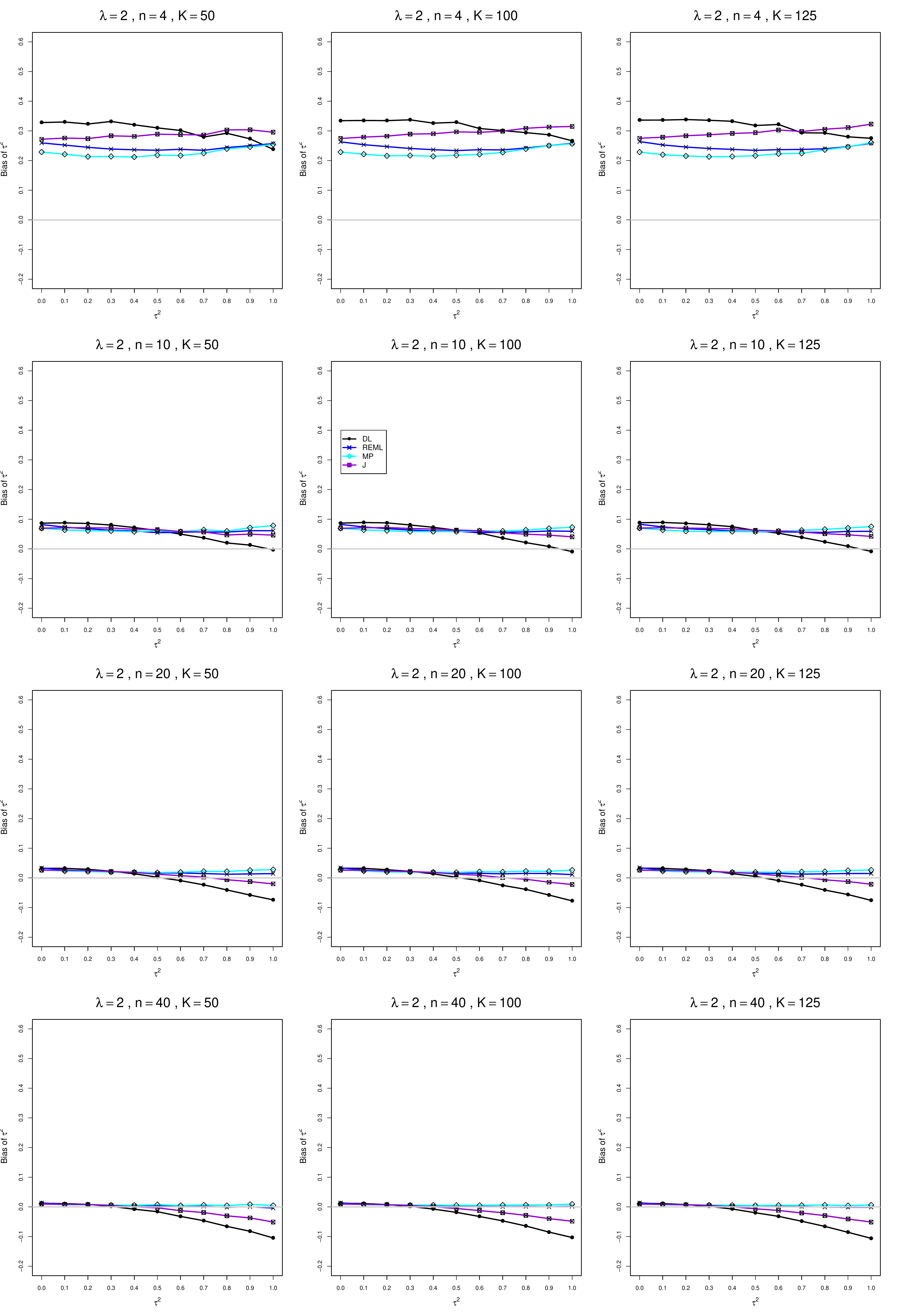}
	\caption{Bias of estimators of between-studies variance $\tau^2$ for $\lambda=2$, $n = 4, \;10, \;20, \;40$, and $K = 50, \;100, \;125$. Usual estimate of $\lambda_i$
		\label{BiasTauRoM2ln_smallN_large_K}}
\end{figure}

\renewcommand{\thefigure}{A3.2.\arabic{figure}}
\setcounter{figure}{0}
\clearpage
\subsection*{A3.2 Coverage of interval estimators of $\tau^2$}
Each figure corresponds to a value of $\lambda \;(= 0, 0.2, 0.5, 1, 2)$, a set of values of $n$ (= 4, 10, 20, 40), and a set of values of $K$ (= 50, 100, 125).\\
Each panel corresponds to a value of $n$ and a value of $K$ and has $\tau^2 = 0.0(0.1)1.0$ on the horizontal axis.\\
The interval estimators of $\tau^2$ are
\begin{itemize}
	\item QP (Q-profile confidence interval)
	\item BJ (Biggerstaff and Jackson interval )
	\item PL (Profile-likelihood interval)
	\item J (Jackson interval)
\end{itemize}

\begin{figure}[t]
	\includegraphics[scale=0.35]{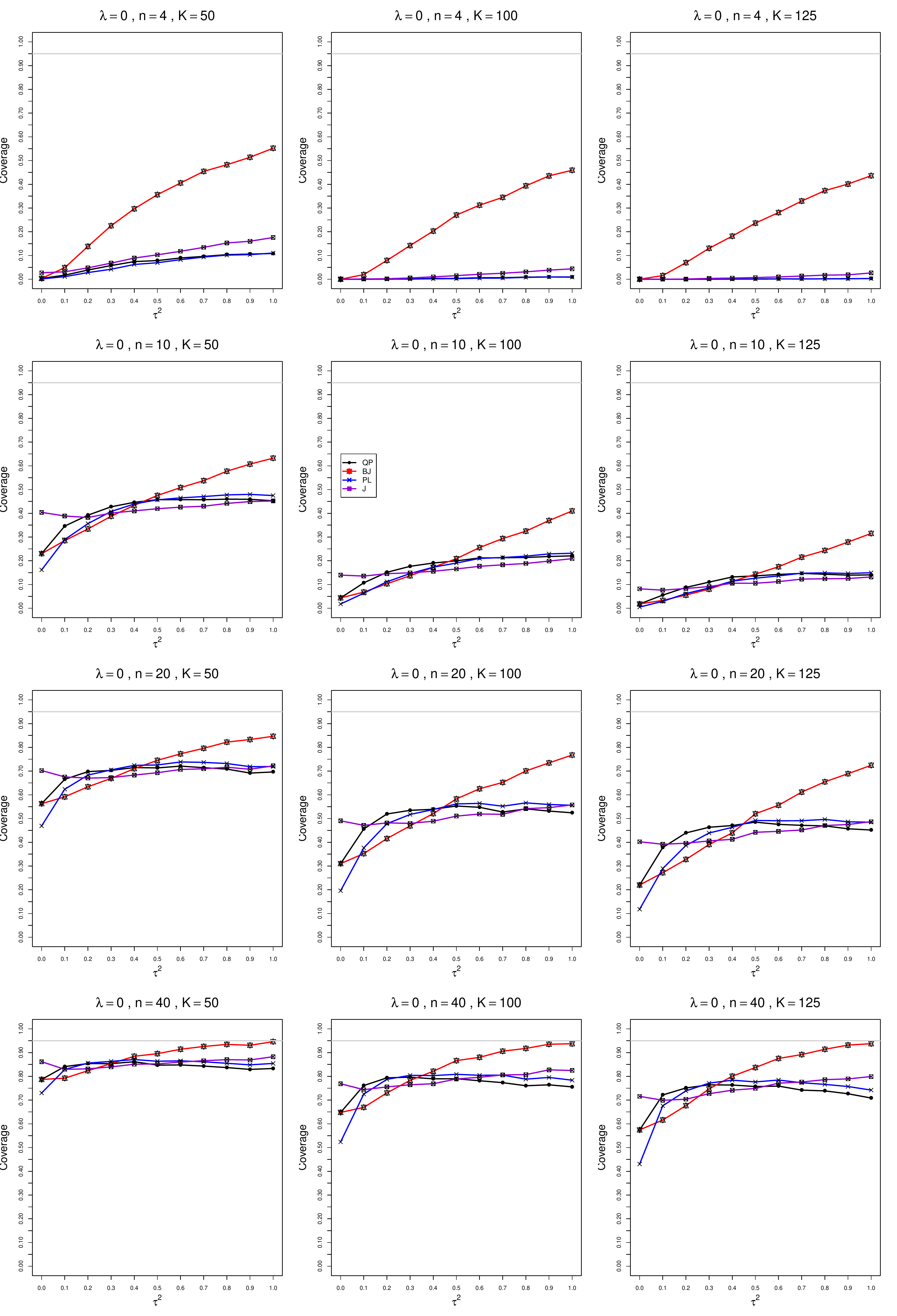}
	\caption{Coverage of 95\% confidence intervals for the between-studies variance $\tau^2$ when $\lambda=0$, $n = 4, \;10, \;20, \;40$, and $K = 50, \;100, \;125$. Usual estimate of $\lambda_i$ \label{CovTauRoM0ln_smallN_large_K}}
\end{figure}

\begin{figure}[t]
	\includegraphics[scale=0.35]{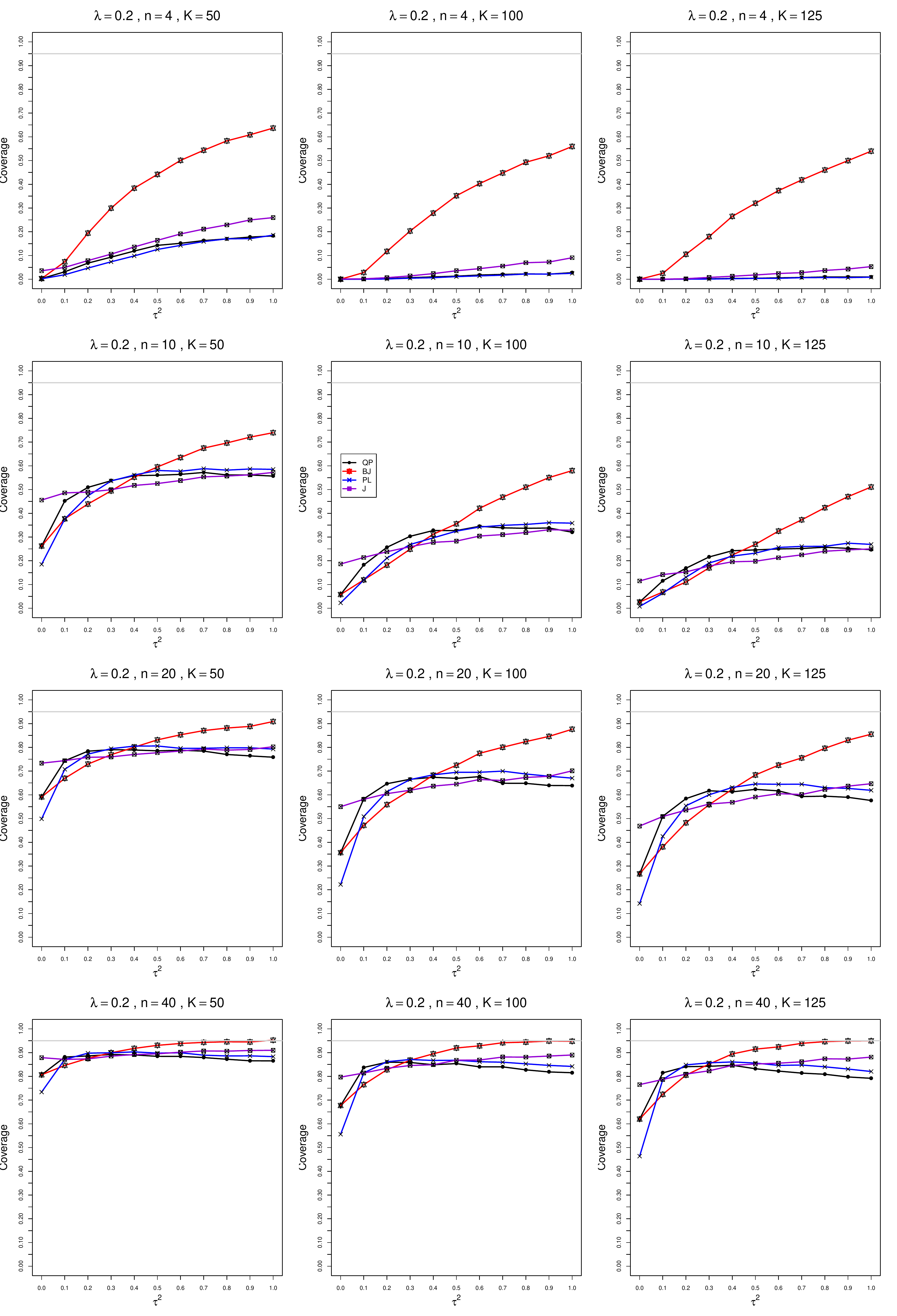}
	\caption{Coverage of 95\% confidence intervals for the between-studies variance $\tau^2$ when $\lambda=0.2$, $n = 4, \;10, \;20, \;40$, and $K = 50, \;100, \;125$. Usual estimate of $\lambda_i$ \label{CovTauRoM02ln_smallN_large_K}}
\end{figure}

\begin{figure}[t]
	\includegraphics[scale=0.35]{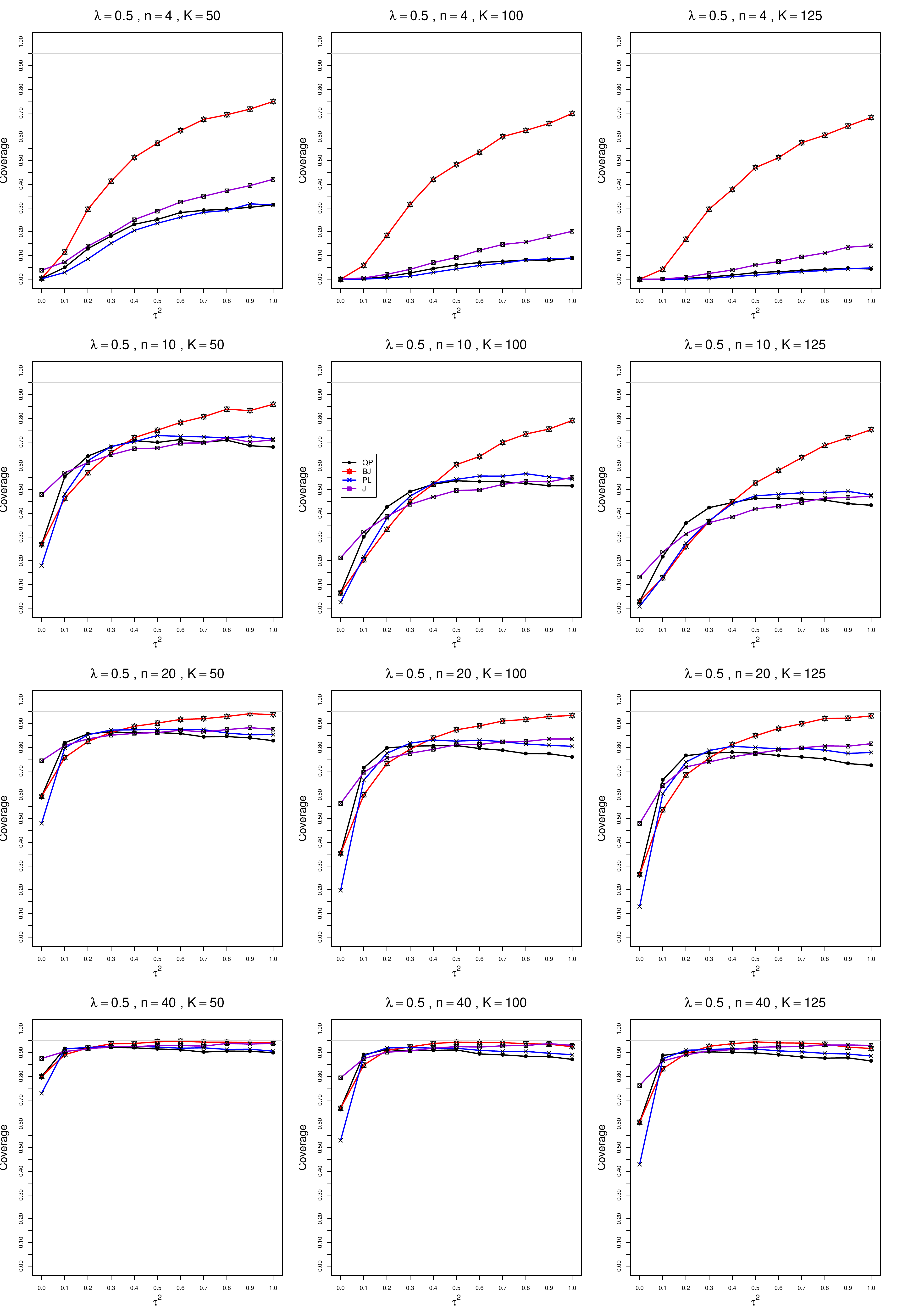}
	\caption{Coverage of 95\% confidence intervals for the between-studies variance $\tau^2$ when $\lambda=0.5$, $n = 4, \;10, \;20, \;40$, and $K = 50, \;100, \;125$. Usual estimate of $\lambda_i$ 		\label{CovTauRoM05ln_smallN_large_K}}
\end{figure}

\begin{figure}[t]
	\includegraphics[scale=0.35]{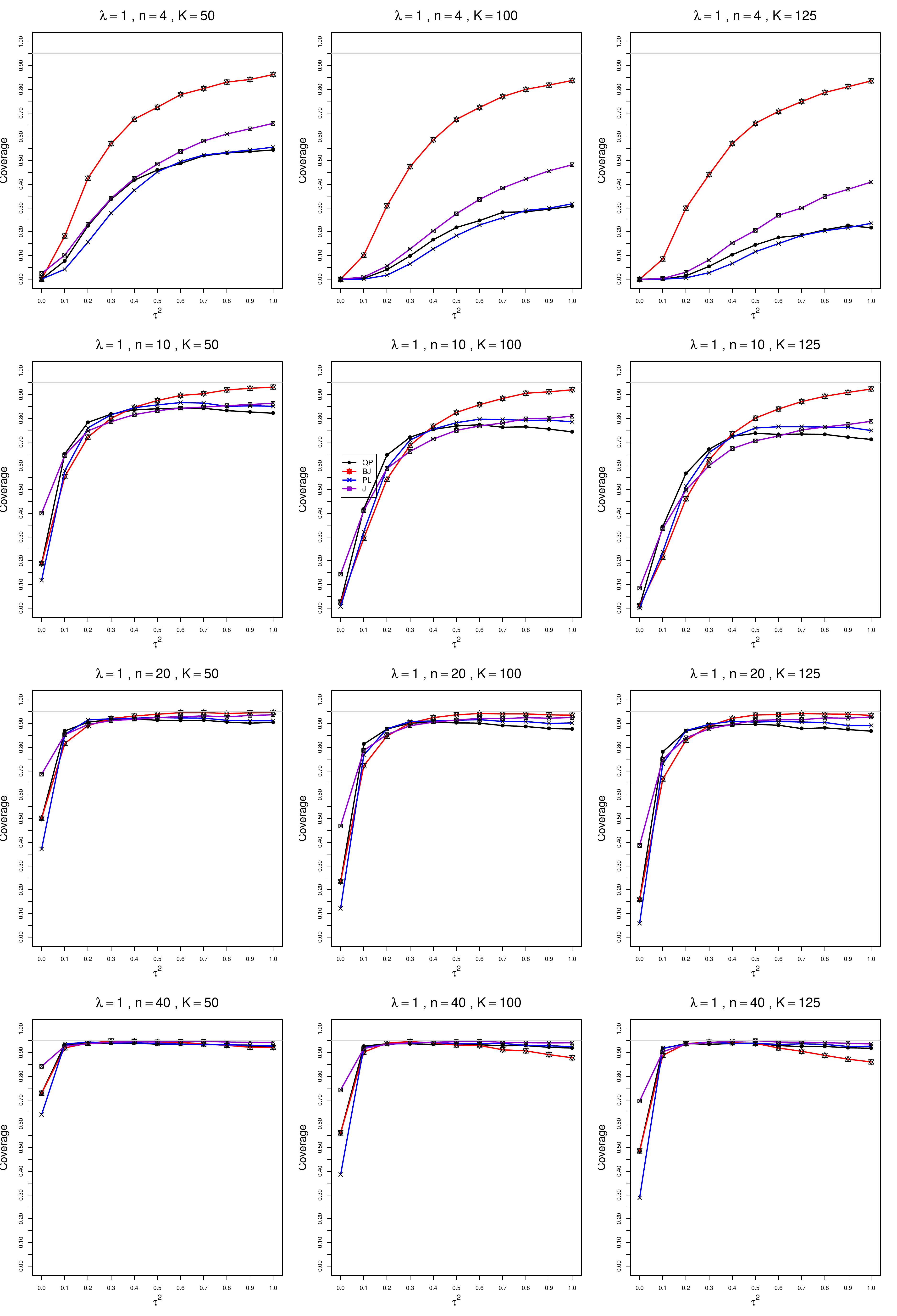}
	\caption{Coverage of 95\% confidence intervals for the between-studies variance $\tau^2$ when $\lambda=1$, $n = 4, \;10, \;20, \;40$, and $K = 50, \;100, \;125$. Usual estimate of $\lambda_i$ 		\label{CovTauRoM1ln_smallN_large_K}}
\end{figure}
\begin{figure}[t]
	\includegraphics[scale=0.35]{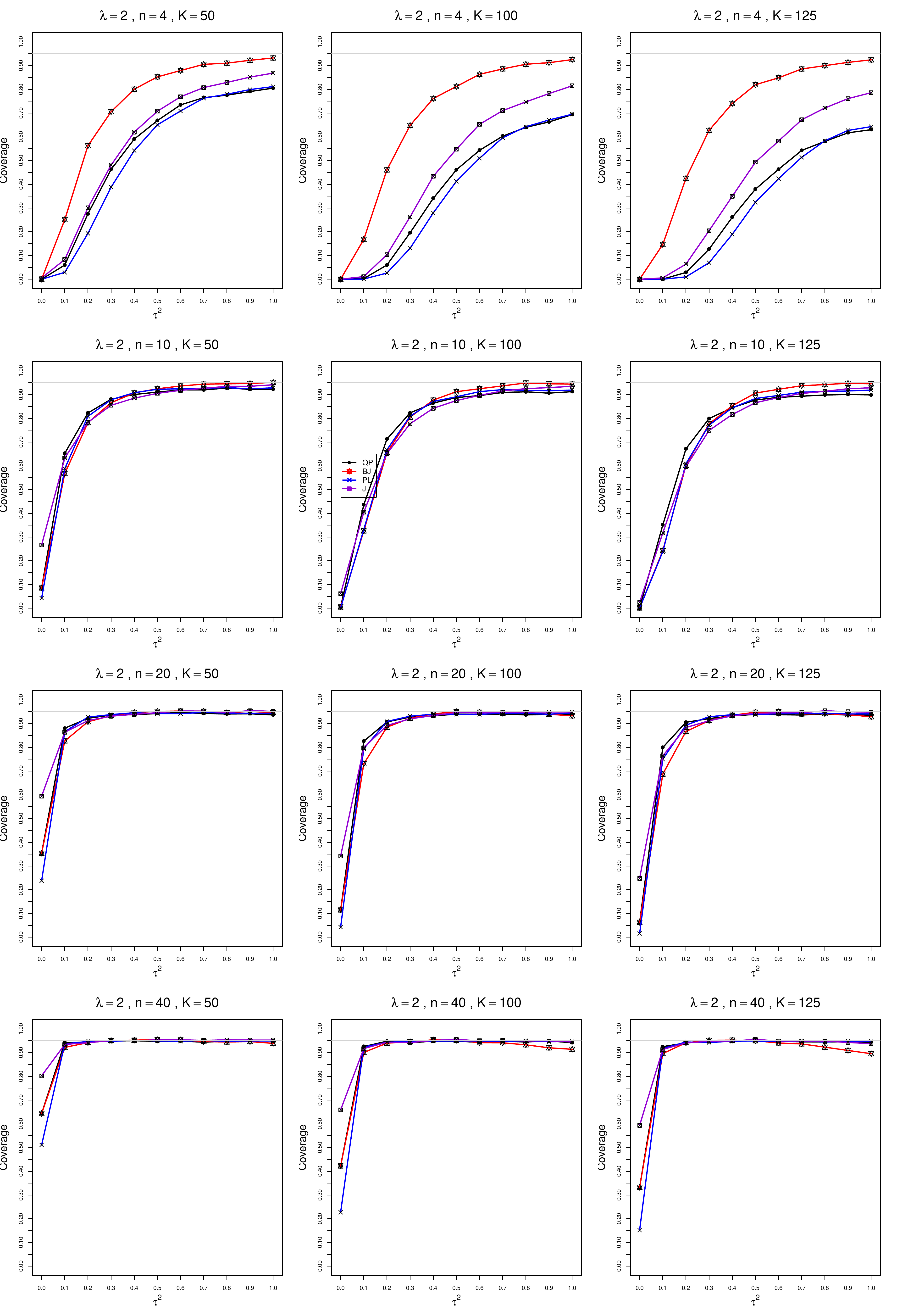}
	\caption{Coverage of 95\% confidence intervals for the between-studies variance $\tau^2$ when $\lambda=2$, $n = 4, \;10, \;20, \;40$, and $K = 50, \;100, \;125$. Usual estimate of $\lambda_i$ 		\label{CovTauRoM2ln_smallN_large_K}}
\end{figure}

\clearpage
\renewcommand{\thefigure}{A4.1.\arabic{figure}}
\setcounter{figure}{0}
\section*{A4. Lognormal model, bias-corrected estimator of $\lambda_i$, $n= 4, 10, 20, 40$, $K=50,100,125$}
\subsection*{A4.1 Bias of point estimators of $\tau^2$}
Each figure corresponds to a value of $\lambda \;(= 0, 0.2, 0.5, 1, 2)$, a set of values of $n$ (= 4, 10, 20, 40), and a set of values of $K$ (= 50, 100, 125).\\
Each panel corresponds to a value of $n$ and a value of $K$ and has $\tau^2 = 0.0(0.1)1.0$ on the horizontal axis.\\
The point estimators of $\tau^2$ are
\begin{itemize}
	\item DL (DerSimonian-Laird)
	\item REML (restricted maximum likelihood)
	\item MP (Mandel-Paule)
	\item J (Jackson)
\end{itemize}

\clearpage

\begin{figure}[t]
	\includegraphics[scale=0.33]{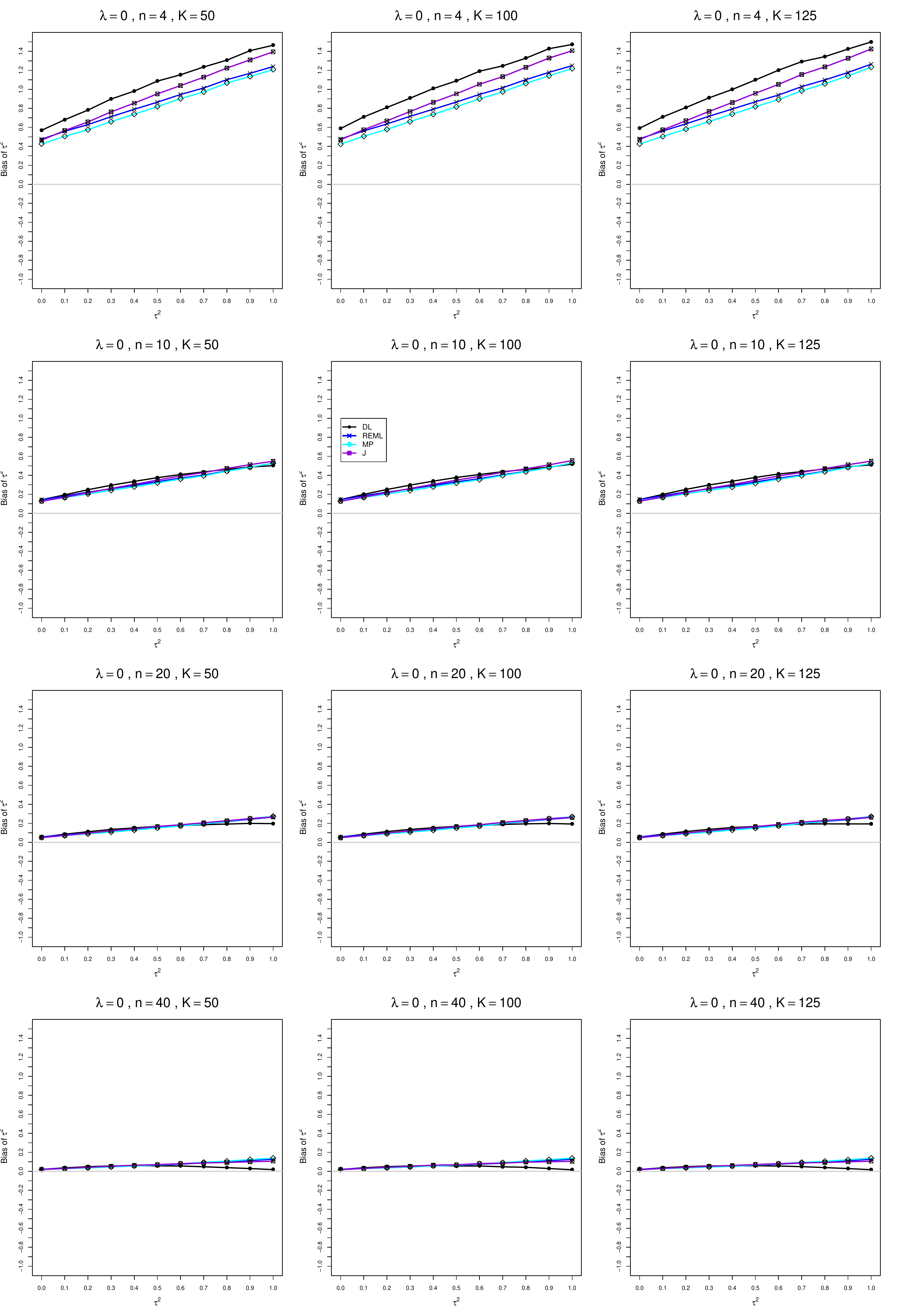}
	\caption{Bias of estimators of between-studies variance $\tau^2$ for $\lambda=0$, $n = 4, \;10, \;20, \;40$, and $K = 50, \;100, \;125$. Bias-corrected estimate of $\lambda_i$
		\label{BiasTauRoM0lnCor_smallN_large_K}}
\end{figure}

\begin{figure}[t]
	\includegraphics[scale=0.33]{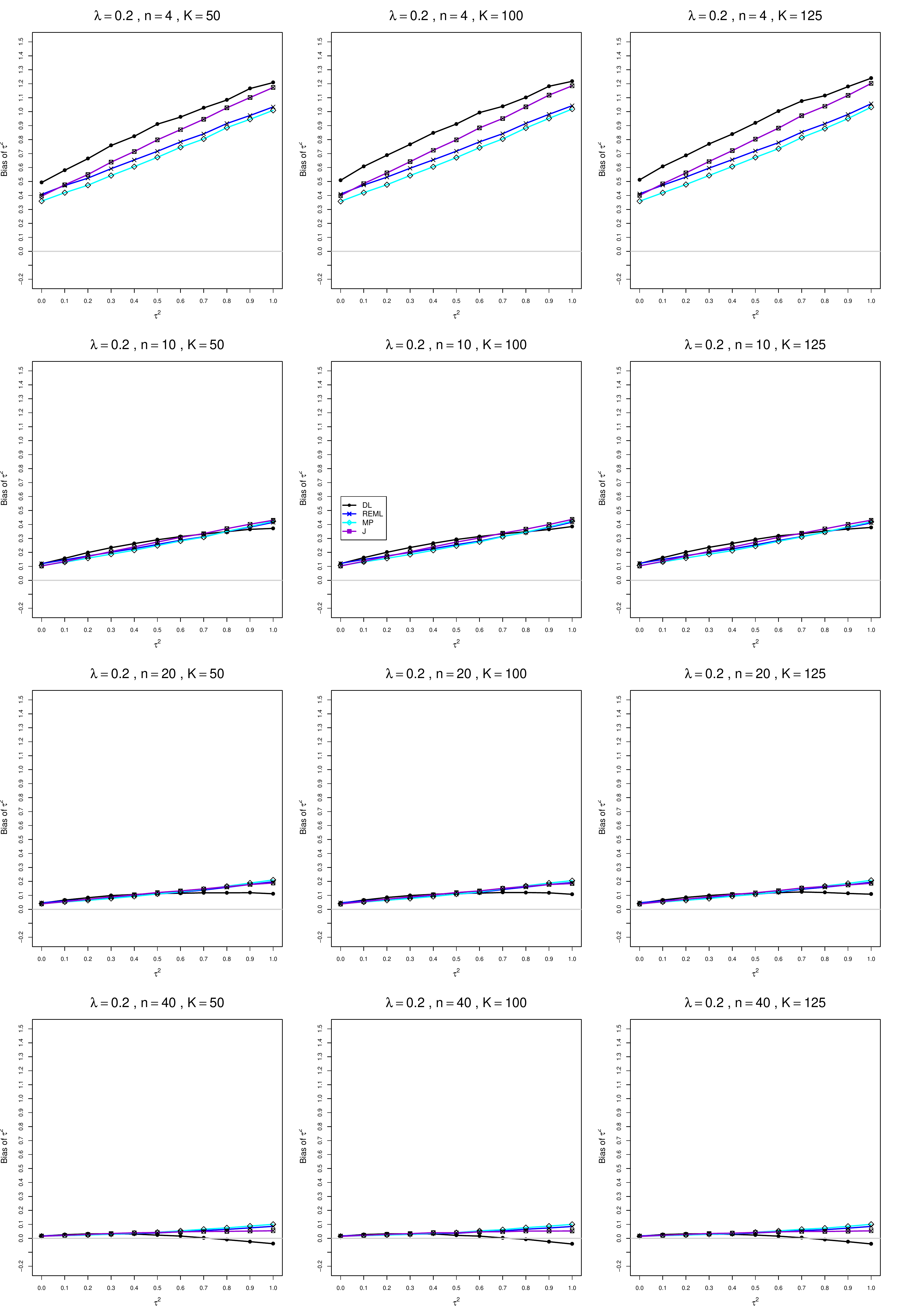}
	\caption{Bias of estimators of between-studies variance $\tau^2$ for $\lambda=0.2$, $n = 4, \;10, \;20, \;40$, and $K = 50, \;100, \;125$. Bias-corrected estimate of $\lambda_i$
		\label{BiasTauRoM02lnCor_smallN_large_K}}
\end{figure}

\begin{figure}[t]
	\includegraphics[scale=0.33]{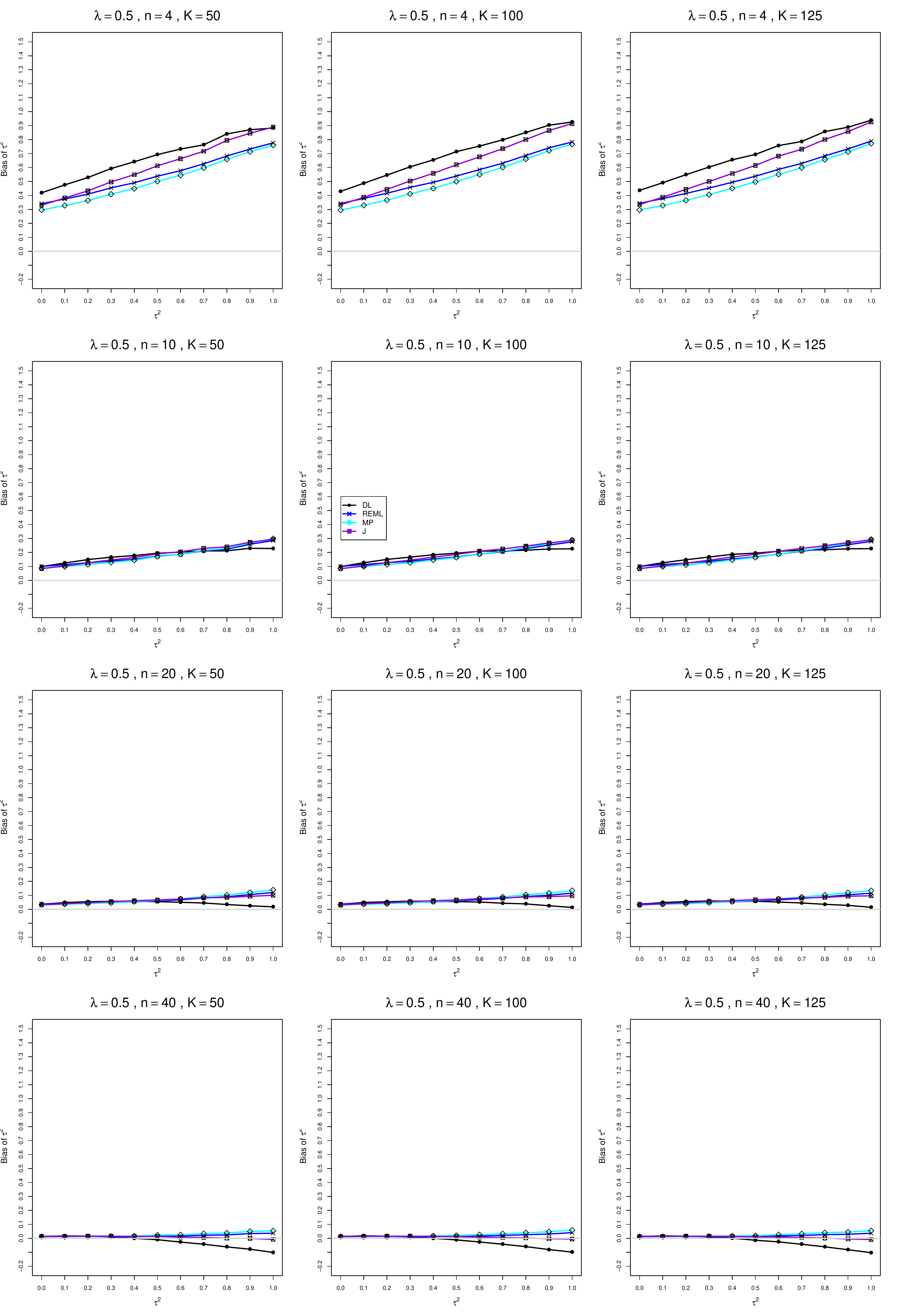}
	\caption{Bias of estimators of between-studies variance $\tau^2$ for $\lambda=0.5$, $n = 4, \;10, \;20, \;40$, and $K = 50, \;100, \;125$. Bias-corrected estimate of $\lambda_i$
		\label{BiasTauRoM05lnCor_smallN_large_K}}
\end{figure}

\begin{figure}[t]
	\includegraphics[scale=0.33]{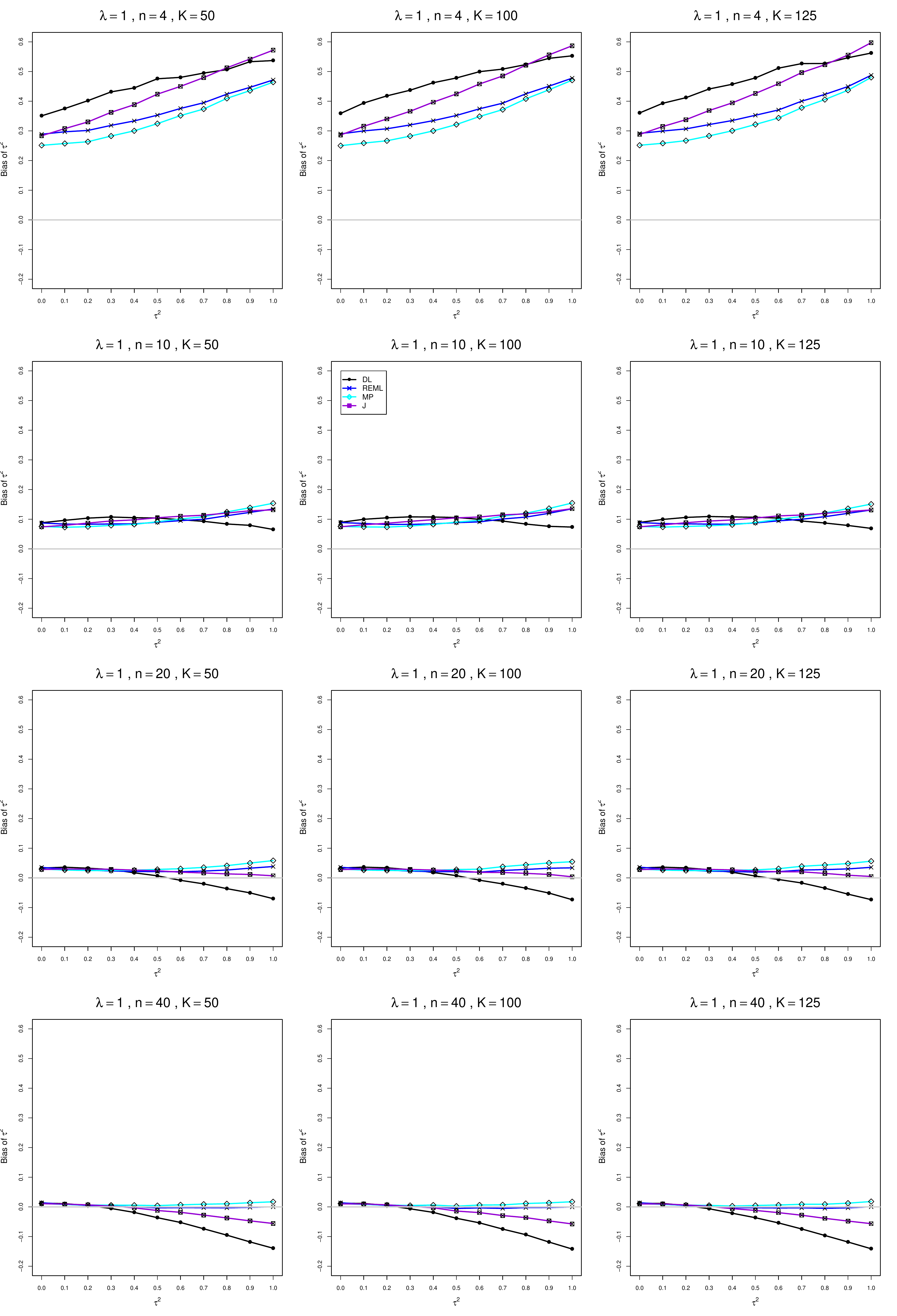}
	\caption{Bias of estimators of between-studies variance $\tau^2$ for $\lambda=1$, $n = 4, \;10, \;20, \;40$, and $K = 50, \;100, \;125$. Bias-corrected estimate of $\lambda_i$
		\label{BiasTauRoM1lnCor_smallN_large_K}}
\end{figure}

\begin{figure}[t]
	\includegraphics[scale=0.33]{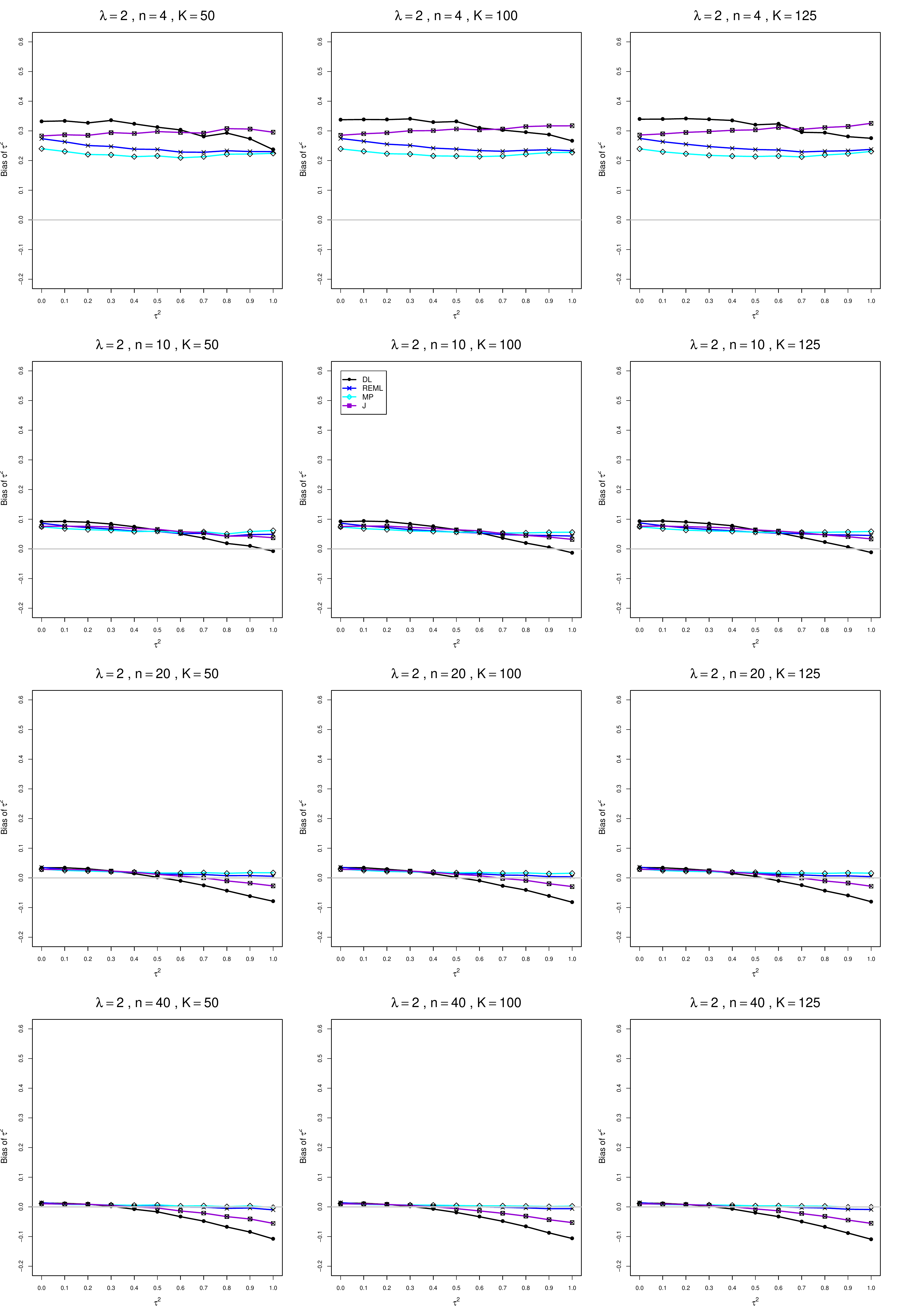}
	\caption{Bias of estimators of between-studies variance $\tau^2$ for $\lambda=2$, $n = 4, \;10, \;20, \;40$, and $K = 50, \;100, \;125$. Bias-corrected estimate of $\lambda_i$
		\label{BiasTauRoM2lnCor_smallN_large_K}}
\end{figure}

\renewcommand{\thefigure}{A4.2.\arabic{figure}}
\setcounter{figure}{0}
\clearpage
\subsection*{A4.2 Coverage of interval estimators of $\tau^2$}
Each figure corresponds to a value of $\lambda \;(= 0, 0.2, 0.5, 1, 2)$, a set of values of $n$ (= 4, 10, 20, 40), and a set of values of $K$ (= 50, 100, 125).\\
Each panel corresponds to a value of $n$ and a value of $K$ and has $\tau^2 = 0.0(0.1)1.0$ on the horizontal axis.\\
The interval estimators of $\tau^2$ are
\begin{itemize}
	\item QP (Q-profile confidence interval)
	\item BJ (Biggerstaff and Jackson interval )
	\item PL (Profile-likelihood interval)
	\item J (Jackson interval)
\end{itemize}

\clearpage

\begin{figure}[t]
	\includegraphics[scale=0.35]{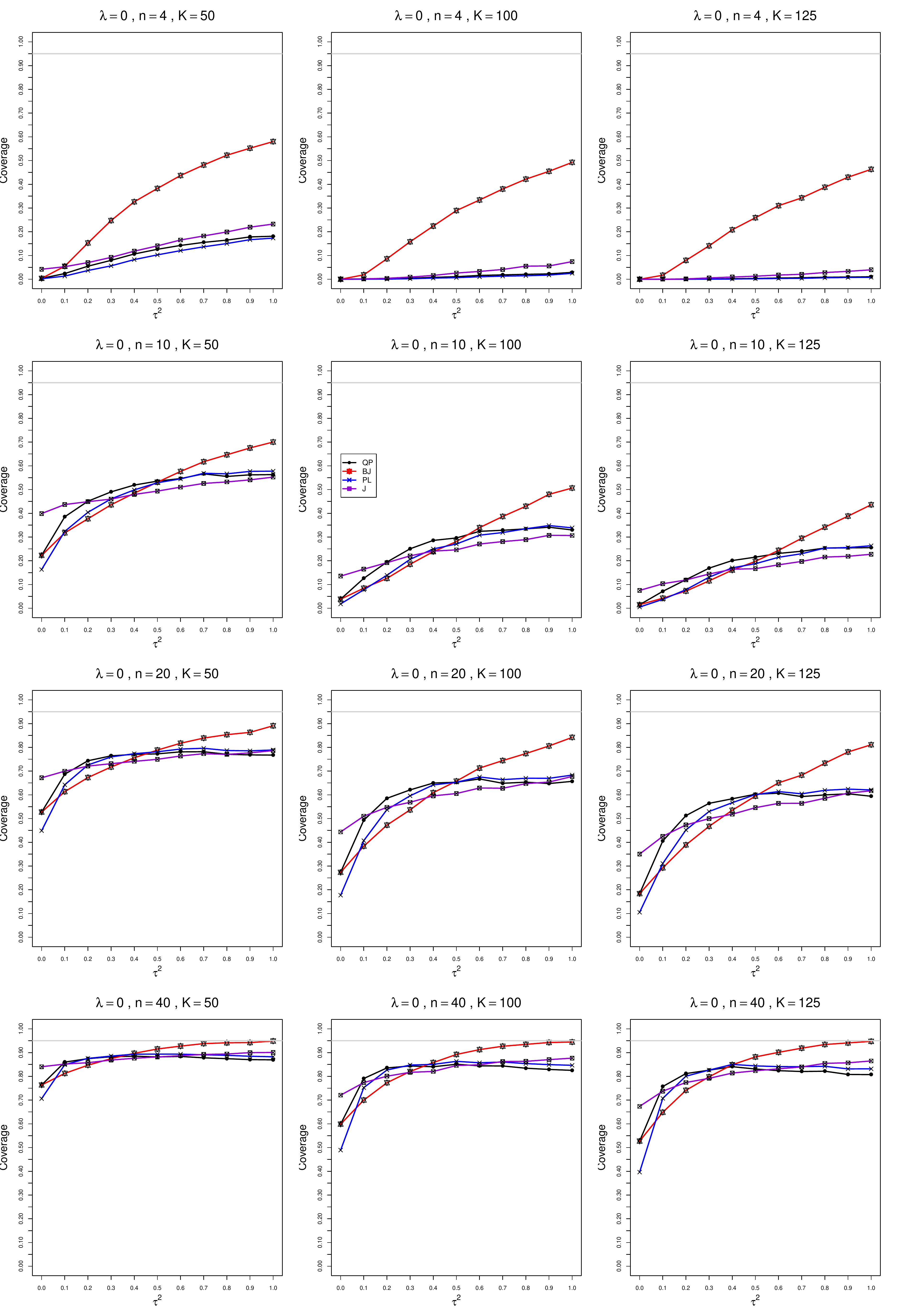}
	\caption{Coverage of 95\% confidence intervals for the between-studies variance $\tau^2$ when $\lambda=0$, $n = 4, \;10, \;20, \;40$, and $K = 50, \;100, \;125$. Bias-corrected estimate of $\lambda_i$
		\label{CovTauRoM0lnCor_smallN_large_K}}
\end{figure}
\begin{figure}[t]
	\includegraphics[scale=0.35]{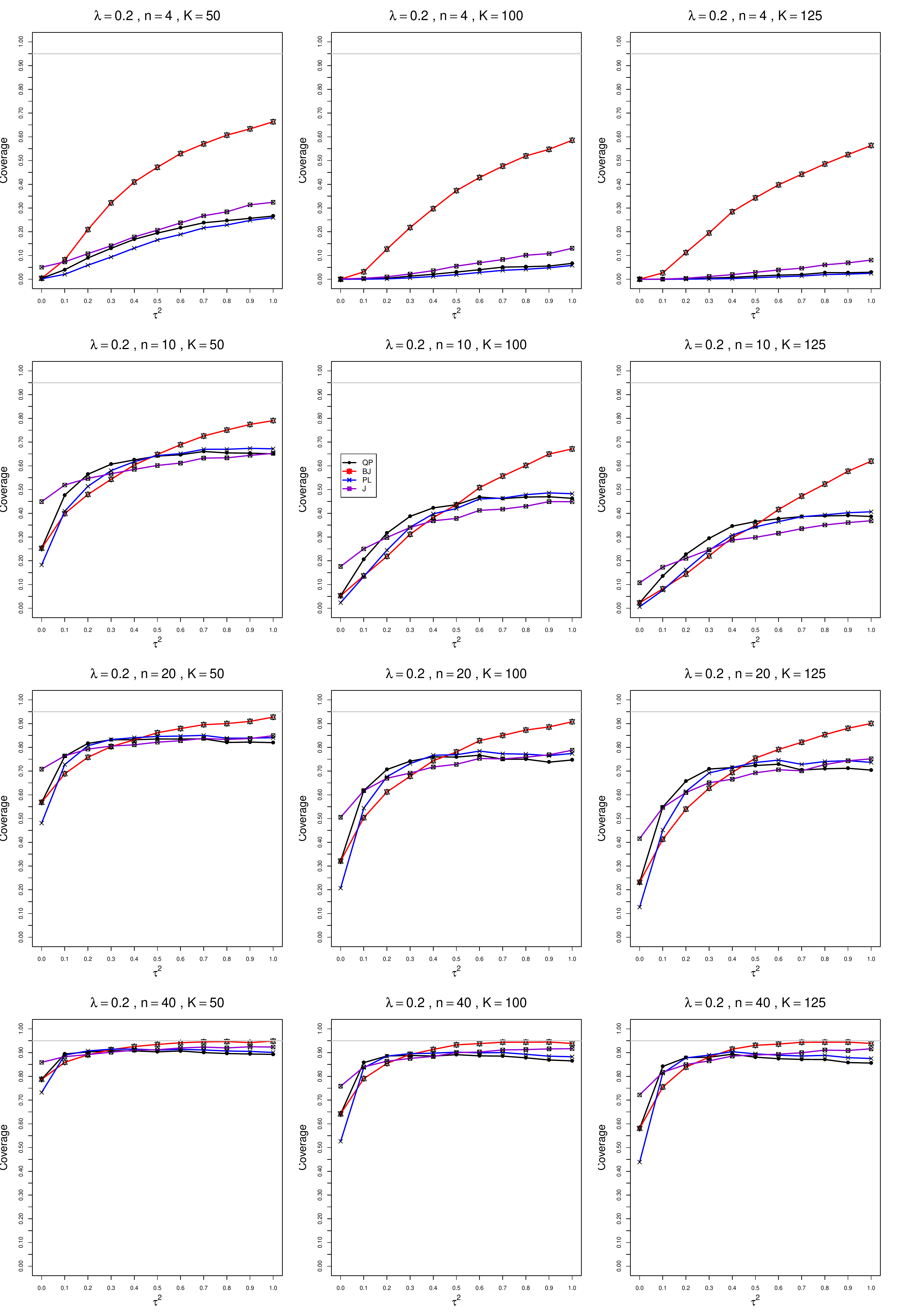}
	\caption{Coverage of 95\% confidence intervals for the between-studies variance $\tau^2$ when $\lambda=0.2$, $n = 4, \;10, \;20, \;40$, and $K = 50, \;100, \;125$. Bias-corrected estimate of $\lambda_i$ 		\label{CovTauRoM02lnCor_smallN_large_K}}
\end{figure}

\begin{figure}[t]
	\includegraphics[scale=0.35]{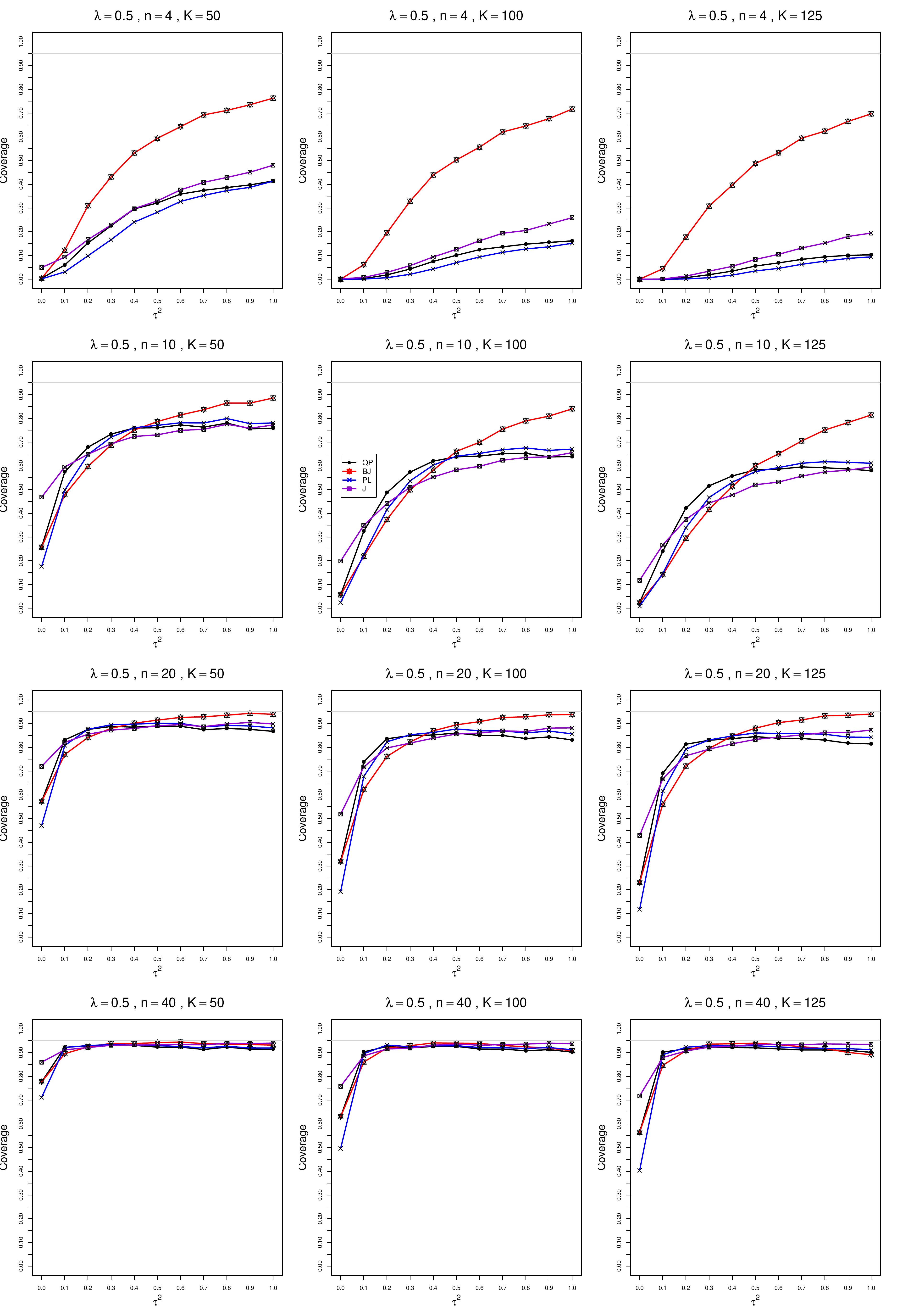}
	\caption{Coverage of 95\% confidence intervals for the between-studies variance $\tau^2$ when $\lambda=0.5$, $n = 4, \;10, \;20, \;40$, and $K = 50, \;100, \;125$. Bias-corrected estimate of $\lambda_i$ 		\label{CovTauRoM05lnCor_smallN_large_K}}
\end{figure}

\begin{figure}[t]
	\includegraphics[scale=0.35]{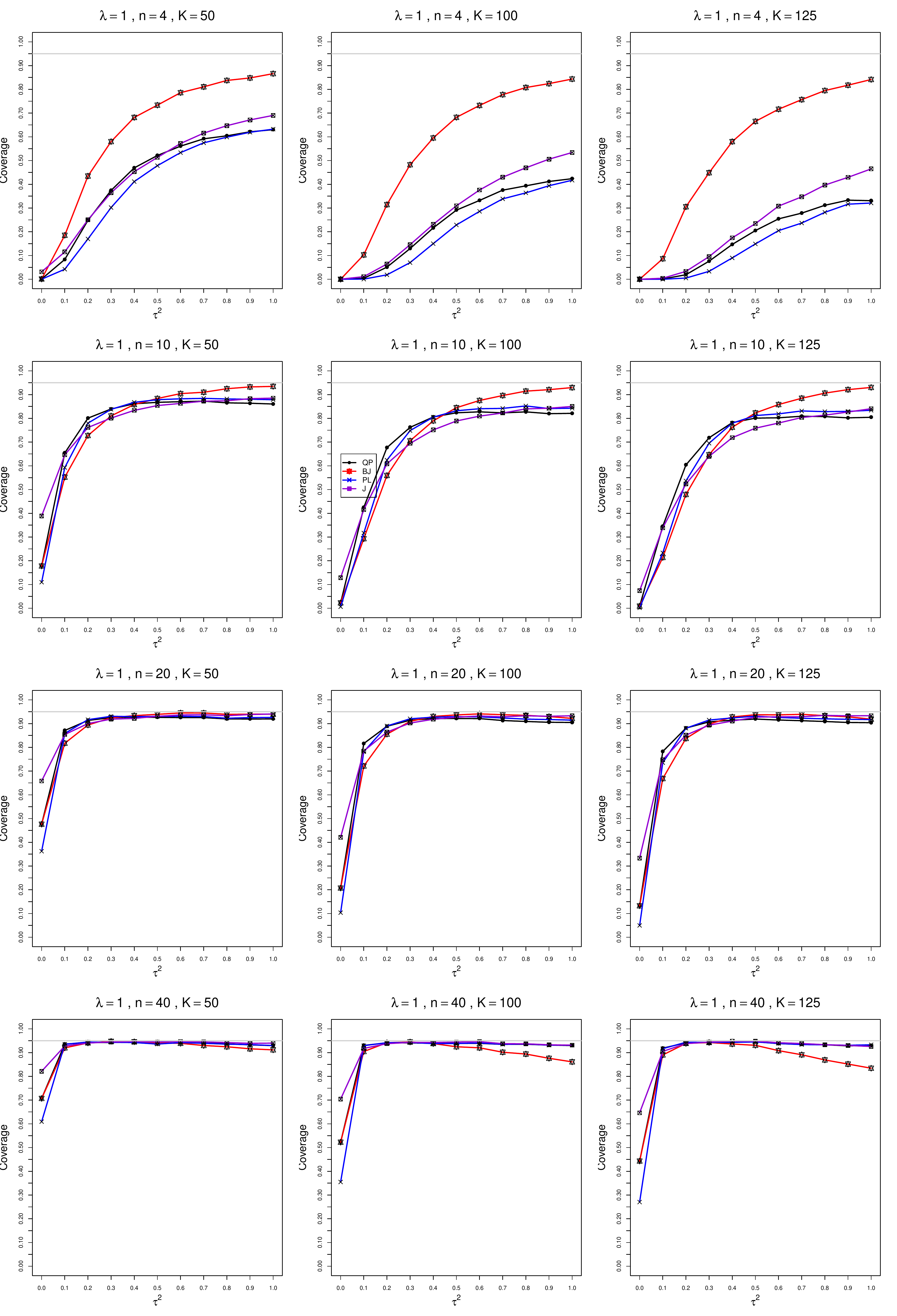}
	\caption{Coverage of 95\% confidence intervals for the between-studies variance $\tau^2$ when $\lambda=1$, $n = 4, \;10, \;20, \;40$, and $K = 50, \;100, \;125$. Bias-corrected estimate of $\lambda_i$ 		\label{CovTauRoM1lnCor_smallN_large_K}}
\end{figure}
\begin{figure}[t]
	\includegraphics[scale=0.35]{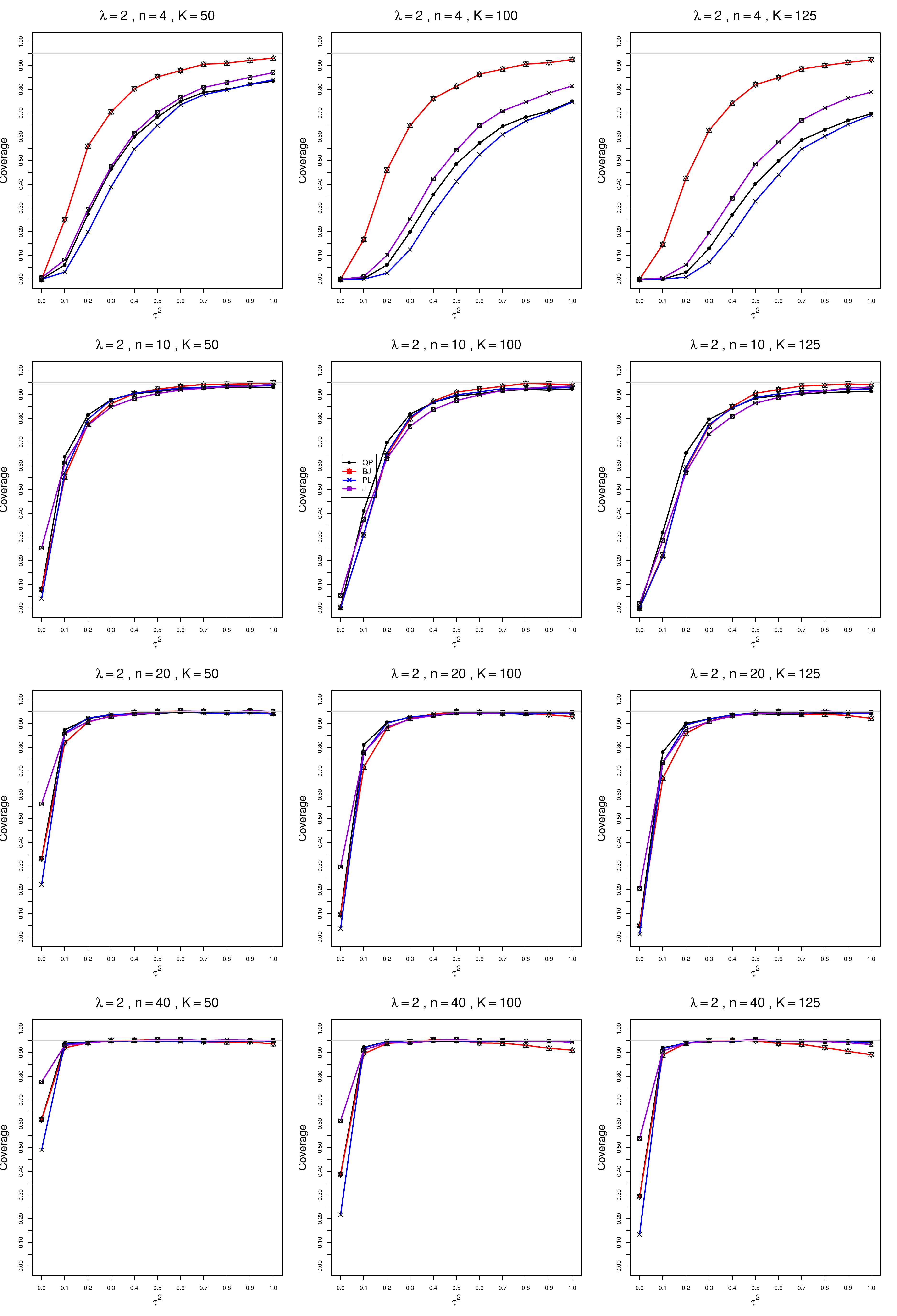}
	\caption{Coverage of 95\% confidence intervals for the between-studies variance $\tau^2$ when $\lambda=2$, $n = 4, \;10, \;20, \;40$, and $K = 50, \;100, \;125$. Bias-corrected estimate of $\lambda_i$
		\label{CovTauRoM2lnCor_smallN_large_K}}
\end{figure}

\clearpage
\renewcommand{\thesection}{B1.\arabic{section}}
\setcounter{section}{0}
\section*{B: Plots of bias and coverage of estimators of $\lambda$, small $n$}
\begin{itemize}
	\item B1. Lognormal model, usual estimator of $\lambda_i$, $K=5,10,30$
	\item B2. Lognormal model, bias-corrected estimator of $\lambda_i$, $K=5,10,30$
	\item B3. Lognormal model, usual estimator of $\lambda_i$, $K=50,100,125$
	\item B4. Lognormal model, bias-corrected estimator of $\lambda_i$, $K=50,100,125$
\end{itemize}

\clearpage
\renewcommand{\thefigure}{B1.1.\arabic{figure}}
\setcounter{figure}{0}
\clearpage
\section*{B1. Lognormal model, usual estimator of $\lambda_i$, $n= 4, 10, 20, 40$, $K=5,10,30$}
\subsection*{B1.1 Bias of point estimators of $\lambda$}
Each figure corresponds to a value of $\lambda \;(= 0, 0.2, 0.5, 1, 2)$, a set of values of $n$ (= 4, 10, 20, 40), and a set of values of $K$ (= 5, 10, 30).\\
Each panel corresponds to a value of $n$ and a value of $K$ and has $\tau^2 = 0.0(0.1)1.0$ on the horizontal axis.\\
The point estimators of $\lambda$ are
\begin{itemize}
	\item DL (DerSimonian-Laird)
	\item REML (restricted maximum likelihood)
	\item MP (Mandel-Paule)
	\item J (Jackson)
	\item SSW (sample-size-weighted)
\end{itemize}

\begin{figure}[t]
	\includegraphics[scale=0.33]{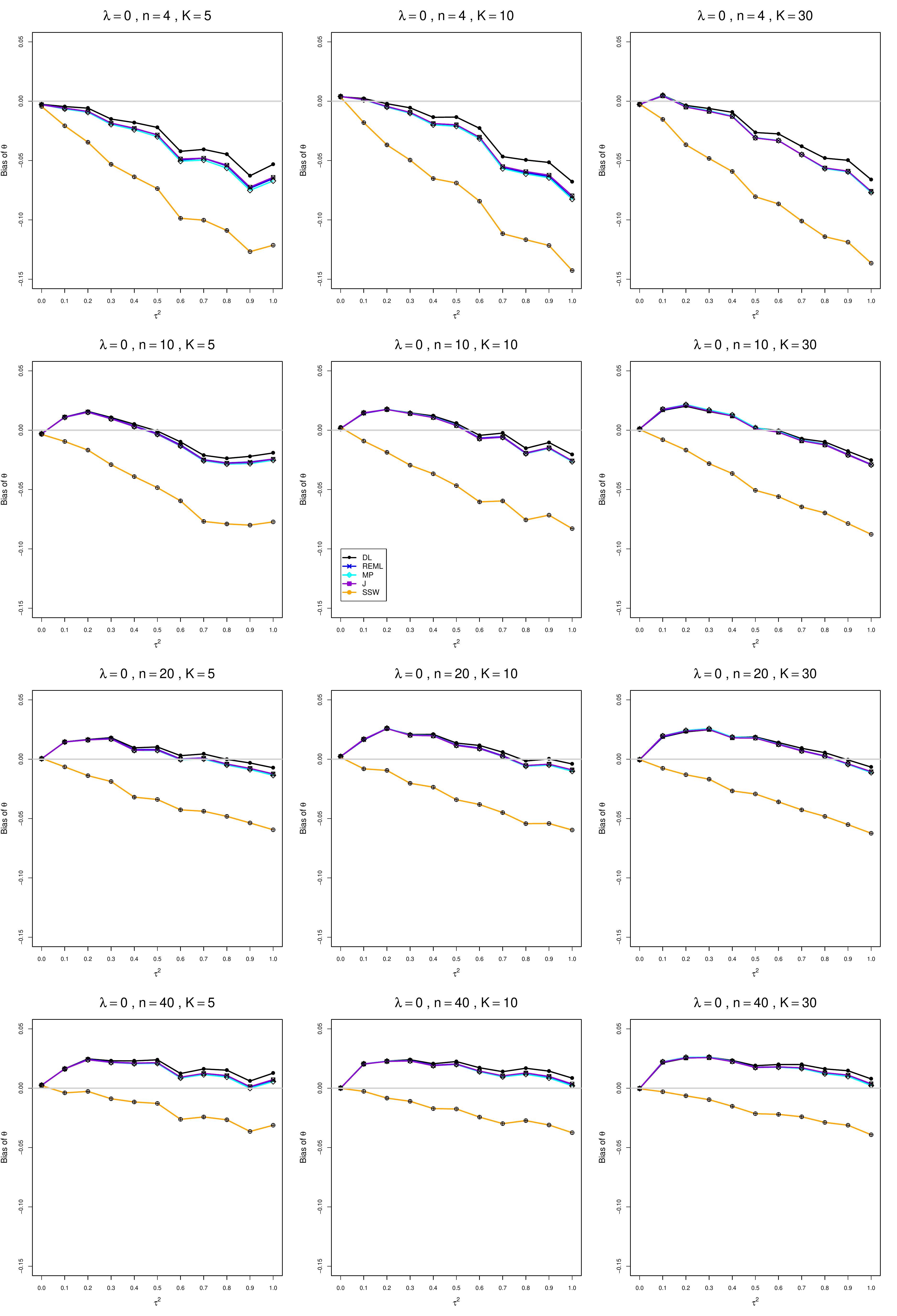}
	\caption{Bias of estimators of $\lambda$ for $\lambda=0$, $n = 4, \;10, \;20, \;40$, and $K = 5, \;10, \;30$. Usual estimate of $\lambda_i$
		\label{BiasThetaRoM0ln_smallN_small_K}}
\end{figure}

\begin{figure}[t]
	\includegraphics[scale=0.33]{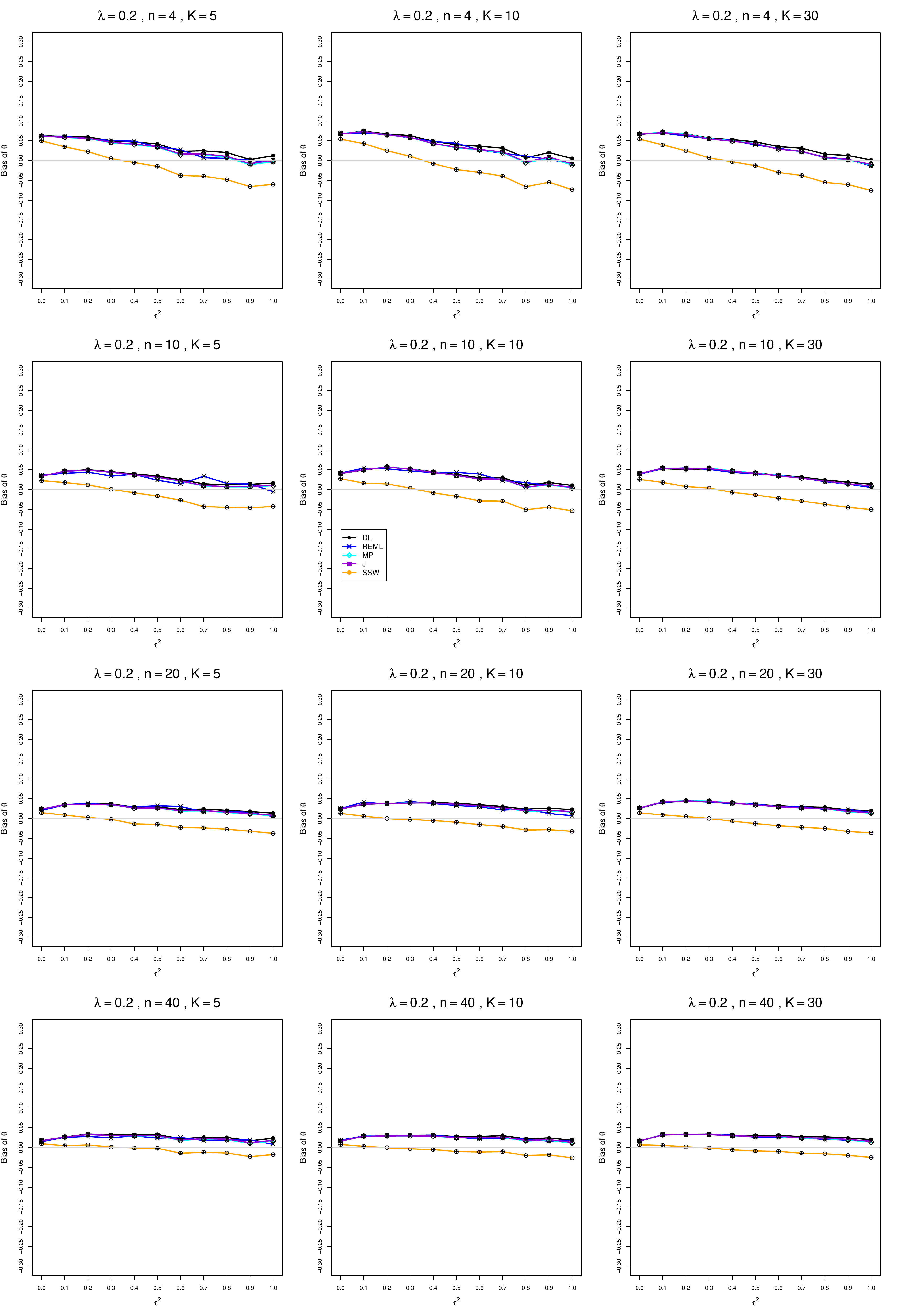}
	\caption{Bias of estimators of $\lambda$ for $\lambda=0.2$, $n = 4, \;10, \;20, \;40$, and $K = 5, \;10, \;30$. Usual estimate of $\lambda_i$ 	
		\label{BiasThetaRoM02ln_smallN_small_K}}
\end{figure}

\begin{figure}[t]
	\includegraphics[scale=0.33]{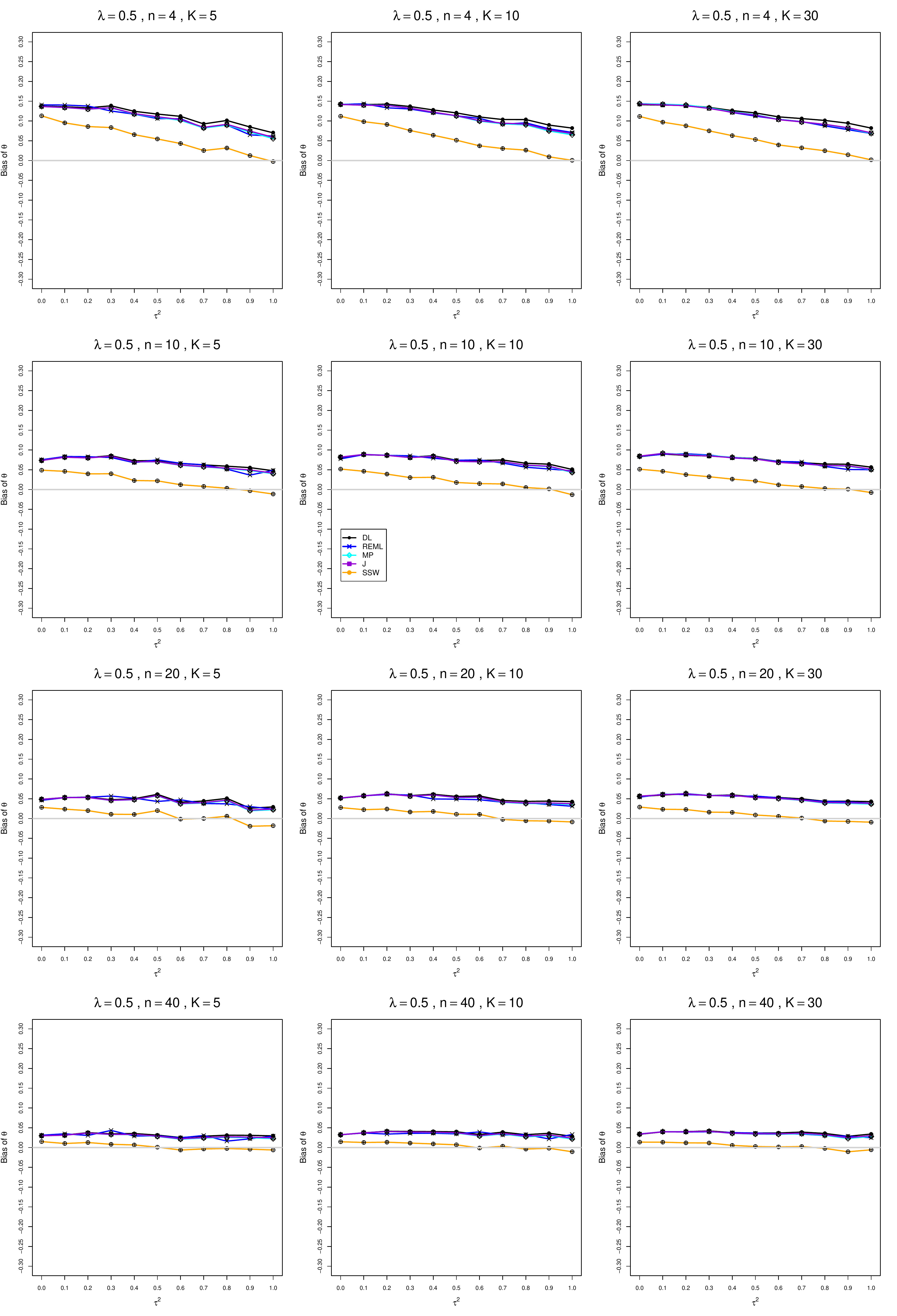}
	\caption{Bias of estimators of $\lambda$ for $\lambda=0.5$, $n = 4, \;10, \;20, \;40$, and $K = 5, \;10, \;30$. Usual estimate of $\lambda_i$
		\label{BiasThetaRoM05ln_smallN_small_K}}
\end{figure}

\begin{figure}[t]
	\includegraphics[scale=0.33]{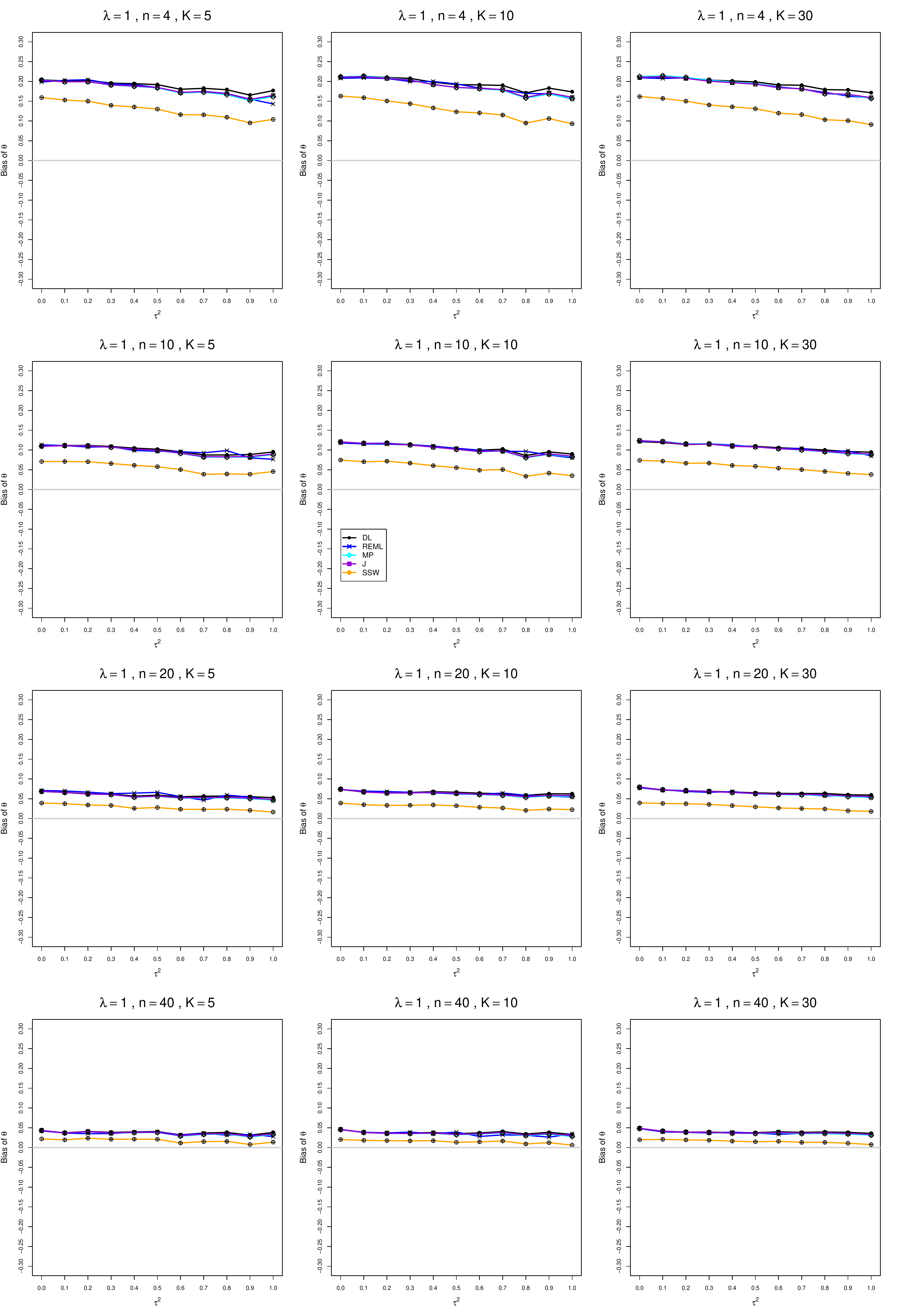}
	\caption{Bias of estimators of $\lambda$ for $\lambda=1$, $n = 4, \;10, \;20, \;40$, and $K = 5, \;10, \;30$. Usual estimate of $\lambda_i$ 		
		\label{BiasThetaRoM1ln_smallN_small_K}}
\end{figure}

\begin{figure}[t]
	\includegraphics[scale=0.33]{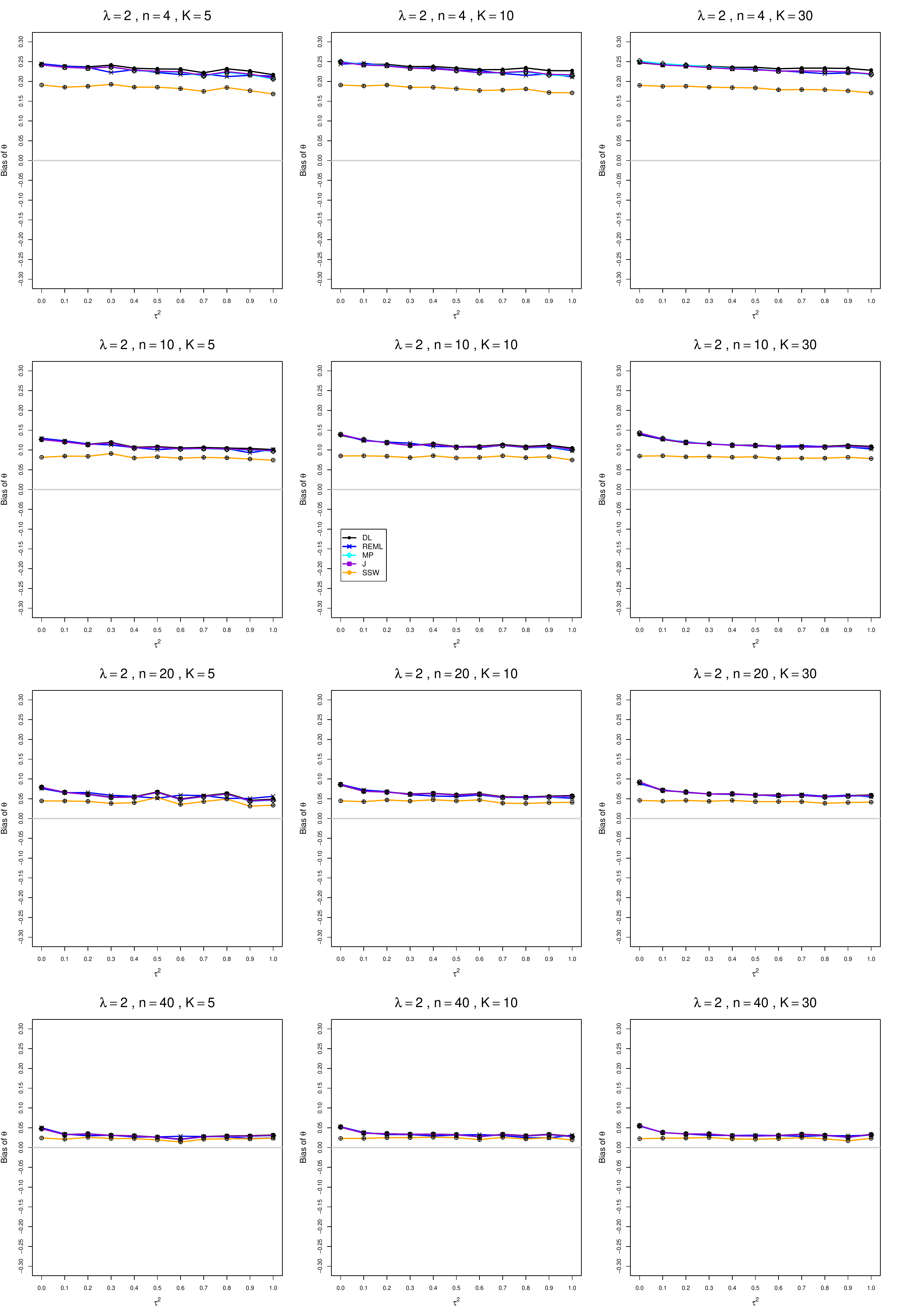}
	\caption{Bias of estimators of $\lambda$ for $\lambda=2$, $n = 4, \;10, \;20, \;40$, and $K = 5, \;10, \;30$. Usual estimate of $\lambda_i$ 		
		\label{BiasThetaRoM2ln_smallN_small_K}}
\end{figure}

\clearpage
\renewcommand{\thefigure}{B1.2.\arabic{figure}}
\setcounter{figure}{0}
\subsection*{B1.2 Coverage of interval estimators of $\lambda$}
Each figure corresponds to a value of $\lambda \;(= 0, 0.2, 0.5, 1, 2)$, a set of values of $n$ (= 4, 10, 20, 40), and a set of values of $K$ (= 5, 10, 30).\\
Each panel corresponds to a value of $n$ and a value of $K$ and has $\tau^2 = 0.0(0.1)1.0$ on the horizontal axis.\\
The interval estimators of $\lambda$ are the companions to the inverse-variance-weighted point estimators
\begin{itemize}
	\item DL (DerSimonian-Laird)
	\item REML (restricted maximum likelihood)
	\item MP (Mandel-Paule)
	\item J (Jackson)
\end{itemize}
and
\begin{itemize}
	\item HKSJ (Hartung-Knapp-Sidik-Jonkman)
	\item HKSJ MP (HKSJ with MP estimator of $\tau^2$)
	\item SSW MP (SSW as center and half-width equal to critical value from $t_{K-1}$ times estimated standard deviation of SSW with $\hat{\tau}^2$ = $\hat{\tau}^2_{MP}$)
\end{itemize}

\begin{figure}[t]
	\includegraphics[scale=0.35]{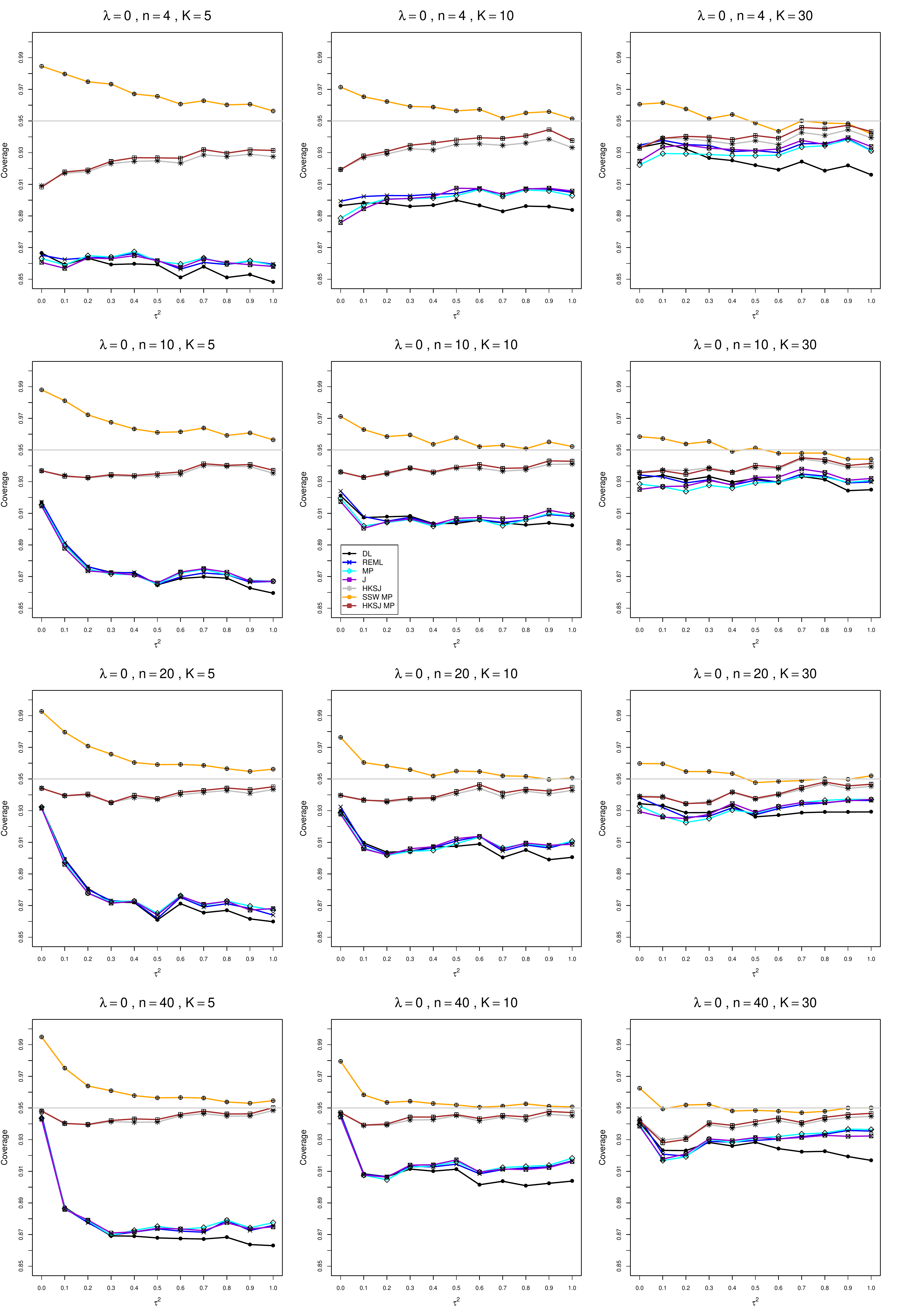}
	\caption{Coverage of 95\% confidence intervals for $\lambda$ when $\lambda=0$, $n = 4, \;10, \;20, \;40$, and $K = 5, \;10, \;30$. Usual estimate of $\lambda_i$
		\label{CovThetaRoM0ln_smallN_small_K}}
\end{figure}

\begin{figure}[t]
	\includegraphics[scale=0.35]{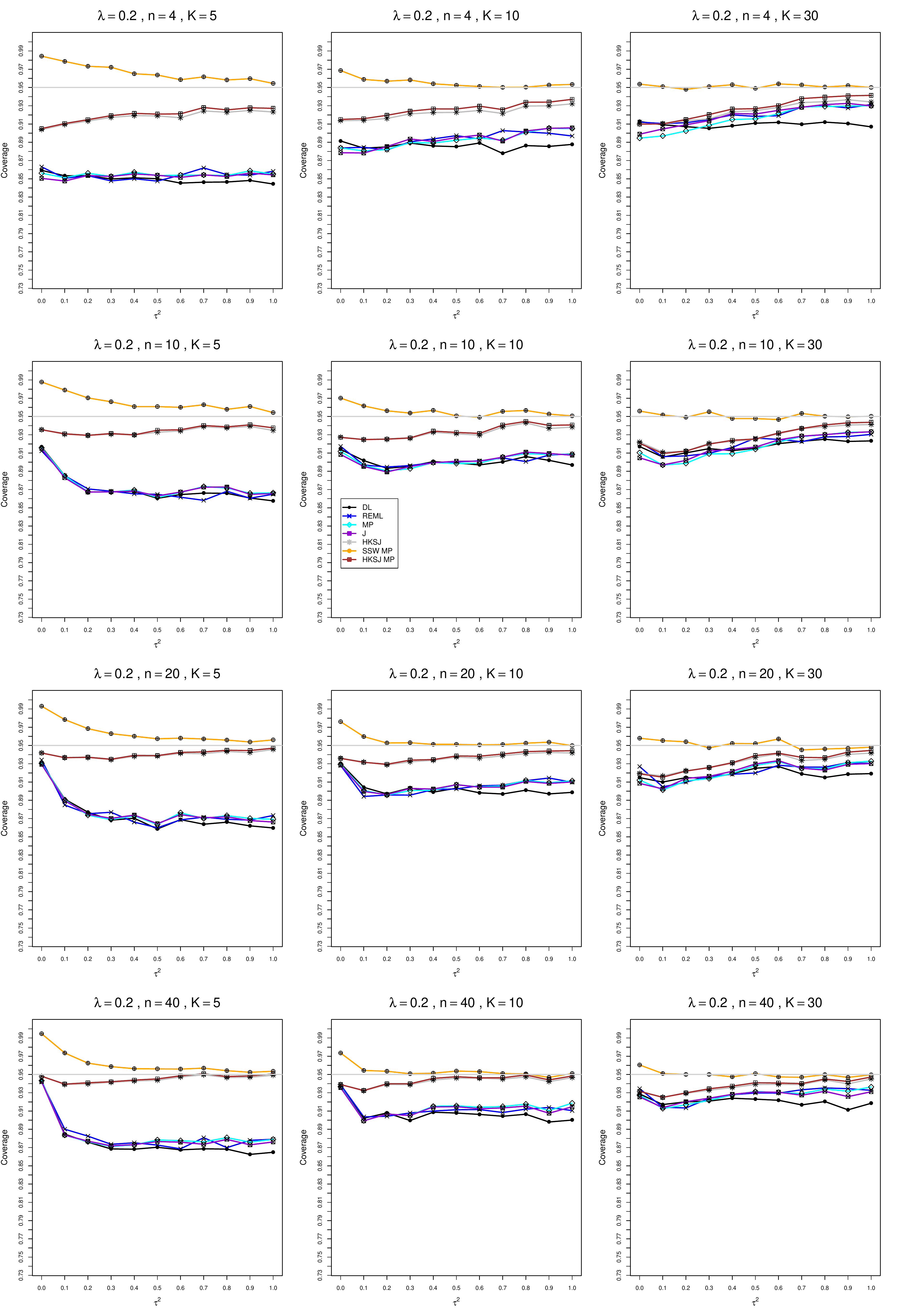}
	\caption{Coverage of 95\% confidence intervals for $\lambda$ when $\lambda=0.2$, $n = 4, \;10, \;20, \;40$, and $K = 5, \;10, \;30$. Usual estimate of $\lambda_i$
		\label{CovThetaRoM02ln_smallN_small_K}}
\end{figure}

\begin{figure}[t]
	\includegraphics[scale=0.35]{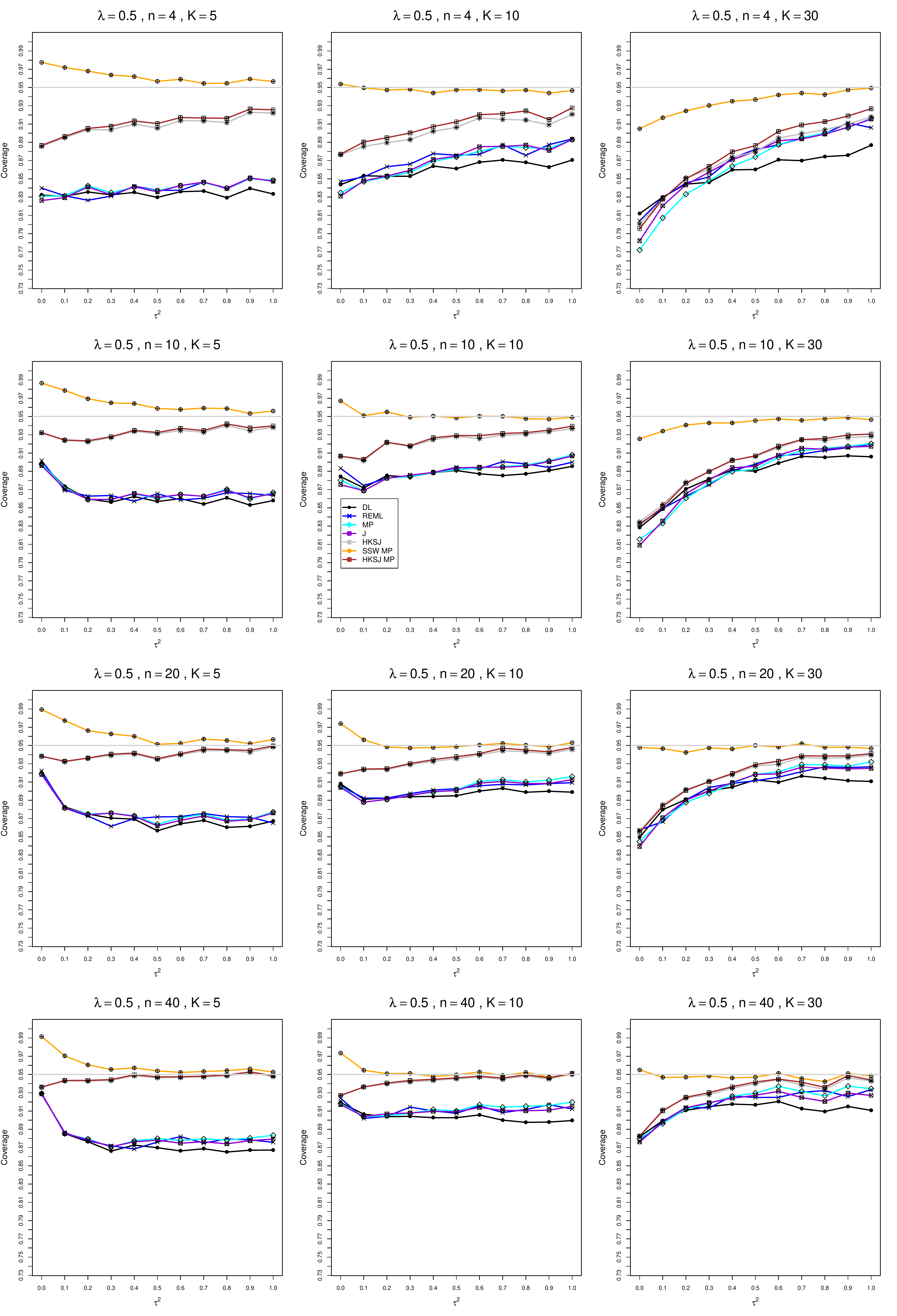}
	\caption{Coverage of 95\% confidence intervals for $\lambda$ when $\lambda=0,5$, $n = 4, \;10, \;20, \;40$, and $K = 5, \;10, \;30$. Usual estimate of $\lambda_i$ 		
		\label{CovThetaRoM05ln_smallN_small_K}}
\end{figure}

\begin{figure}[t]
	\includegraphics[scale=0.35]{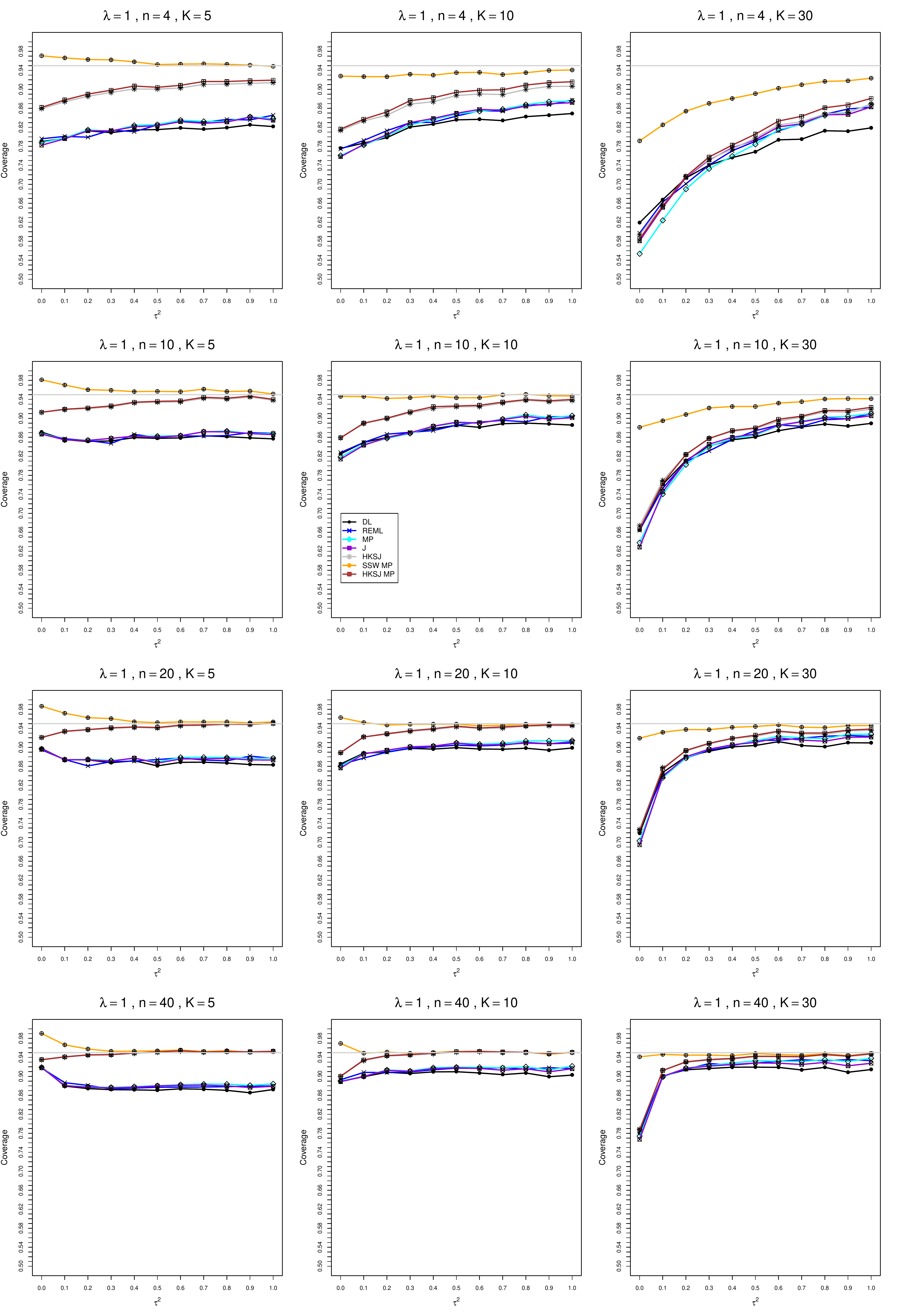}
	\caption{Coverage of 95\% confidence intervals for $\lambda$ when $\lambda=1$, $n = 4, \;10, \;20, \;40$, and $K = 5, \;10, \;30$. Usual estimate of $\lambda_i$ 	
		\label{CovThetaRoM1ln_smallN_small_K}}
\end{figure}
\begin{figure}[t]
	\includegraphics[scale=0.35]{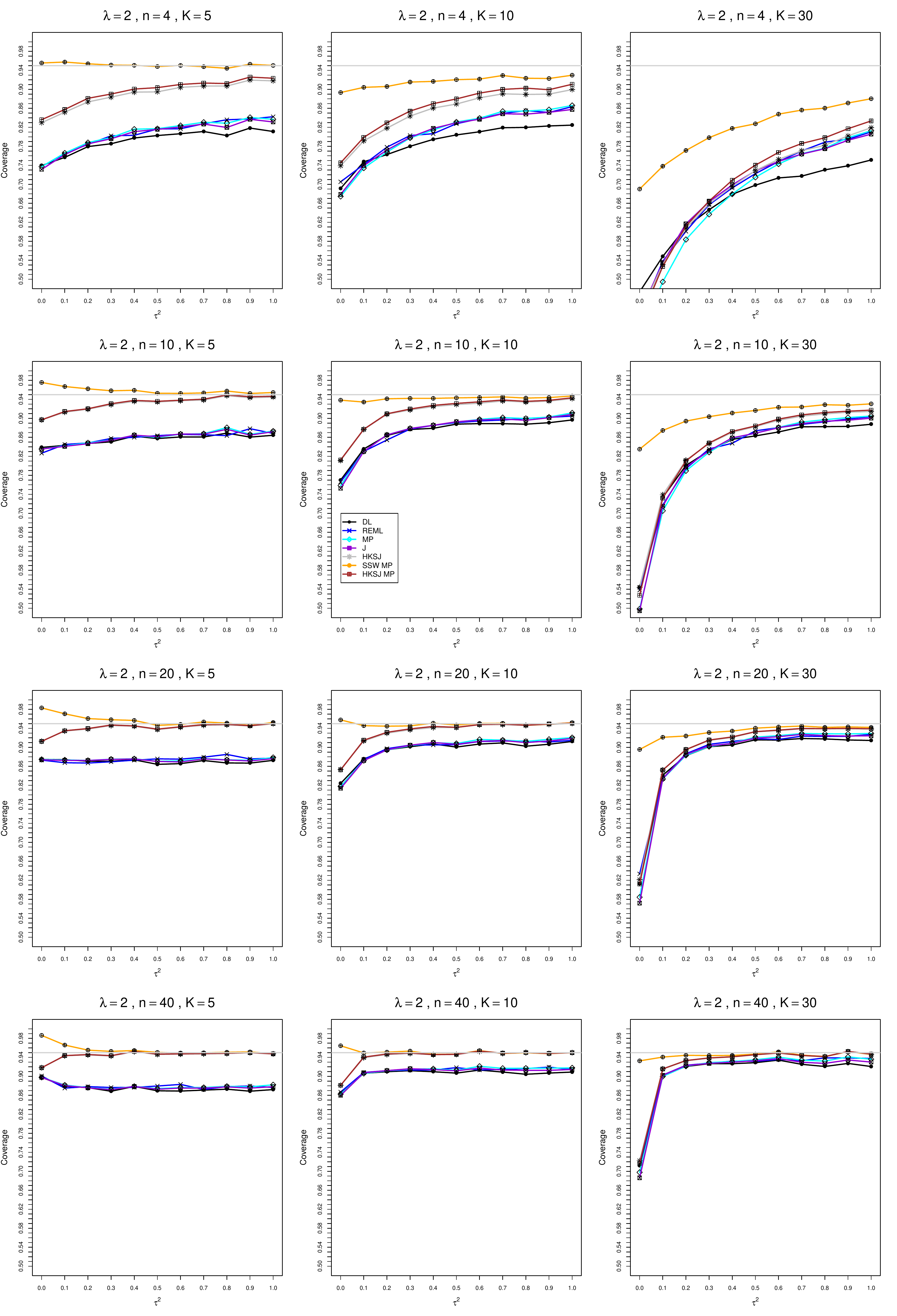}
	\caption{Coverage of 95\% confidence intervals for $\lambda$ when $\lambda=2$, $n = 4, \;10, \;20, \;40$, and $K = 5, \;10, \;30$. Usual estimate of $\lambda_i$ 		
		\label{CovThetaRoM2ln_smallN_small_K}}
\end{figure}

\clearpage
\setcounter{figure}{0}
\renewcommand{\thefigure}{B2.1.\arabic{figure}}

\section*{B2. Lognormal model, bias-corrected estimator of $\lambda_i$, $n= 4, 10, 20, 40$, $K=5,10,30$}

\subsection*{B2.1 Bias of point estimators of $\lambda$}
Each figure corresponds to a value of $\lambda \;(= 0, 0.2, 0.5, 1, 2)$, a set of values of $n$ (= 4, 10, 20, 40), and a set of values of $K$ (= 5, 10, 30).\\
Each panel corresponds to a value of $n$ and a value of $K$ and has $\tau^2 = 0.0(0.1)1.0$ on the horizontal axis.\\
The point estimators of $\lambda$ are
\begin{itemize}
	\item DL (DerSimonian-Laird)
	\item REML (restricted maximum likelihood)
	\item MP (Mandel-Paule)
	\item J (Jackson)
	\item SSW (sample-size-weighted)
\end{itemize}

\begin{figure}[t]
	\includegraphics[scale=0.33]{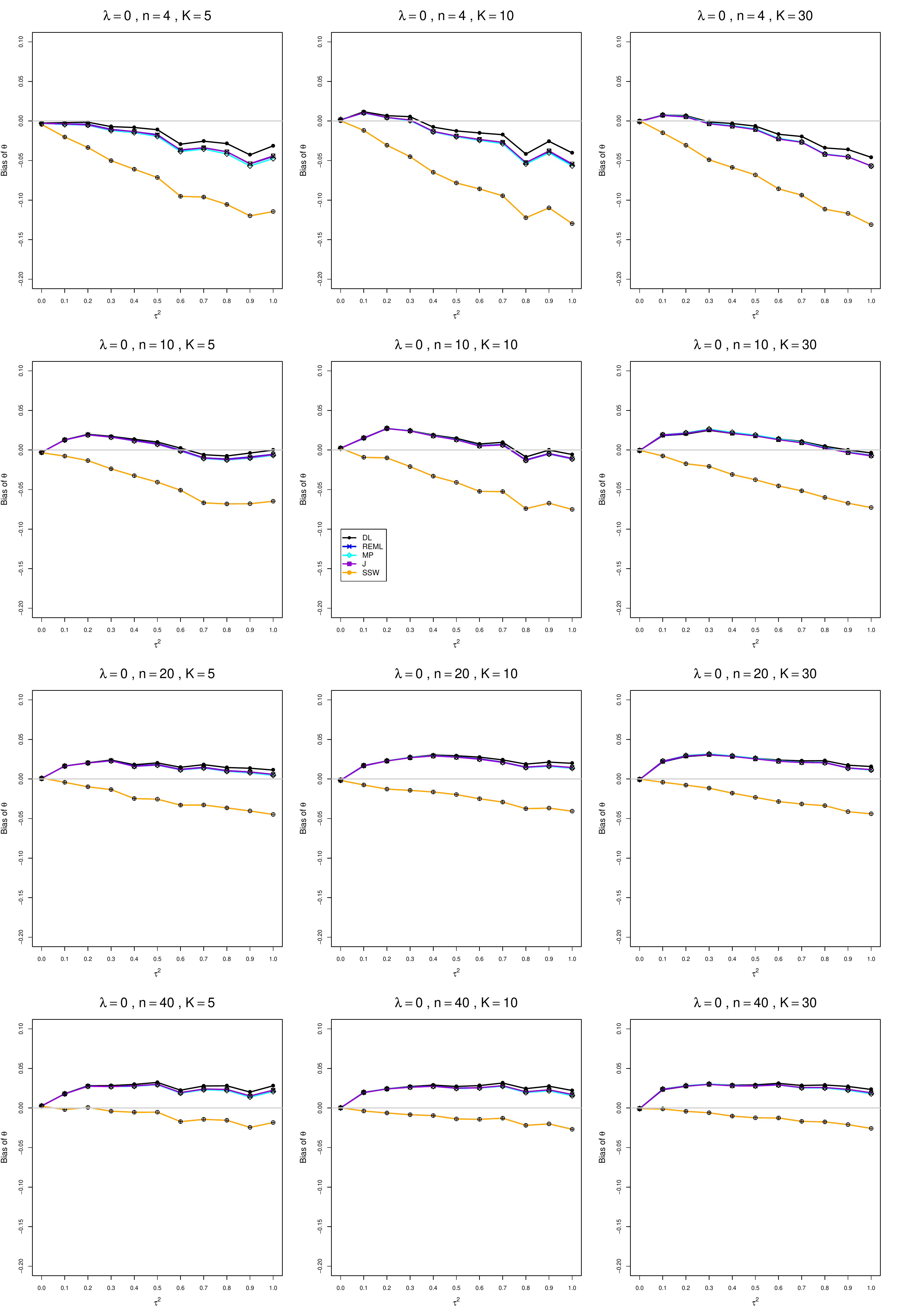}
	\caption{Bias of estimators of $\lambda$ for $\lambda=0$, $n = 4, \;10, \;20, \;40$, and $K = 5, \;10, \;30$. Bias-corrected estimate of $\lambda_i$
		\label{BiasThetaRoM0lnCor_smallN_small_K}}
\end{figure}

\begin{figure}[t]
	\includegraphics[scale=0.33]{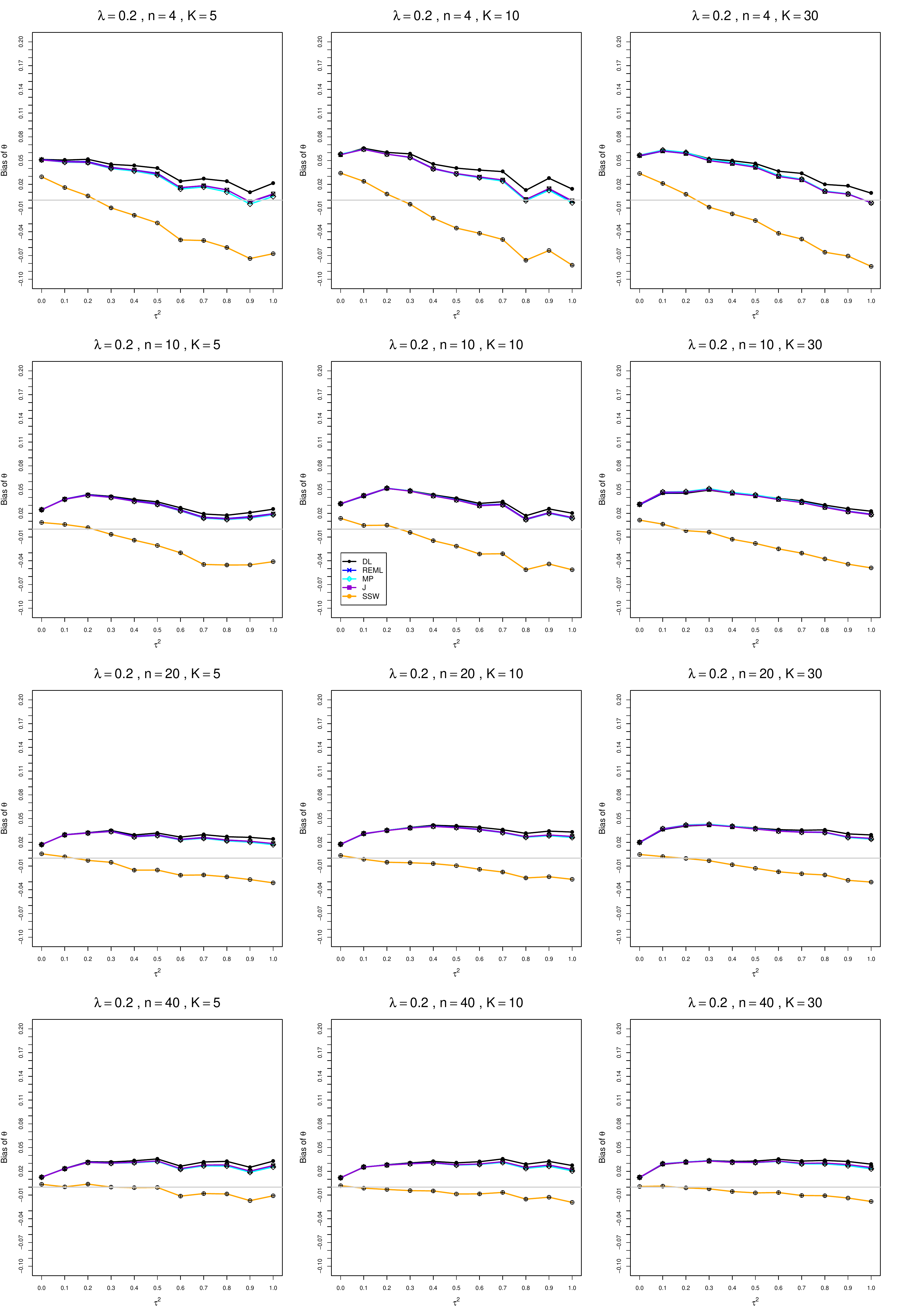}
	\caption{Bias of estimators of $\lambda$ for $\lambda=0.2$, $n = 4, \;10, \;20, \;40$, and $K = 5, \;10, \;30$. Bias-corrected estimate of $\lambda_i$
		\label{BiasThetaRoM02lnCor_smallN_small_K}}
\end{figure}

\begin{figure}[t]
	\includegraphics[scale=0.33]{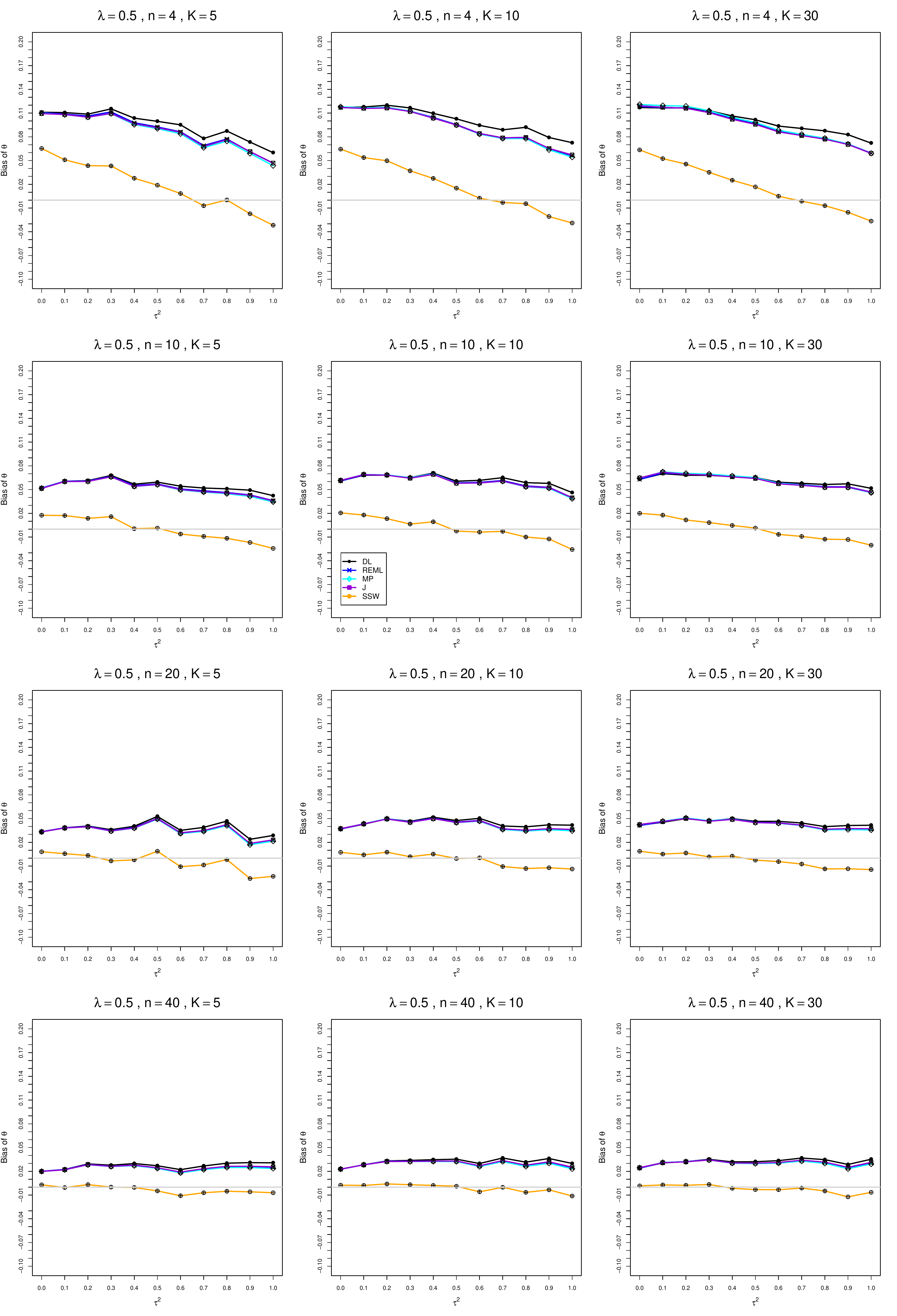}
	\caption{Bias of estimators of $\lambda$ for $\lambda=0.5$, $n = 4, \;10, \;20, \;40$, and $K = 5, \;10, \;30$. Bias-corrected estimate of $\lambda_i$
		\label{BiasThetaRoM05lnCor_smallN_small_K}}
\end{figure}

\begin{figure}[t]
	\includegraphics[scale=0.33]{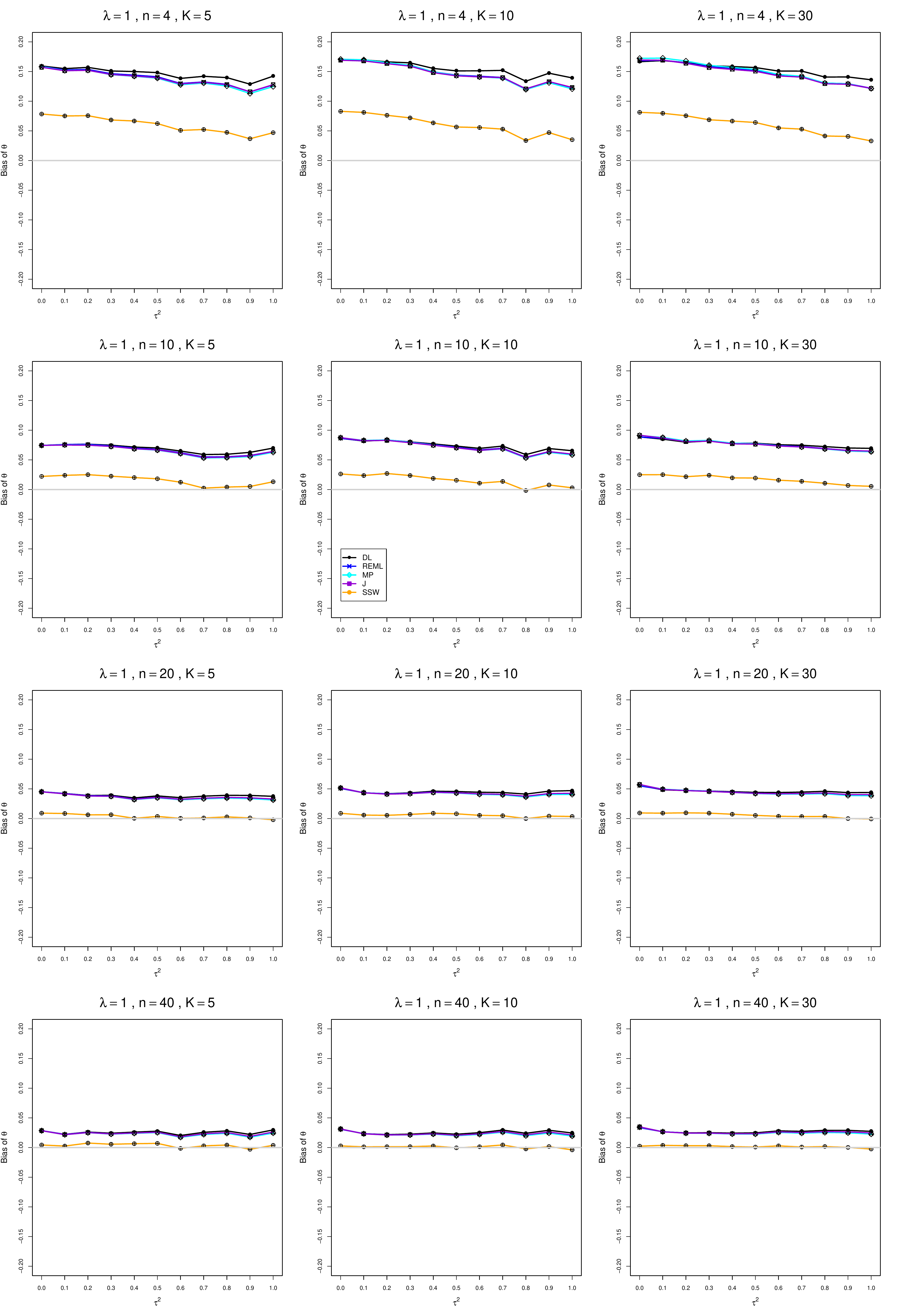}
	\caption{Bias of estimators of $\lambda$ for $\lambda=1$, $n = 4, \;10, \;20, \;40$, and $K = 5, \;10, \;30$. Bias-corrected estimate of $\lambda_i$
		\label{BiasThetaRoM1lnCor_smallN_small_K}}
\end{figure}

\begin{figure}[t]
	\includegraphics[scale=0.33]{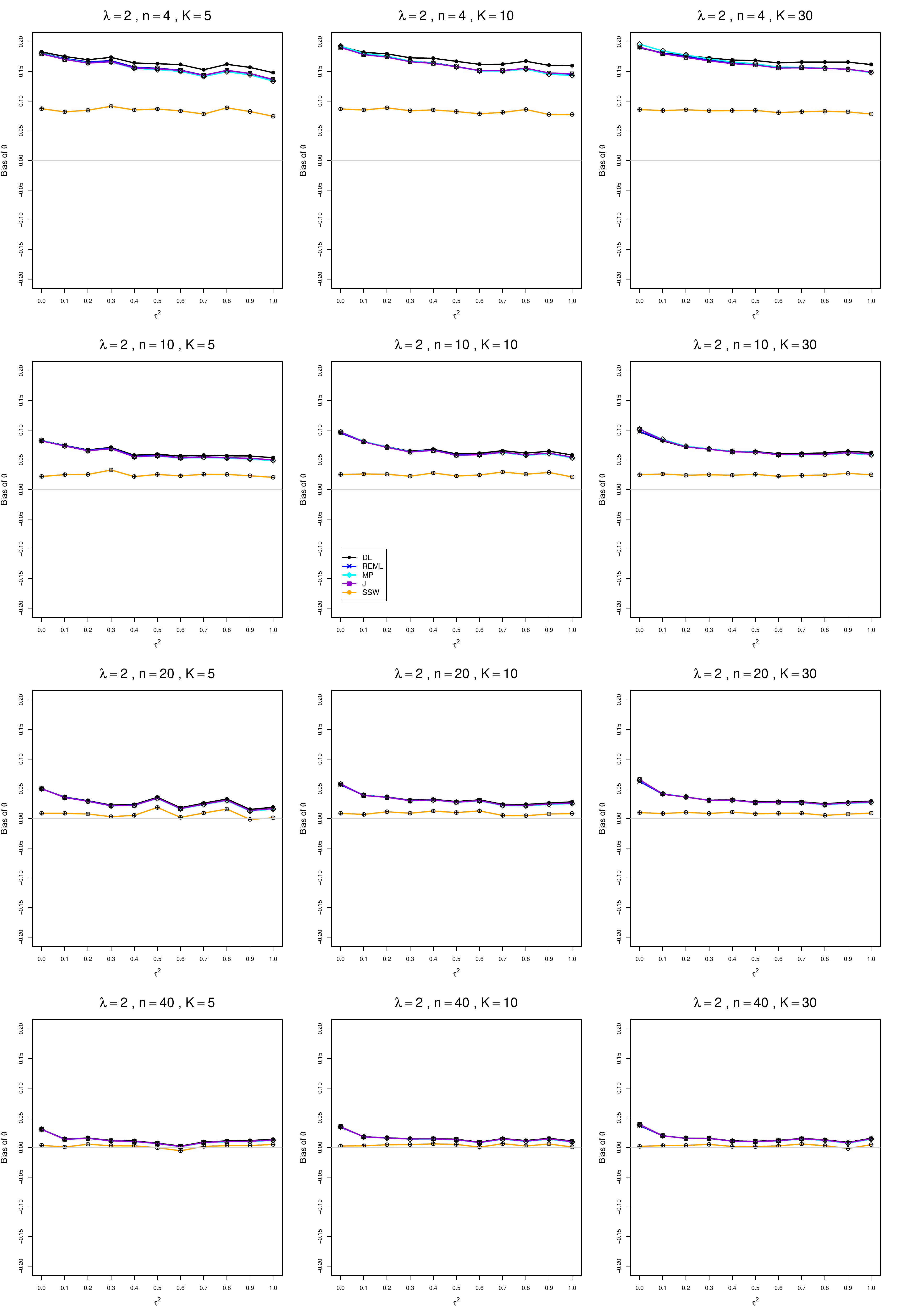}
	\caption{Bias of estimators of $\lambda$ for $\lambda=2$, $n = 4, \;10, \;20, \;40$, and $K = 5, \;10, \;30$. Bias-corrected estimate of $\lambda_i$
		\label{BiasThetaRoM2lnCor_smallN_small_K}}
\end{figure}

\clearpage
\renewcommand{\thefigure}{B2.2.\arabic{figure}}
\setcounter{figure}{0}
\subsection*{B2.2 Coverage of interval estimators of $\lambda$}
Each figure corresponds to a value of $\lambda \;(= 0, 0.2, 0.5, 1, 2)$, a set of values of $n$ (= 4, 10, 20, 40), and a set of values of $K$ (= 5, 10, 30).\\
Each panel corresponds to a value of $n$ and a value of $K$ and has $\tau^2 = 0.0(0.1)1.0$ on the horizontal axis.\\
The interval estimators of $\lambda$ are the companions to the inverse-variance-weighted point estimators
\begin{itemize}
	\item DL (DerSimonian-Laird)
	\item REML (restricted maximum likelihood)
	\item MP (Mandel-Paule)
	\item J (Jackson)
\end{itemize}
and
\begin{itemize}
	\item HKSJ (Hartung-Knapp-Sidik-Jonkman)
	\item HKSJ MP (HKSJ with MP estimator of $\tau^2$)
	\item SSW MP (SSW as center and half-width equal to critical value from $t_{K-1}$ times estimated standard deviation of SSW with $\hat{\tau}^2$ = $\hat{\tau}^2_{MP}$)
\end{itemize}

\begin{figure}[t]
	\includegraphics[scale=0.35]{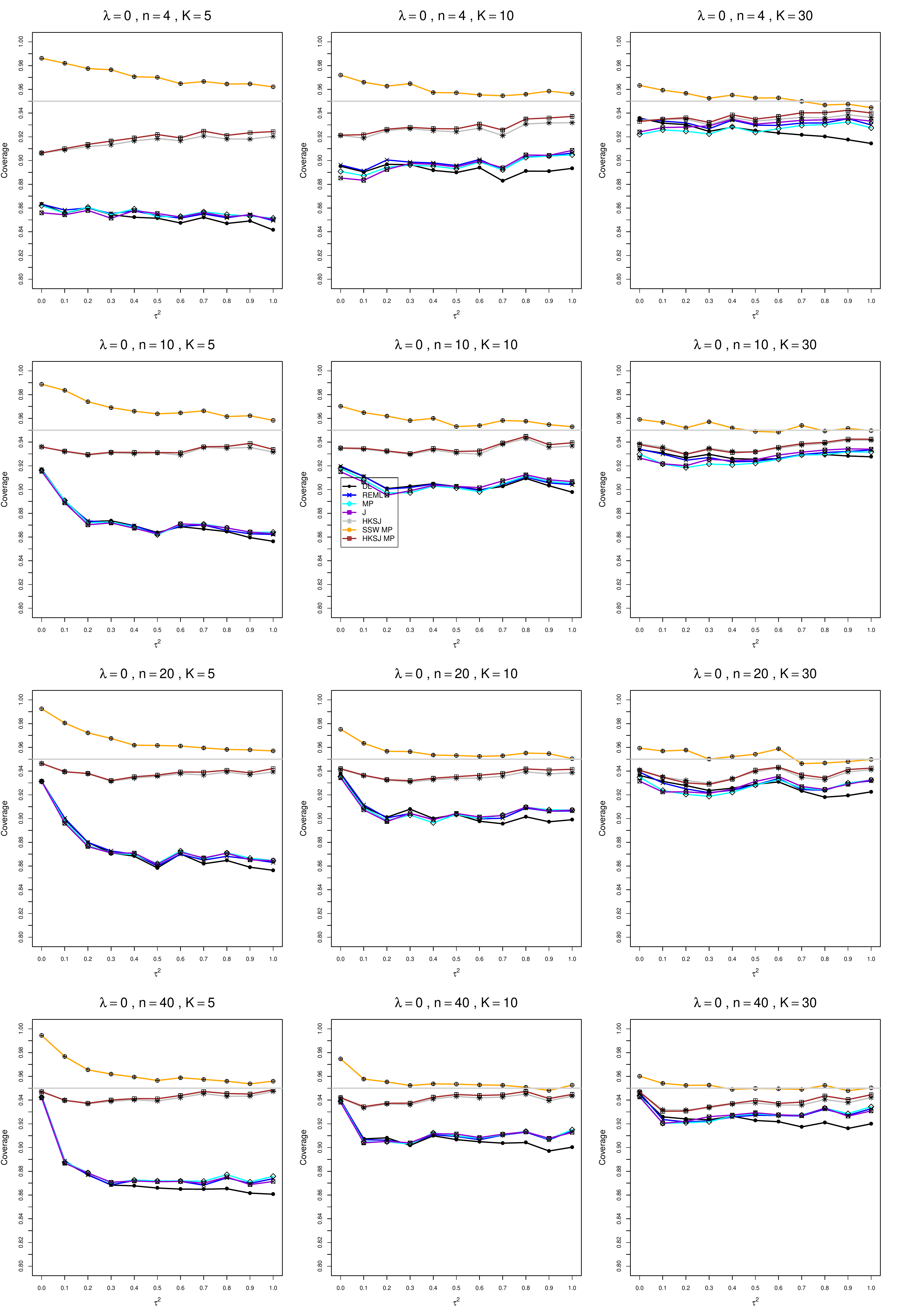}
	\caption{Coverage of 95\% confidence intervals for $\lambda$ when $\lambda=0$, $n = 4, \;10, \;20, \;40$, and $K = 5, \;10, \;30$. Bias-corrected estimate of $\lambda_i$ 		
		\label{CovThetaRoM0lnCor_smallN_small_K}}
\end{figure}
\begin{figure}[t]
	\includegraphics[scale=0.35]{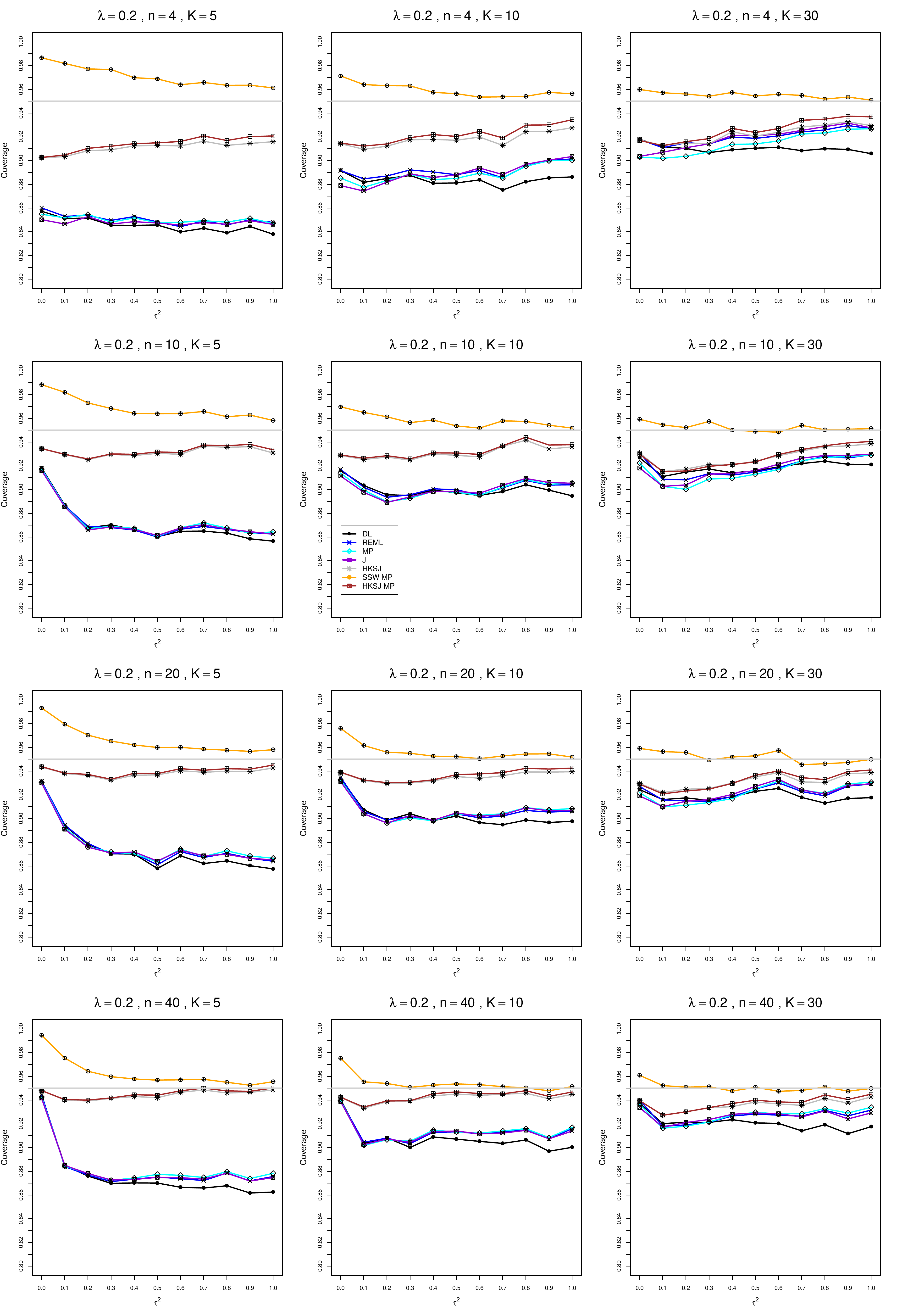}
	\caption{Coverage of 95\% confidence intervals for $\lambda$ when $\lambda=0.2$, $n = 4, \;10, \;20, \;40$, and $K = 5, \;10, \;30$. Bias-corrected estimate of $\lambda_i$ 		
		\label{CovThetaRoM02lnCor_smallN_small_K}}
\end{figure}

\begin{figure}[t]
	\includegraphics[scale=0.35]{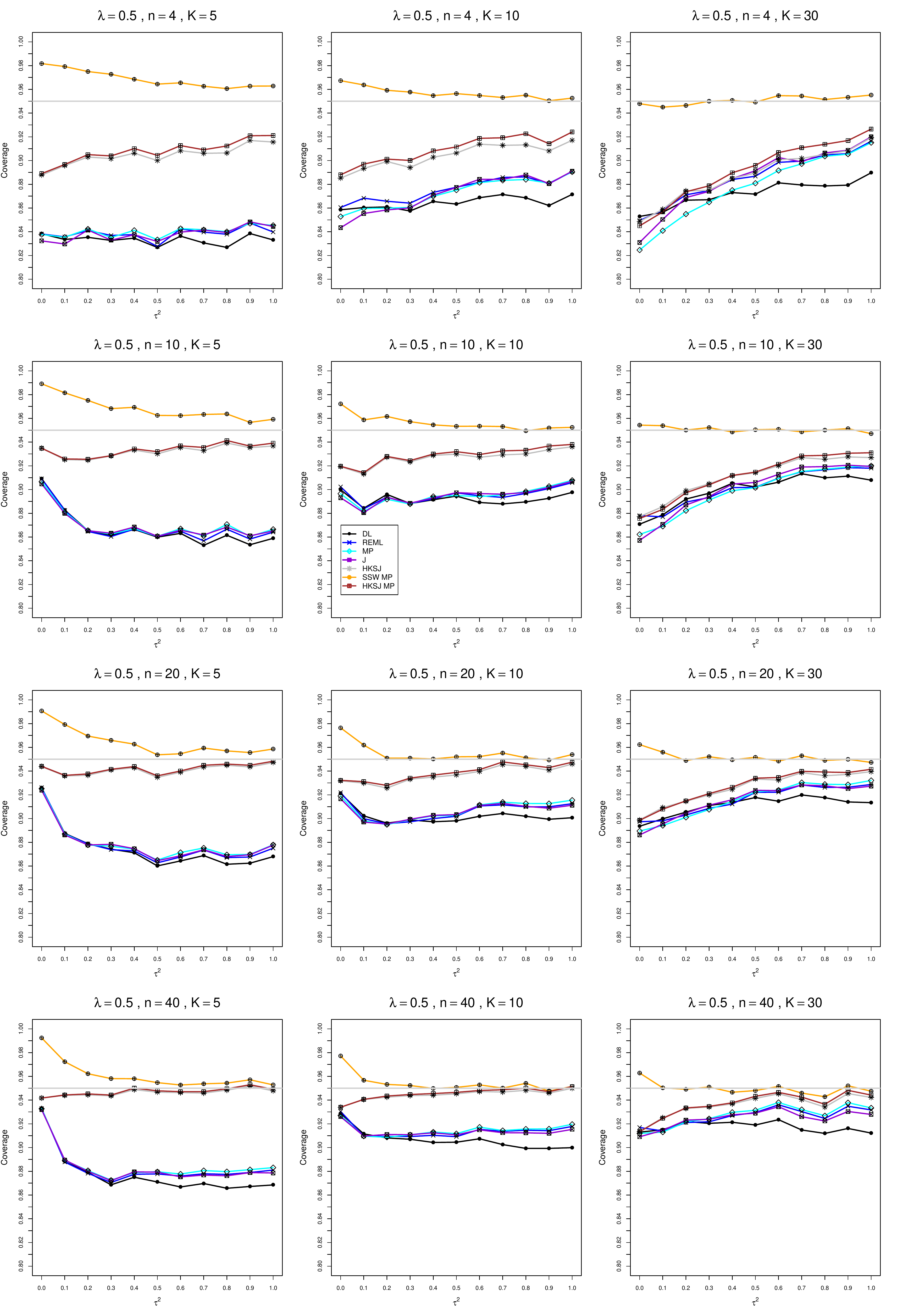}
	\caption{Coverage of 95\% confidence intervals for $\lambda$ when $\lambda=0.5$, $n = 4, \;10, \;20, \;40$, and $K = 5, \;10, \;30$. Bias-corrected estimate of $\lambda_i$ 		
		\label{CovThetaRoM05lnCor_smallN_small_K}}
\end{figure}

\begin{figure}[t]
	\includegraphics[scale=0.35]{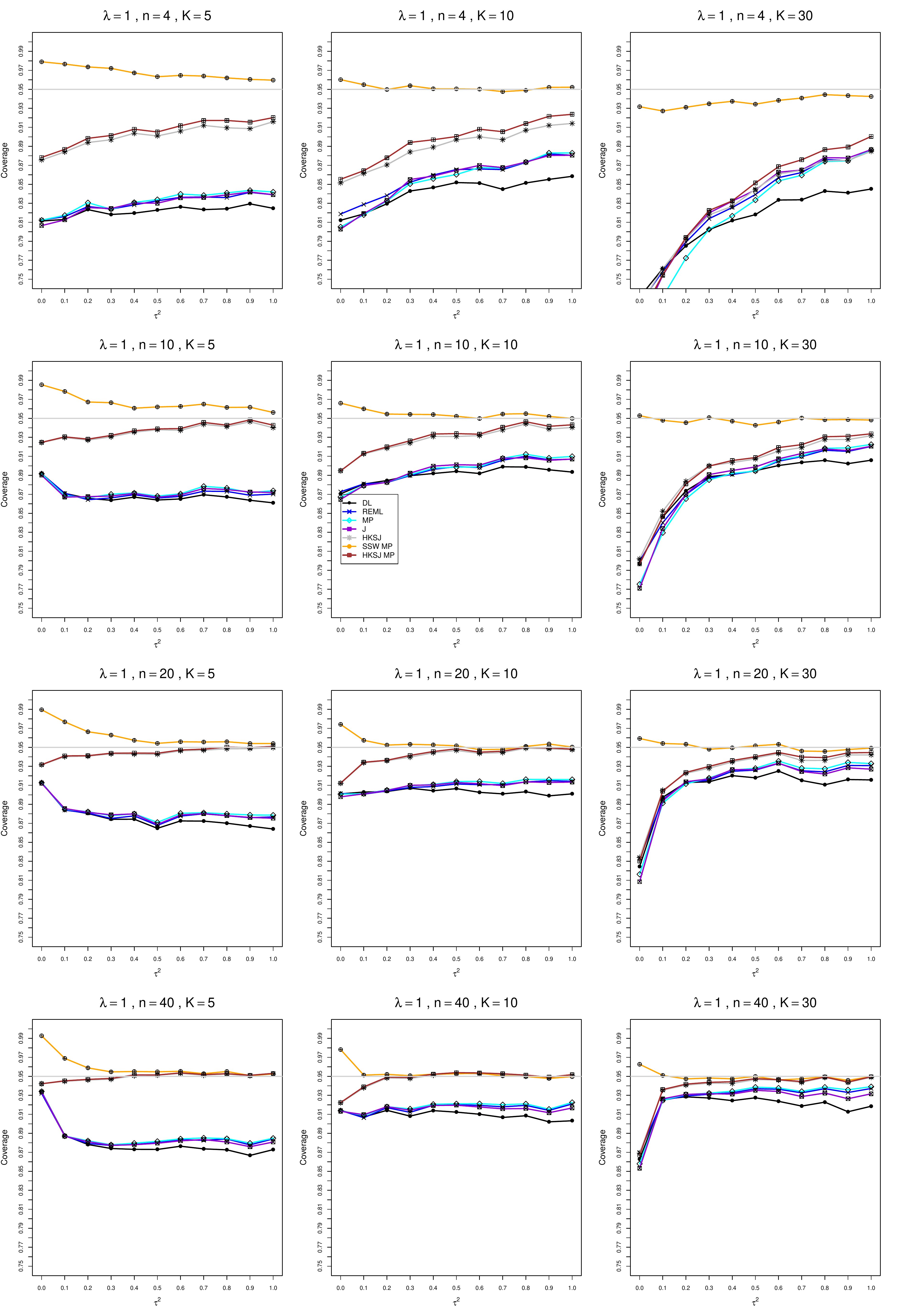}
	\caption{Coverage of 95\% confidence intervals for $\lambda$ when $\lambda=1$, $n = 4, \;10, \;20, \;40$, and $K = 5, \;10, \;30$. Bias-corrected estimate of $\lambda_i$ 		
		\label{CovThetaRoM1lnCor_smallN_small_K}}
\end{figure}
\begin{figure}[t]
	\includegraphics[scale=0.35]{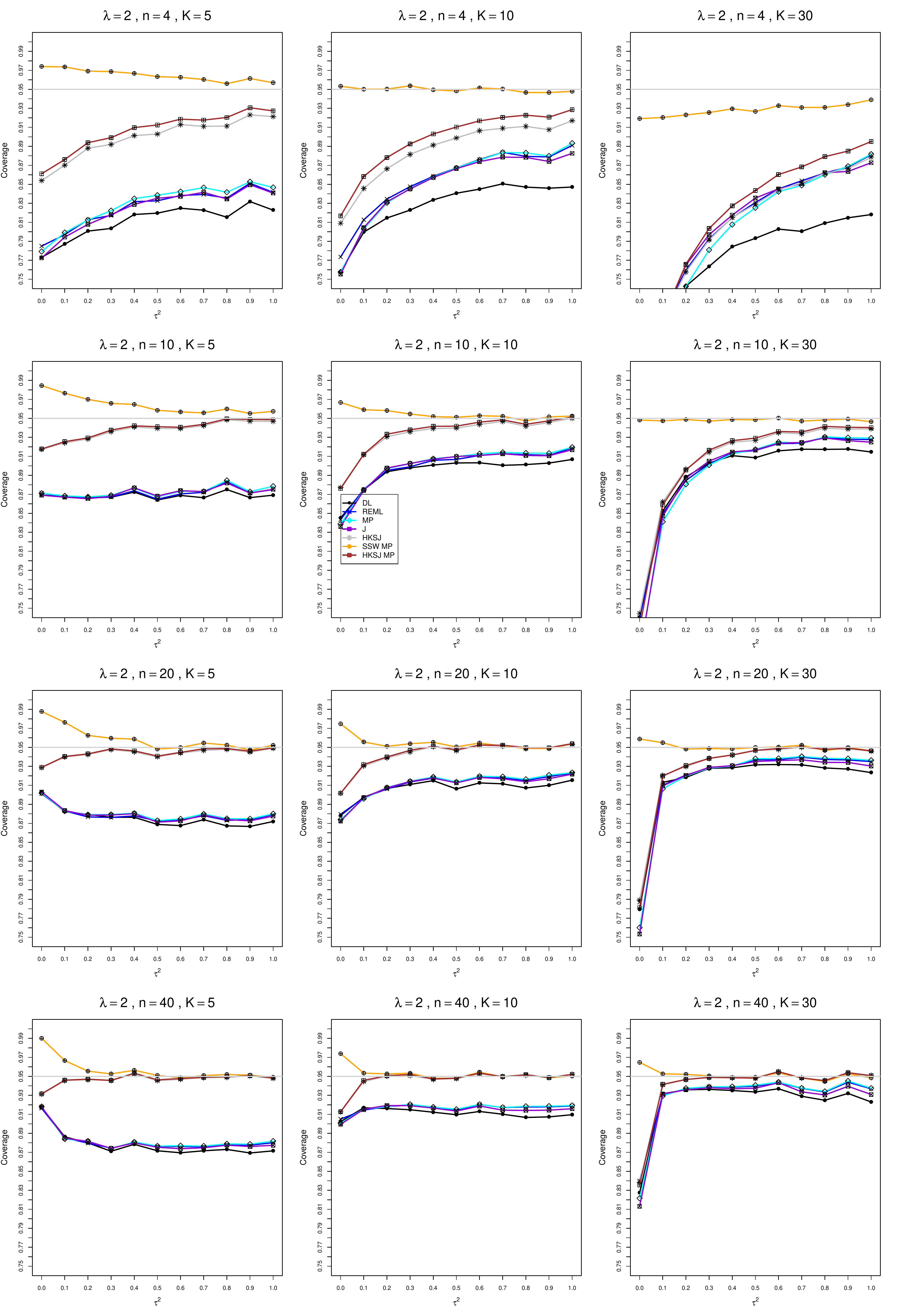}
	\caption{Coverage of 95\% confidence intervals for $\lambda$ when $\lambda=2$, $n = 4, \;10, \;20, \;40$, and $K = 5, \;10, \;30$. Bias-corrected estimate of $\lambda_i$ 		\label{CovThetaRoM2lnCor_smallN_small_K}}
\end{figure}

\renewcommand{\thefigure}{B3.1.\arabic{figure}}
\setcounter{figure}{0}

\clearpage
\section*{B3. Lognormal model, usual estimator of $\lambda_i$, $n= 4, 10, 20, 40$, $K=50,100,125$}
\subsection*{B3.1 Bias of point estimators of $\lambda$}
Each figure corresponds to a value of $\lambda \;(= 0, 0.2, 0.5, 1, 2)$, a set of values of $n$ (= 4, 10, 20, 40), and a set of values of $K$ (= 50, 100, 125).\\
Each panel corresponds to a value of $n$ and a value of $K$ and has $\tau^2 = 0.0(0.1)1.0$ on the horizontal axis.\\
The point estimators of $\lambda$ are
\begin{itemize}
	\item DL (DerSimonian-Laird)
	\item REML (restricted maximum likelihood)
	\item MP (Mandel-Paule)
	\item J (Jackson)
	\item SSW (sample-size-weighted)
\end{itemize}

\begin{figure}[t]
	\includegraphics[scale=0.33]{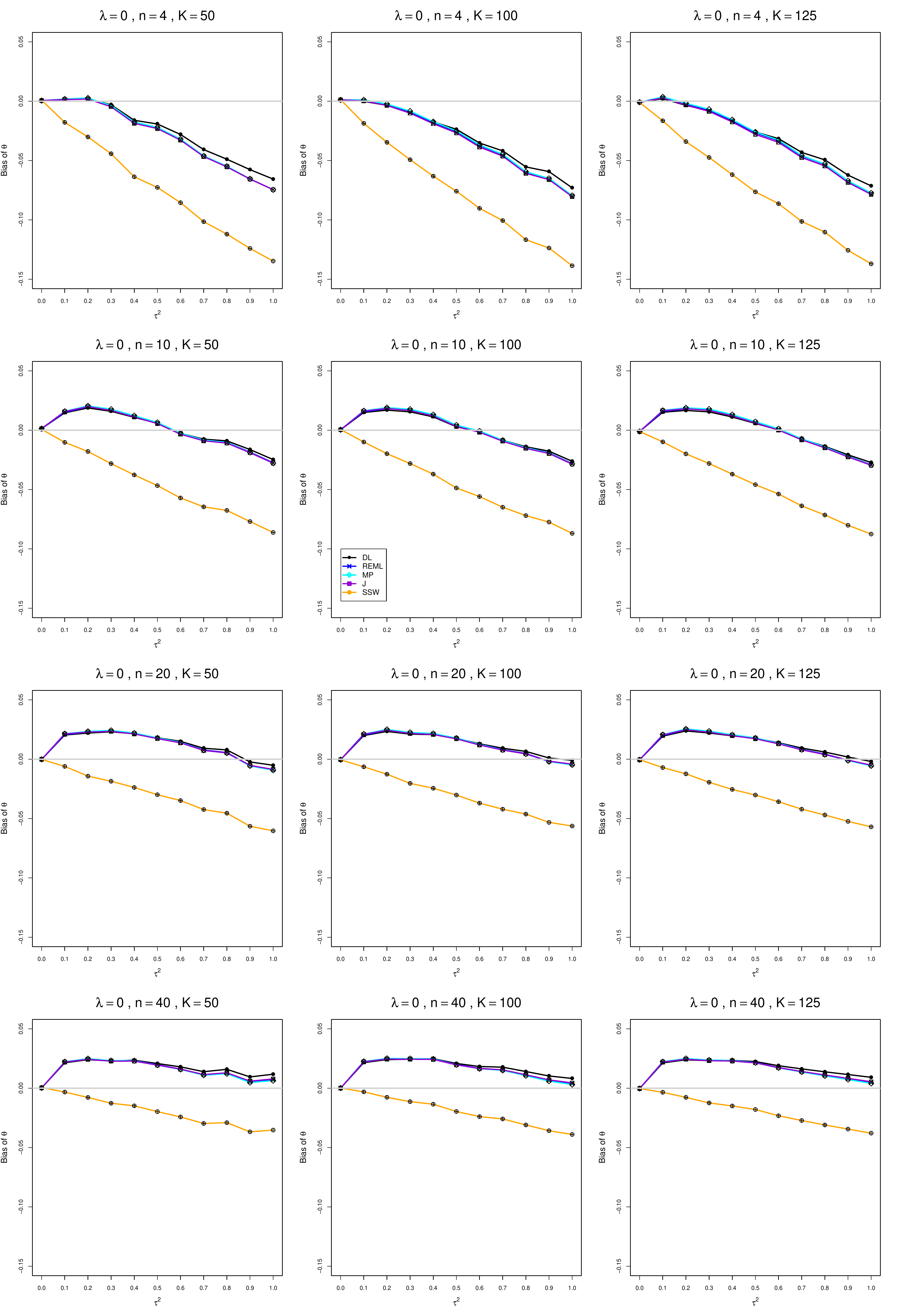}
	\caption{Bias of estimators of $\lambda$ for $\lambda=0$, $n = 4, \;10, \;20, \;40$, and $K = 50, \;100, \;125$. Usual estimate of $\lambda_i$
		\label{BiasThetaRoM0ln_smallN_large_K}}
\end{figure}

\begin{figure}[t]
	\includegraphics[scale=0.33]{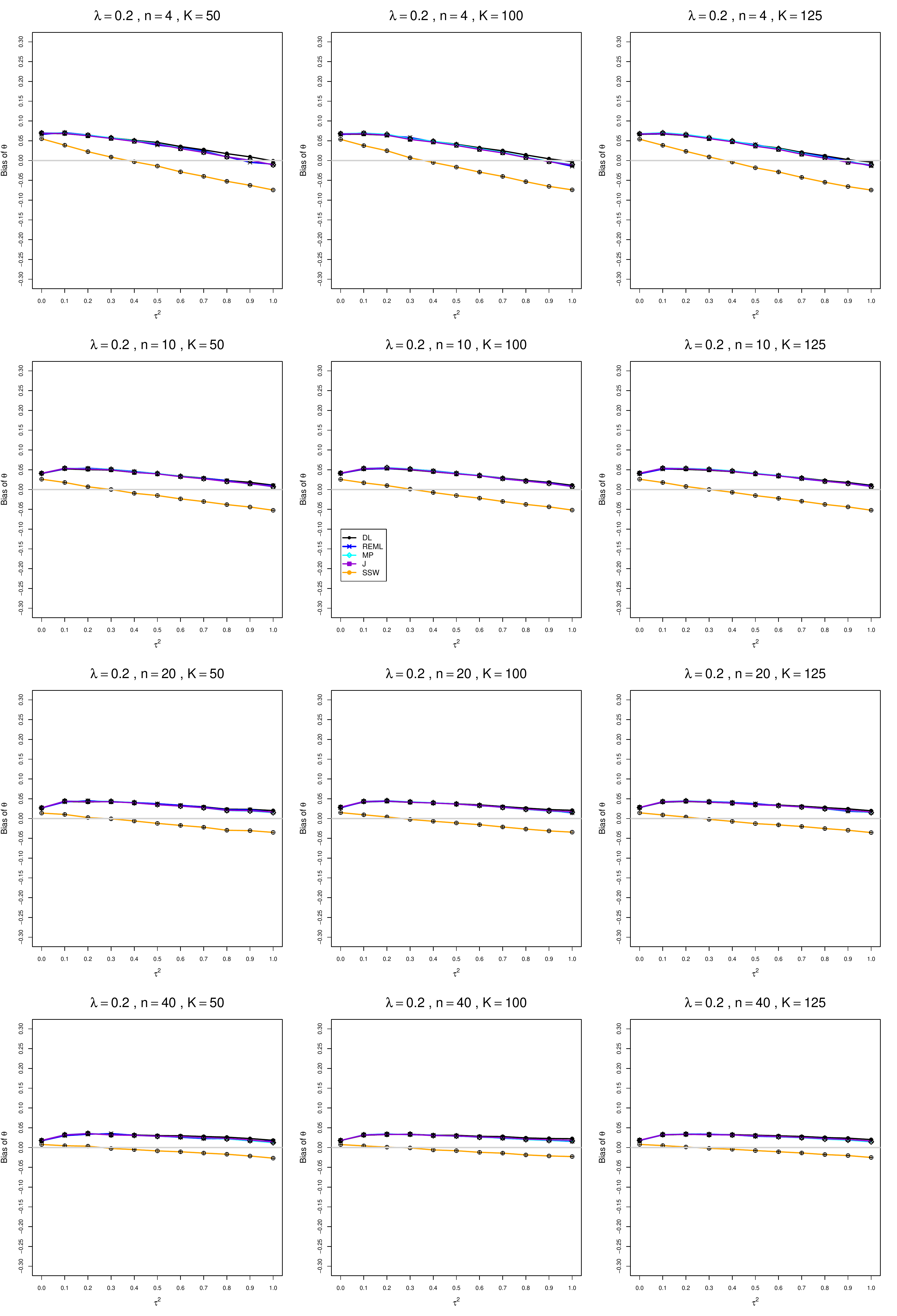}
	\caption{Bias of estimators of $\lambda$ for $\lambda=0.2$, $n = 4, \;10, \;20, \;40$, and $K = 50, \;100, \;125$. Usual estimate of $\lambda_i$ 	
		\label{BiasThetaRoM02ln_smallN_large_K}}
\end{figure}

\begin{figure}[t]
	\includegraphics[scale=0.33]{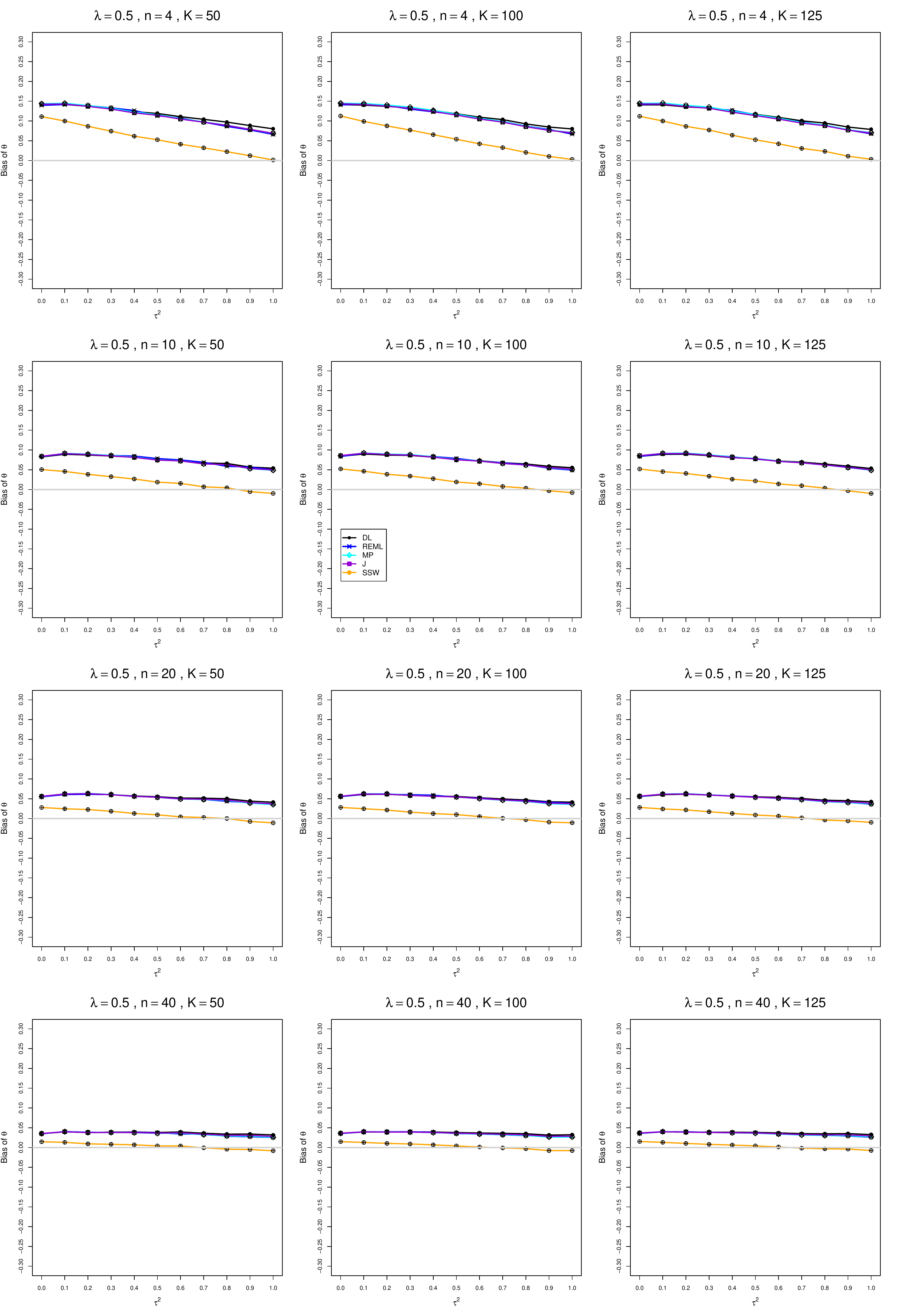}
	\caption{Bias of estimators of $\lambda$ for $\lambda=0.5$, $n = 4, \;10, \;20, \;40$, and $K = 50, \;100, \;125$. Usual estimate of $\lambda_i$
		\label{BiasThetaRoM05ln_smallN_large_K}}
\end{figure}

\begin{figure}[t]
	\includegraphics[scale=0.33]{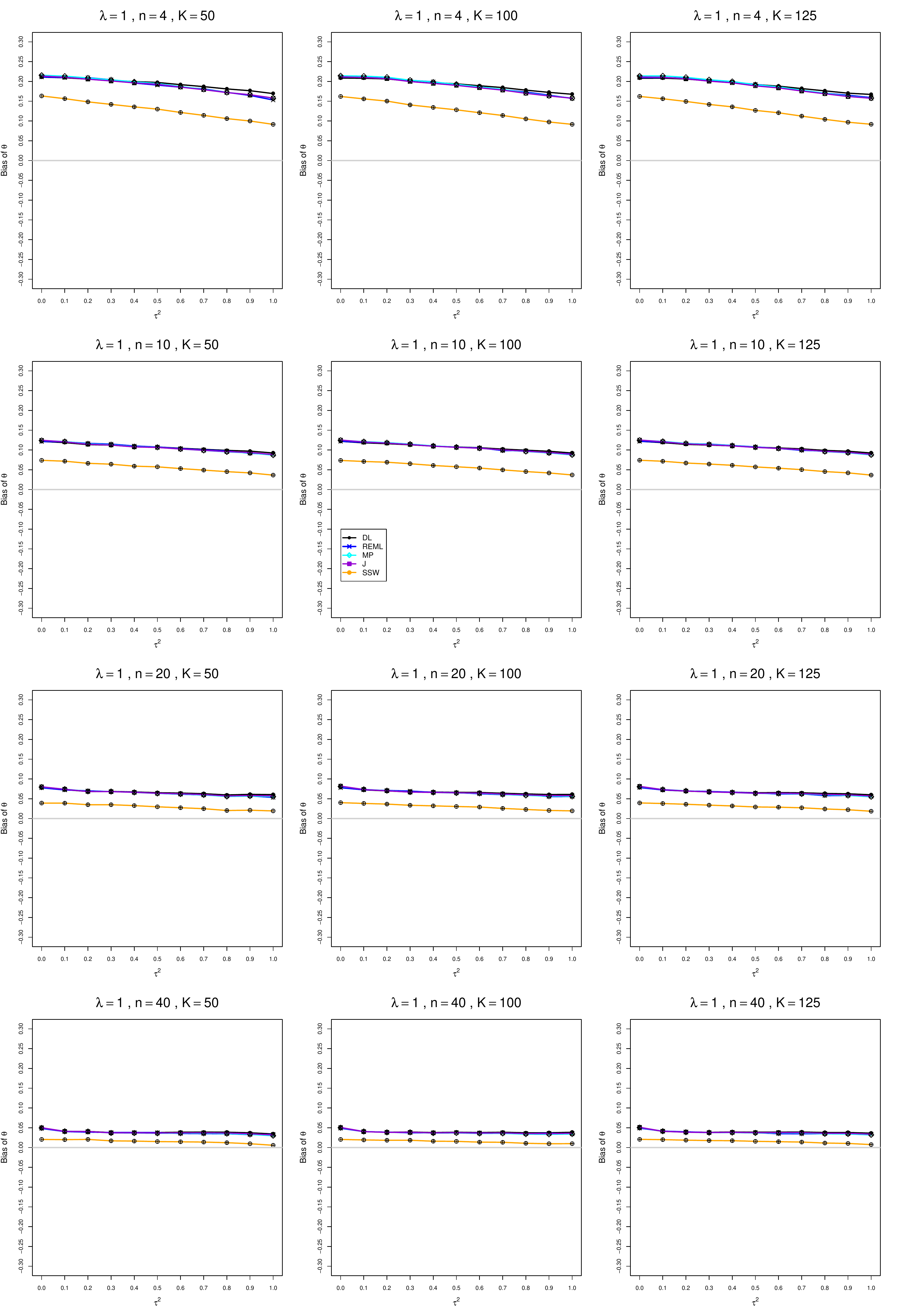}
	\caption{Bias of estimators of $\lambda$ for $\lambda=1$, $n = 4, \;10, \;20, \;40$, and $K = 50, \;100, \;125$. Usual estimate of $\lambda_i$ 		
		\label{BiasThetaRoM1ln_smallN_large_K}}
\end{figure}

\begin{figure}[t]
	\includegraphics[scale=0.33]{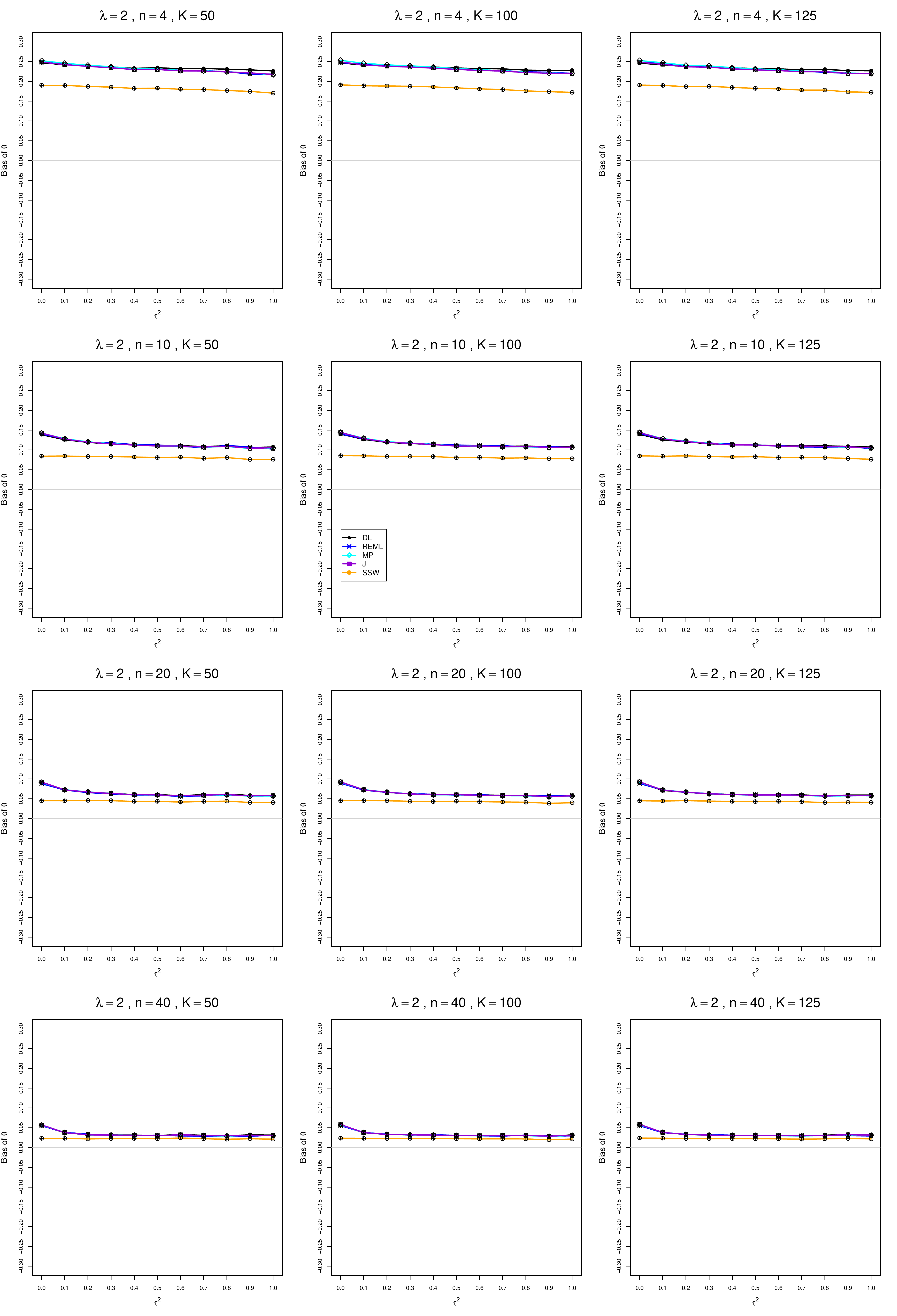}
	\caption{Bias of estimators of $\lambda$ for $\lambda=2$, $n = 4, \;10, \;20, \;40$, and $K = 50, \;100, \;125$. Usual estimate of $\lambda_i$ 		
		\label{BiasThetaRoM2ln_smallN_large_K}}
\end{figure}

\clearpage
\renewcommand{\thefigure}{B3.2.\arabic{figure}}
\setcounter{figure}{0}
\subsection*{B3.2 Coverage of interval estimators of $\lambda$}
Each figure corresponds to a value of $\lambda \;(= 0, 0.2, 0.5, 1, 2)$, a set of values of $n$ (= 4, 10, 20, 40), and a set of values of $K$ (= 50, 100, 125).\\
Each panel corresponds to a value of $n$ and a value of $K$ and has $\tau^2 = 0.0(0.1)1.0$ on the horizontal axis.\\
The interval estimators of $\lambda$ are the companions to the inverse-variance-weighted point estimators
\begin{itemize}
	\item DL (DerSimonian-Laird)
	\item REML (restricted maximum likelihood)
	\item MP (Mandel-Paule)
	\item J (Jackson)
\end{itemize}
and
\begin{itemize}
	\item HKSJ (Hartung-Knapp-Sidik-Jonkman)
	\item HKSJ MP (HKSJ with MP estimator of $\tau^2$)
	\item SSW MP (SSW as center and half-width equal to critical value from $t_{K-1}$ times estimated standard deviation of SSW with $\hat{\tau}^2$ = $\hat{\tau}^2_{MP}$)
\end{itemize}

\begin{figure}[t]
	\includegraphics[scale=0.35]{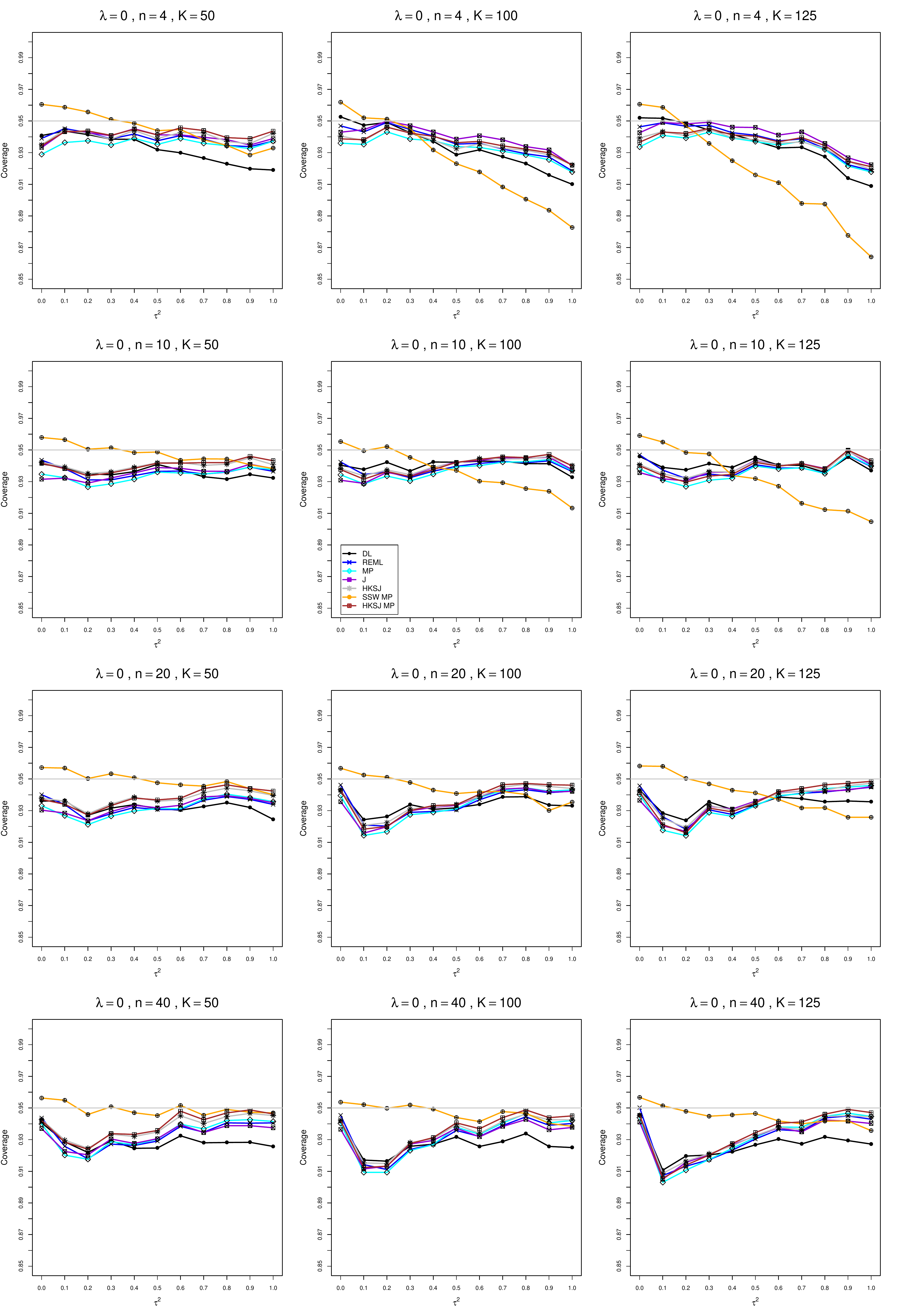}
	\caption{Coverage of 95\% confidence intervals for $\lambda$ when $\lambda=0$, $n = 4, \;10, \;20, \;40$, and $K = 50, \;100, \;125$. Usual estimate of $\lambda_i$
		\label{CovThetaRoM0ln_smallN_large_K}}
\end{figure}

\begin{figure}[t]
	\includegraphics[scale=0.35]{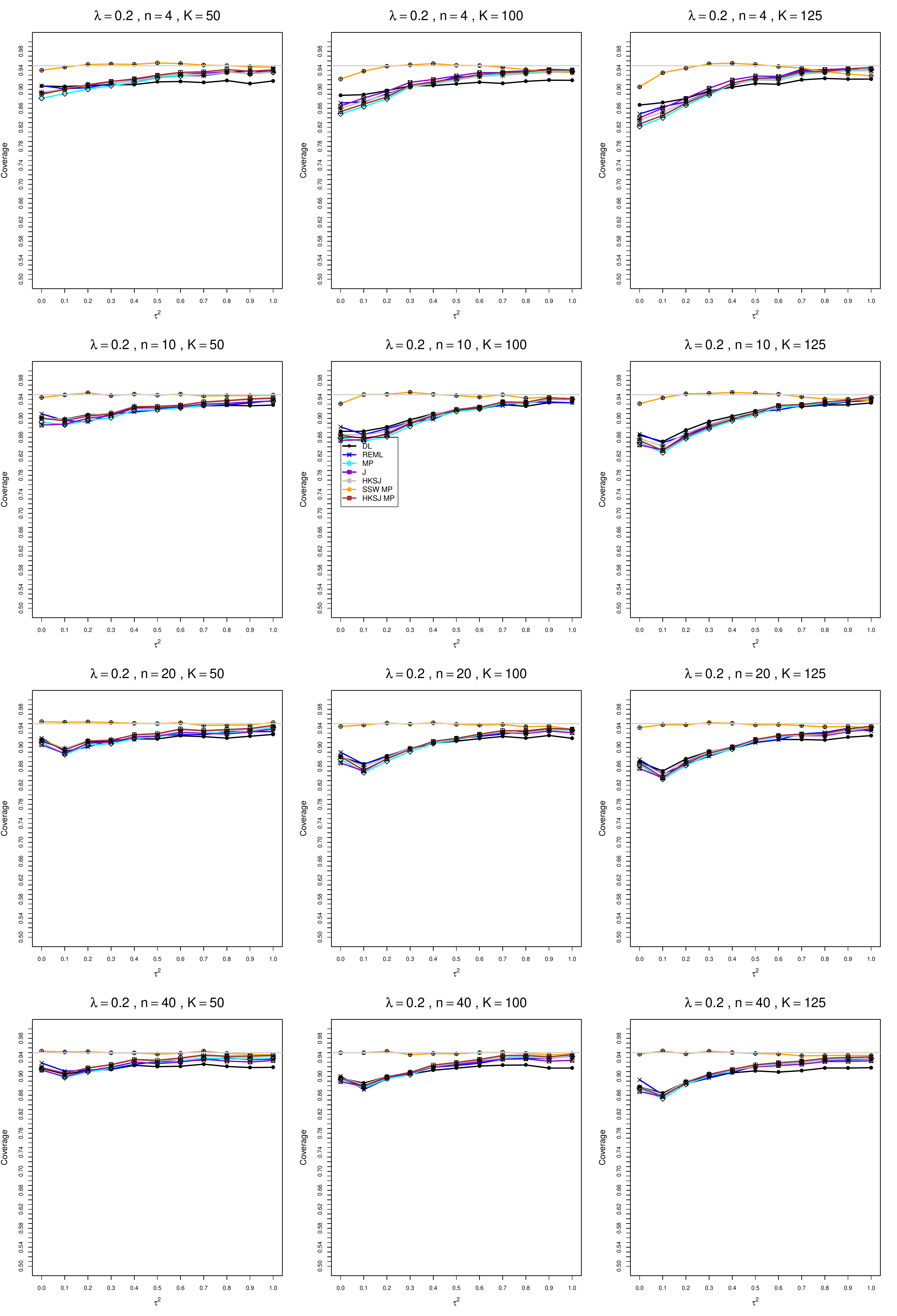}
	\caption{Coverage of 95\% confidence intervals for $\lambda$ when $\lambda=0.2$, $n = 4, \;10, \;20, \;40$, and $K = 50, \;100, \;125$. Usual estimate of $\lambda_i$
		\label{CovThetaRoM02ln_smallN_large_K}}
\end{figure}

\begin{figure}[t]
	\includegraphics[scale=0.35]{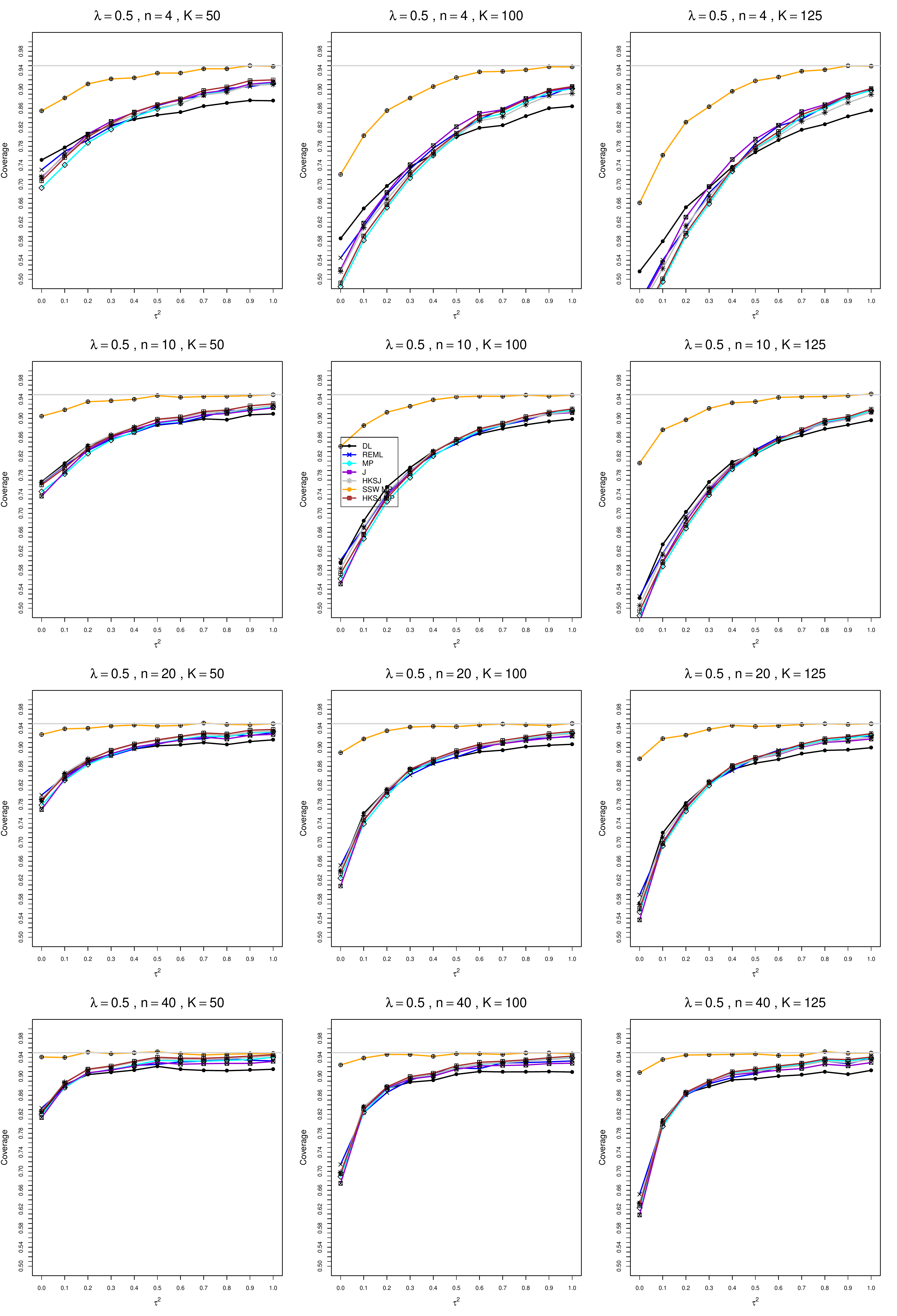}
	\caption{Coverage of 95\% confidence intervals for $\lambda$ when $\lambda=0.5$, $n = 4, \;10, \;20, \;40$, and $K = 50, \;100, \;125$. Usual estimate of $\lambda_i$
		\label{CovThetaRoM05ln_smallN_large_K}}
\end{figure}

\begin{figure}[t]
	\includegraphics[scale=0.35]{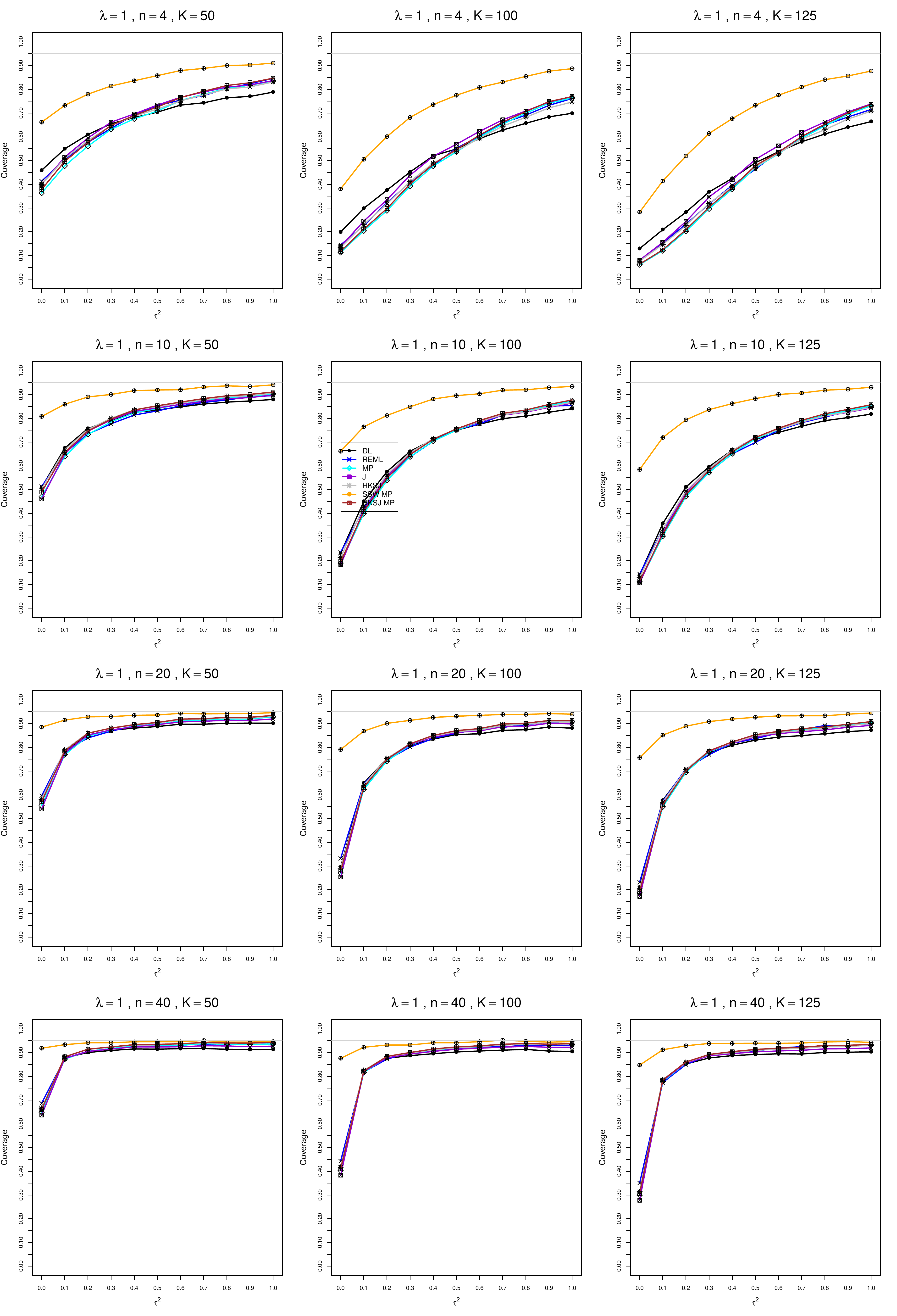}
	\caption{Coverage of 95\% confidence intervals for $\lambda$ when $\lambda=1$, $n = 4, \;10, \;20, \;40$, and $K = 50, \;100, \;125$. Usual estimate of $\lambda_i$
		\label{CovThetaRoM1ln_smallN_large_K}}
\end{figure}
\begin{figure}[t]
	\includegraphics[scale=0.35]{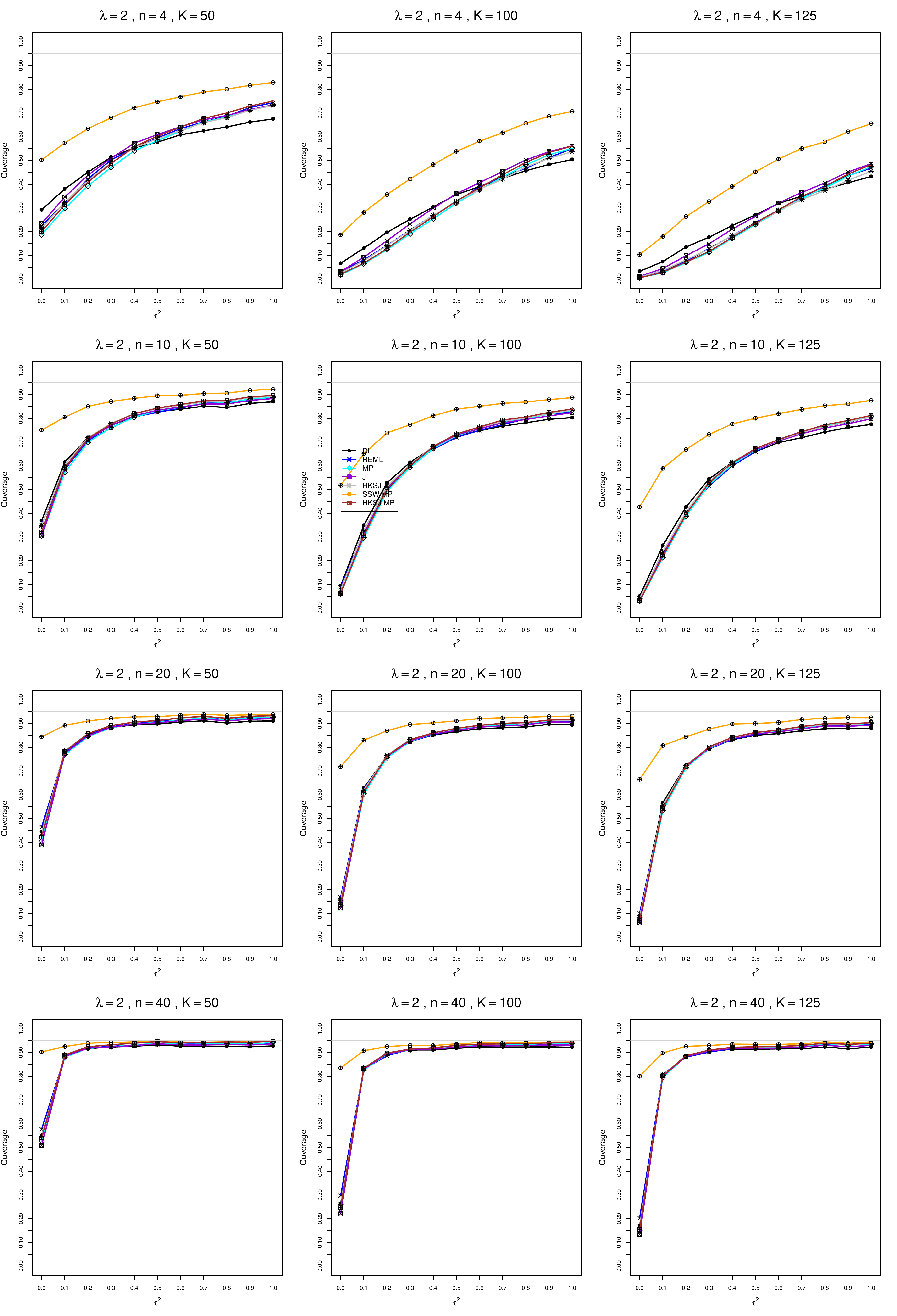}
	\caption{Coverage of 95\% confidence intervals for $\lambda$ when $\lambda=2$, $n = 4, \;10, \;20, \;40$, and $K = 50, \;100, \;125$. Usual estimate of $\lambda_i$ 	
		\label{CovThetaRoM2ln_smallN_large_K}}
\end{figure}

\clearpage
\setcounter{figure}{0}
\renewcommand{\thefigure}{B4.1.\arabic{figure}}

\section*{B4. Lognormal model, bias-corrected estimator of $\lambda_i$, $n= 4, 10, 20, 40$, $K=50,100,125$}

\subsection*{B4.1 Bias of point estimators of $\lambda$}
Each figure corresponds to a value of $\lambda \;(= 0, 0.2, 0.5, 1, 2)$, a set of values of $n$ (= 4, 10, 20, 40), and a set of values of $K$ (= 50, 100, 125).\\
Each panel corresponds to a value of $n$ and a value of $K$ and has $\tau^2 = 0.0(0.1)1.0$ on the horizontal axis.\\
The point estimators of $\lambda$ are
\begin{itemize}
	\item DL (DerSimonian-Laird)
	\item REML (restricted maximum likelihood)
	\item MP (Mandel-Paule)
	\item J (Jackson)
	\item SSW (sample-size-weighted)
\end{itemize}

\begin{figure}[t]
	\includegraphics[scale=0.33]{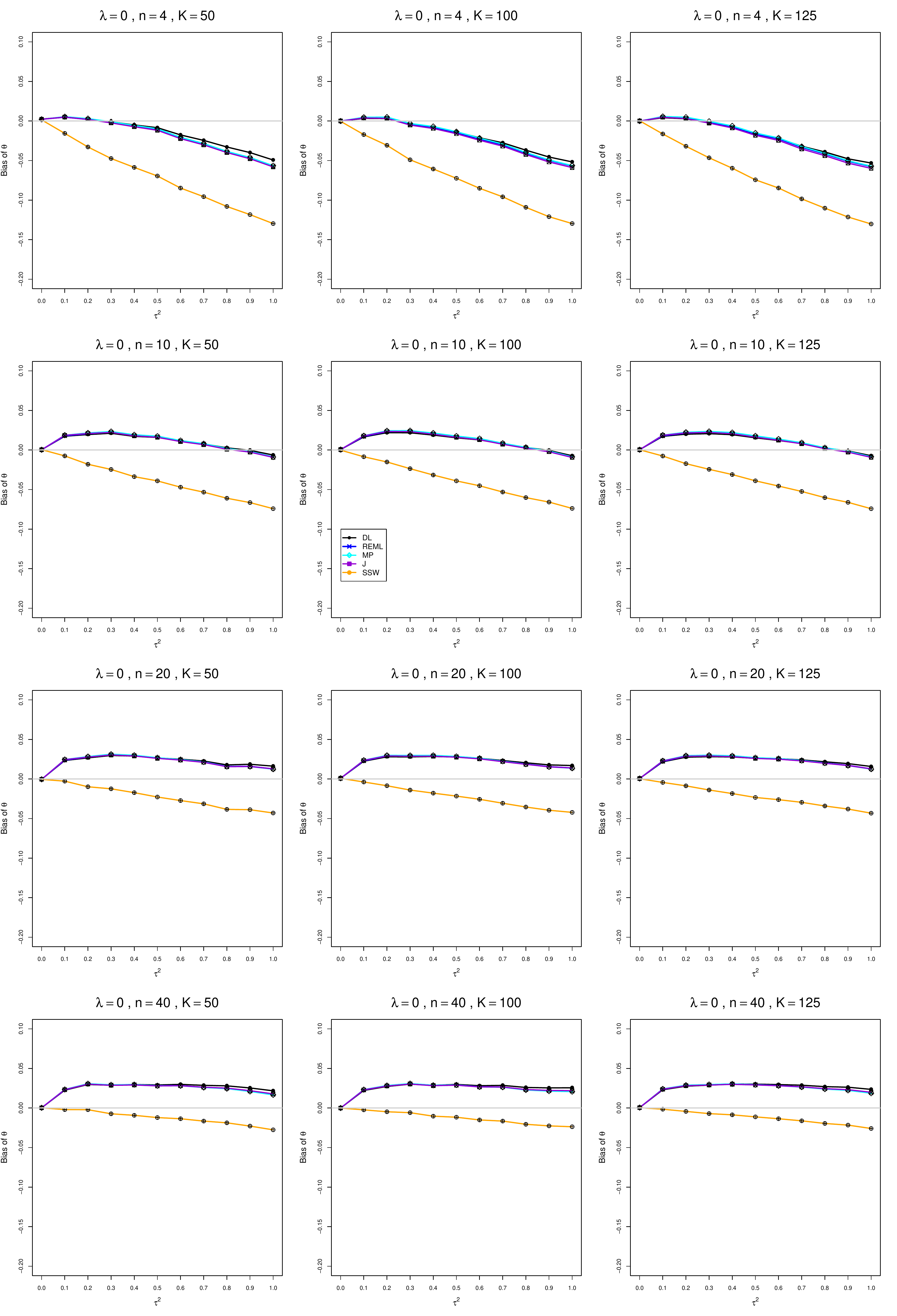}
	\caption{Bias of estimators of $\lambda$ for $\lambda=0$, $n = 4, \;10, \;20, \;40$, and $K = 50, \;100, \;125$. Bias-corrected estimate of $\lambda_i$
		\label{BiasThetaRoM0lnCor_smallN_large_K}}
\end{figure}

\begin{figure}[t]
	\includegraphics[scale=0.33]{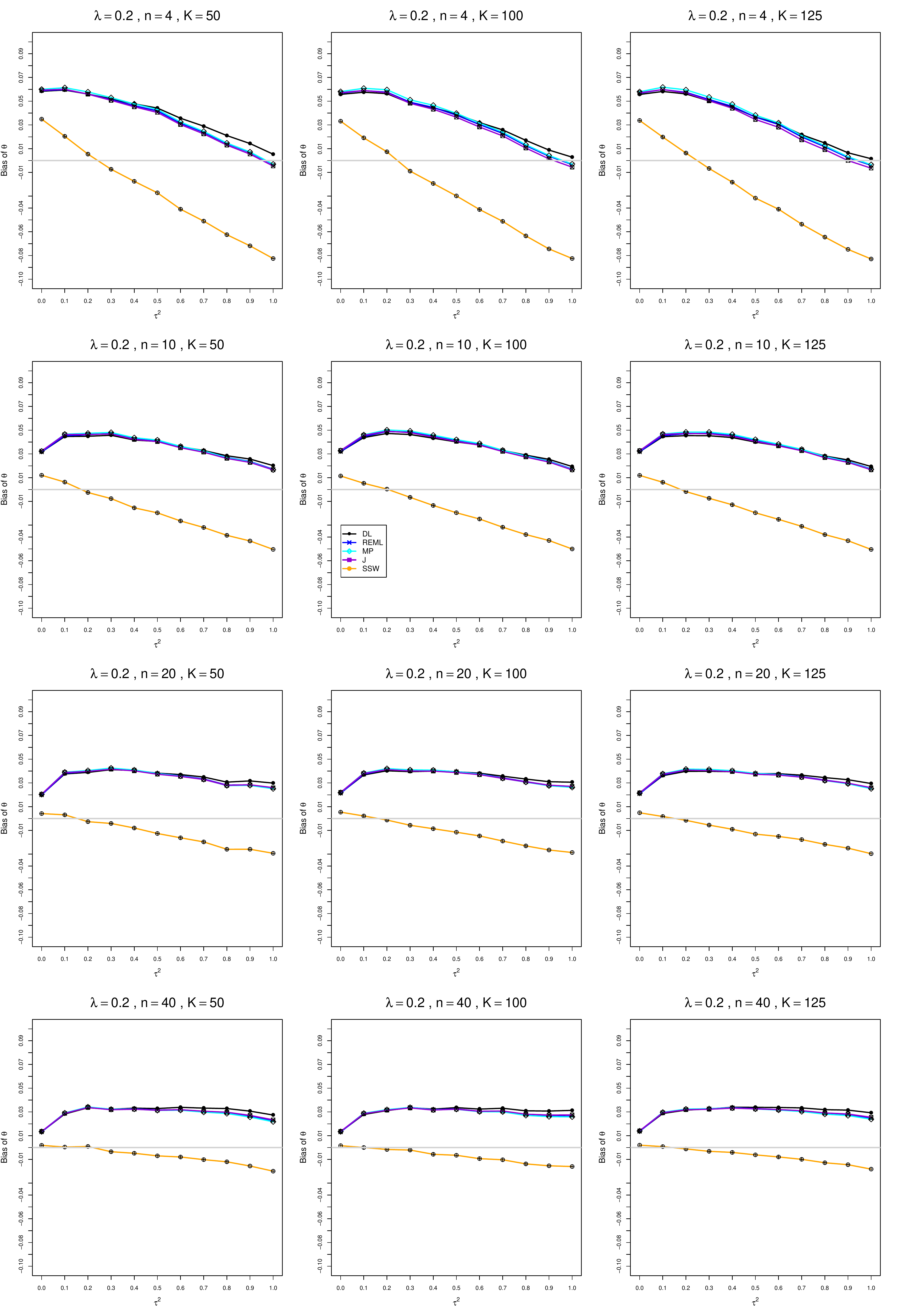}
	\caption{Bias of estimators of $\lambda$ for $\lambda=0.2$, $n = 4, \;10, \;20, \;40$, and $K = 50, \;100, \;125$. Bias-corrected estimate of $\lambda_i$
		\label{BiasThetaRoM02lnCor_smallN_large_K}}
\end{figure}

\begin{figure}[t]
	\includegraphics[scale=0.33]{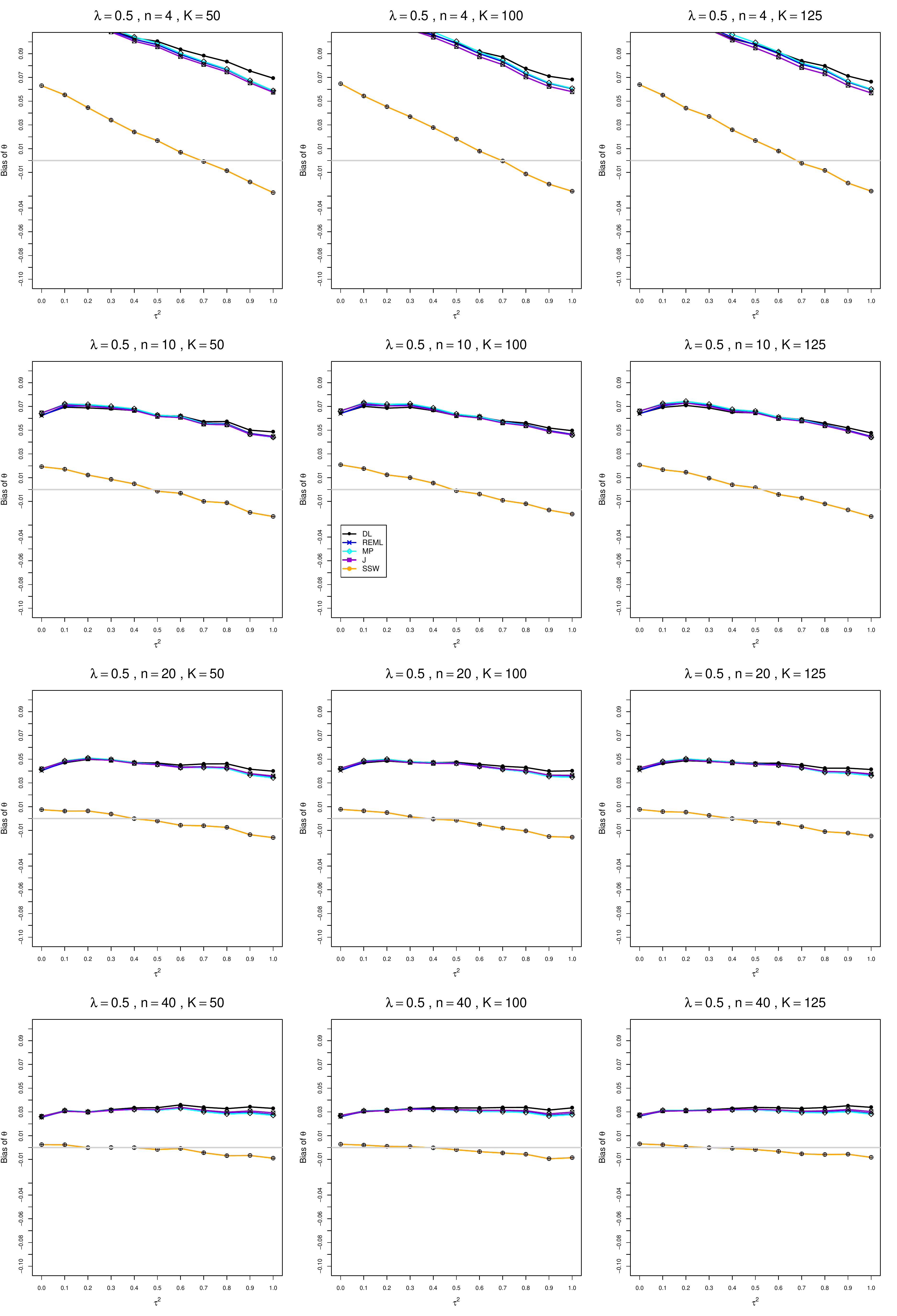}
	\caption{Bias of estimators of $\lambda$ for $\lambda=0.5$, $n = 4, \;10, \;20, \;40$, and $K = 50, \;100, \;125$. Bias-corrected estimate of $\lambda_i$
		\label{BiasThetaRoM05lnCor_smallN_large_K}}
\end{figure}

\begin{figure}[t]
	\includegraphics[scale=0.33]{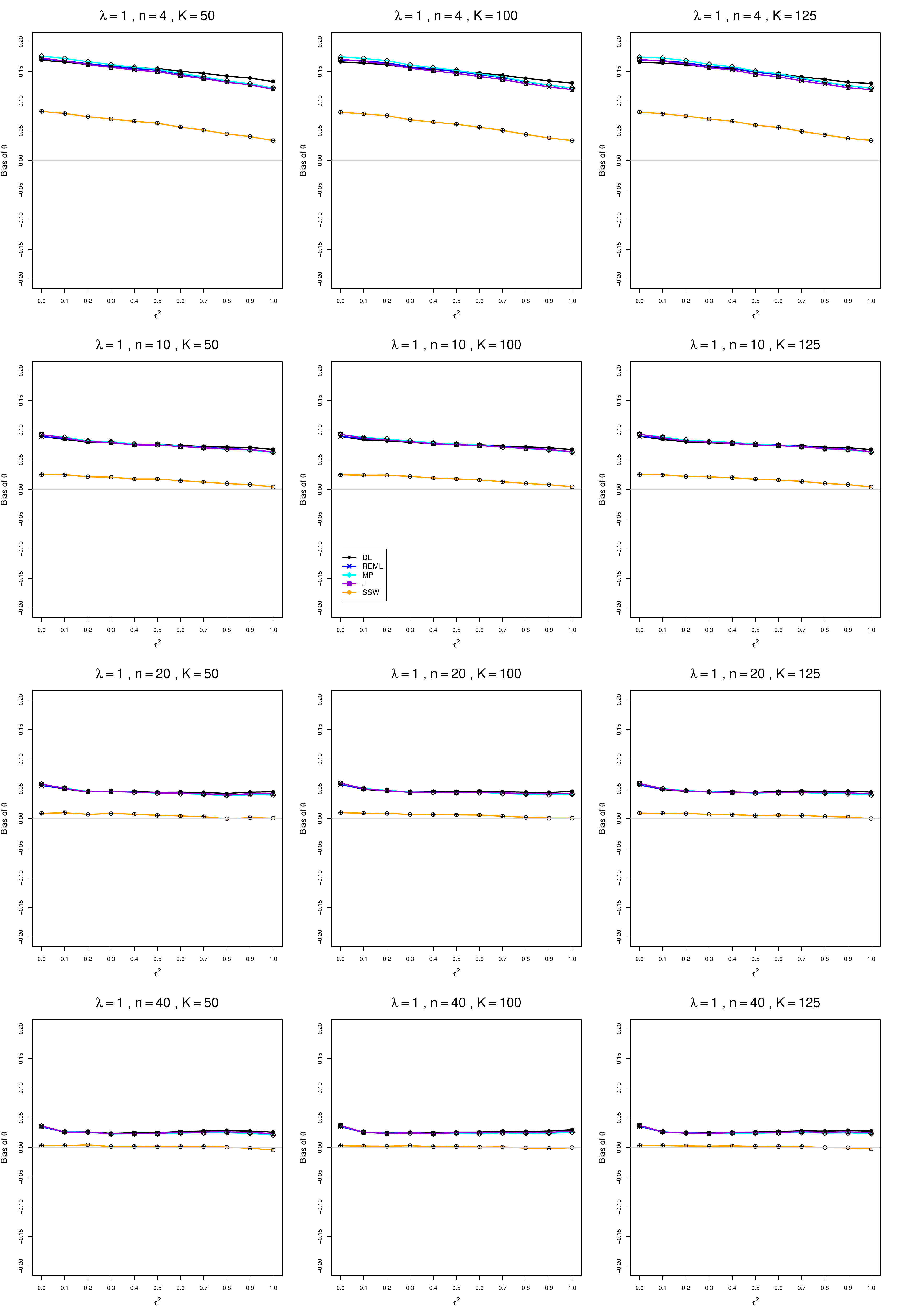}
	\caption{Bias of estimators of $\lambda$ for $\lambda=1$, $n = 4, \;10, \;20, \;40$, and $K = 50, \;100, \;125$. Bias-corrected estimate of $\lambda_i$
		\label{BiasThetaRoM1lnCor_smallN_large_K}}
\end{figure}

\begin{figure}[t]
	\includegraphics[scale=0.33]{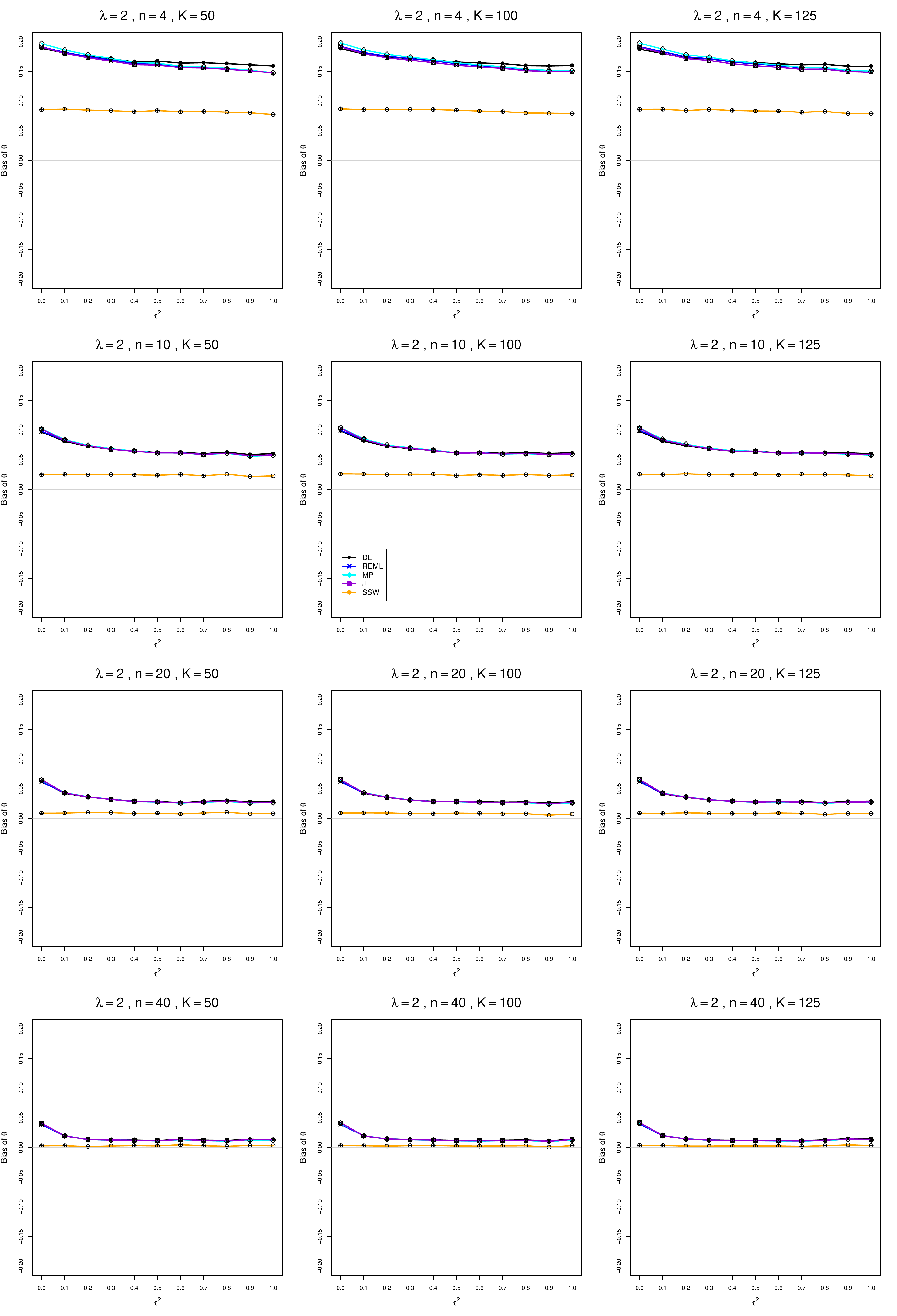}
	\caption{Bias of estimators of $\lambda$ for $\lambda=2$, $n = 4, \;10, \;20, \;40$, and $K = 50, \;100, \;125$. Bias-corrected estimate of $\lambda_i$
		\label{BiasThetaRoM2lnCor_smallN_large_K}}
\end{figure}

\clearpage
\renewcommand{\thefigure}{B4.2.\arabic{figure}}
\setcounter{figure}{0}
\subsection*{B4.2 Coverage of interval estimators of $\lambda$}
Each figure corresponds to a value of $\lambda \;(= 0, 0.2, 0.5, 1, 2)$, a set of values of $n$ (= 4, 10, 20, 40), and a set of values of $K$ (= 50, 100, 125).\\
Each panel corresponds to a value of $n$ and a value of $K$ and has $\tau^2 = 0.0(0.1)1.0$ on the horizontal axis.\\
The interval estimators of $\lambda$ are the companions to the inverse-variance-weighted point estimators
\begin{itemize}
	\item DL (DerSimonian-Laird)
	\item REML (restricted maximum likelihood)
	\item MP (Mandel-Paule)
	\item J (Jackson)
\end{itemize}
and
\begin{itemize}
	\item HKSJ (Hartung-Knapp-Sidik-Jonkman)
	\item HKSJ MP (HKSJ with MP estimator of $\tau^2$)
	\item SSW MP (SSW as center and half-width equal to critical value from $t_{K-1}$ times estimated standard deviation of SSW with $\hat{\tau}^2$ = $\hat{\tau}^2_{MP}$)
\end{itemize}

\begin{figure}[t]
	\includegraphics[scale=0.35]{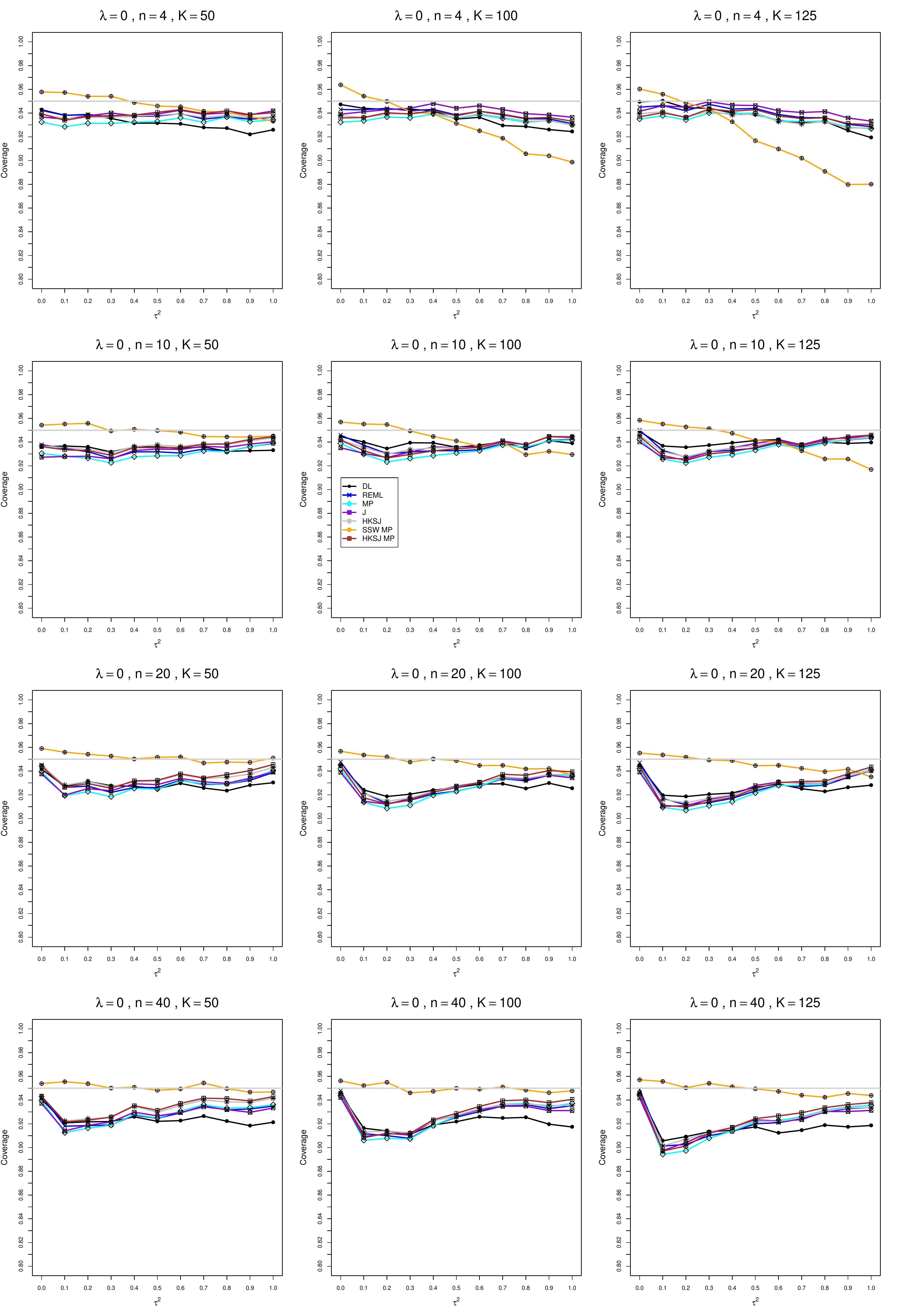}
	\caption{Coverage of 95\% confidence intervals for $\lambda$ when $\lambda=0$, $n = 4, \;10, \;20, \;40$, and $K = 50, \;100, \;125$. Bias-corrected estimate of $\lambda_i$ 		\label{CovThetaRoM0lnCor_smallN_large_K}}
\end{figure}
\begin{figure}[t]
	\includegraphics[scale=0.35]{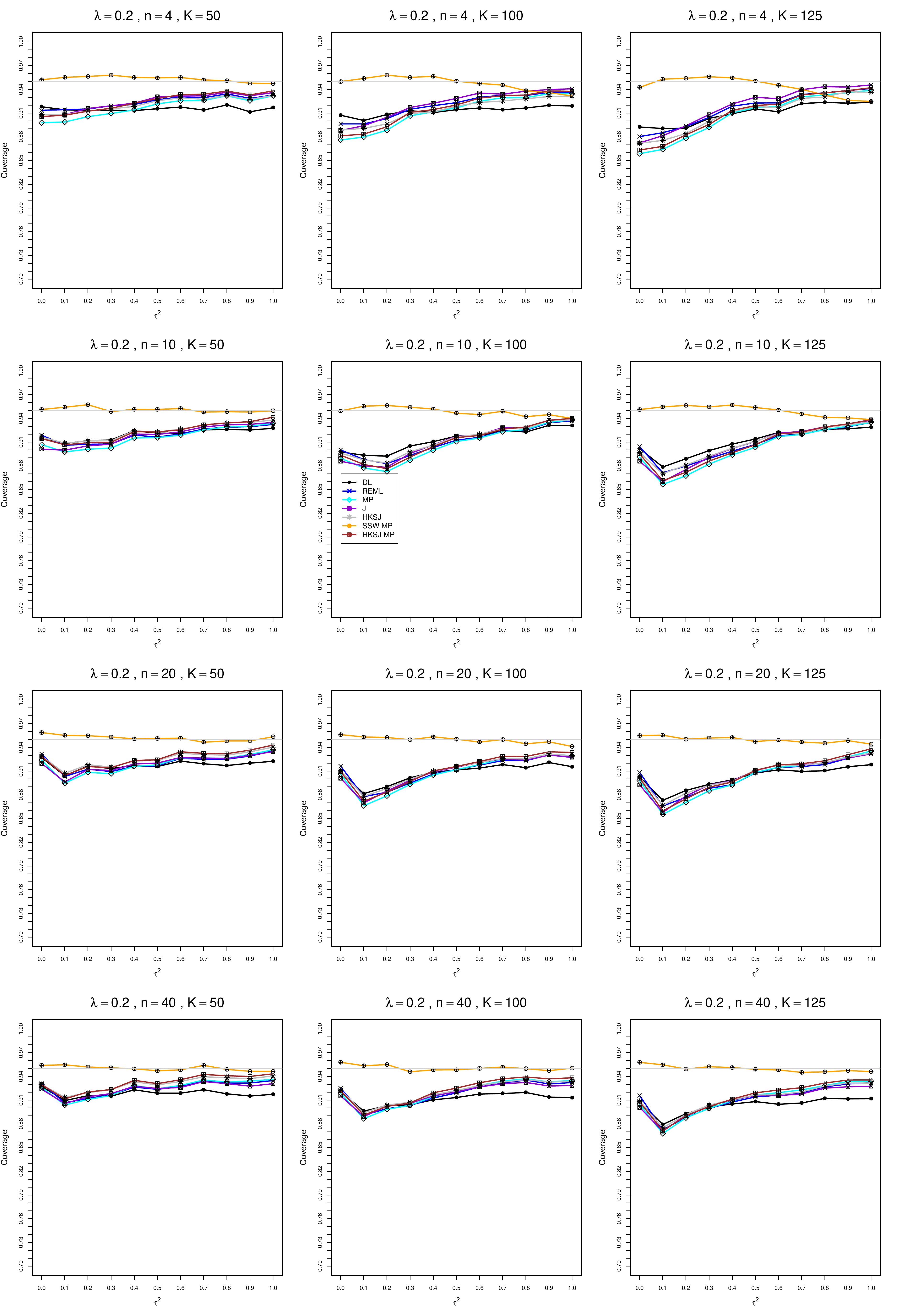}
	\caption{Coverage of 95\% confidence intervals for $\lambda$ when $\lambda=0.2$, $n = 4, \;10, \;20, \;40$, and $K = 50, \;100, \;125$. Bias-corrected estimate of $\lambda_i$ 		\label{CovThetaRoM02lnCor_smallN_large_K}}
\end{figure}

\begin{figure}[t]
	\includegraphics[scale=0.35]{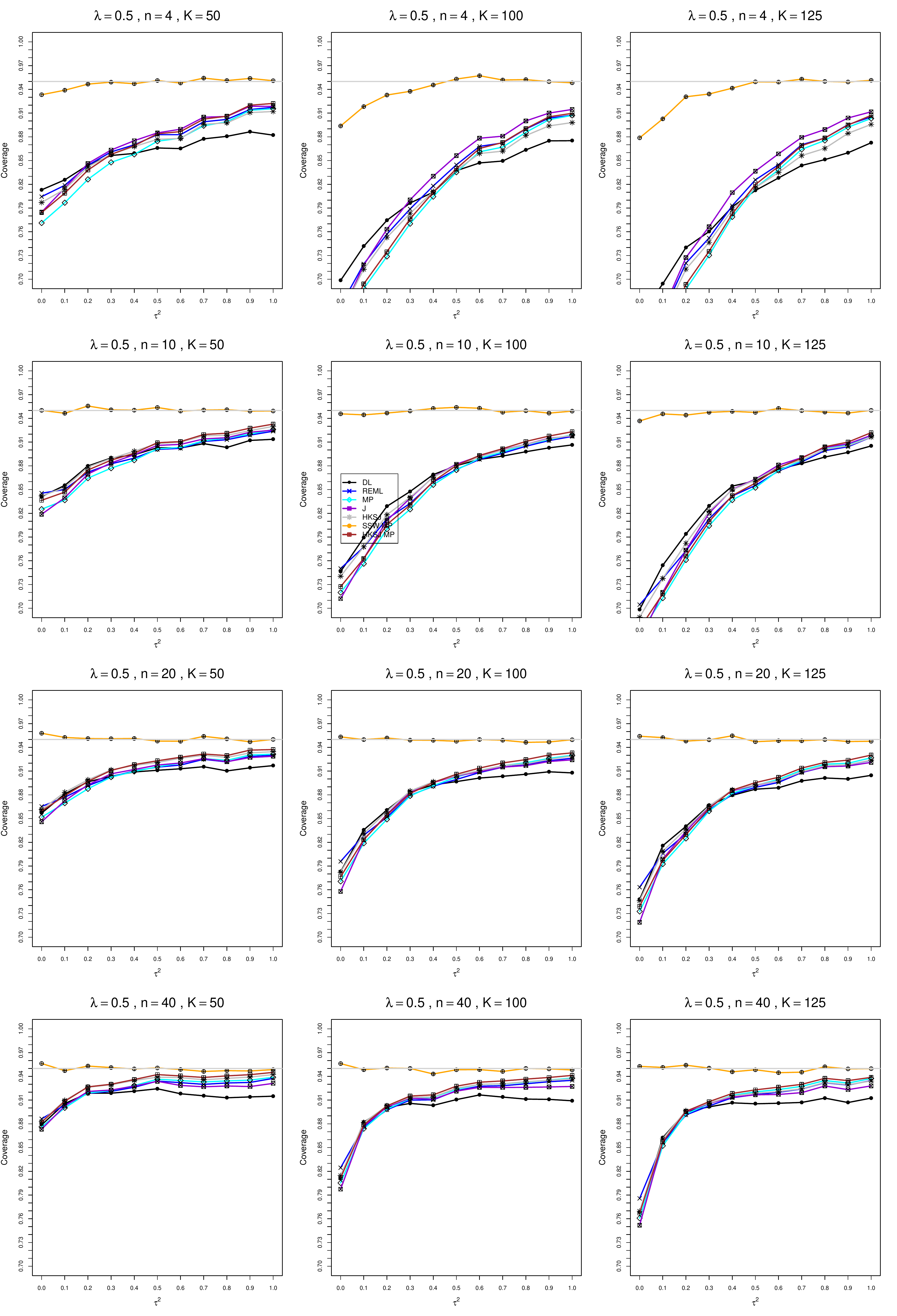}
	\caption{Coverage of 95\% confidence intervals for $\lambda$ when $\lambda=0.5$, $n = 4, \;10, \;20, \;40$, and $K = 50, \;100, \;125$. Bias-corrected estimate of $\lambda_i$ 		\label{CovThetaRoM05lnCor_smallN_large_K}}
\end{figure}

\begin{figure}[t]
	\includegraphics[scale=0.35]{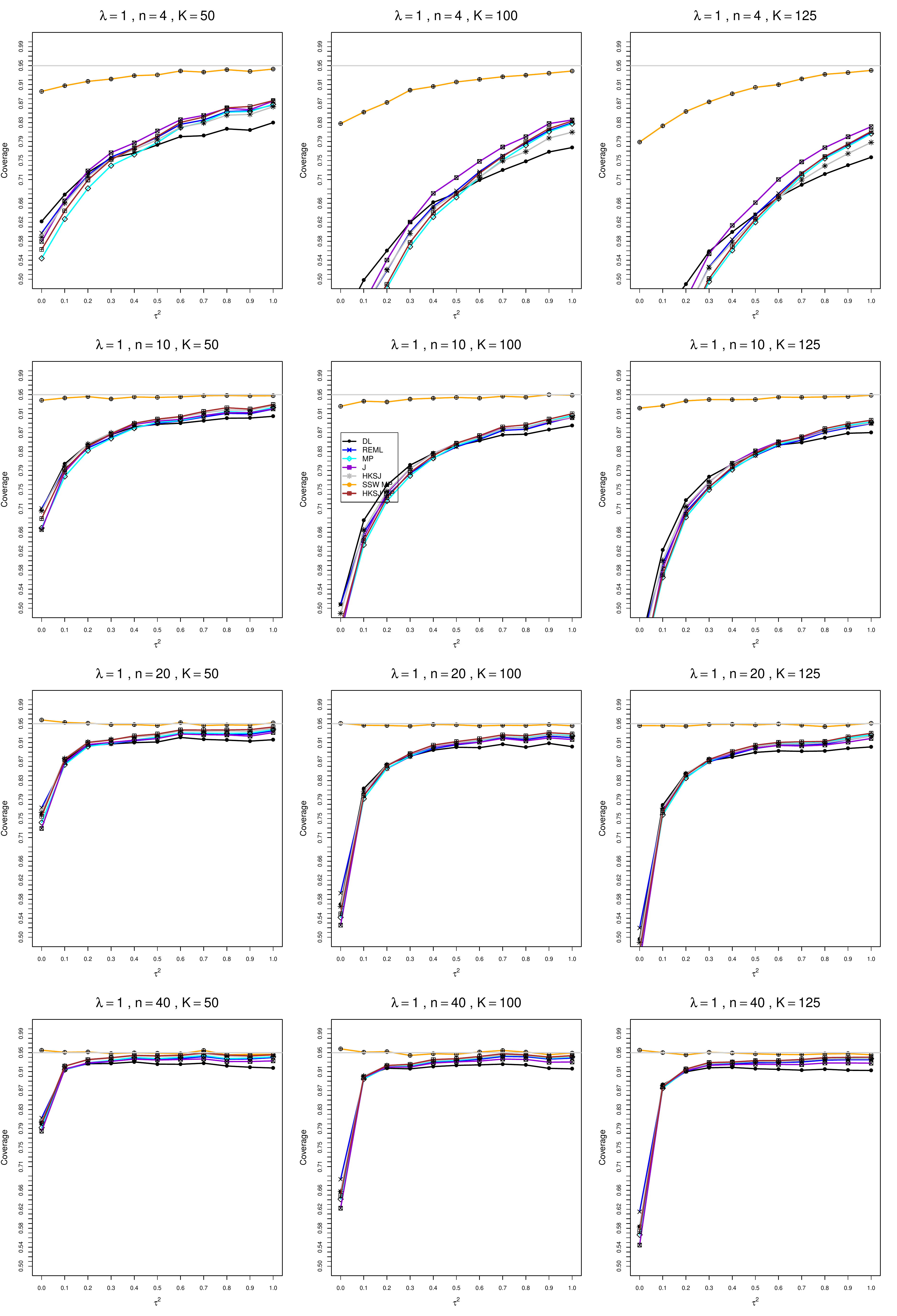}
	\caption{Coverage of 95\% confidence intervals for $\lambda$ when $\lambda=1$, $n = 4, \;10, \;20, \;40$, and $K = 50, \;100, \;125$. Bias-corrected estimate of $\lambda_i$		\label{CovThetaRoM1lnCor_smallN_large_K}}
\end{figure}
\begin{figure}[t]
	\includegraphics[scale=0.35]{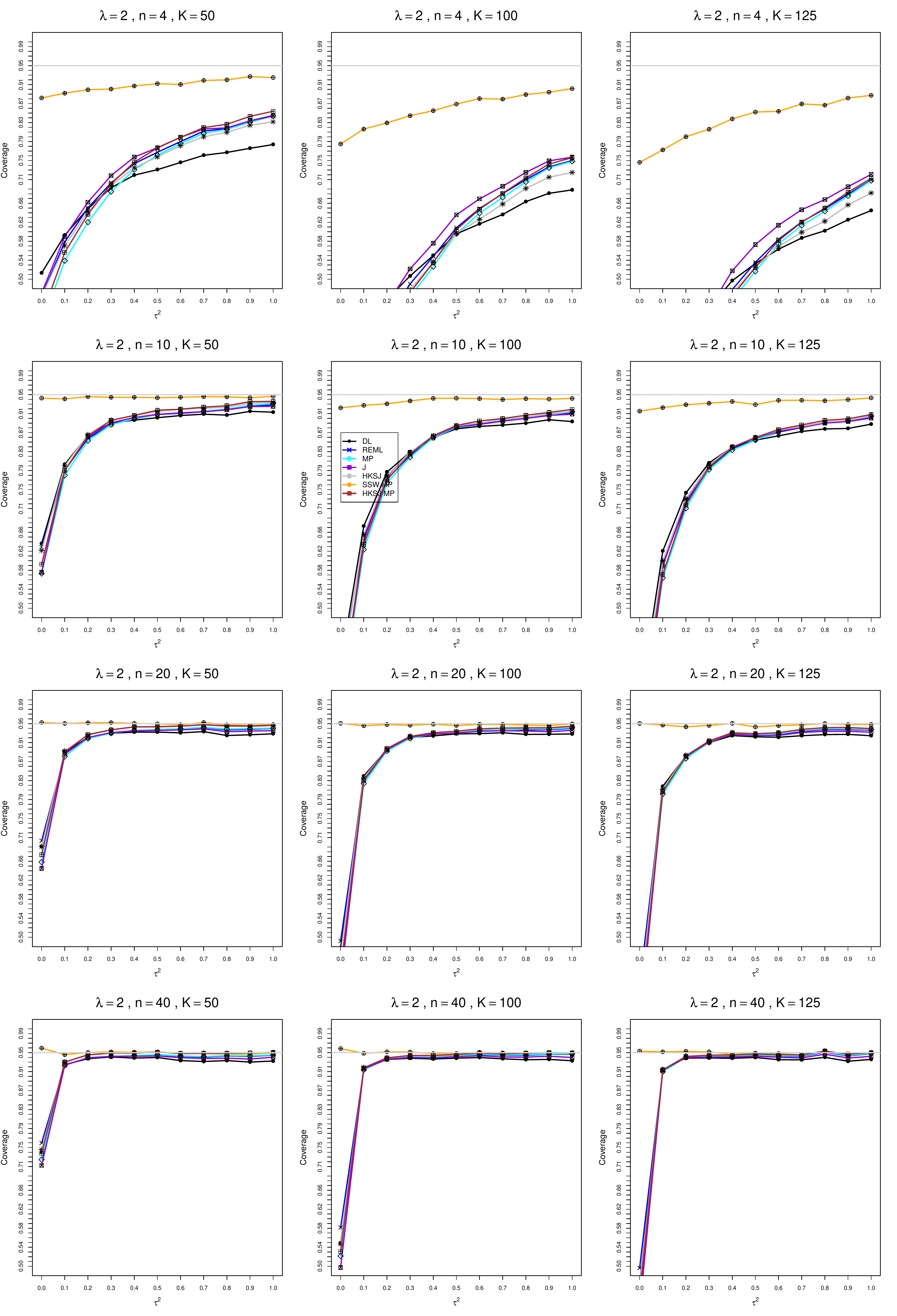}
	\caption{Coverage of 95\% confidence intervals for $\lambda$ when $\lambda=2$, $n = 4, \;10, \;20, \;40$, and $K = 50, \;100, \;125$. Bias-corrected estimate of $\lambda_i$ 		\label{CovThetaRoM2lnCor_smallN_large_K}}
\end{figure}

\clearpage
\renewcommand{\thefigure}{C1.\arabic{figure}}
\renewcommand{\thesection}{C1.\arabic{section}}
\setcounter{section}{0}
\setcounter{figure}{0}

\section*{C: Plots of bias and coverage of estimators of $\tau^2$, large $n$}

\begin{itemize}
	\item C1. Lognormal model, usual estimator of $\lambda_i$, $K=5,10,30$
	\item C2. Lognormal model, bias-corrected estimator of $\lambda_i$, $K=5,10,30$
	\item C3. Lognormal model, usual estimator of $\lambda_i$, $K=50,100,125$
	\item C4. Lognormal model, bias-corrected estimator of $\lambda_i$, $K=50,100,125$
\end{itemize}

\clearpage

\section*{C1. Lognormal model, usual estimator of $\lambda_i$, $n= 100,250,640,1000$, $K=5,10,30$}
\subsection*{C1.1 Bias of point estimators of $\tau^2$}
Each figure corresponds to a value of $\lambda \;(= 0, 0.2, 0.5, 1, 2)$, a set of values of $n$ (= 100, 250, 640, 1000), and a set of values of $K$ (= 5, 10, 30).\\
Each panel corresponds to a value of $n$ and a value of $K$ and has $\tau^2 = 0.0(0.1)1.0$ on the horizontal axis.\\
The point estimators of $\tau^2$ are
\begin{itemize}
	\item DL (DerSimonian-Laird)
	\item REML (restricted maximum likelihood)
	\item MP (Mandel-Paule)
	\item J (Jackson)
\end{itemize}
\clearpage

\setcounter{figure}{0}
\renewcommand{\thefigure}{C1.1.\arabic{figure}}
\begin{figure}[t]
	\includegraphics[scale=0.33]{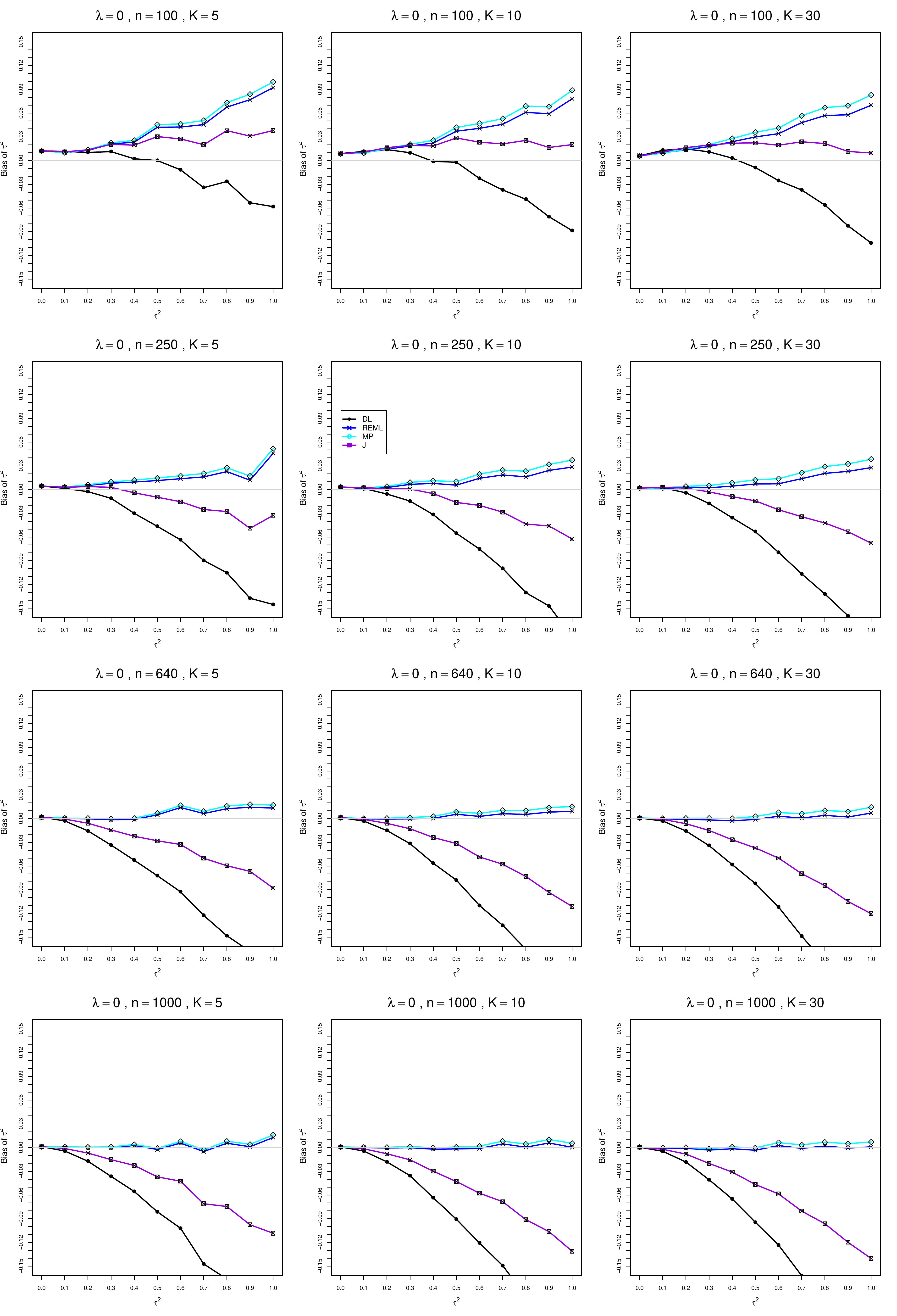}
	\caption{Bias of estimators of between-studies variance $\tau^2$ for $\lambda=0$, $n = 100, \;250, \;640, \;1000$, and $K = 5, \;10, \;30$. Usual estimate of $\lambda_i$
		\label{BiasTauRoM0ln_largeN_small_K}}
\end{figure}

\begin{figure}[t]
	\includegraphics[scale=0.33]{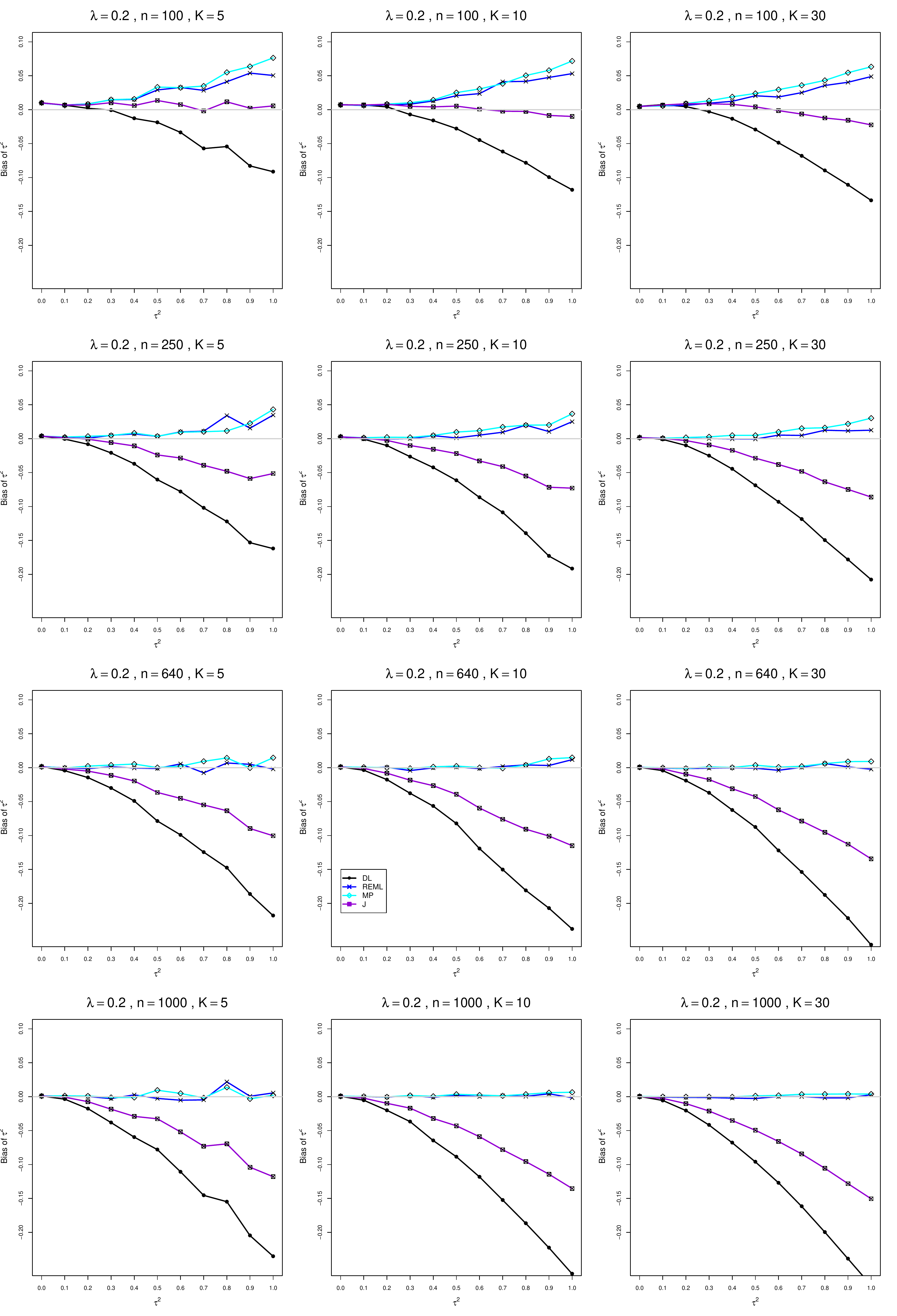}
	\caption{Bias of estimators of between-studies variance $\tau^2$ for $\lambda=0.2$, $n = 100, \;250, \;640, \;1000$, and $K = 5, \;10, \;30$. Usual estimate of $\lambda_i$
		\label{BiasTauRoM02ln_largeN_small_K}}
\end{figure}

\begin{figure}[t]
	\includegraphics[scale=0.33]{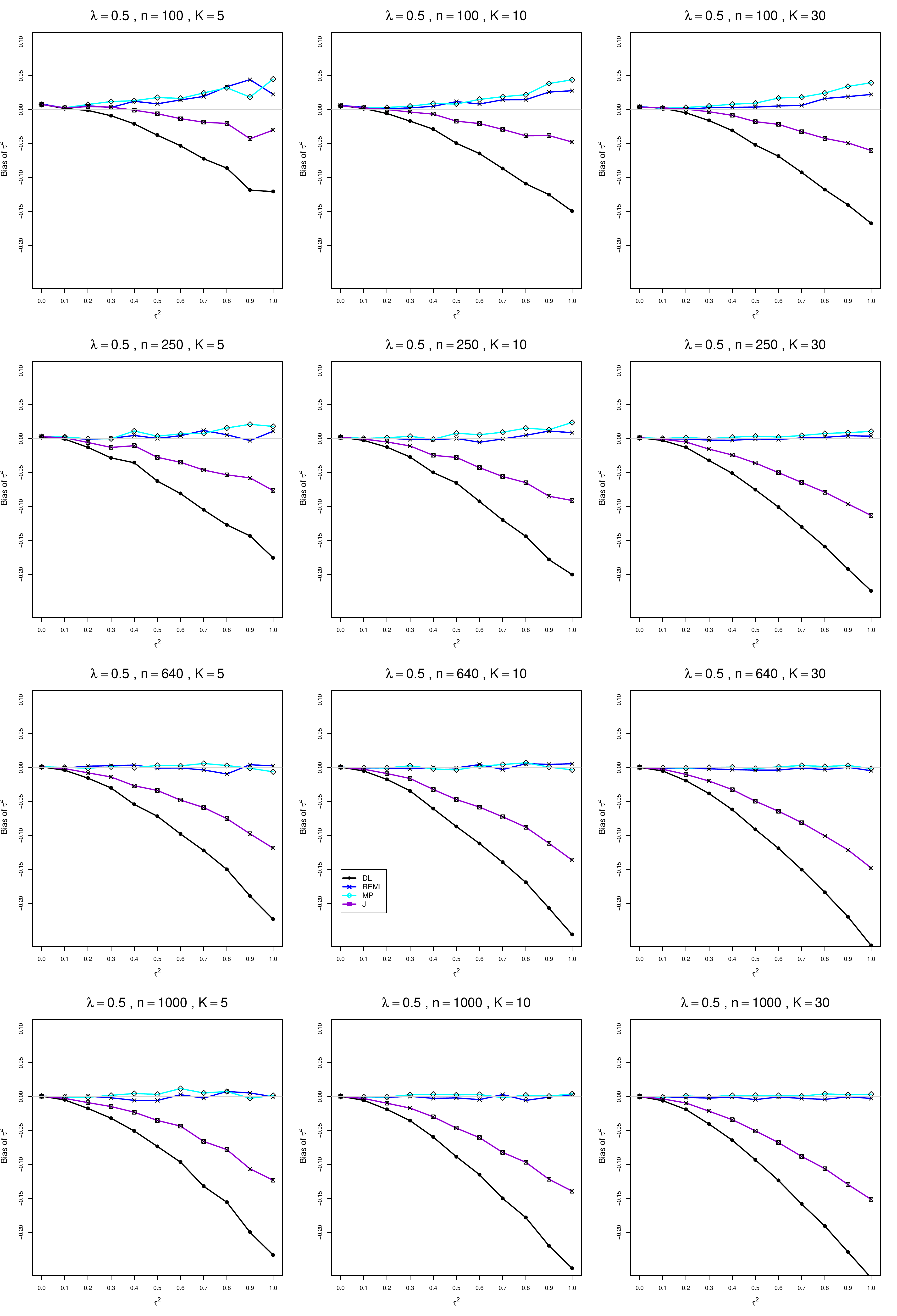}
	\caption{Bias of estimators of between-studies variance $\tau^2$ for $\lambda=0.5$, $n = 100, \;250, \;640, \;1000$, and $K = 5, \;10, \;30$. Usual estimate of $\lambda_i$
		\label{BiasTauRoM05ln_largeN_small_K}}
\end{figure}

\begin{figure}[t]
	\includegraphics[scale=0.33]{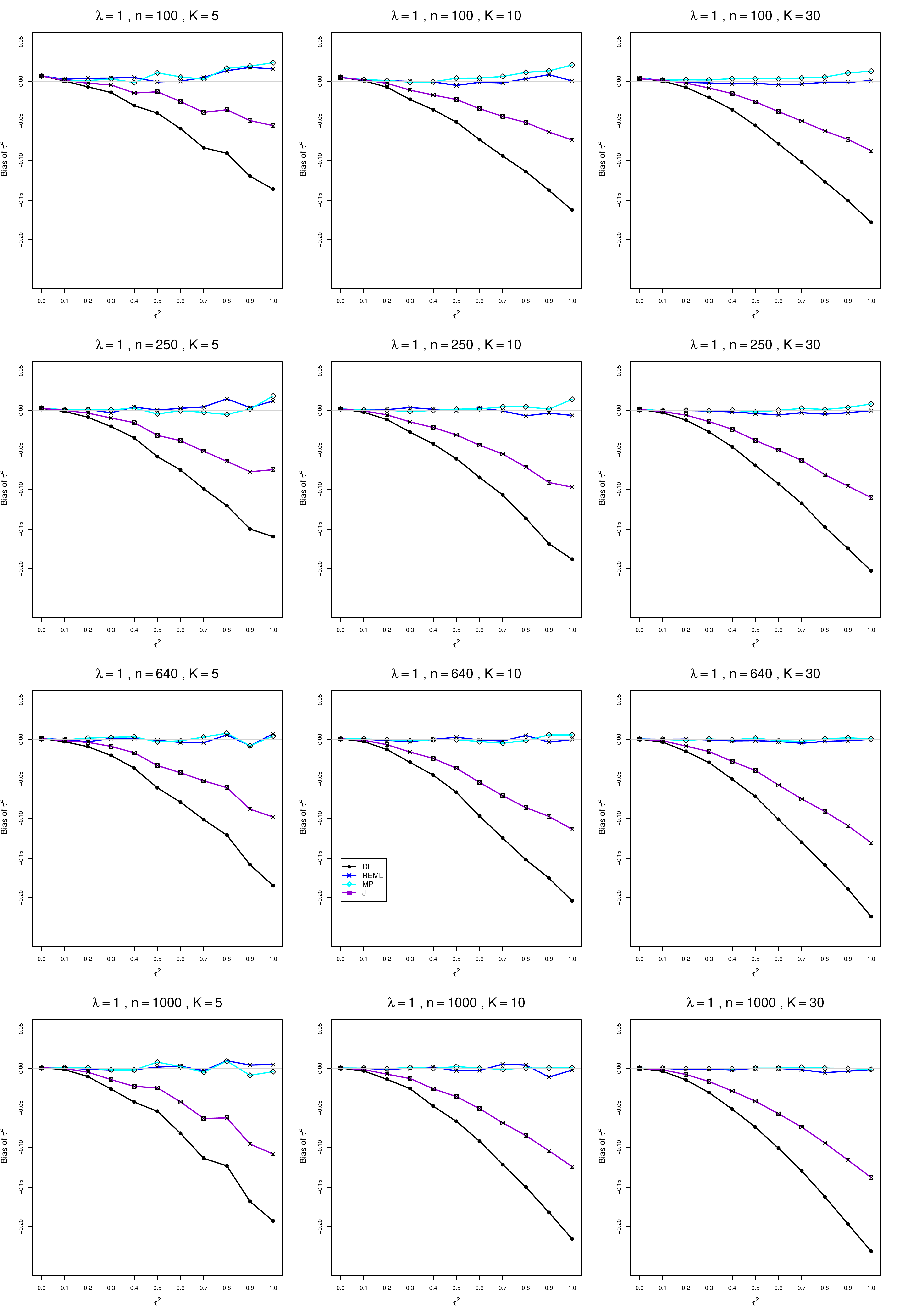}
	\caption{Bias of estimators of between-studies variance $\tau^2$ for $\lambda=1$, $n = 100, \;250, \;640, \;1000$, and $K = 5, \;10, \;30$. Usual estimate of $\lambda_i$
		\label{BiasTauRoM1ln_largeN_small_K}}
\end{figure}

\begin{figure}[t]
	\includegraphics[scale=0.33]{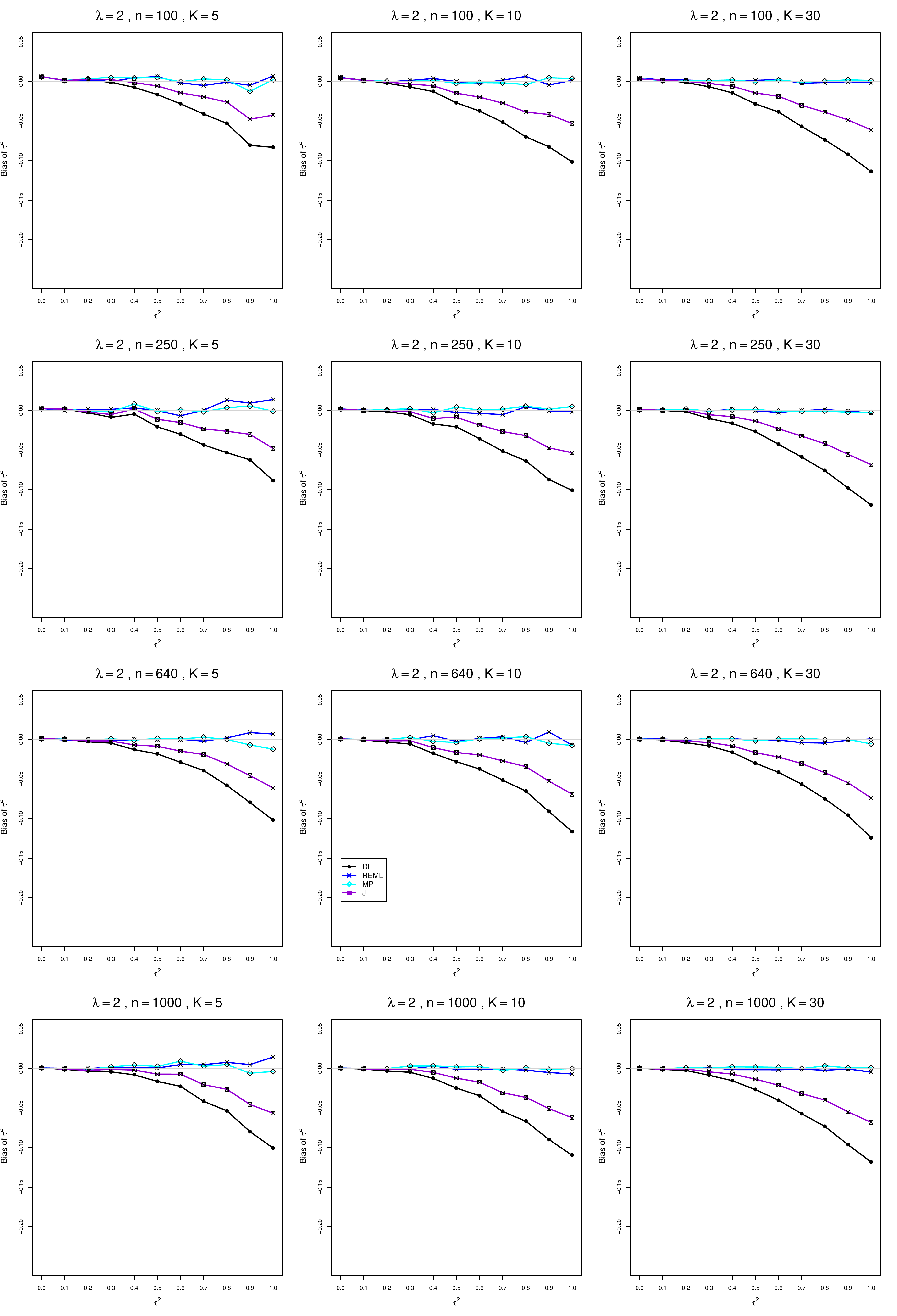}
	\caption{Bias of estimators of between-studies variance $\tau^2$ for $\lambda=2$, $n = 100, \;250, \;640, \;1000$, and $K = 5, \;10, \;30$. Usual estimate of $\lambda_i$
		\label{BiasTauRoM2ln_largeN_small_K}}
\end{figure}

\clearpage
\subsection*{C1.2 Coverage of interval estimators of $\tau^2$}
Each figure corresponds to a value of $\lambda \;(= 0, 0.2, 0.5, 1, 2)$, a set of values of $n$ (= 100, 250, 640, 1000), and a set of values of $K$ (= 5, 10, 30).\\
Each panel corresponds to a value of $n$ and a value of $K$ and has $\tau^2 = 0.0(0.1)1.0$ on the horizontal axis.\\
The interval estimators of $\tau^2$ are
\begin{itemize}
	\item QP (Q-profile confidence interval)
	\item BJ (Biggerstaff and Jackson interval)
	\item PL (Profile-likelihood interval)
	\item J (Jackson interval)
\end{itemize}

\setcounter{figure}{0}
\renewcommand{\thefigure}{C1.2.\arabic{figure}}
\begin{figure}[t]
	\includegraphics[scale=0.35]{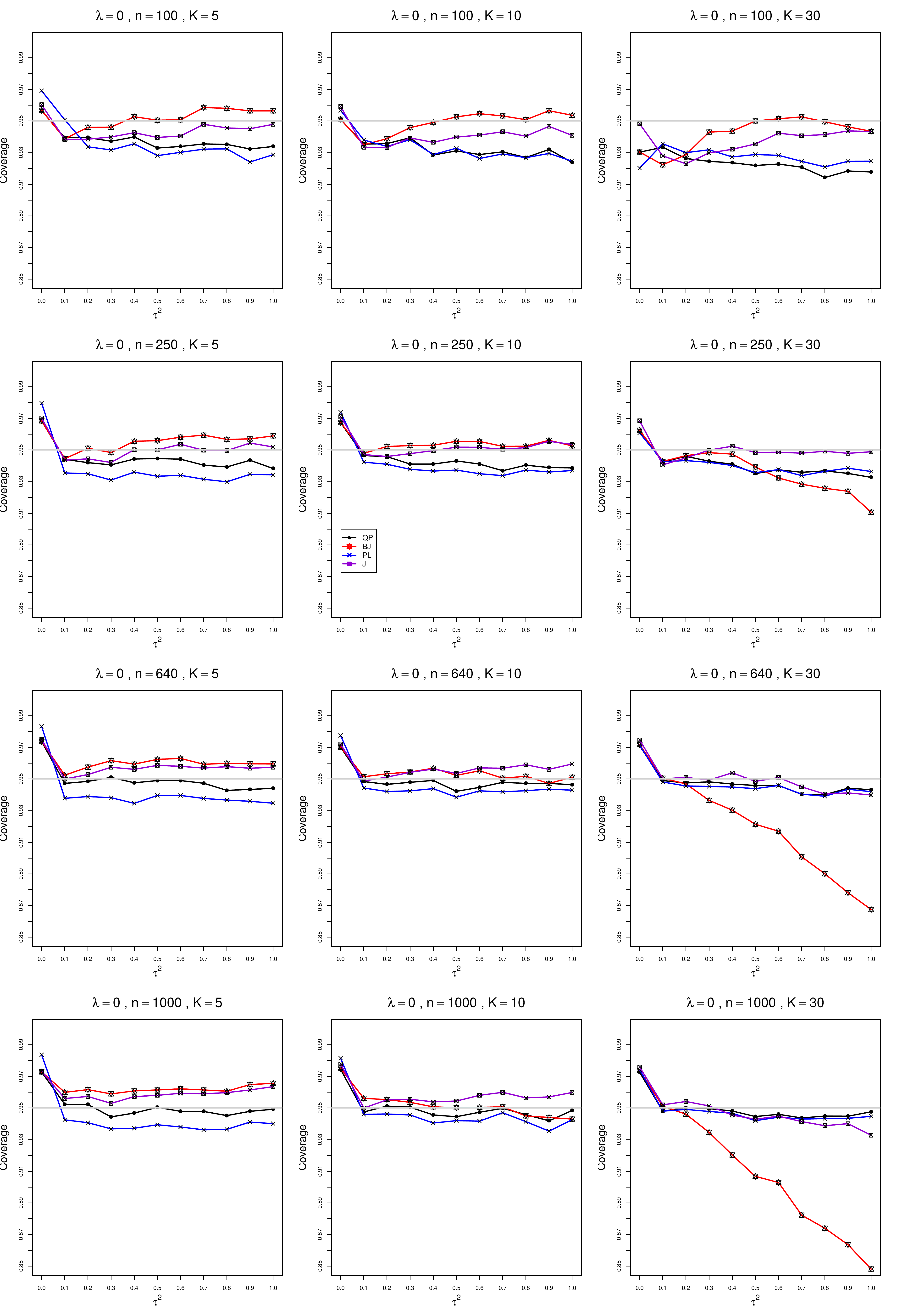}
	\caption{Coverage of 95\% confidence intervals for the between-studies variance $\tau^2$ when $\lambda=0$, $n = 100, \;250, \;640, \;1000$, and $K = 5, \;10, \;30$. Usual estimate of $\lambda_i$
		\label{CovTauRoM0ln_largeN_small_K}}
\end{figure}

\begin{figure}[t]
	\includegraphics[scale=0.35]{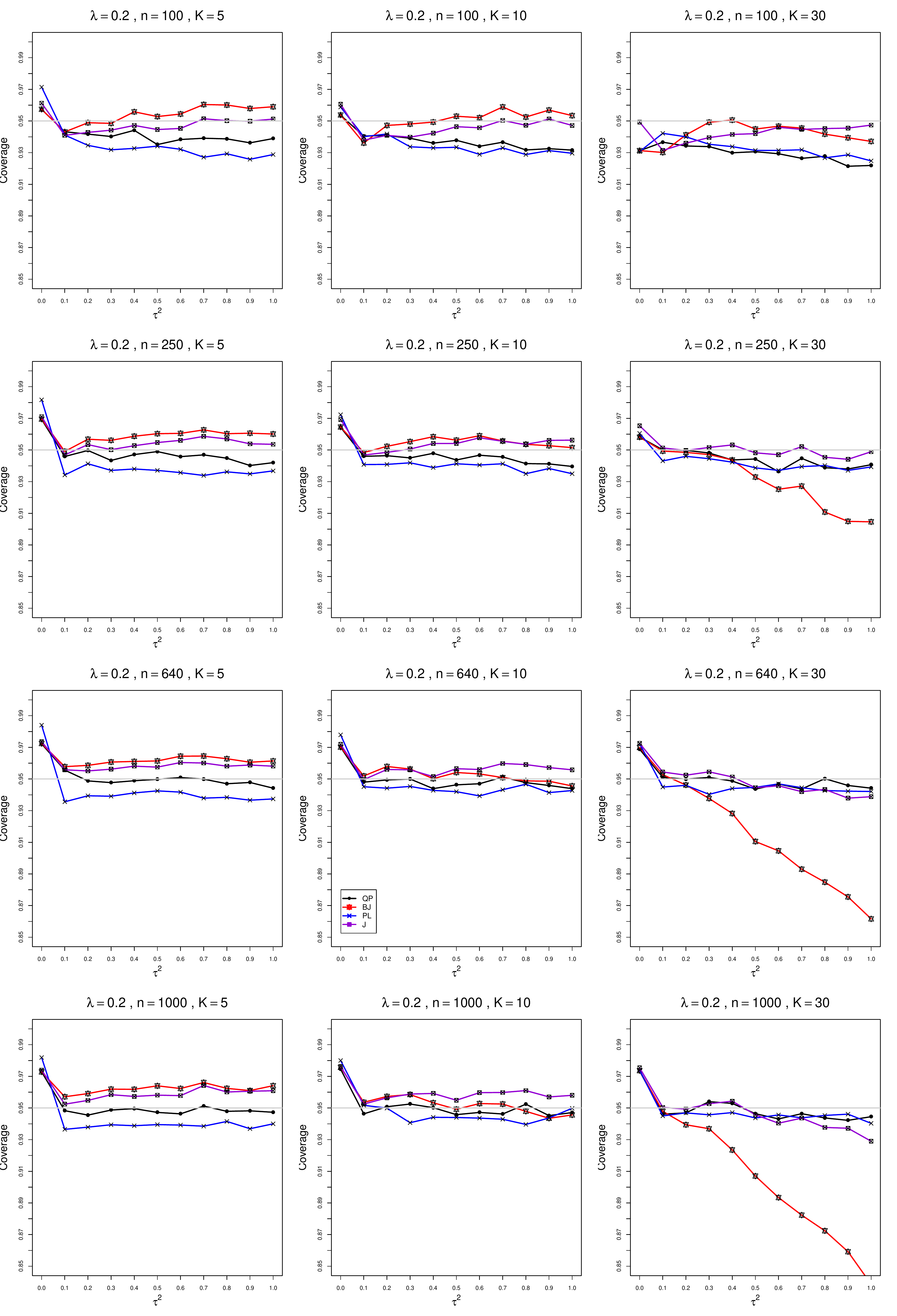}
	\caption{Coverage of 95\% confidence intervals for the between-studies variance $\tau^2$ when $\lambda=0.2$, $n = 100, \;250, \;640, \;1000$, and $K = 5, \;10, \;30$. Usual estimate of $\lambda_i$
		\label{CovTauRoM02ln_largeN_small_K}}
\end{figure}

\begin{figure}[t]
	\includegraphics[scale=0.35]{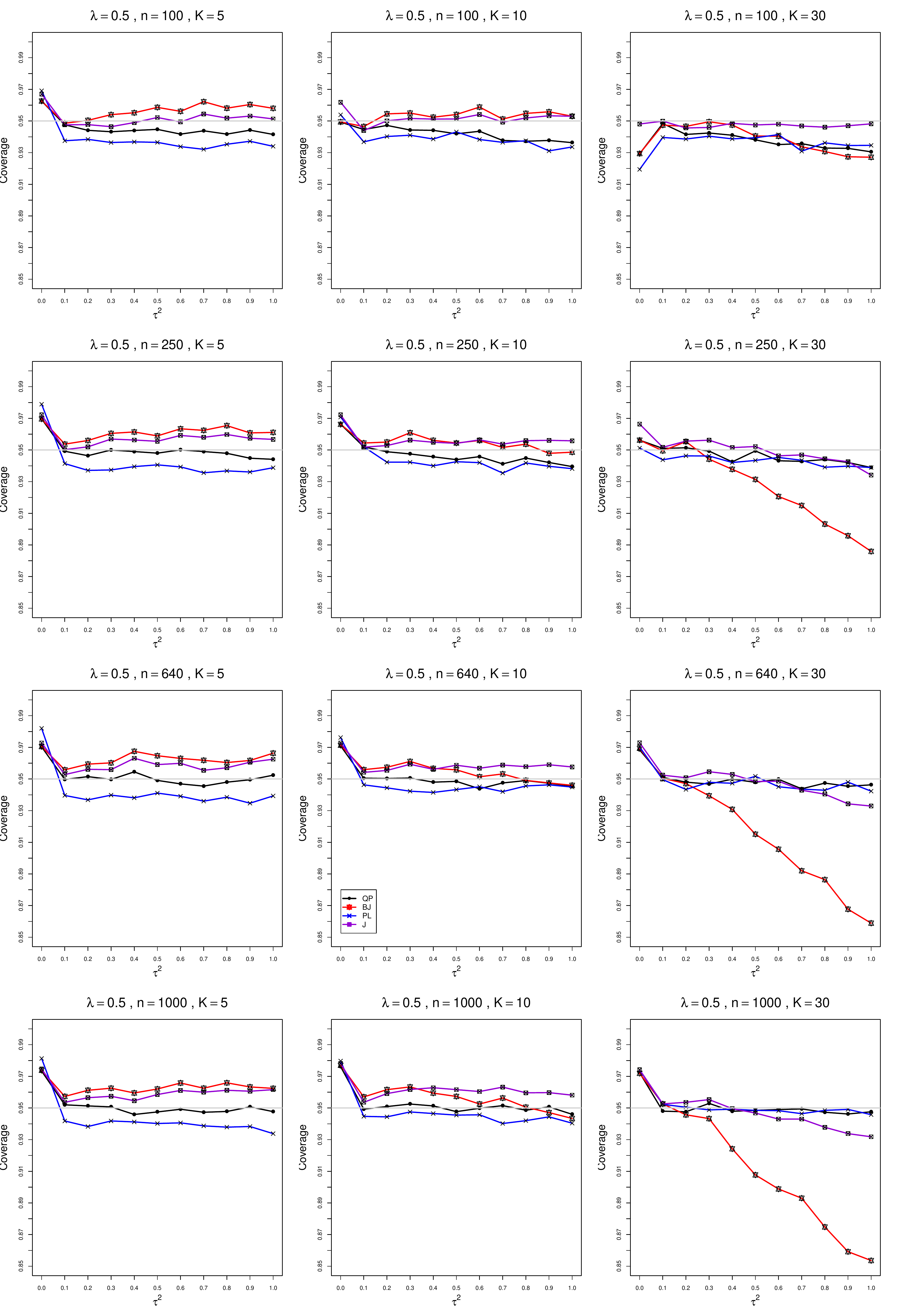}
	\caption{Coverage of 95\% confidence intervals for the between-studies variance $\tau^2$ when $\lambda=0.5$, $n = 100, \;250, \;640, \;1000$, and $K = 5, \;10, \;30$. Usual estimate of $\lambda_i$.
		\label{CovTauRoM05ln_largeN_small_K}}
\end{figure}

\begin{figure}[t]
	\includegraphics[scale=0.35]{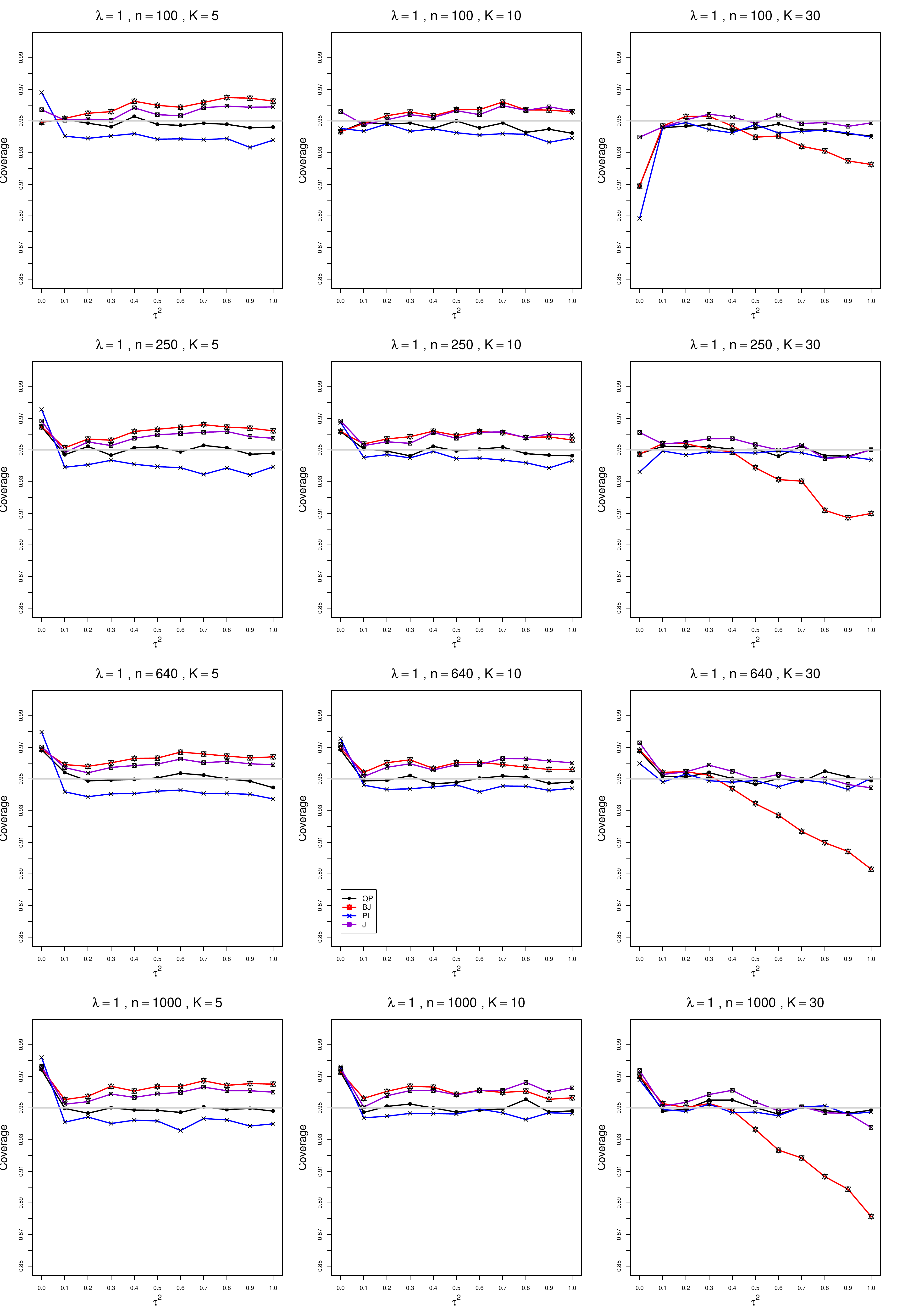}
	\caption{Coverage of 95\% confidence intervals for the between-studies variance $\tau^2$ when $\lambda=1$, $n = 100, \;250, \;640, \;1000$, and $K = 5, \;10, \;30$. Usual estimate of $\lambda_i$
		\label{CovTauRoM1ln_largeN_small_K}}
\end{figure}

\begin{figure}[t]
	\includegraphics[scale=0.35]{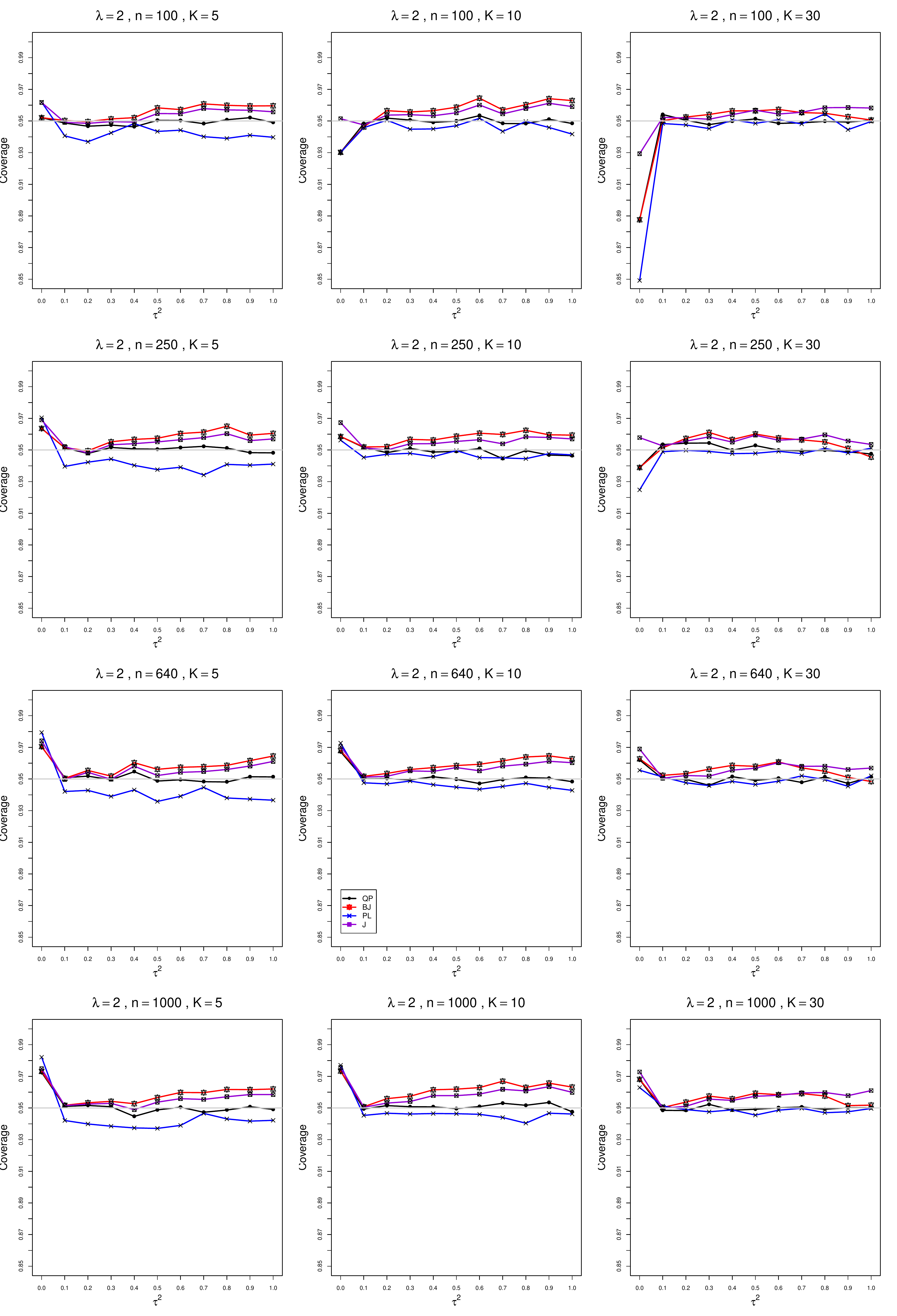}
	\caption{Coverage of 95\% confidence intervals for the between-studies variance $\tau^2$ when $\lambda=2$, $n = 100, \;250, \;640, \;1000$, and $K = 5, \;10, \;30$. Usual estimate of $\lambda_i$.
		\label{CovTauRoM2ln_largeN_small_K}}
\end{figure}

\clearpage

\section*{C2. Lognormal model, bias-corrected estimator of $\lambda_i$, $n= 100, 250, 640, 1000$, $K=5,10,30$}
\subsection*{C2.1 Bias of point estimators of $\tau^2$}
Each figure corresponds to a value of $\lambda \;(= 0, 0.2, 0.5, 1, 2)$, a set of values of $n$ (= 100, 250, 640, 1000), and a set of values of $K$ (= 5, 10, 30).\\
Each panel corresponds to a value of $n$ and a value of $K$ and has $\tau^2 = 0.0(0.1)1.0$ on the horizontal axis.\\
The point estimators of $\tau^2$ are
\begin{itemize}
	\item DL (DerSimonian-Laird)
	\item REML (restricted maximum likelihood)
	\item MP (Mandel-Paule)
	\item J (Jackson)
\end{itemize}

\clearpage
\setcounter{figure}{0}
\renewcommand{\thefigure}{C2.1.\arabic{figure}}
\begin{figure}[t]
	\includegraphics[scale=0.33]{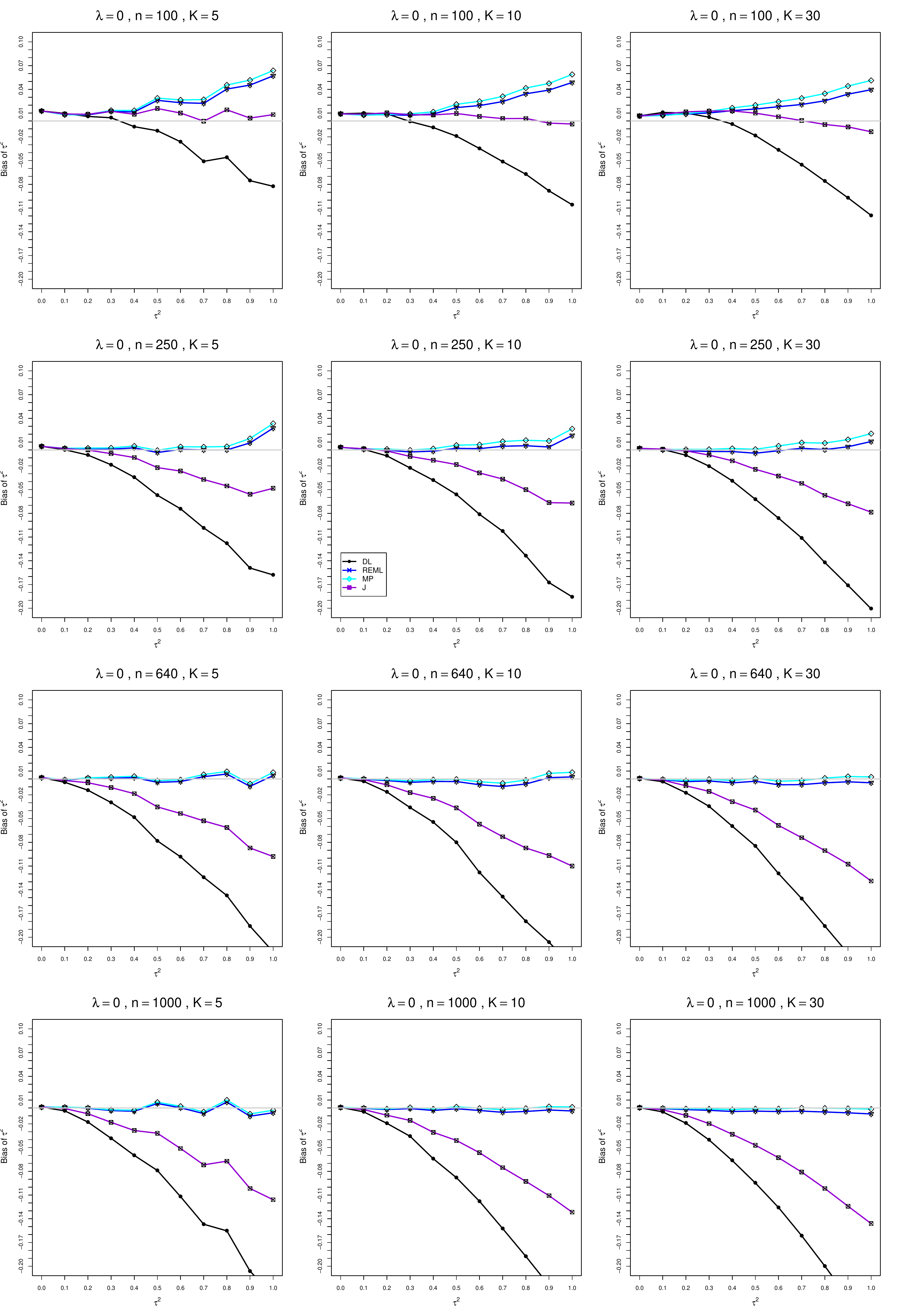}
	\caption{Bias of estimators of between-studies variance $\tau^2$ for $\lambda=0$, $n = 100, \;250, \;640, \;1000$, and $K = 5, \;10, \;30$. Bias-corrected estimate of $\lambda_i$
		\label{BiasTauRoM0lnCor_largeN_small_K}}
\end{figure}

\begin{figure}[t]
	\includegraphics[scale=0.33]{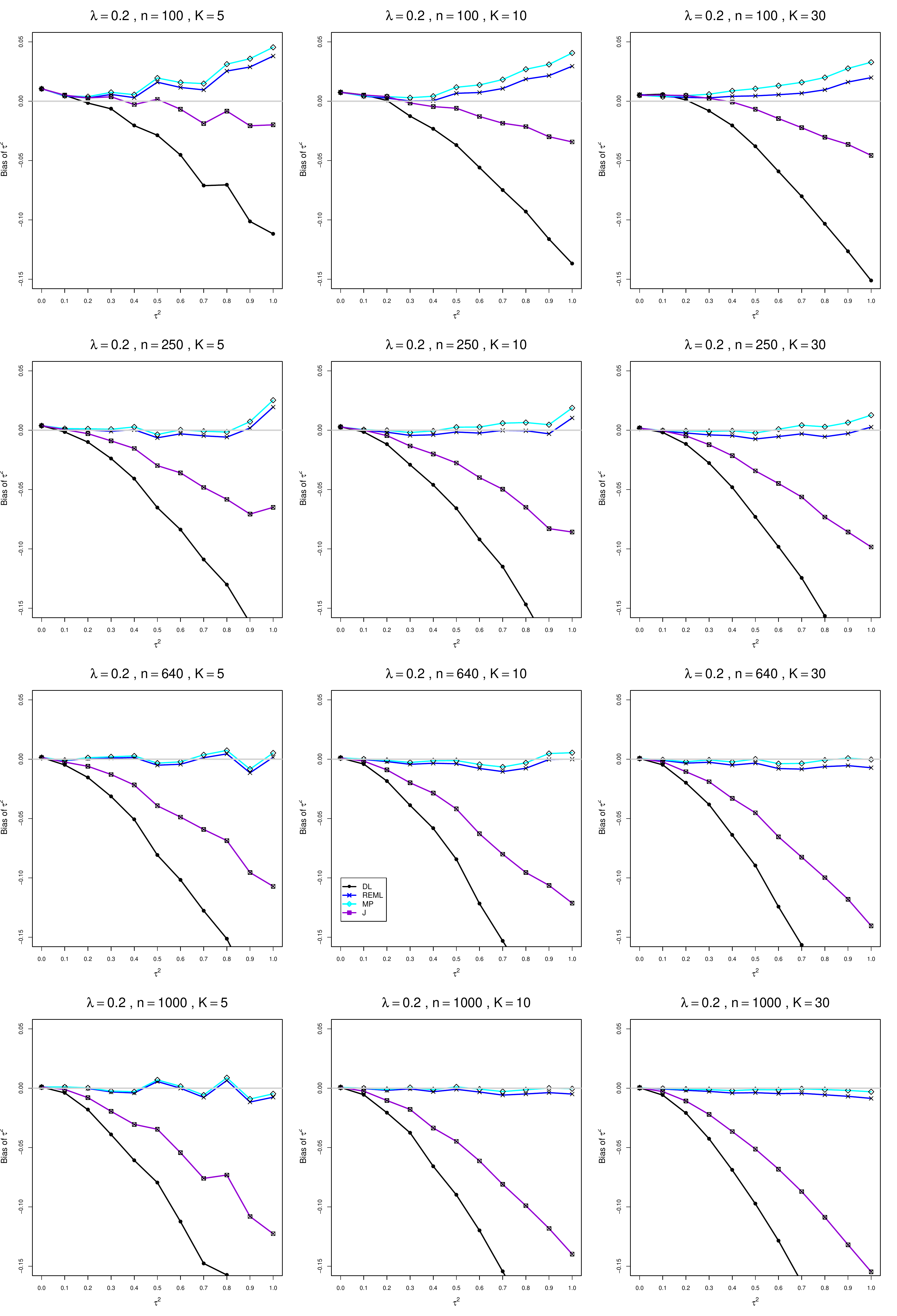}
	\caption{Bias of estimators of between-studies variance $\tau^2$ for $\lambda=0.2$, $n = 100, \;250, \;640, \;1000$, and $K = 5, \;10, \;30$. Bias-corrected estimate of $\lambda_i$
		\label{BiasTauRoM02lnCor_largeN_small_K}}
\end{figure}

\begin{figure}[t]
	\includegraphics[scale=0.33]{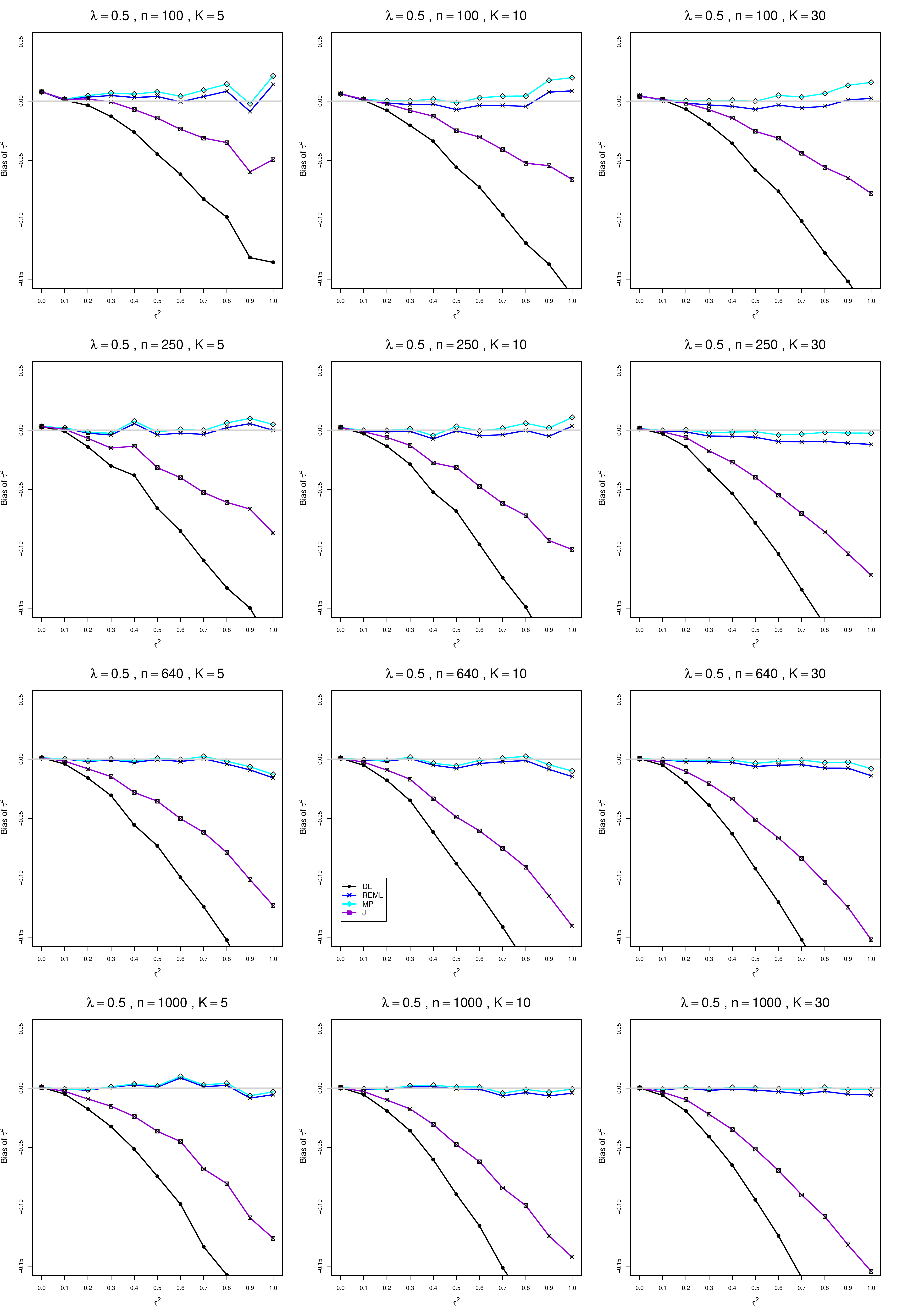}
	\caption{Bias of estimators of between-studies variance $\tau^2$ for $\lambda=0.5$, $n = 100, \;250, \;640, \;1000$, and $K = 5, \;10, \;30$. Bias-corrected estimate of $\lambda_i$
		\label{BiasTauRoM05lnCor_largeN_small_K}}
\end{figure}

\begin{figure}[t]
	\includegraphics[scale=0.33]{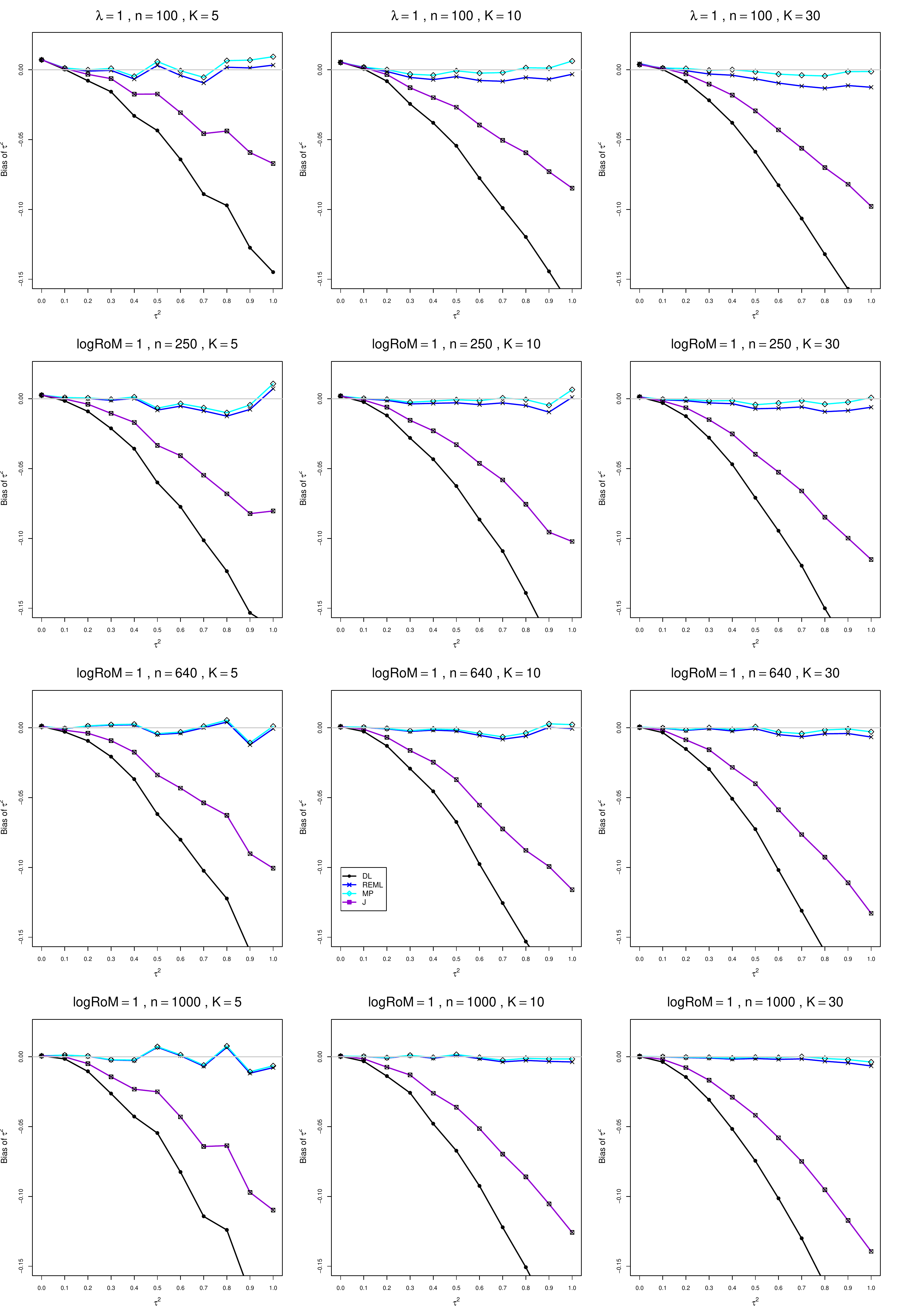}
	\caption{Bias of estimators of between-studies variance $\tau^2$ for $\lambda=1$, $n = 100, \;250, \;640, \;1000$, and $K = 5, \;10, \;30$. Bias-corrected estimate of $\lambda_i$
		\label{BiasTauRoM1lnCor_largeN_small_K}}
\end{figure}

\begin{figure}[t]
	\includegraphics[scale=0.33]{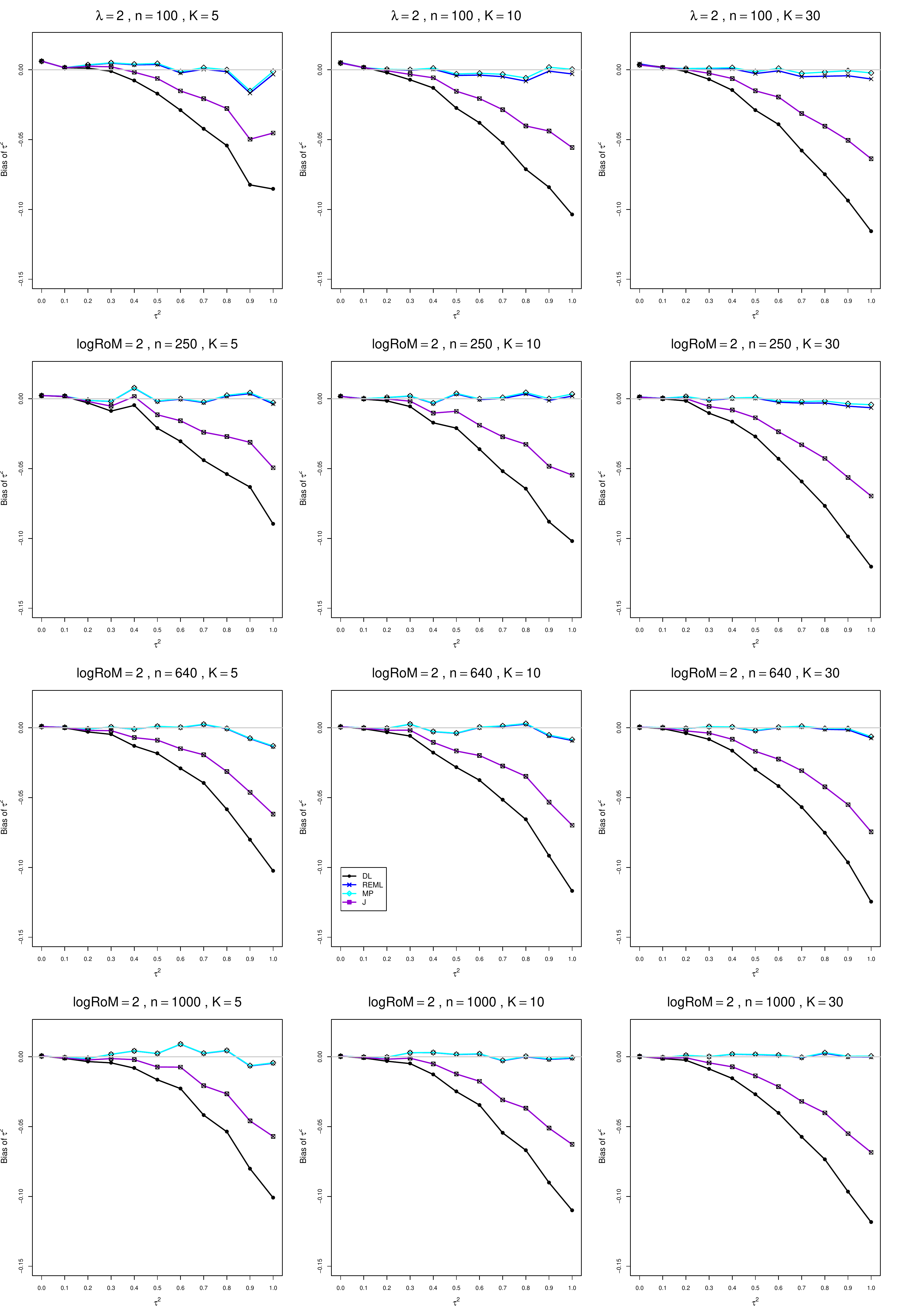}
	\caption{Bias of estimators of between-studies variance $\tau^2$ for $\lambda=2$, $n = 100, \;250, \;640, \;1000$, and $K = 5, \;10, \;30$. Bias-corrected estimate of $\lambda_i$
		\label{BiasTauRoM2lnCor_largeN_small_K}}
\end{figure}

\clearpage
\subsection*{C2.2 Coverage of interval estimators of $\tau^2$}
Each figure corresponds to a value of $\lambda \;(= 0, 0.2, 0.5, 1, 2)$, a set of values of $n$ (= 100, 250, 640, 1000), and a set of values of $K$ (= 5, 10, 30).\\
Each panel corresponds to a value of $n$ and a value of $K$ and has $\tau^2 = 0.0(0.1)1.0$ on the horizontal axis.\\
The interval estimators of $\tau^2$ are
\begin{itemize}
	\item QP (Q-profile confidence interval)
	\item BJ (Biggerstaff and Jackson interval )
	\item PL (Profile-likelihood interval)
	\item J (Jackson interval)
\end{itemize}

\setcounter{figure}{0}
\renewcommand{\thefigure}{C2.2.\arabic{figure}}
\begin{figure}[t]
	\includegraphics[scale=0.35]{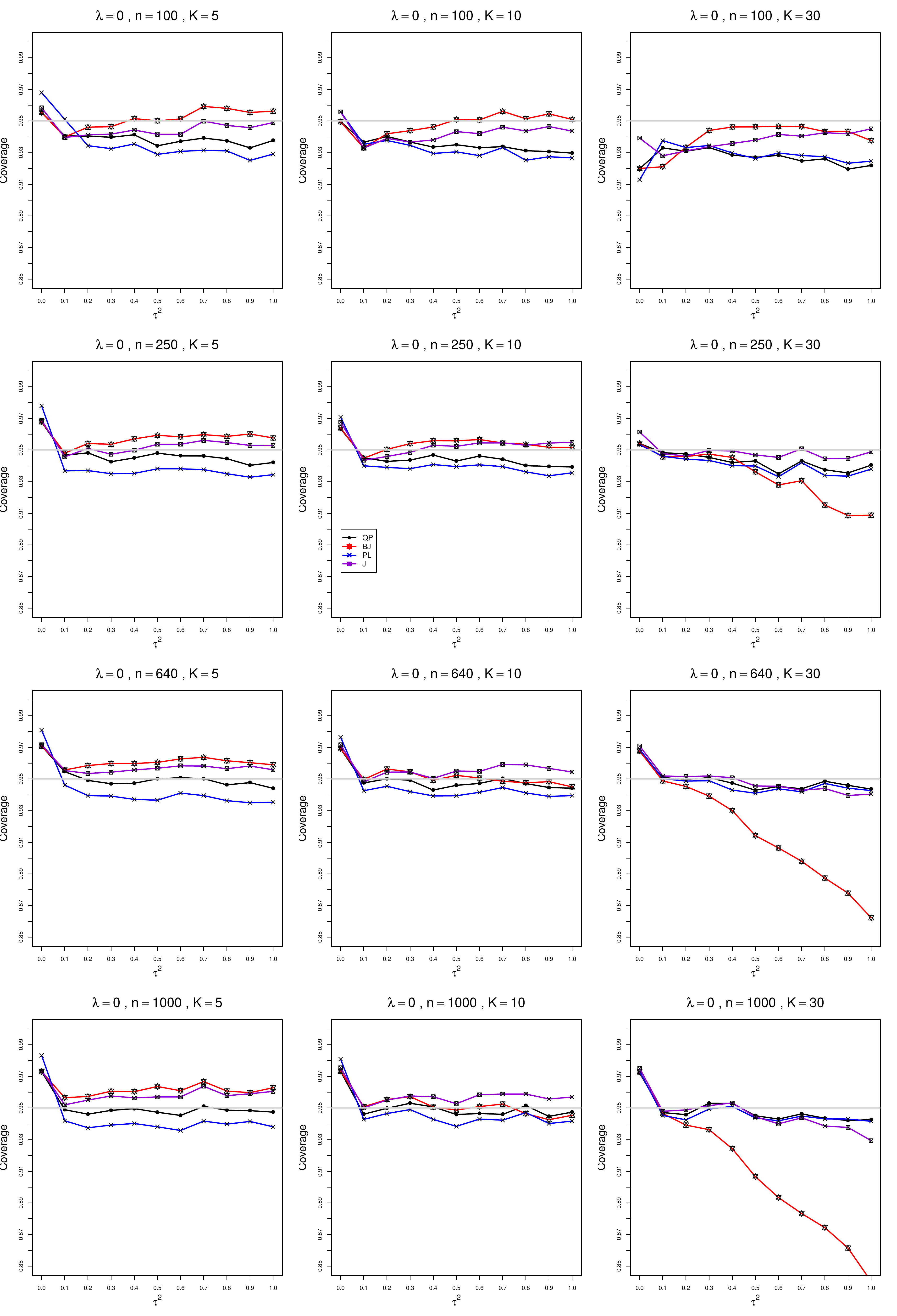}
	\caption{Coverage of 95\% confidence intervals for the between-studies variance $\tau^2$ when $\lambda=0$, $n = 100, \;250, \;640, \;1000$, and $K = 5, \;10, \;30$. Bias-corrected estimate of $\lambda_i$ 		\label{CovTauRoM0lnCor_largeN_small_K}}
\end{figure}

\begin{figure}[t]
	\includegraphics[scale=0.35]{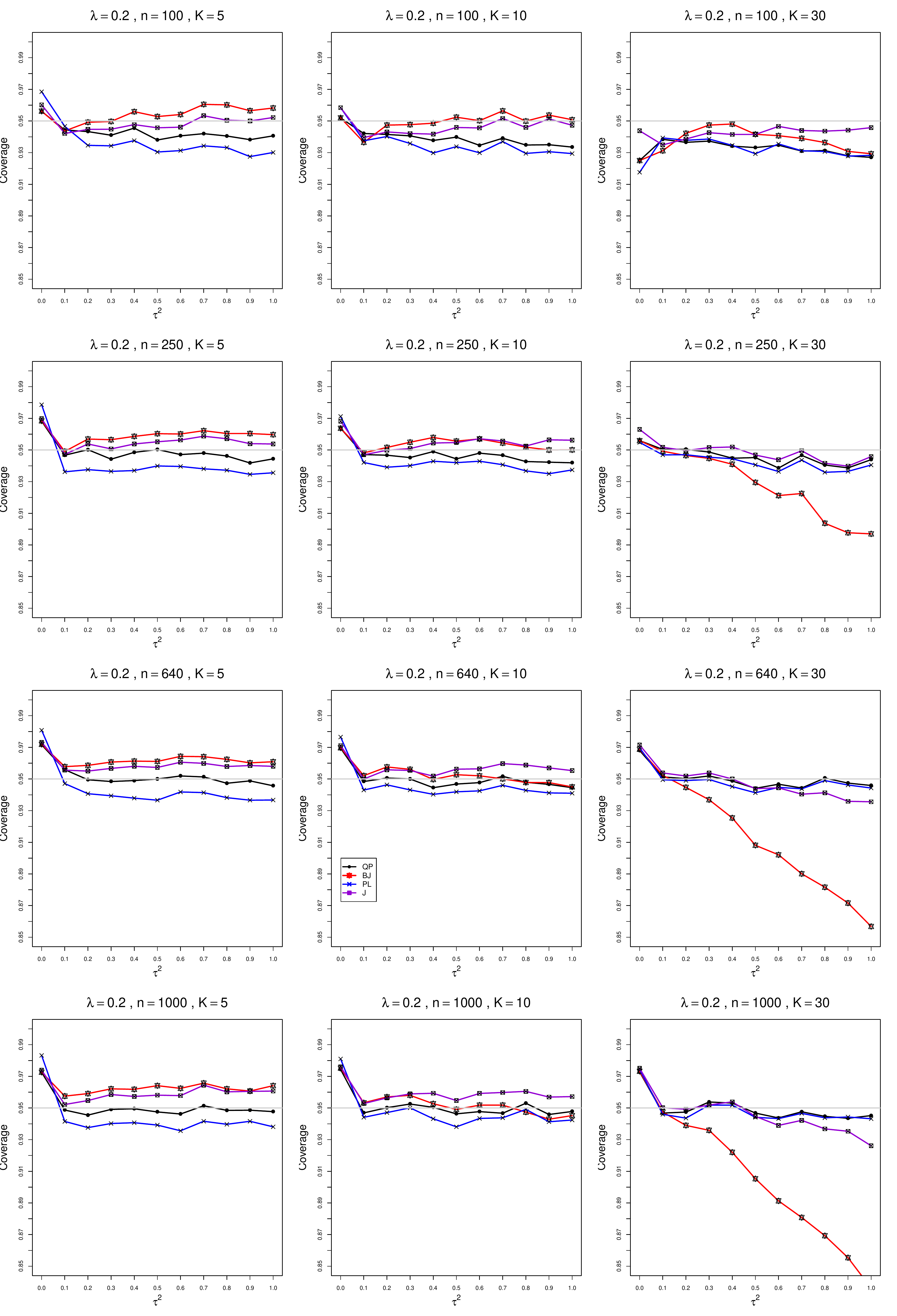}
	\caption{Coverage of 95\% confidence intervals for the between-studies variance $\tau^2$ when $\lambda=0.2$, $n = 100, \;250, \;640, \;1000$, and $K = 5, \;10, \;30$. Bias-corrected estimate of $\lambda_i$ 		\label{CovTauRoM02lnCor_largeN_small_K}}
\end{figure}

\begin{figure}[t]
	\includegraphics[scale=0.35]{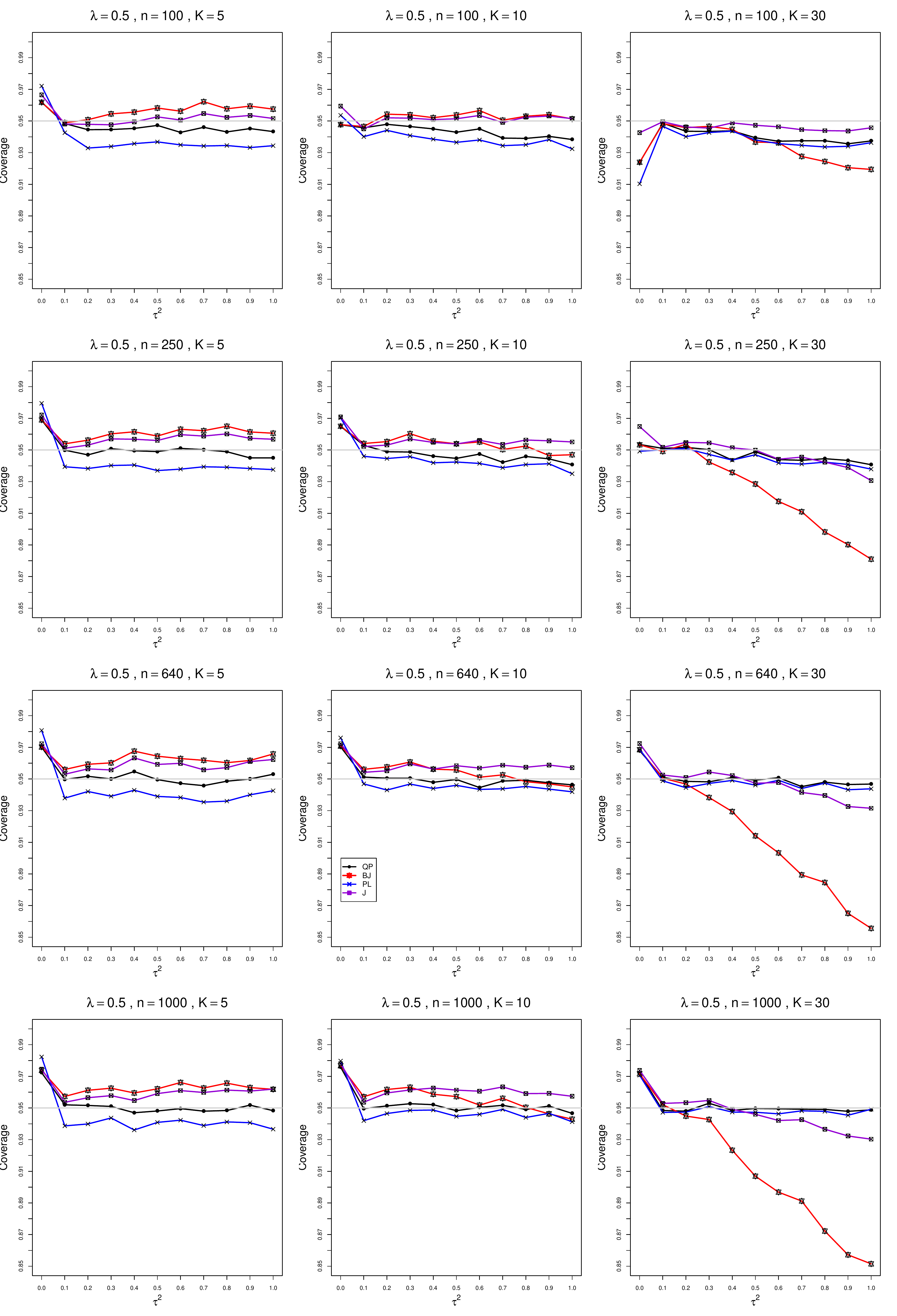}
	\caption{Coverage of 95\% confidence intervals for the between-studies variance $\tau^2$ when $\lambda=0.5$, $n = 100, \;250, \;640, \;1000$, and $K = 5, \;10, \;30$. Bias-corrected estimate of $\lambda_i$ 		\label{CovTauRoM05lnCor_largeN_small_K}}
\end{figure}

\begin{figure}[t]
	\includegraphics[scale=0.35]{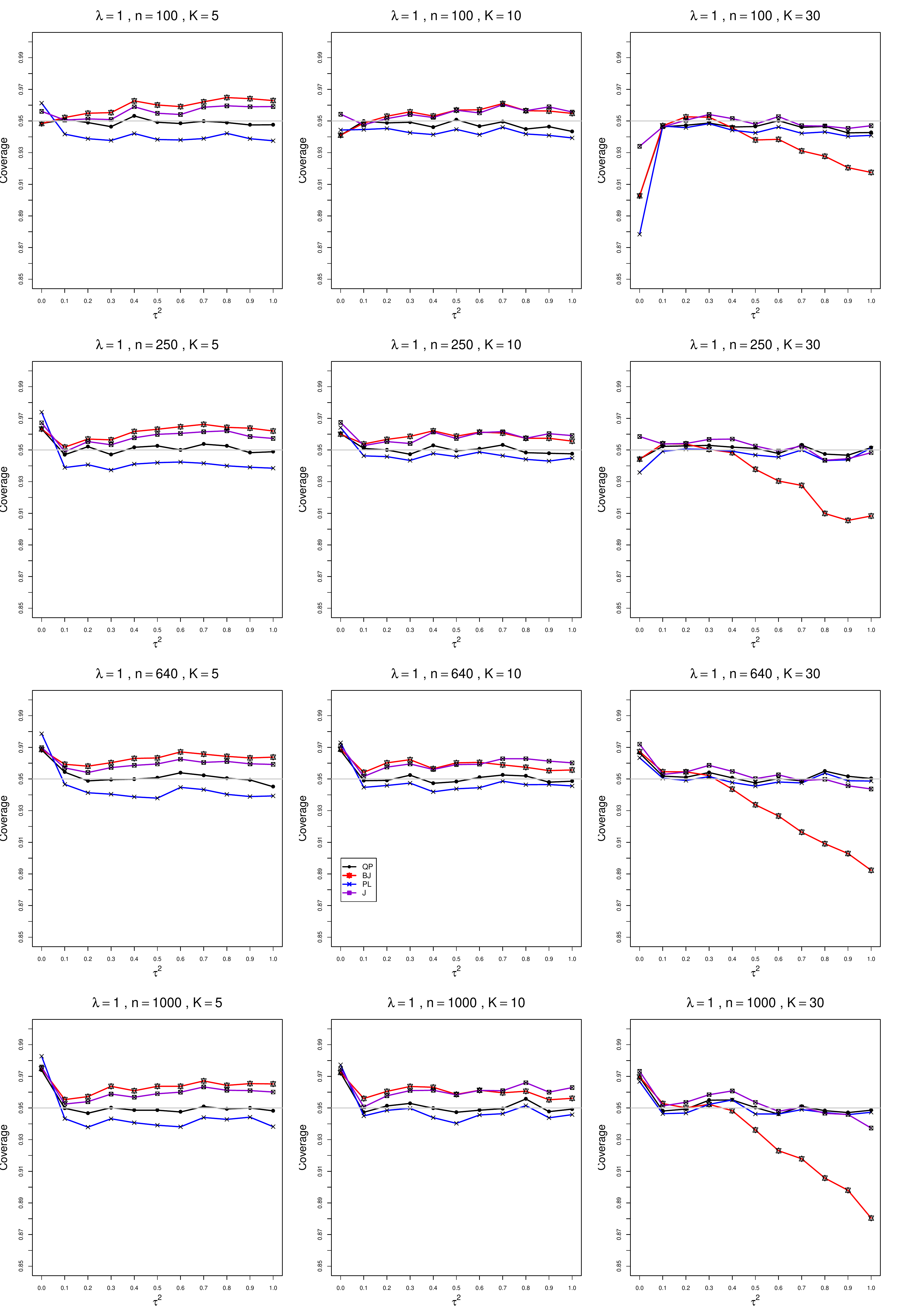}
	\caption{Coverage of 95\% confidence intervals for the between-studies variance $\tau^2$ when $\lambda=1$, $n = 100, \;250, \;640, \;1000$, and $K = 5, \;10, \;30$. Bias-corrected estimate of $\lambda_i$ 		\label{CovTauRoM1lnCor_largeN_small_K}}
\end{figure}

\begin{figure}[t]
	\includegraphics[scale=0.35]{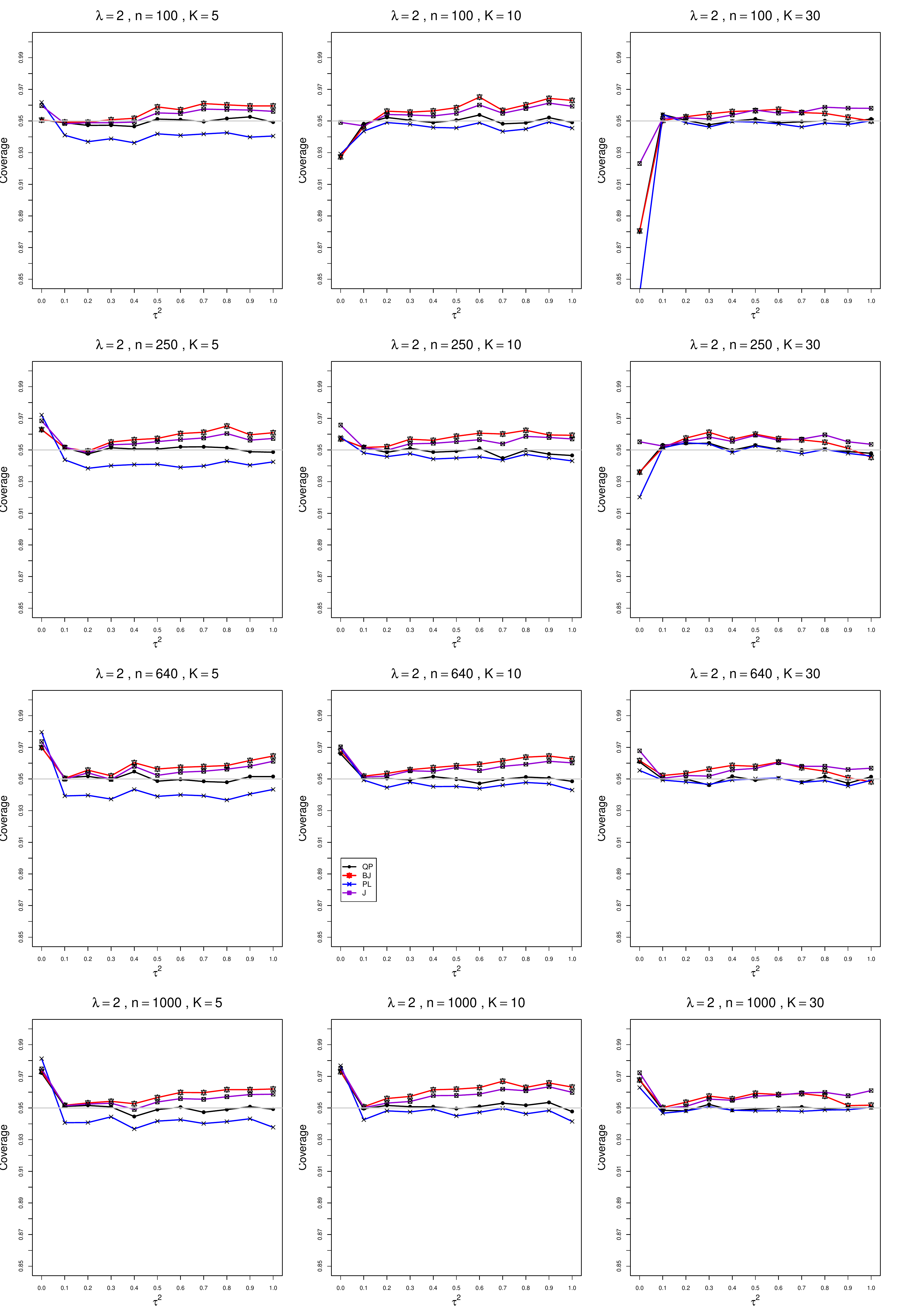}
	\caption{Coverage of 95\% confidence intervals for the between-studies variance $\tau^2$ when $\lambda=2$, $n = 100, \;250, \;640, \;1000$, and $K = 5, \;10, \;30$. Bias-corrected estimate of $\lambda_i$ 		\label{CovTauRoM2lnCor_largeN_small_K}}
\end{figure}

\clearpage

\section*{C3. Lognormal model, usual estimator of $\lambda_i$, $n= 100,250,640,1000$, $K=50,100,125$}
\subsection*{C3.1 Bias of point estimators of $\tau^2$}
Each figure corresponds to a value of $\lambda \;(= 0, 0.2, 0.5, 1, 2)$, a set of values of $n$ (= 100, 250, 640, 1000), and a set of values of $K$ (= 50, 100, 125).\\
Each panel corresponds to a value of $n$ and a value of $K$ and has $\tau^2 = 0.0(0.1)1.0$ on the horizontal axis.\\
The point estimators of $\tau^2$ are
\begin{itemize}
	\item DL (DerSimonian-Laird)
	\item REML (restricted maximum likelihood)
	\item MP (Mandel-Paule)
	\item J (Jackson)
\end{itemize}
\clearpage

\setcounter{figure}{0}
\renewcommand{\thefigure}{C3.1.\arabic{figure}}
\begin{figure}[t]
	\includegraphics[scale=0.33]{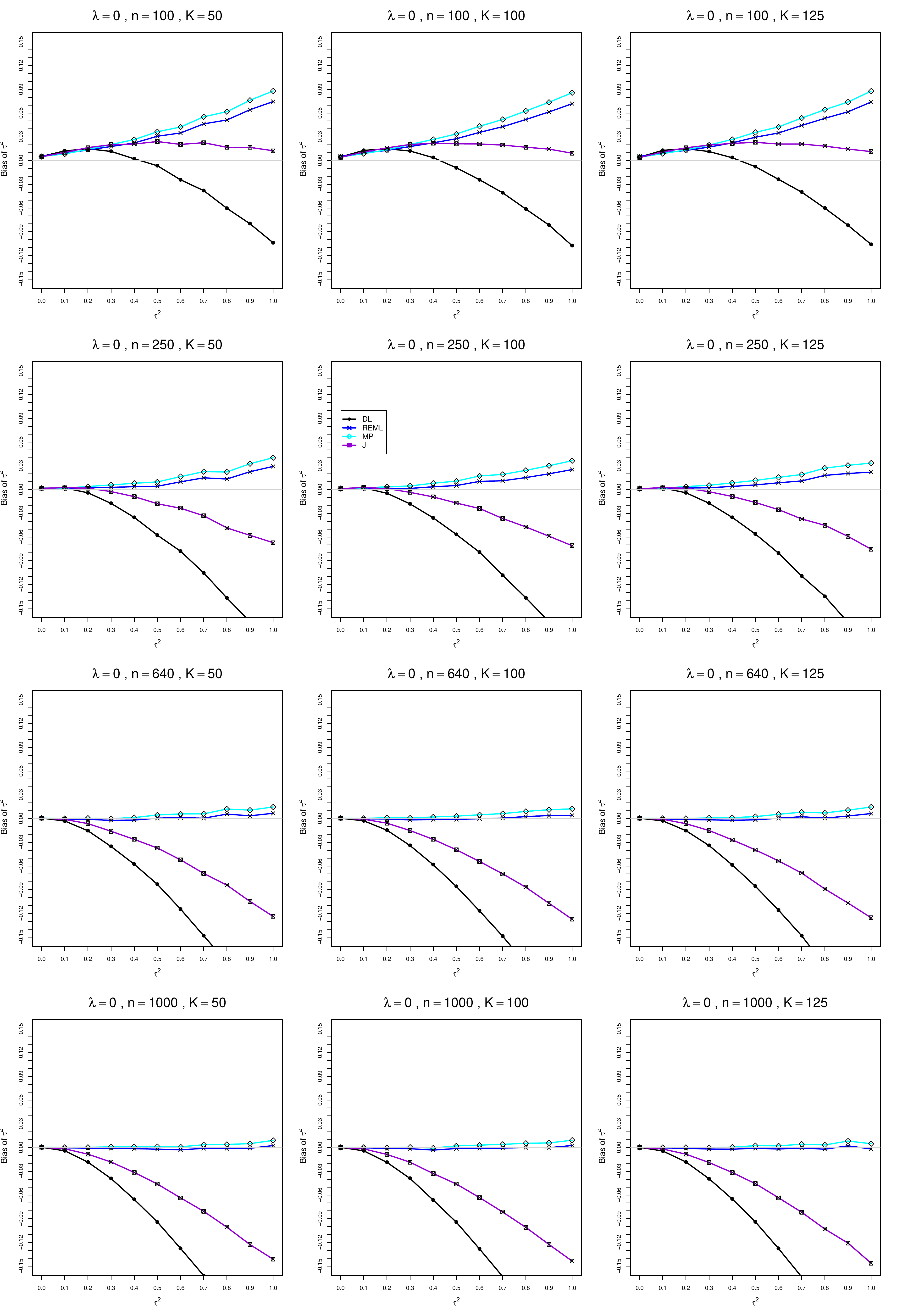}
	\caption{Bias of estimators of between-studies variance $\tau^2$ for $\lambda=0$, $n = 100, \;250, \;640, \;1000$, and $K = 50, \;100, \;125$. Usual estimate of $\lambda_i$
		\label{BiasTauRoM0ln_largeN_large_K}}
\end{figure}

\begin{figure}[t]
	\includegraphics[scale=0.33]{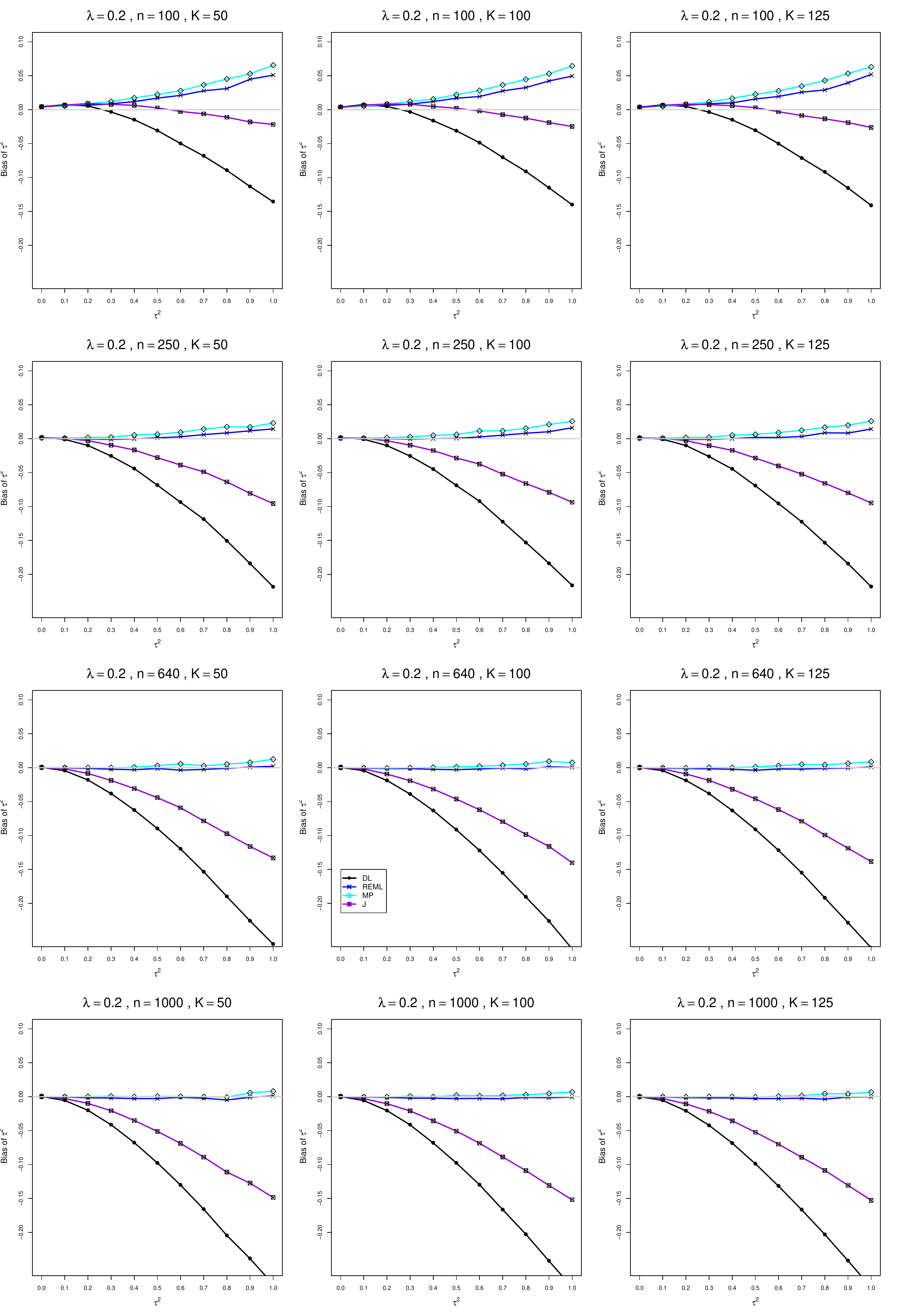}
	\caption{Bias of estimators of between-studies variance $\tau^2$ for $\lambda=0.2$, $n = 100, \;250, \;640, \;1000$, and $K = 50, \;100, \;125$. Usual estimate of $\lambda_i$
		\label{BiasTauRoM02ln_largeN_large_K}}
\end{figure}

\begin{figure}[t]
	\includegraphics[scale=0.33]{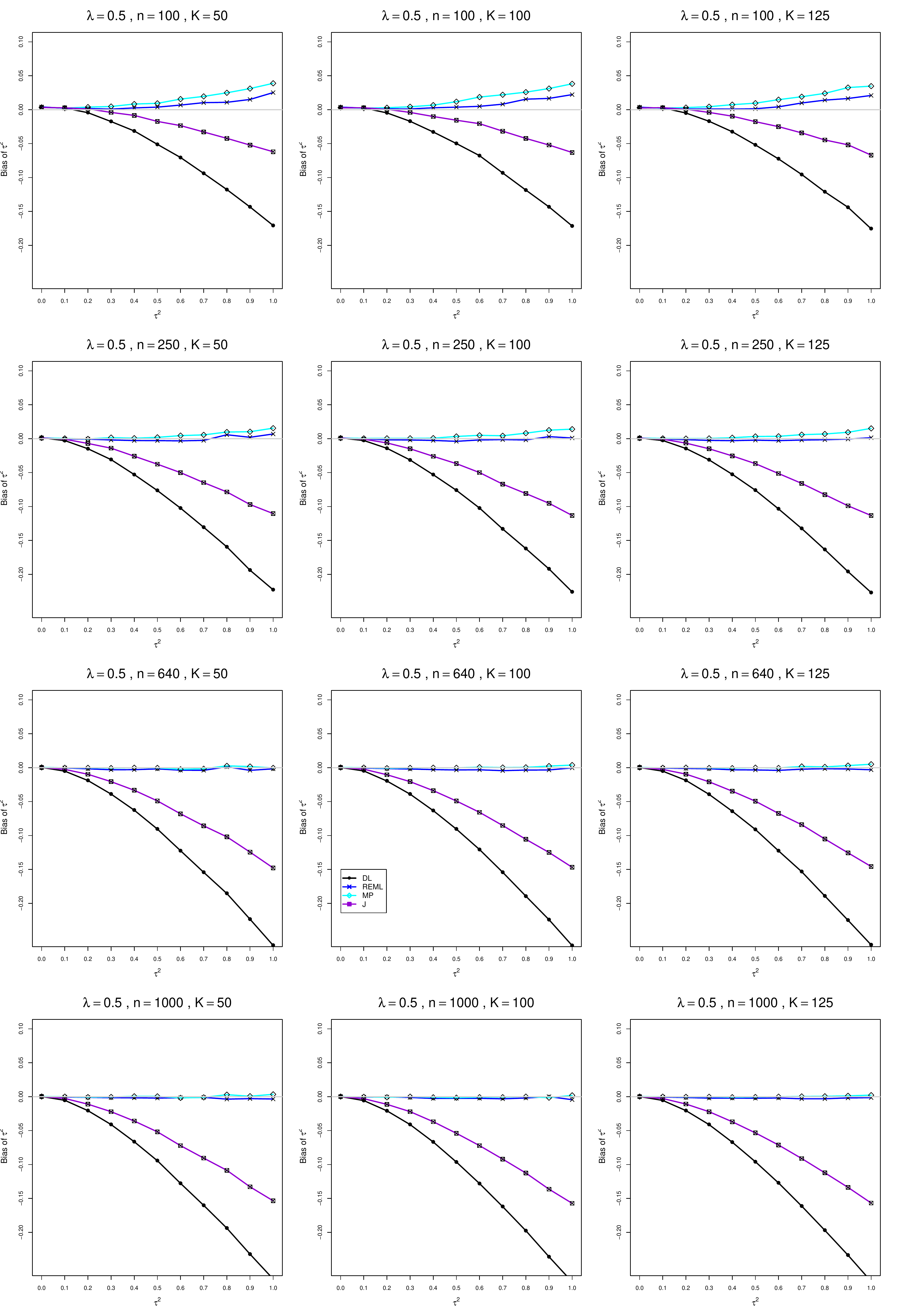}
	\caption{Bias of estimators of between-studies variance $\tau^2$ for $\lambda=0.5$, $n = 100, \;250, \;640, \;1000$, and $K = 50, \;100, \;125$. Usual estimate of $\lambda_i$
		\label{BiasTauRoM05ln_largeN_large_K}}
\end{figure}

\begin{figure}[t]
	\includegraphics[scale=0.33]{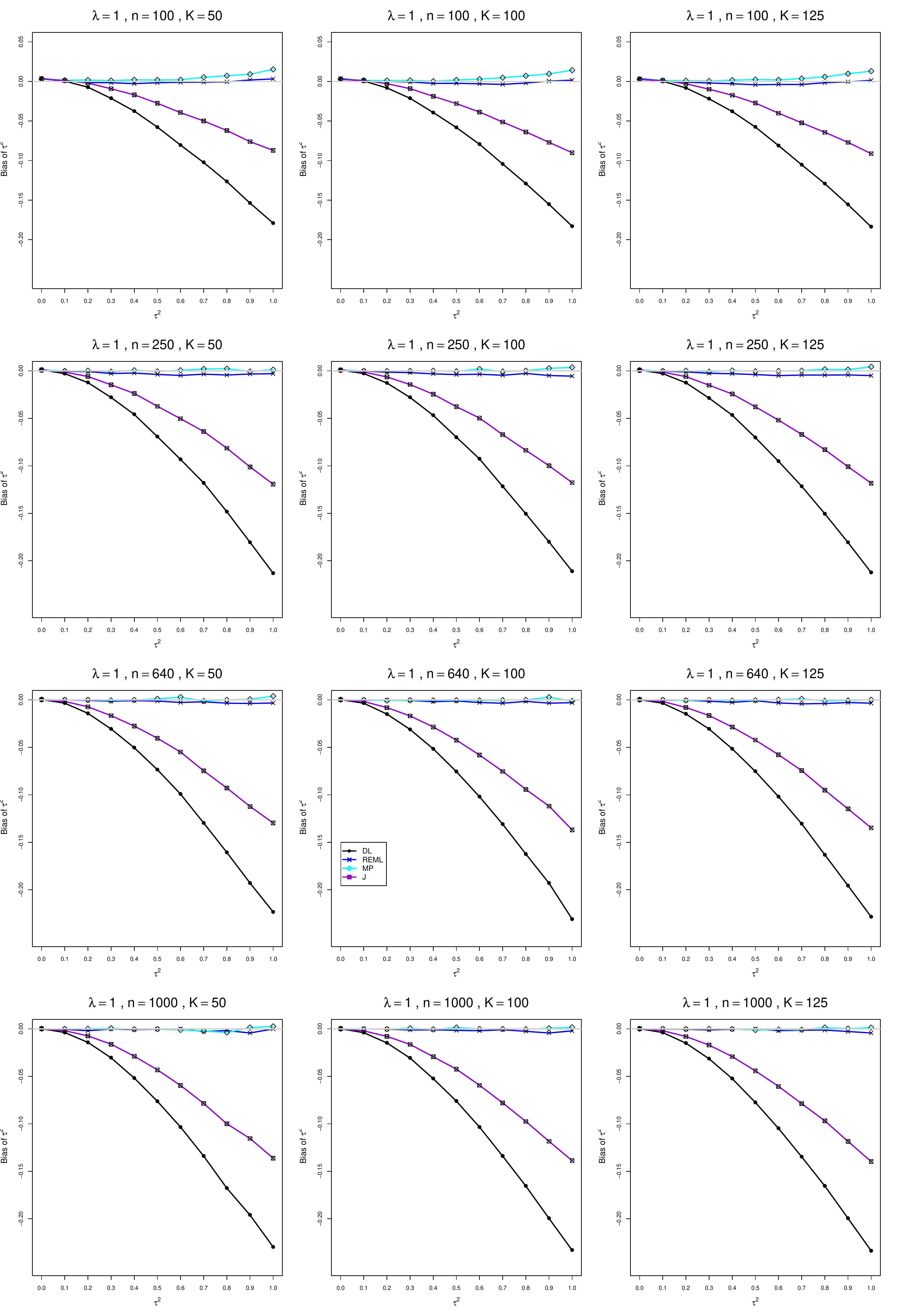}
	\caption{Bias of estimators of between-studies variance $\tau^2$ for $\lambda=1$, $n = 100, \;250, \;640, \;1000$, and $K = 50, \;100, \;125$. Usual estimate of $\lambda_i$
		\label{BiasTauRoM1ln_largeN_large_K}}
\end{figure}

\begin{figure}[t]
	\includegraphics[scale=0.33]{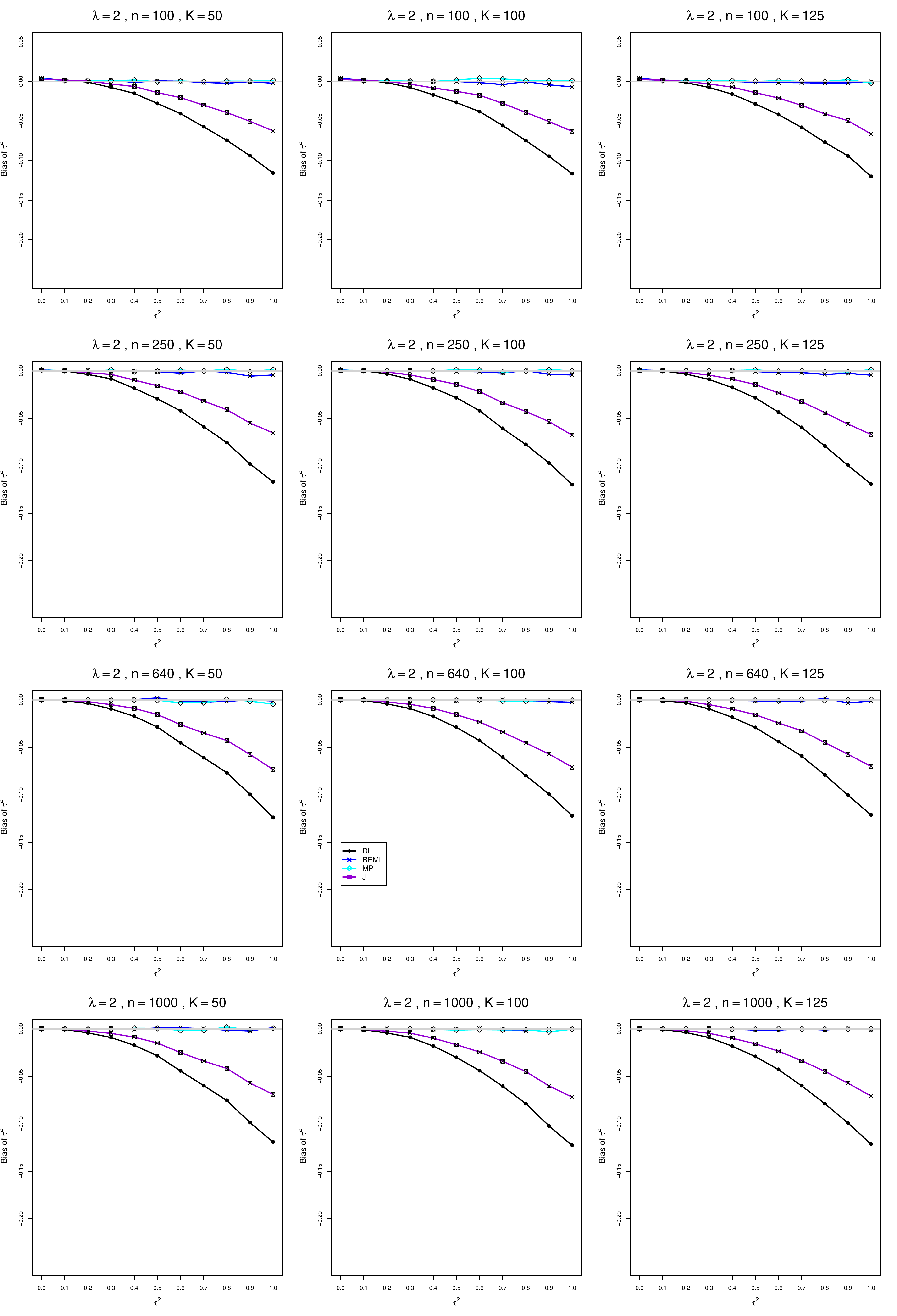}
	\caption{Bias of estimators of between-studies variance $\tau^2$ for $\lambda=2$, $n = 100, \;250, \;640, \;1000$, and $K = 50, \;100, \;125$. Usual estimate of $\lambda_i$
		\label{BiasTauRoM2ln_largeN_large_K}}
\end{figure}

\clearpage
\subsection*{C3.2 Coverage of interval estimators of $\tau^2$}
Each figure corresponds to a value of $\lambda \;(= 0, 0.2, 0.5, 1, 2)$, a set of values of $n$ (= 100, 250, 640, 1000), and a set of values of $K$ (= 50, 100, 125).\\
Each panel corresponds to a value of $n$ and a value of $K$ and has $\tau^2 = 0.0(0.1)1.0$ on the horizontal axis.\\
The interval estimators of $\tau^2$ are
\begin{itemize}
	\item QP (Q-profile confidence interval)
	\item BJ (Biggerstaff and Jackson interval )
	\item PL (Profile-likelihood interval)
	\item J (Jackson interval)
\end{itemize}

\setcounter{figure}{0}
\renewcommand{\thefigure}{C3.2.\arabic{figure}}
\begin{figure}[t]
	\includegraphics[scale=0.35]{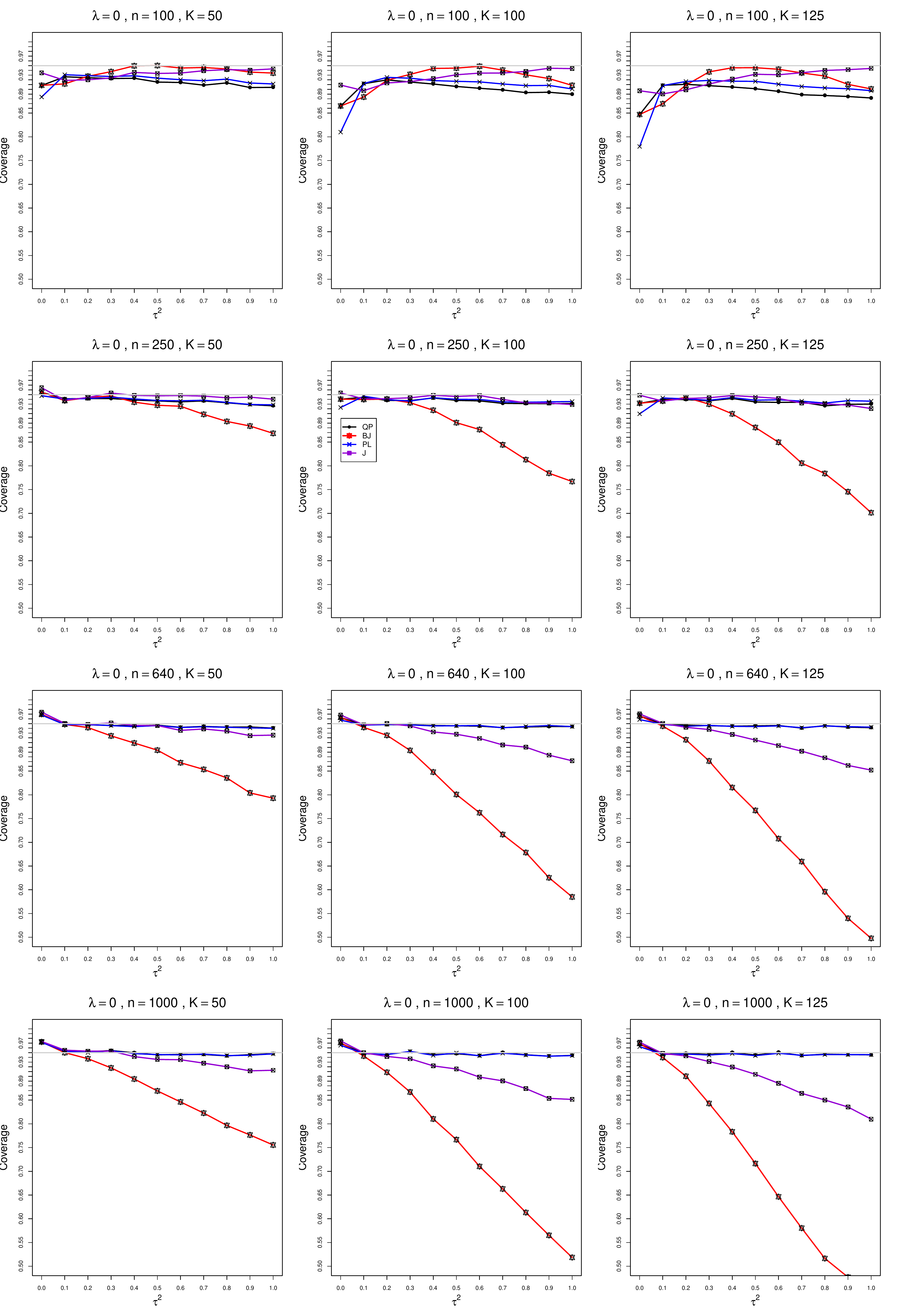}
	\caption{Coverage of 95\% confidence intervals for the between-studies variance $\tau^2$ when $\lambda=0$, $n = 100, \;250, \;640, \;1000$, and $K = 50, \;100, \;125$. Usual estimate of $\lambda_i$
		\label{CovTauRoM0ln_largeN_large_K}}
\end{figure}

\begin{figure}[t]
	\includegraphics[scale=0.35]{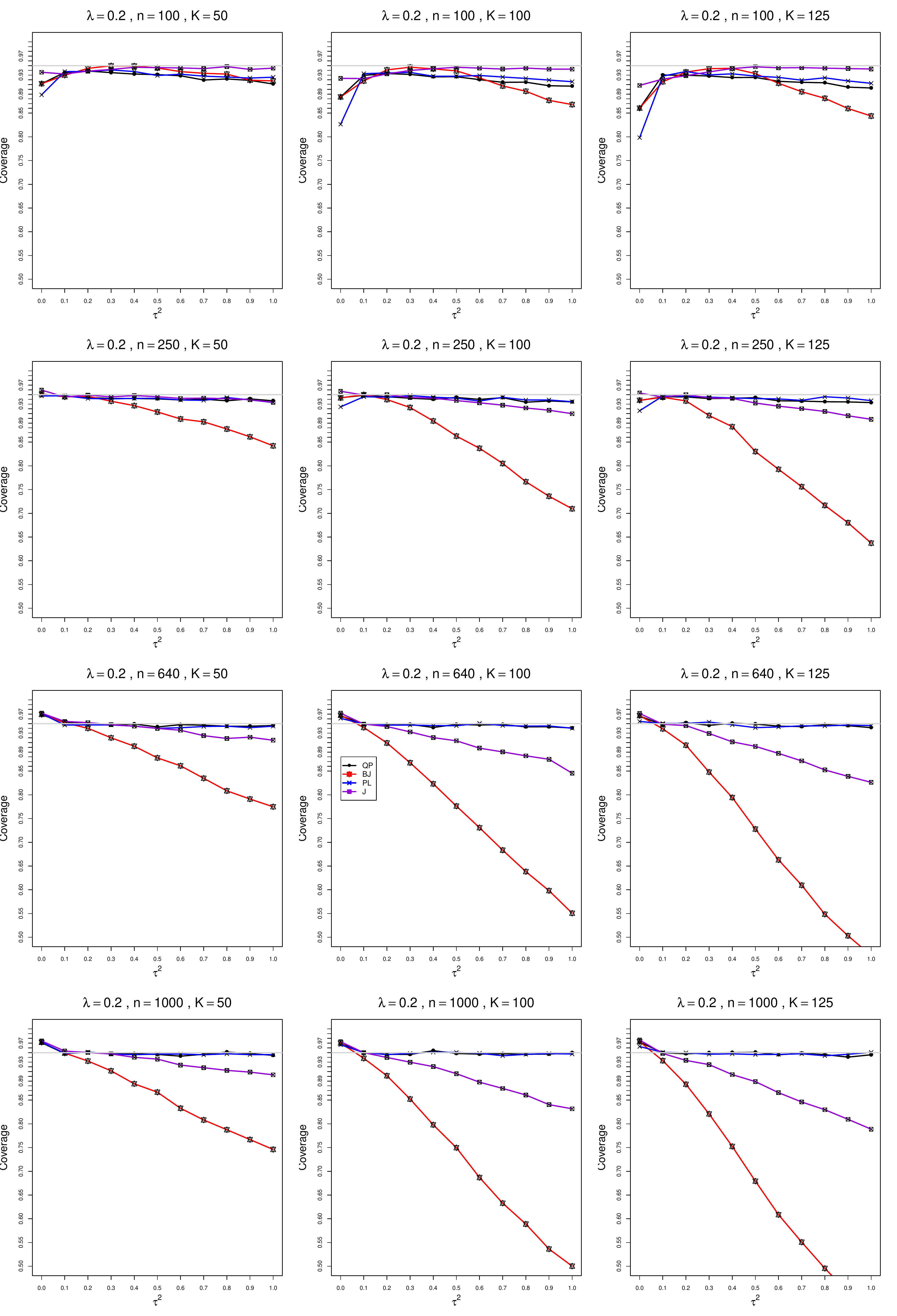}
	\caption{Coverage of 95\% confidence intervals for the between-studies variance $\tau^2$ when $\lambda=0.2$, $n = 100, \;250, \;640, \;1000$, and $K = 50, \;100, \;125$. Usual estimate of $\lambda_i$
		\label{CovTauRoM02ln_largeN_large_K}}
\end{figure}

\begin{figure}[t]
	\includegraphics[scale=0.35]{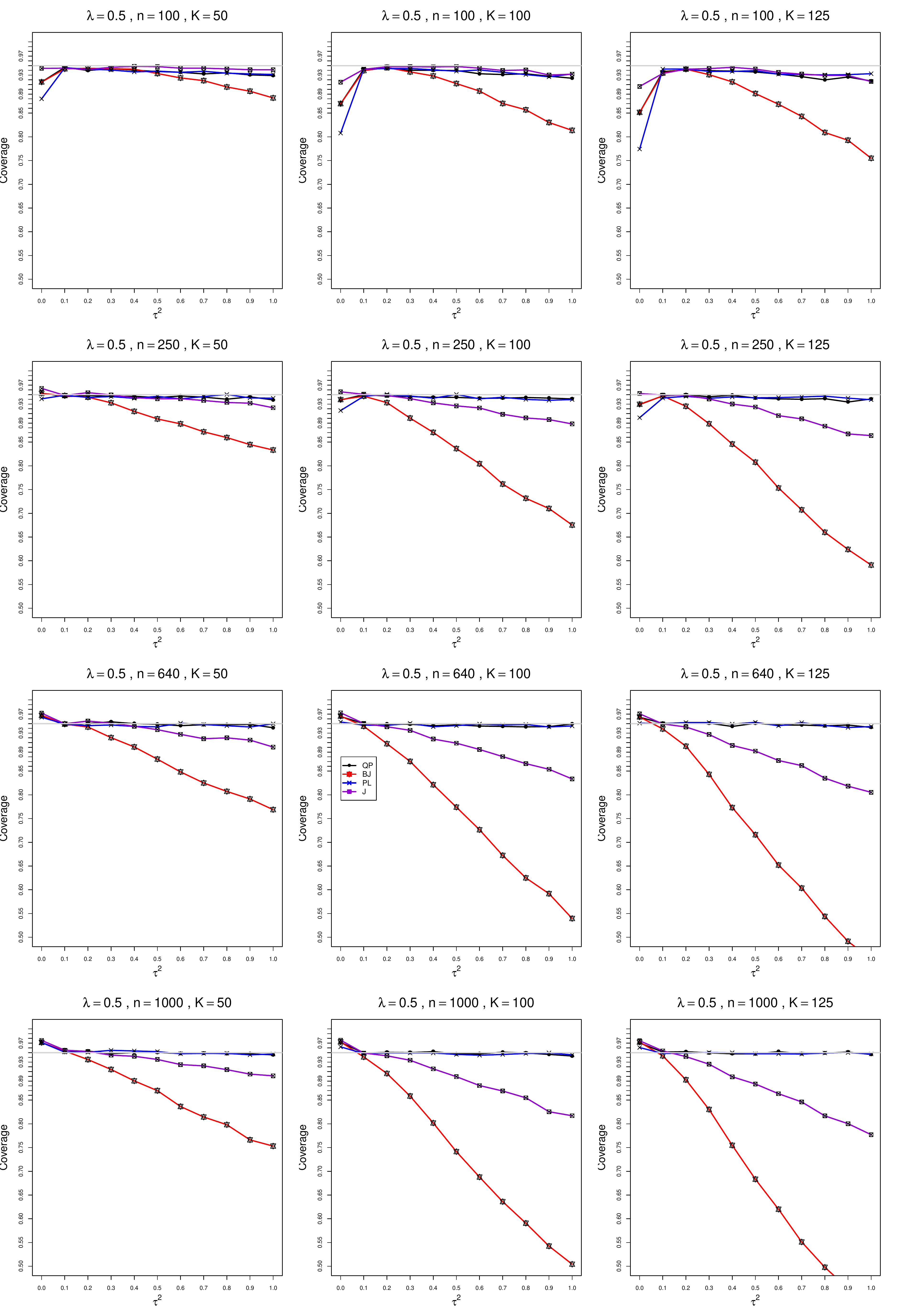}
	\caption{Coverage of 95\% confidence intervals for the between-studies variance $\tau^2$ when $\lambda=0.5$, $n = 100, \;250, \;640, \;1000$, and $K = 50, \;100, \;125$. Usual estimate of $\lambda_i$
		\label{CovTauRoM05ln_largeN_large_K}}
\end{figure}

\begin{figure}[t]
	\includegraphics[scale=0.35]{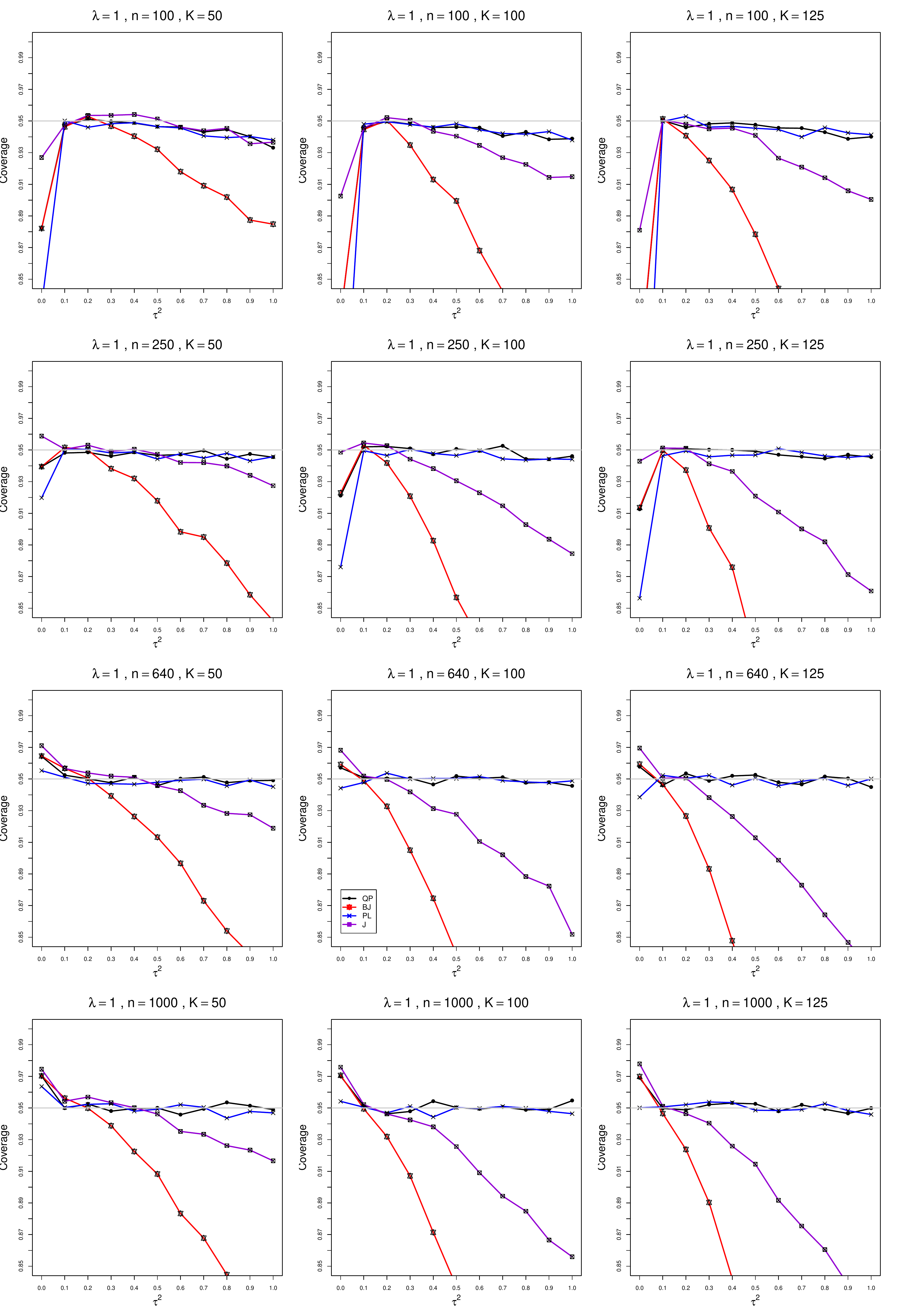}
	\caption{Coverage of 95\% confidence intervals for the between-studies variance $\tau^2$ when $\lambda=1$, $n = 100, \;250, \;640, \;1000$, and $K = 50, \;100, \;125$. Usual estimate of $\lambda_i$
		\label{CovTauRoM1ln_largeN_large_K}}
\end{figure}

\begin{figure}[t]
	\includegraphics[scale=0.35]{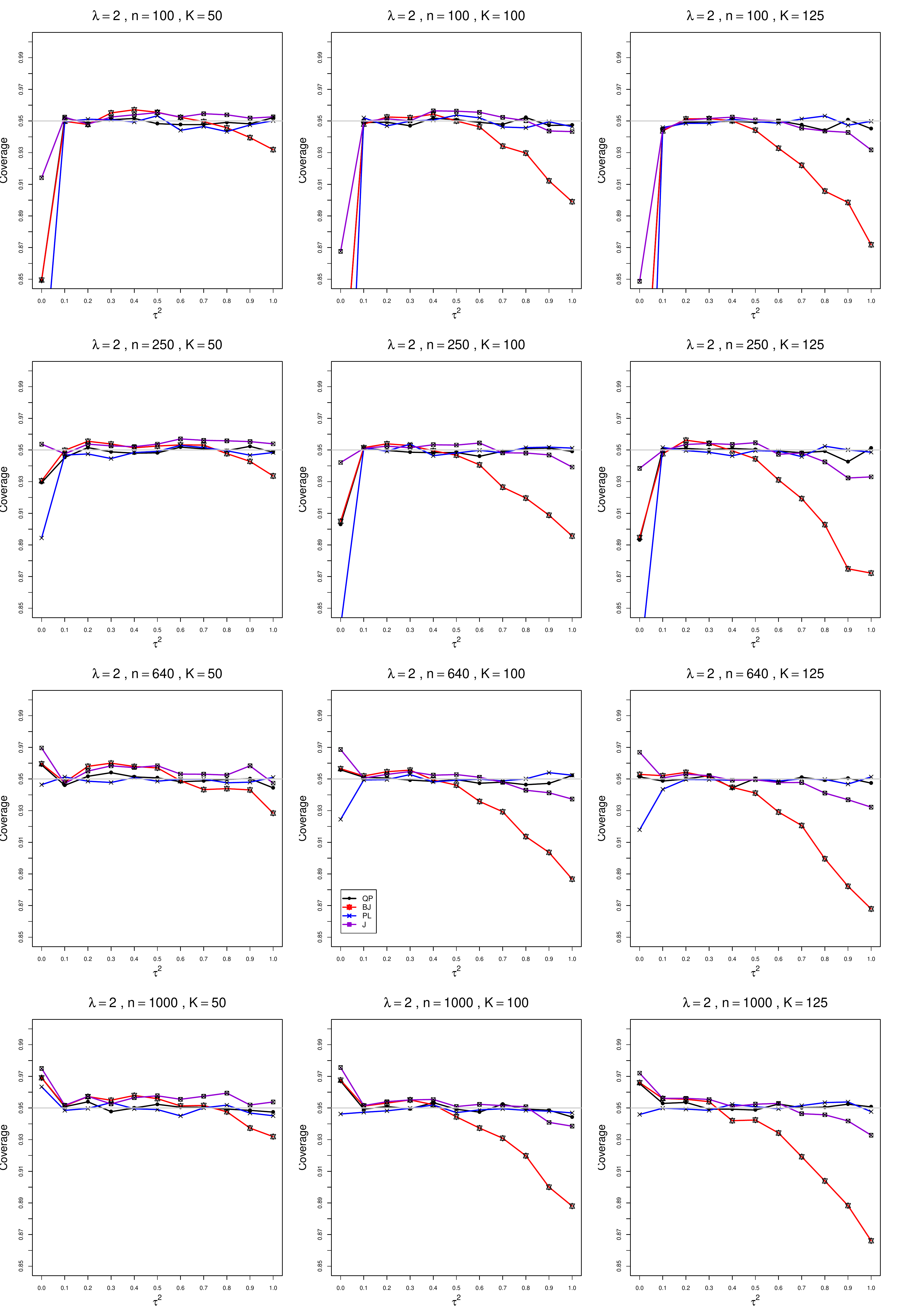}
	\caption{Coverage of 95\% confidence intervals for the between-studies variance $\tau^2$ when $\lambda=2$, $n = 100, \;250, \;640, \;1000$, and $K = 50, \;100, \;125$. Usual estimate of $\lambda_i$
		\label{CovTauRoM2ln_largeN_large_K}}
\end{figure}

\clearpage

\section*{C4. Lognormal model, bias-corrected estimator of $\lambda_i$, $n= 100, 250, 640, 1000$, $K=50,100,125$}
\subsection*{C4.1 Bias of point estimators of $\tau^2$}
Each figure corresponds to a value of $\lambda \;(= 0, 0.2, 0.5, 1, 2)$, a set of values of $n$ (= 100, 250, 640, 1000), and a set of values of $K$ (= 50, 100, 125).\\
Each panel corresponds to a value of $n$ and a value of $K$ and has $\tau^2 = 0.0(0.1)1.0$ on the horizontal axis.\\
The point estimators of $\tau^2$ are
\begin{itemize}
	\item DL (DerSimonian-Laird)
	\item REML (restricted maximum likelihood)
	\item MP (Mandel-Paule)
	\item J (Jackson)
\end{itemize}

\clearpage
\setcounter{figure}{0}
\renewcommand{\thefigure}{C4.1.\arabic{figure}}
\begin{figure}[t]
	\includegraphics[scale=0.33]{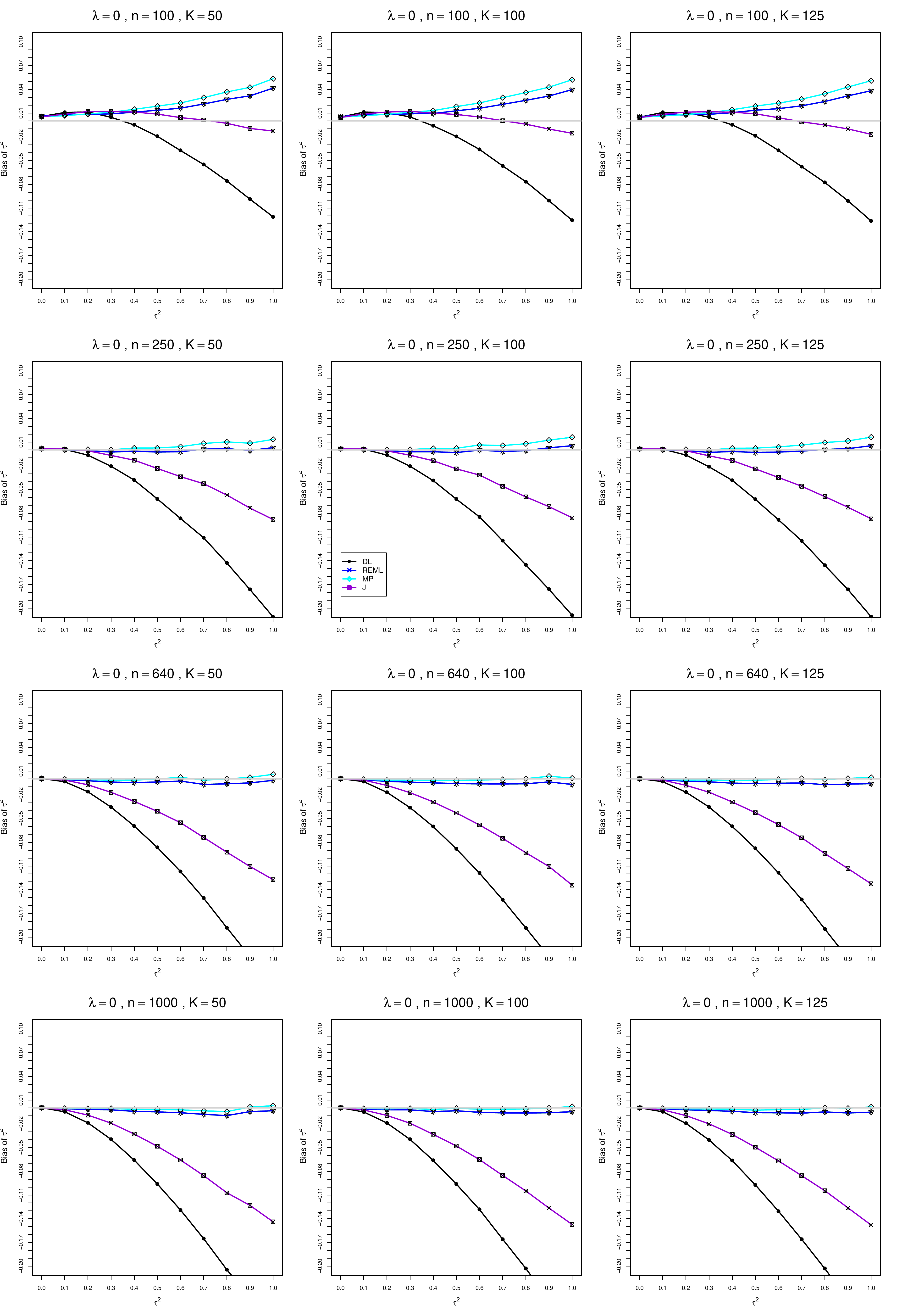}
	\caption{Bias of estimators of between-studies variance $\tau^2$ for $\lambda=0$, $n = 100, \;250, \;640, \;1000$, and $K = 50, \;100, \;125$. Bias-corrected estimate of $\lambda_i$
		\label{BiasTauRoM0lnCor_largeN_large_K}}
\end{figure}

\begin{figure}[t]
	\includegraphics[scale=0.33]{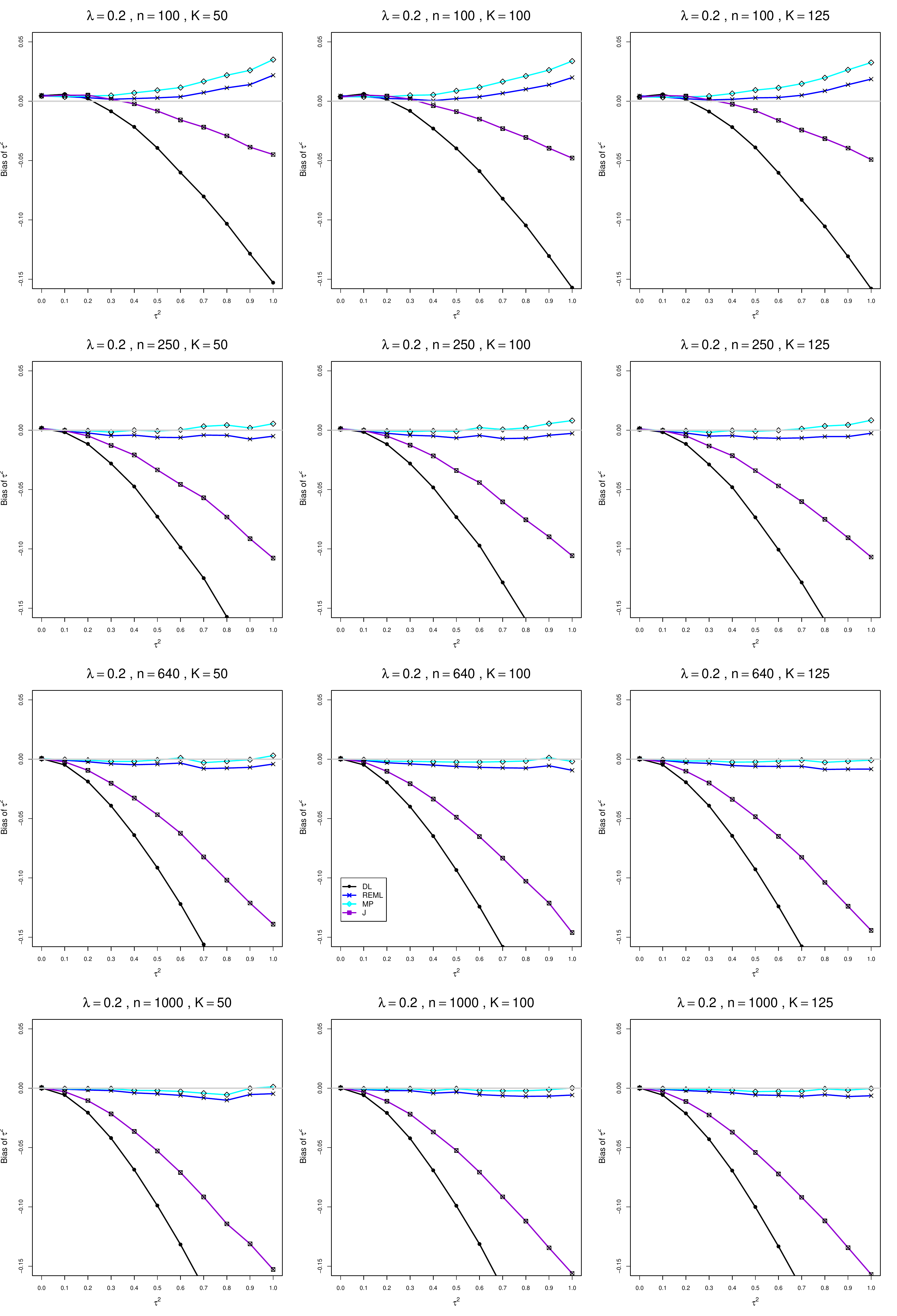}
	\caption{Bias of estimators of between-studies variance $\tau^2$ for $\lambda=0.2$, $n = 100, \;250, \;640, \;1000$, and $K = 50, \;100, \;125$. Bias-corrected estimate of $\lambda_i$
		\label{BiasTauRoM02lnCor_largeN_large_K}}
\end{figure}

\begin{figure}[t]
	\includegraphics[scale=0.33]{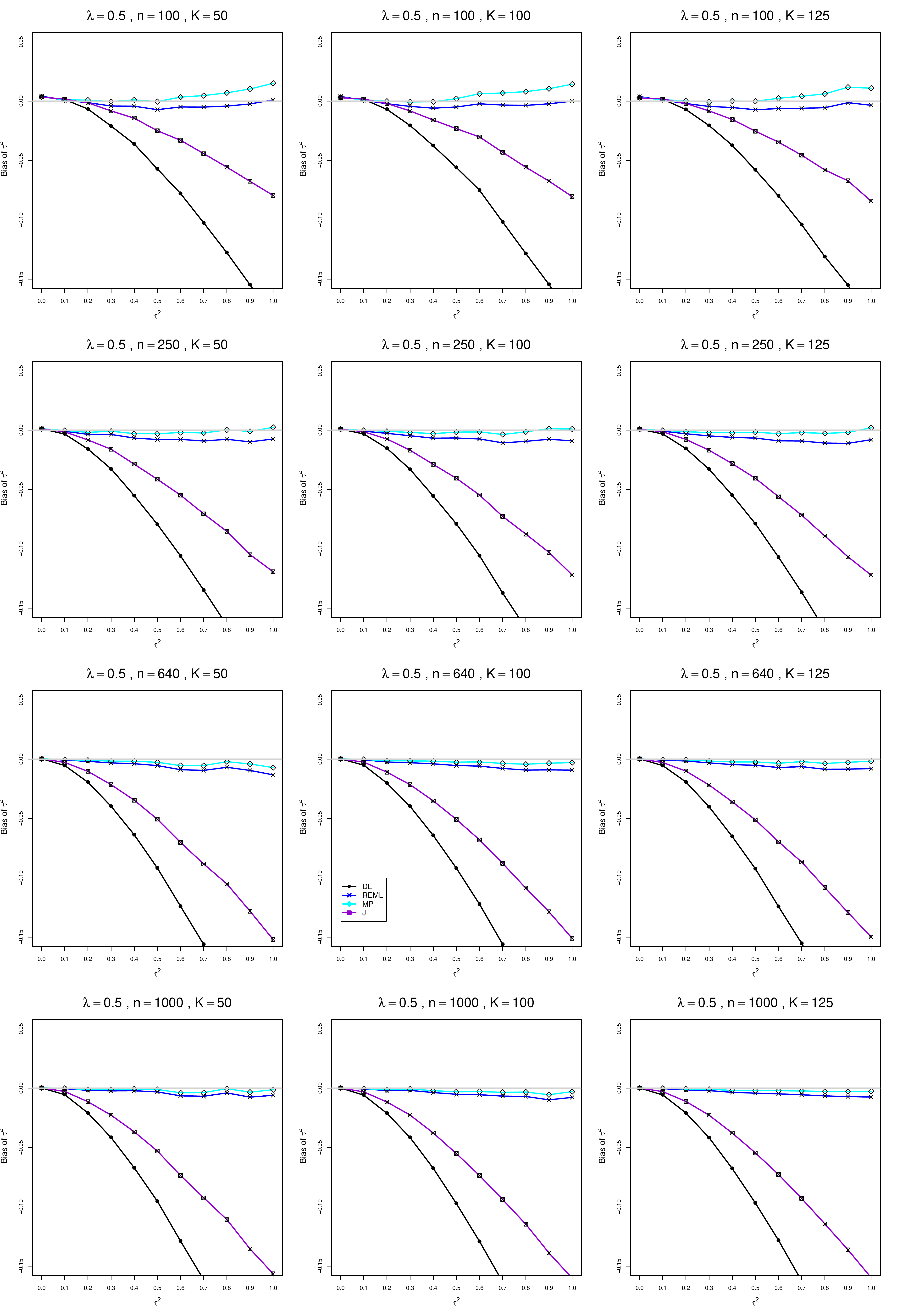}
	\caption{Bias of estimators of between-studies variance $\tau^2$ for $\lambda=0.5$, $n = 100, \;250, \;640, \;1000$, and $K = 50, \;100, \;125$. Bias-corrected estimate of $\lambda_i$
		\label{BiasTauRoM05lnCor_largeN_large_K}}
\end{figure}

\begin{figure}[t]
	\includegraphics[scale=0.33]{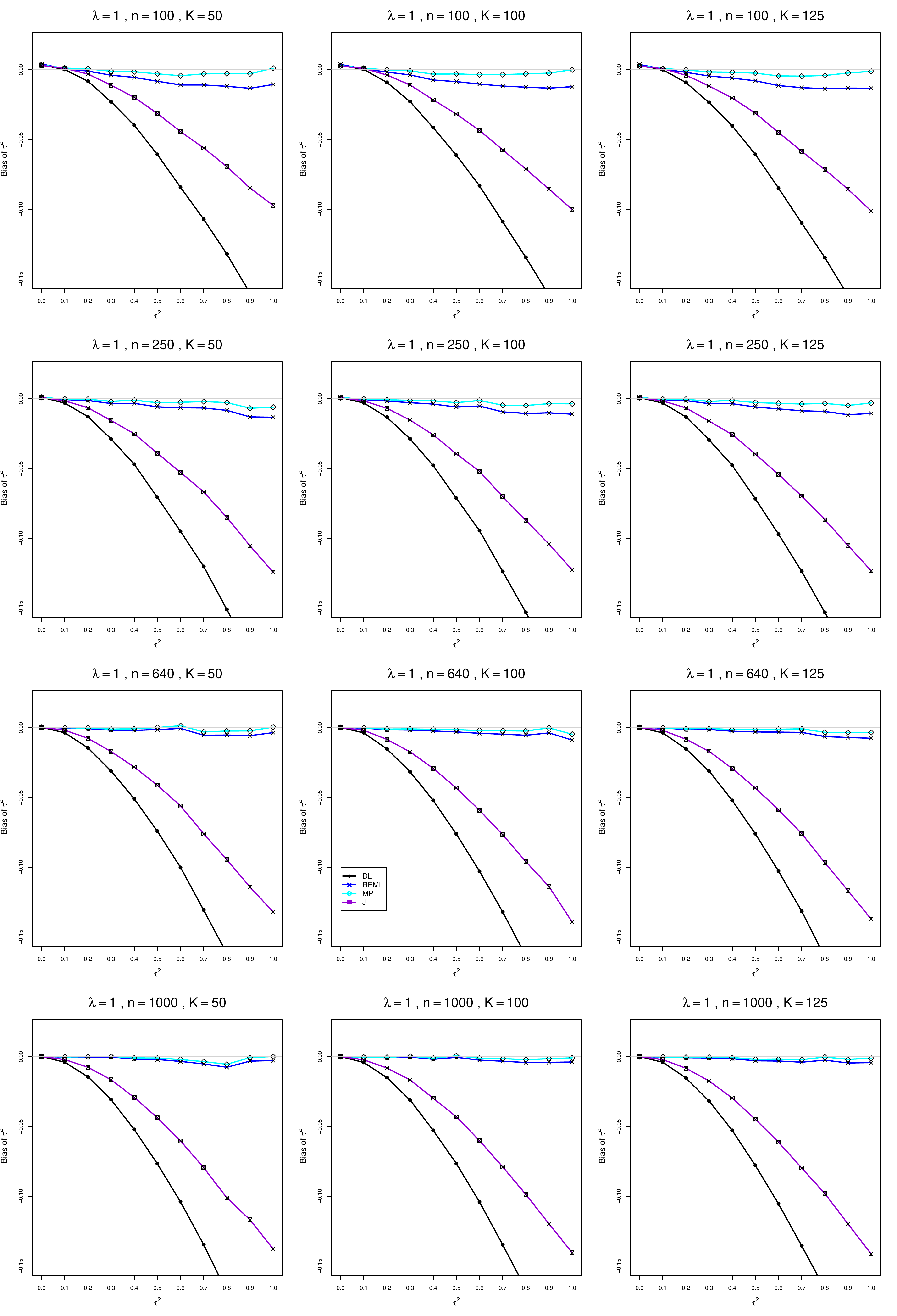}
	\caption{Bias of estimators of between-studies variance $\tau^2$ for $\lambda=1$, $n = 100, \;250, \;640, \;1000$, and $K = 50, \;100, \;125$. Bias-corrected estimate of $\lambda_i$
		\label{BiasTauRoM1lnCor_largeN_large_K}}
\end{figure}

\begin{figure}[t]
	\includegraphics[scale=0.33]{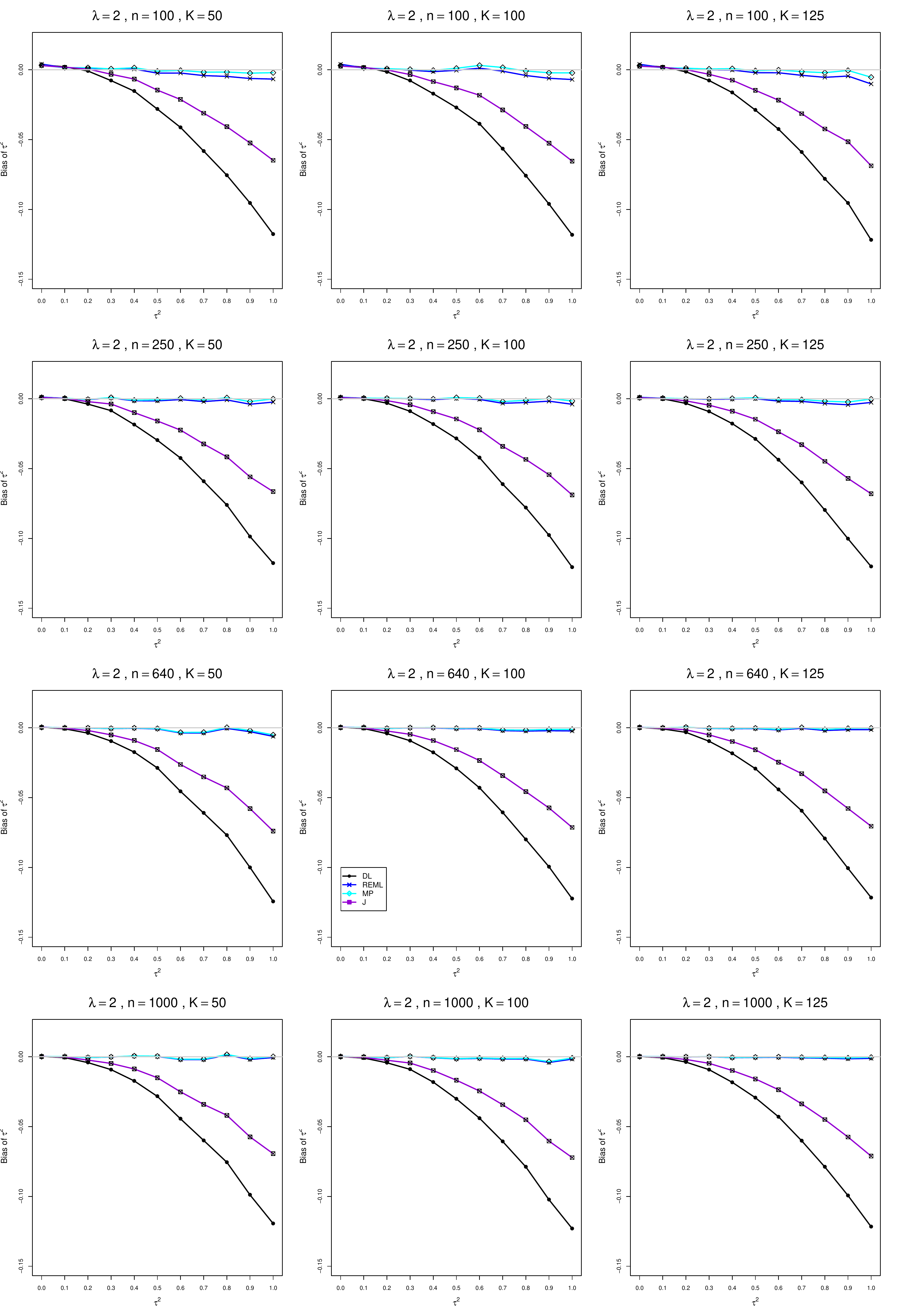}
	\caption{Bias of estimators of between-studies variance $\tau^2$ for $\lambda=2$, $n = 100, \;250, \;640, \;1000$, and $K = 50, \;100, \;125$. Bias-corrected estimate of $\lambda_i$
		\label{BiasTauRoM2lnCor_largeN_large_K}}
\end{figure}

\clearpage
\subsection*{C4.2 Coverage of interval estimators of $\tau^2$}
Each figure corresponds to a value of $\lambda \;(= 0, 0.2, 0.5, 1, 2)$, a set of values of $n$ (= 100, 250, 640, 1000), and a set of values of $K$ (= 50, 100, 125).\\
Each panel corresponds to a value of $n$ and a value of $K$ and has $\tau^2 = 0.0(0.1)1.0$ on the horizontal axis.\\
The interval estimators of $\tau^2$ are
\begin{itemize}
	\item QP (Q-profile confidence interval)
	\item BJ (Biggerstaff and Jackson interval )
	\item PL (Profile-likelihood interval)
	\item J (Jackson interval)
\end{itemize}

\setcounter{figure}{0}
\renewcommand{\thefigure}{C4.2.\arabic{figure}}
\begin{figure}[t]
	\includegraphics[scale=0.35]{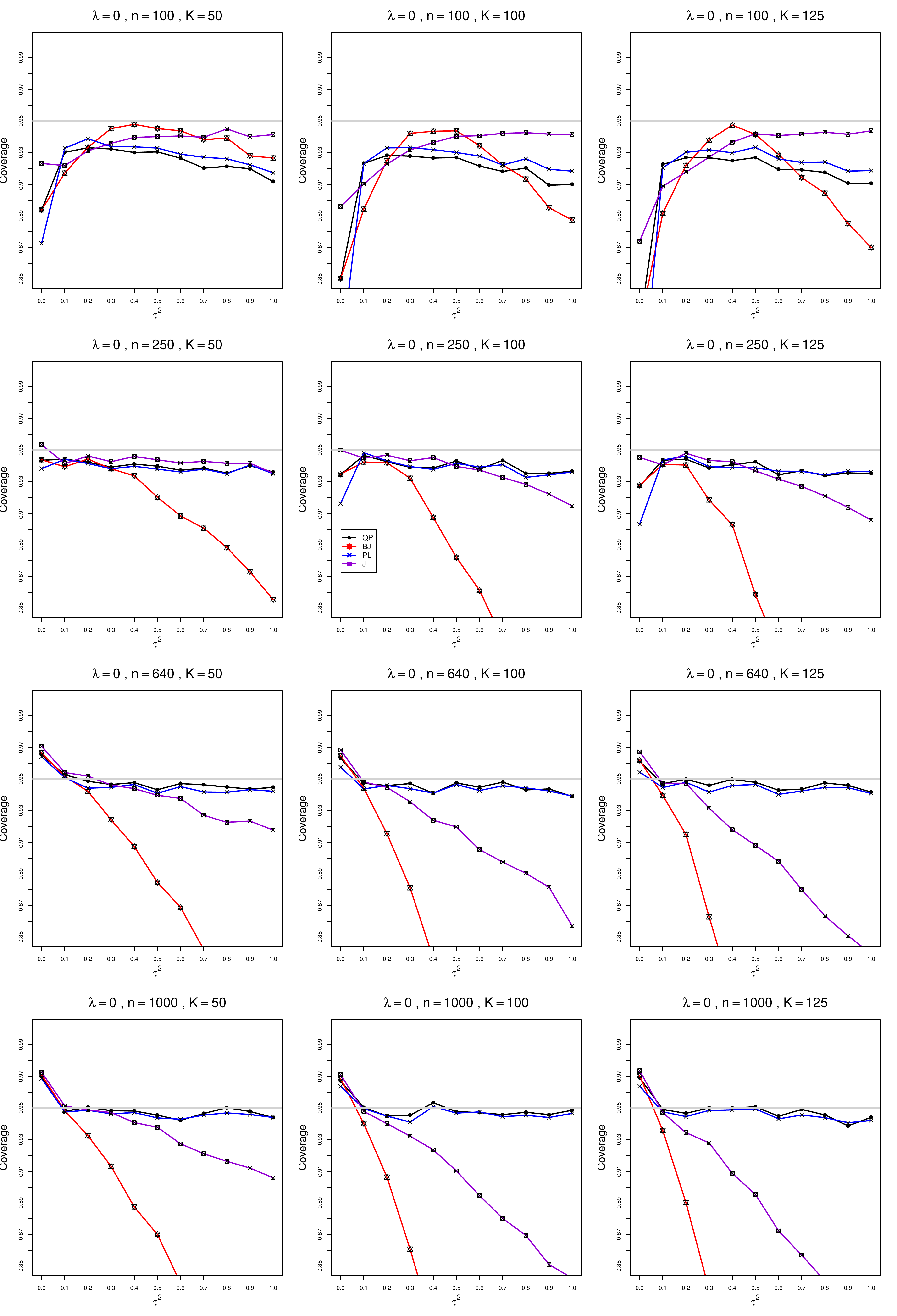}
	\caption{Coverage of 95\% confidence intervals for the between-studies variance $\tau^2$ when $\lambda=0$, $n = 100, \;250, \;640, \;1000$, and $K = 50, \;100, \;125$. Bias-corrected estimate of $\lambda_i$ 		\label{CovTauRoM0lnCor_largeN_large_K}}
\end{figure}

\begin{figure}[t]
	\includegraphics[scale=0.35]{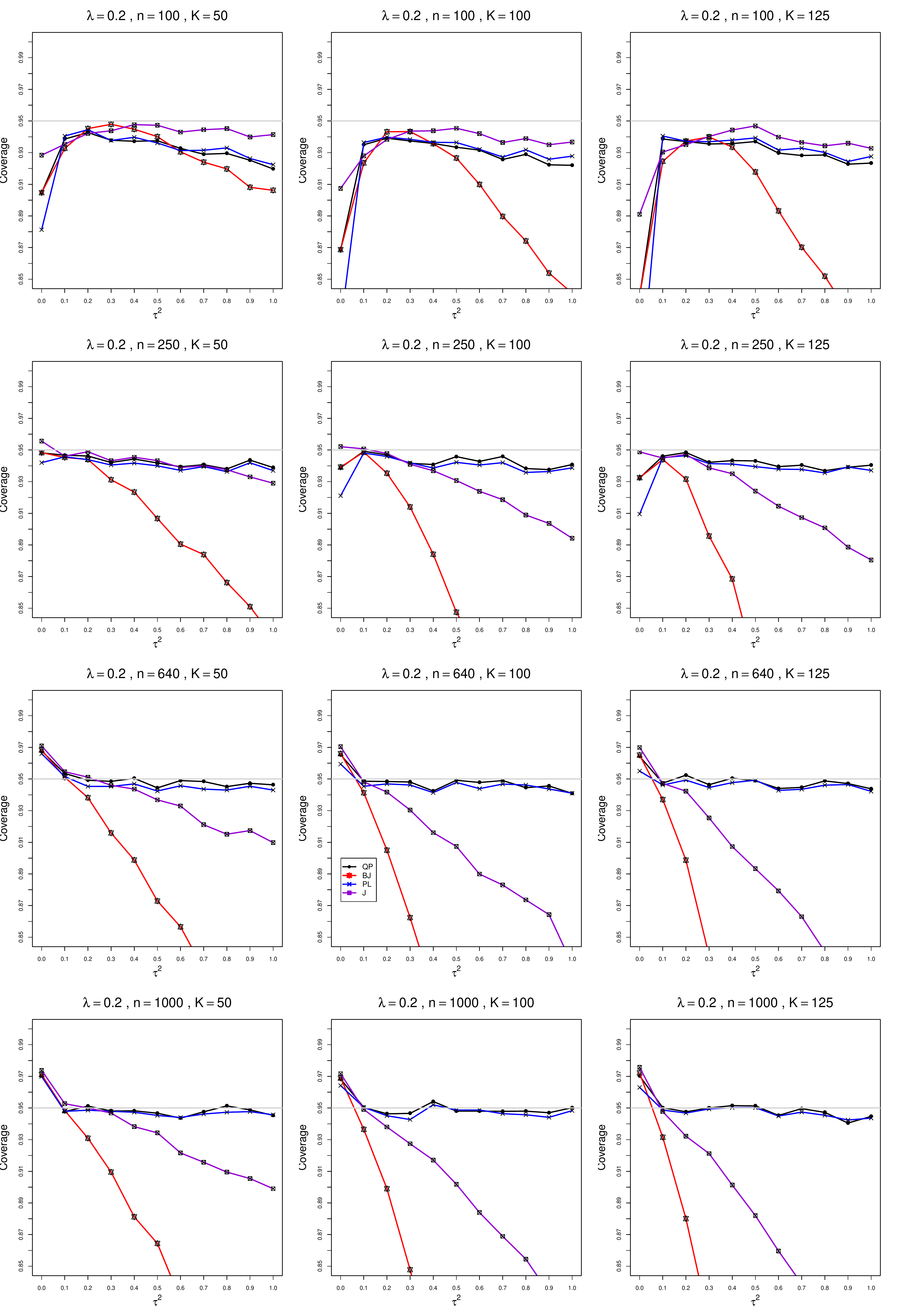}
	\caption{Coverage of 95\% confidence intervals for the between-studies variance $\tau^2$ when $\lambda=0.2$, $n = 100, \;250, \;640, \;1000$, and $K = 50, \;100, \;125$. Bias-corrected estimate of $\lambda_i$ 		\label{CovTauRoM02lnCor_largeN_large_K}}
\end{figure}

\begin{figure}[t]
	\includegraphics[scale=0.35]{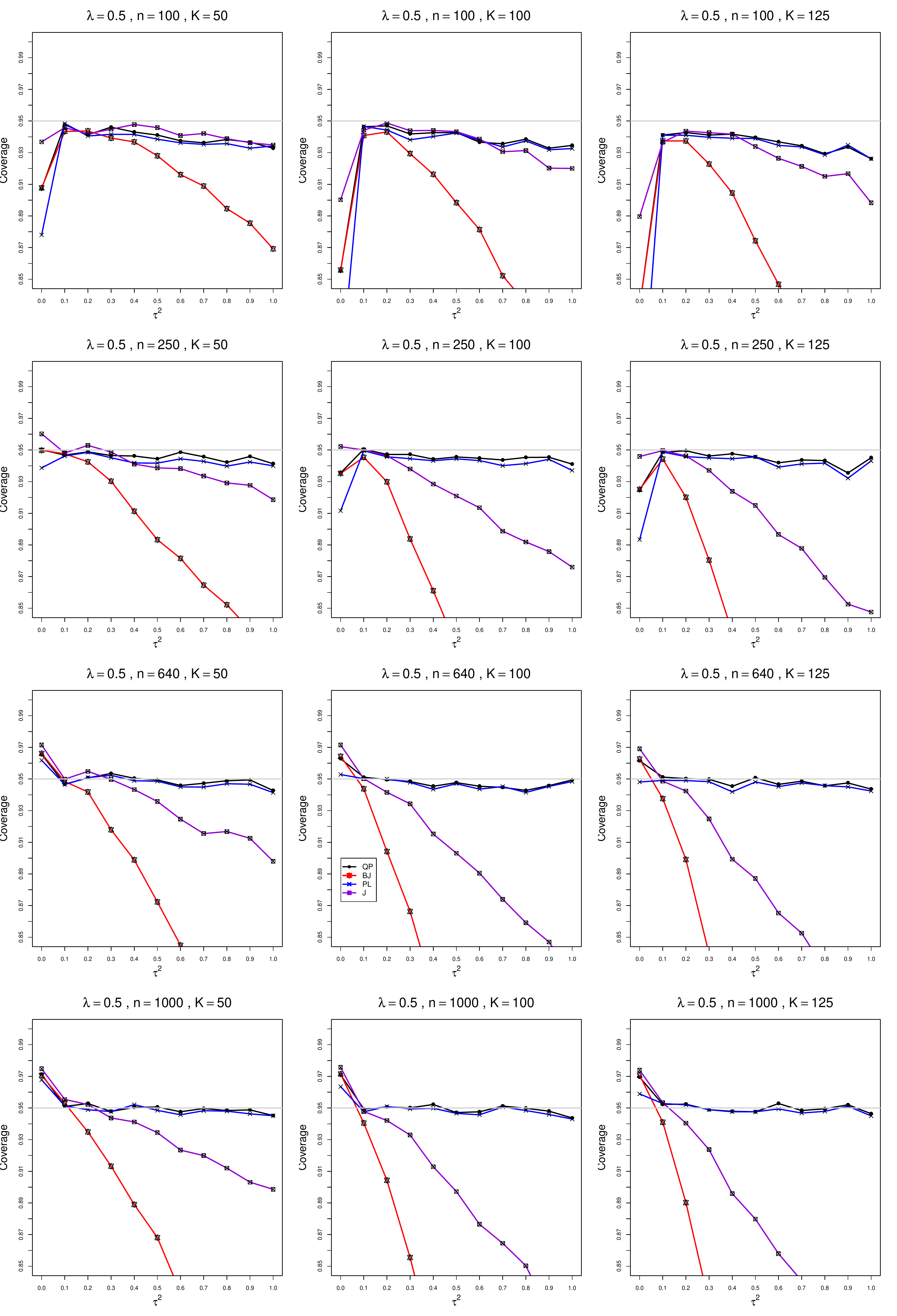}
	\caption{Coverage of 95\% confidence intervals for the between-studies variance $\tau^2$ when $\lambda=0.5$, $n = 100, \;250, \;640, \;1000$, and $K = 50, \;100, \;125$. Bias-corrected estimate of $\lambda_i$ 		\label{CovTauRoM05lnCor_largeN_large_K}}
\end{figure}

\begin{figure}[t]
	\includegraphics[scale=0.35]{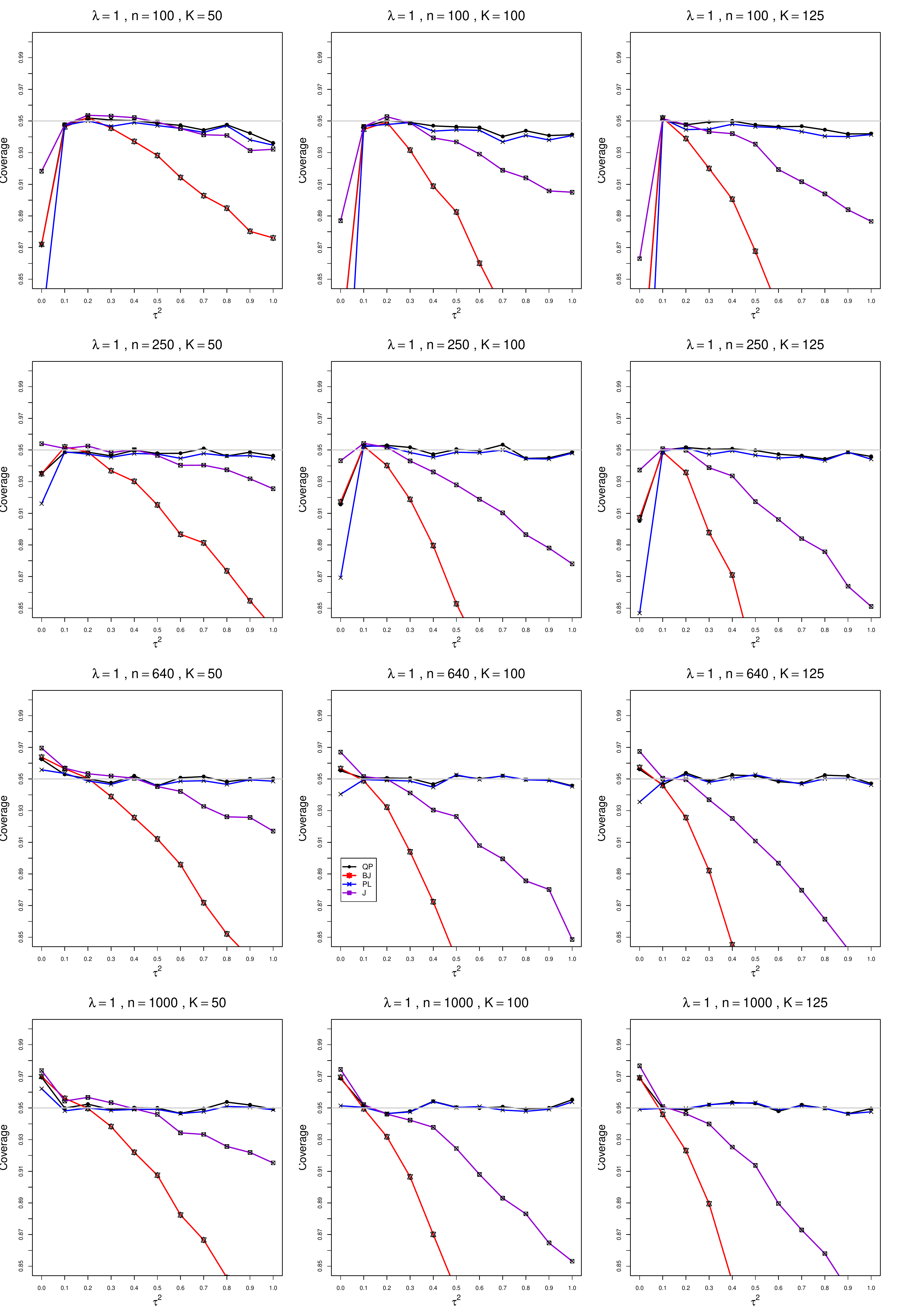}
	\caption{Coverage of 95\% confidence intervals for the between-studies variance $\tau^2$ when $\lambda=1$, $n = 100, \;250, \;640, \;1000$, and $K = 50, \;100, \;125$. Bias-corrected estimate of $\lambda_i$ 		\label{CovTauRoM1lnCor_largeN_large_K}}
\end{figure}

\begin{figure}[t]
	\includegraphics[scale=0.35]{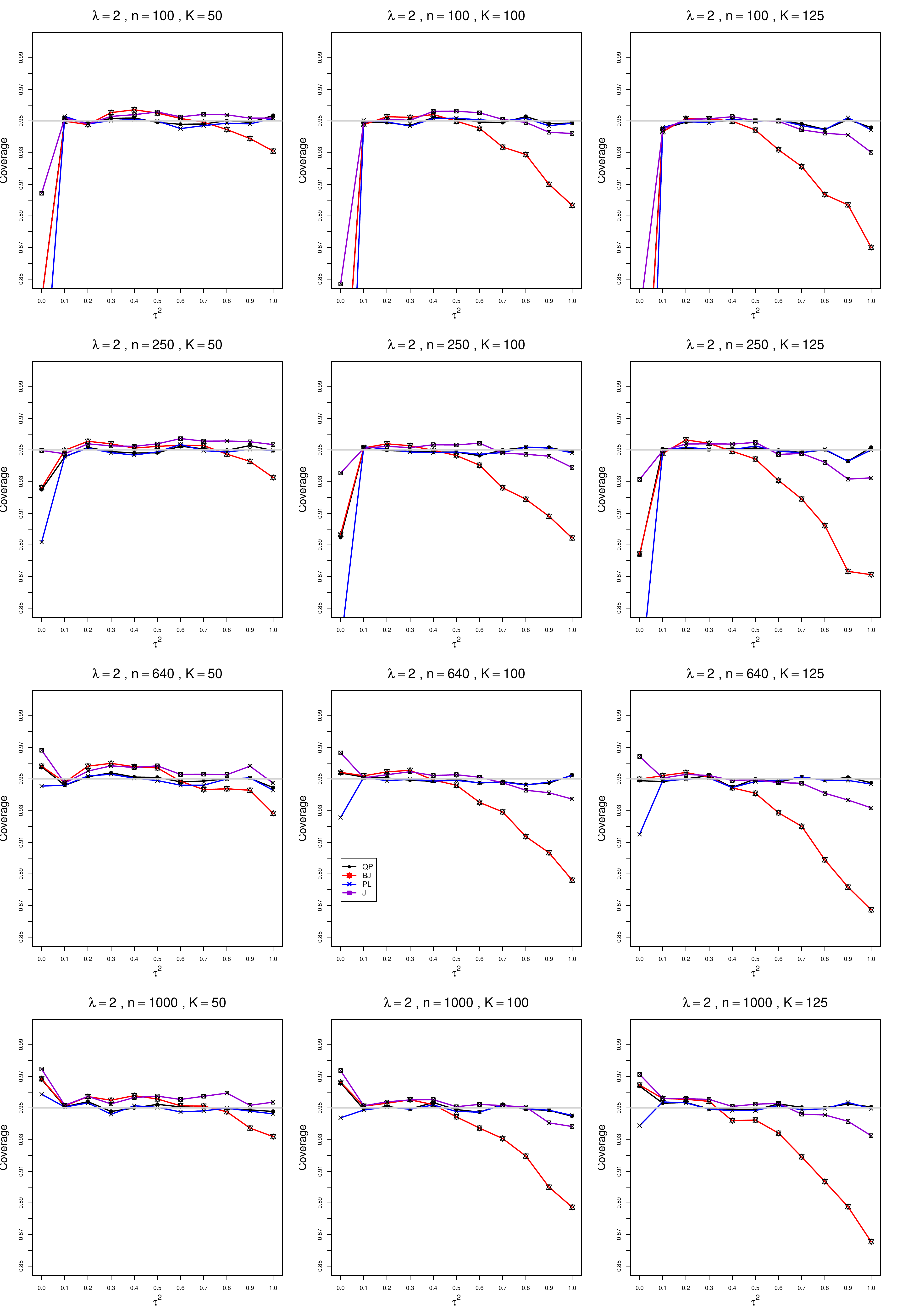}
	\caption{Coverage of 95\% confidence intervals for the between-studies variance $\tau^2$ when $\lambda=2$, $n = 100, \;250, \;640, \;1000$, and $K = 50, \;100, \;125$. Bias-corrected estimate of $\lambda_i$ 		\label{CovTauRoM2lnCor_largeN_large_K}}
\end{figure}

\clearpage
\renewcommand{\thesection}{D1.\arabic{section}}
\setcounter{section}{0}
\section*{D: Plots of bias and coverage of estimators of $\lambda$, large $n$}

\begin{itemize}
	\item D1. Lognormal model, usual estimator of $\lambda_i$, $K=5,10,30$
	\item D2. Lognormal model, bias-corrected estimator of $\lambda_i$, $K=5,10,30$
	\item D3. Lognormal model, usual estimator of $\lambda_i$, $K=50,100,125$
	\item D4. Lognormal model, bias-corrected estimator of $\lambda_i$, $K=50,100,125$
\end{itemize}

\clearpage

\section*{D1. Lognormal model, usual estimator of $\lambda_i$, $n= 100, 250, 640, 1000$, $K=5,10,30$}
\subsection*{D1.1 Bias of point estimators of $\lambda$}
Each figure corresponds to a value of $\lambda \;(= 0, 0.2, 0.5, 1, 2)$, a set of values of $n$ (= 100, 250, 640, 1000), and a set of values of $K$ (= 5, 10, 30).\\
Each panel corresponds to a value of $n$ and a value of $K$ and has $\tau^2 = 0.0(0.1)1.0$ on the horizontal axis.\\
The point estimators of $\lambda$ are
\begin{itemize}
	\item DL (DerSimonian-Laird)
	\item REML (restricted maximum likelihood)
	\item MP (Mandel-Paule)
	\item J (Jackson)
	\item SSW (sample-size-weighted)
\end{itemize}

\clearpage
\setcounter{figure}{0}
\renewcommand{\thefigure}{D1.1.\arabic{figure}}

\begin{figure}[t]
	\includegraphics[scale=0.33]{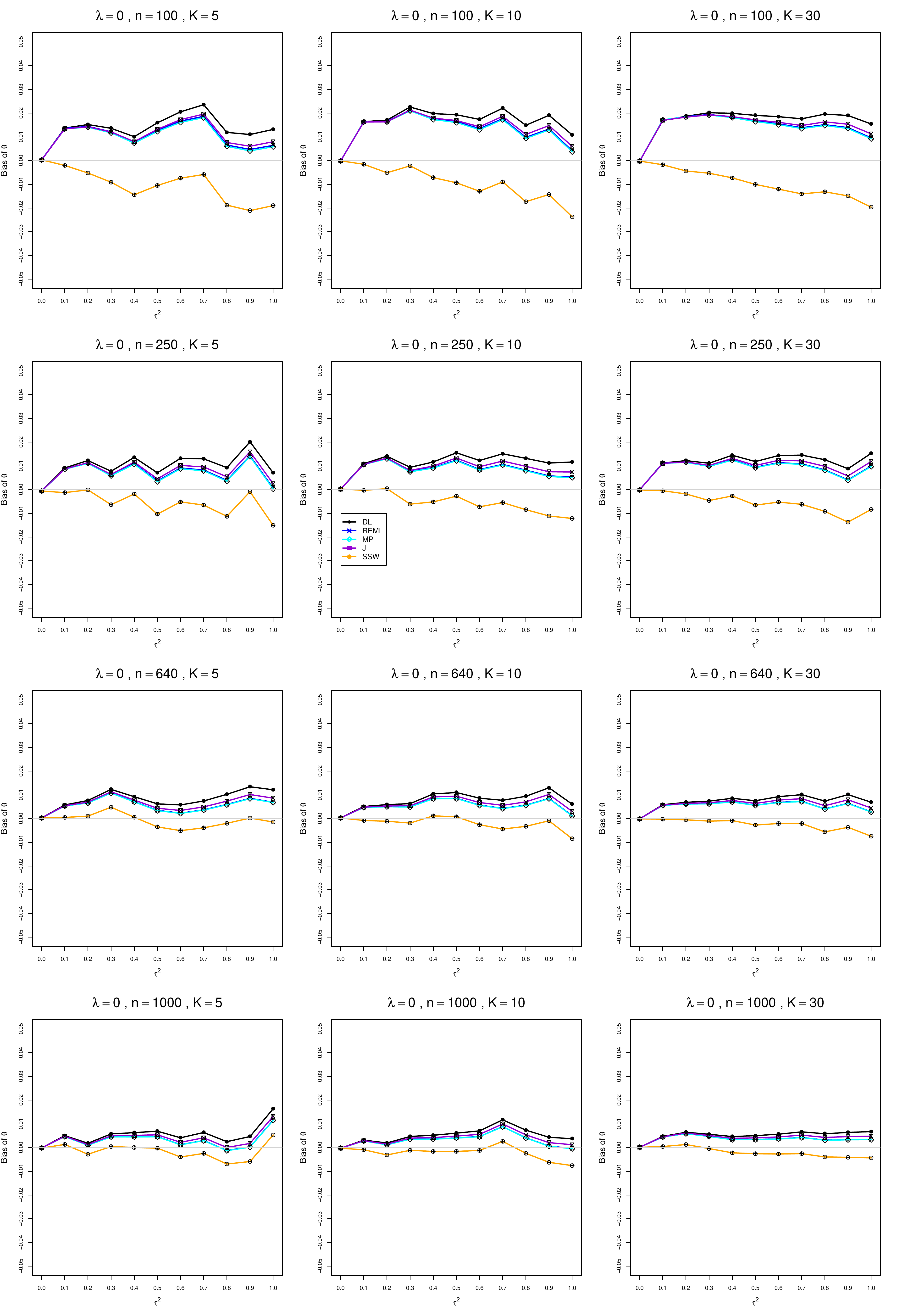}
	\caption{Bias of estimators of $\lambda$ for $\lambda=0$, $n = 100, \;250, \;640, \;1000$, and $K = 5, \;10, \;30$. Usual estimate of $\lambda_i$
		\label{BiasThetaRoM0ln_largeN_small_K}}
\end{figure}

\begin{figure}[t]
	\includegraphics[scale=0.33]{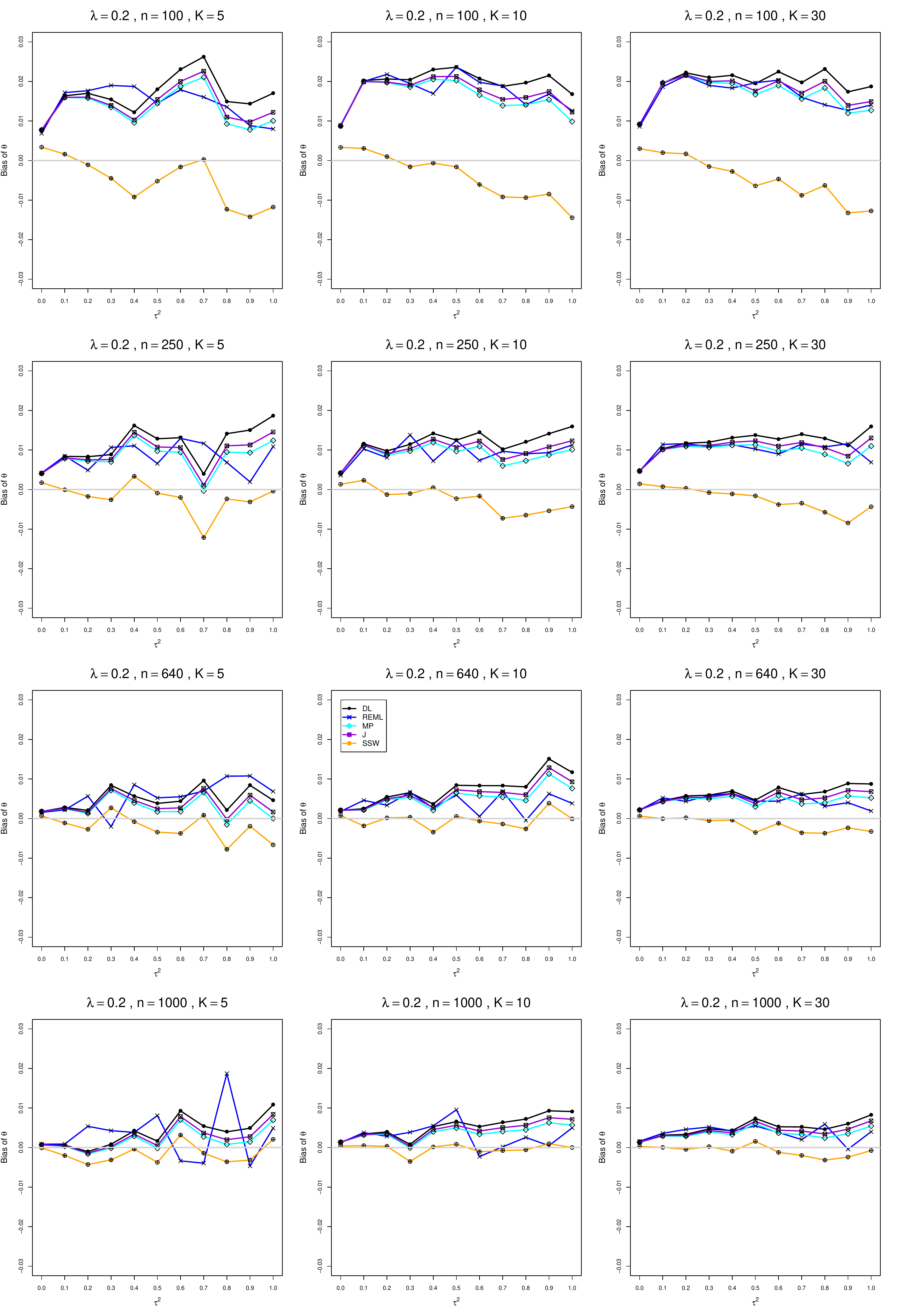}
	\caption{Bias of estimators of $\lambda$ for $\lambda=0.2$, $n = 100, \;250, \;640, \;1000$, and $K = 5, \;10, \;30$. Usual estimate of $\lambda_i$
		\label{BiasThetaRoM02ln_largeN_small_K}}
\end{figure}

\begin{figure}[t]
	\includegraphics[scale=0.33]{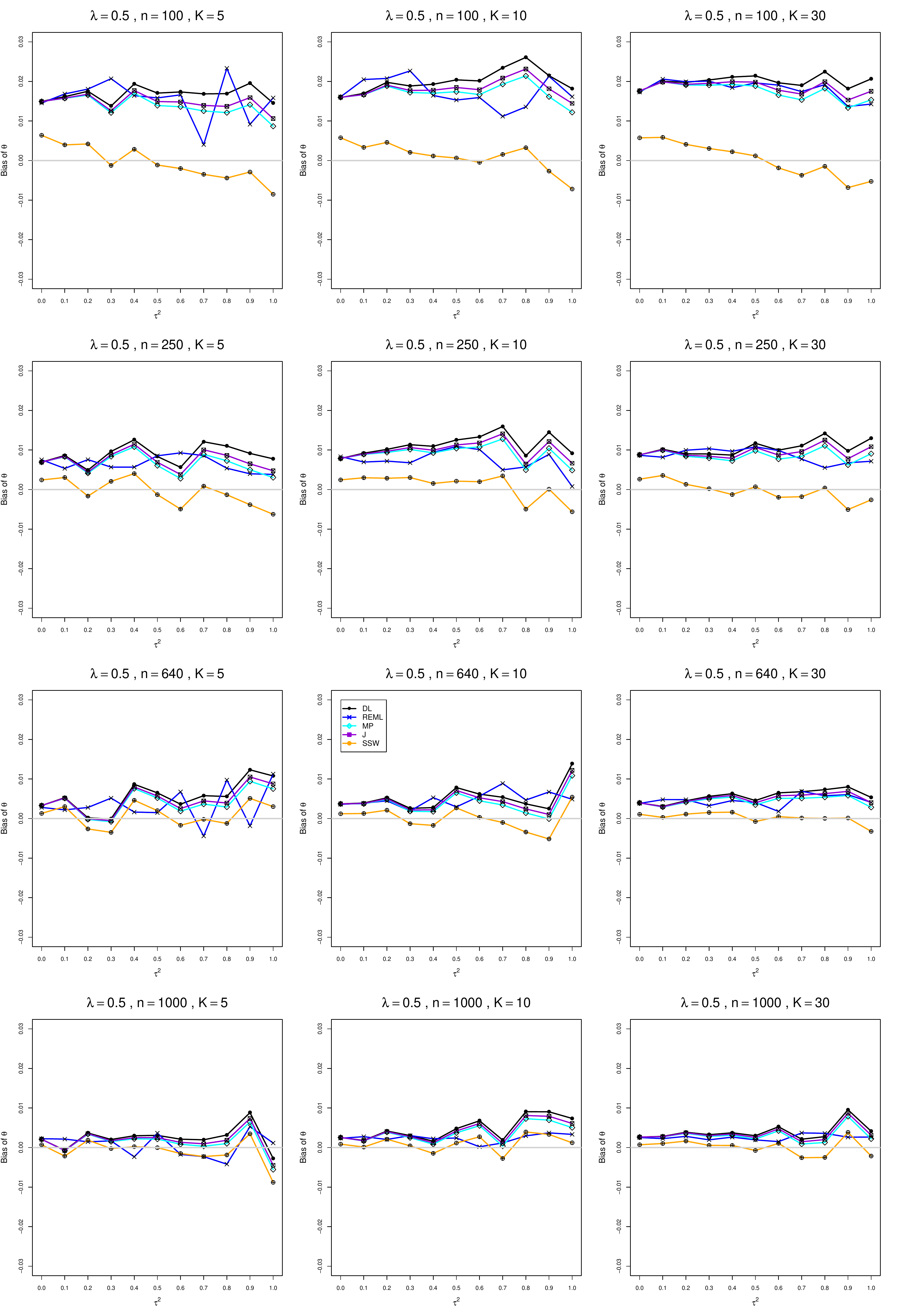}
	\caption{Bias of estimators of $\lambda$ for $\lambda=0.5$, $n = 100, \;250, \;640, \;1000$, and $K = 5, \;10, \;30$. Usual estimate of $\lambda_i$
		\label{BiasThetaRoM05ln_largeN_small_K}}
\end{figure}

\begin{figure}[t]
	\includegraphics[scale=0.33]{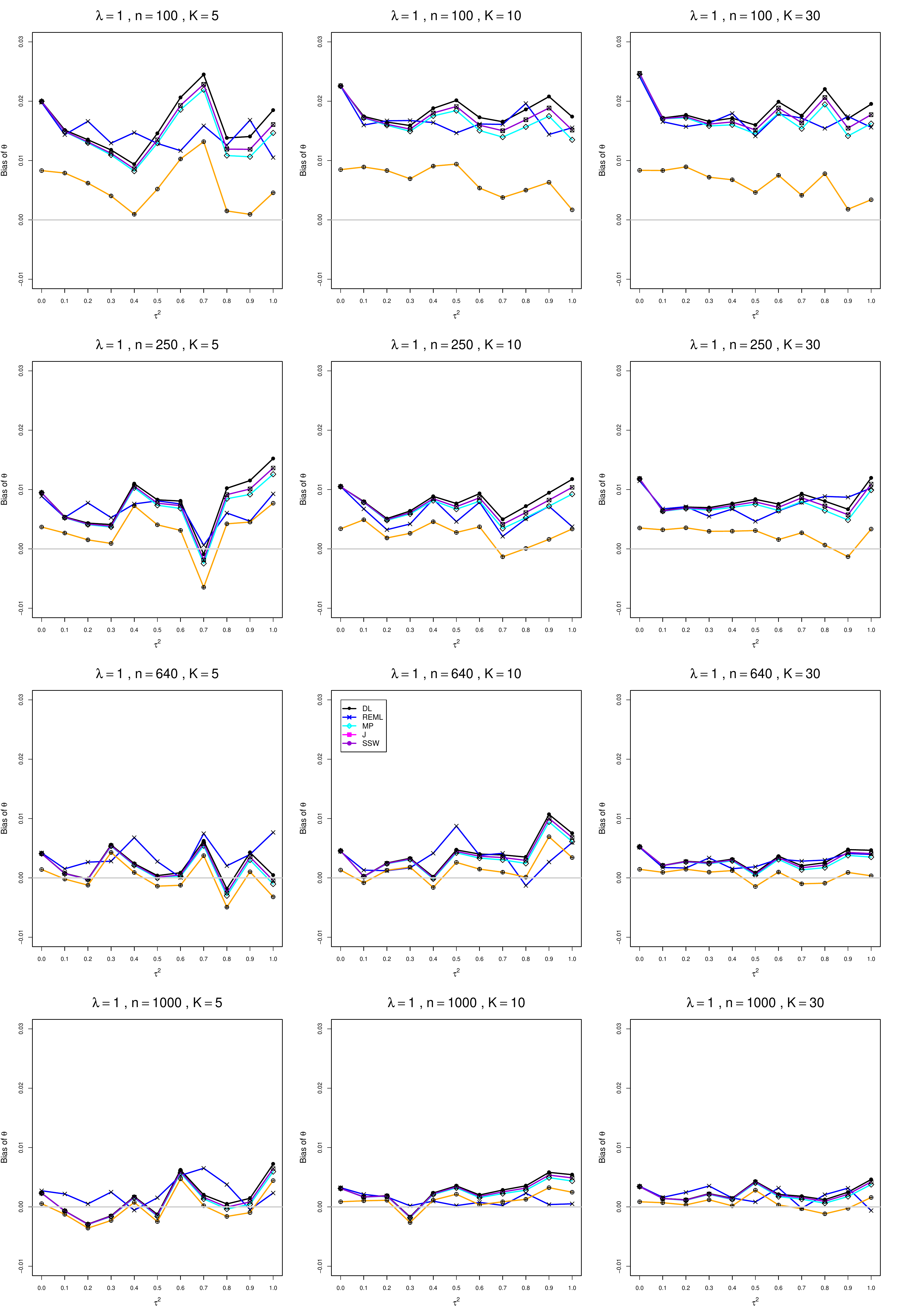}
	\caption{Bias of estimators of $\lambda$ for $\lambda=1$, $n = 100, \;250, \;640, \;1000$, and $K = 5, \;10, \;30$. Usual estimate of $\lambda_i$
		\label{BiasThetaRoM1ln_largeN_small_K}}
\end{figure}

\begin{figure}[t]
	\includegraphics[scale=0.33]{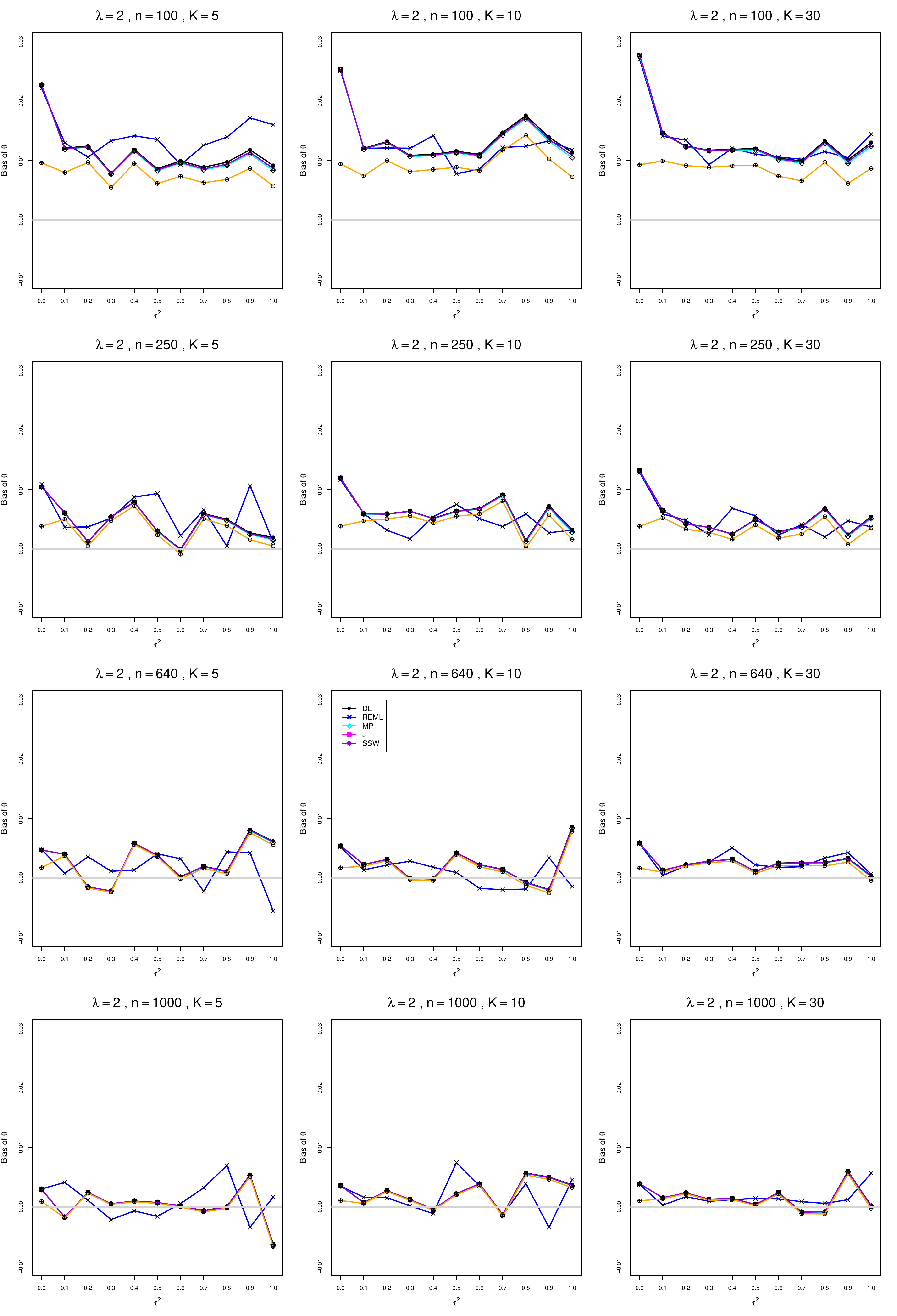}
	\caption{Bias of estimators of $\lambda$ for $\lambda=2$, $n = 100, \;250, \;640, \;1000$, and $K = 5, \;10, \;30$. Usual estimate of $\lambda_i$
		\label{BiasThetaRoM2ln_largeN_small_K}}
\end{figure}

\clearpage
\subsection*{D1.2 Coverage of interval estimators of $\lambda$}
Each figure corresponds to a value of $\lambda \;(= 0, 0.2, 0.5, 1, 2)$, a set of values of $n$ (= 100, 250, 640, 1000), and a set of values of $K$ (= 5, 10, 30).\\
Each panel corresponds to a value of $n$ and a value of $K$ and has $\tau^2 = 0.0(0.1)1.0$ on the horizontal axis.\\
The interval estimators of $\lambda$ are the companions to the inverse-variance-weighted point estimators
\begin{itemize}
	\item DL (DerSimonian-Laird)
	\item REML (restricted maximum likelihood)
	\item MP (Mandel-Paule)
	\item J (Jackson)
\end{itemize}
and
\begin{itemize}
	\item HKSJ (Hartung-Knapp-Sidik-Jonkman)
	\item HKSJ MP (HKSJ with MP estimator of $\tau^2$)
	\item SSW MP (SSW as center and half-width equal to critical value from $t_{K-1}$ times estimated standard deviation of SSW with $\hat{\tau}^2$ = $\hat{\tau}^2_{MP}$)
\end{itemize}

\clearpage
\setcounter{figure}{0}
\renewcommand{\thefigure}{D1.2.\arabic{figure}}
\begin{figure}[t]
	\includegraphics[scale=0.35]{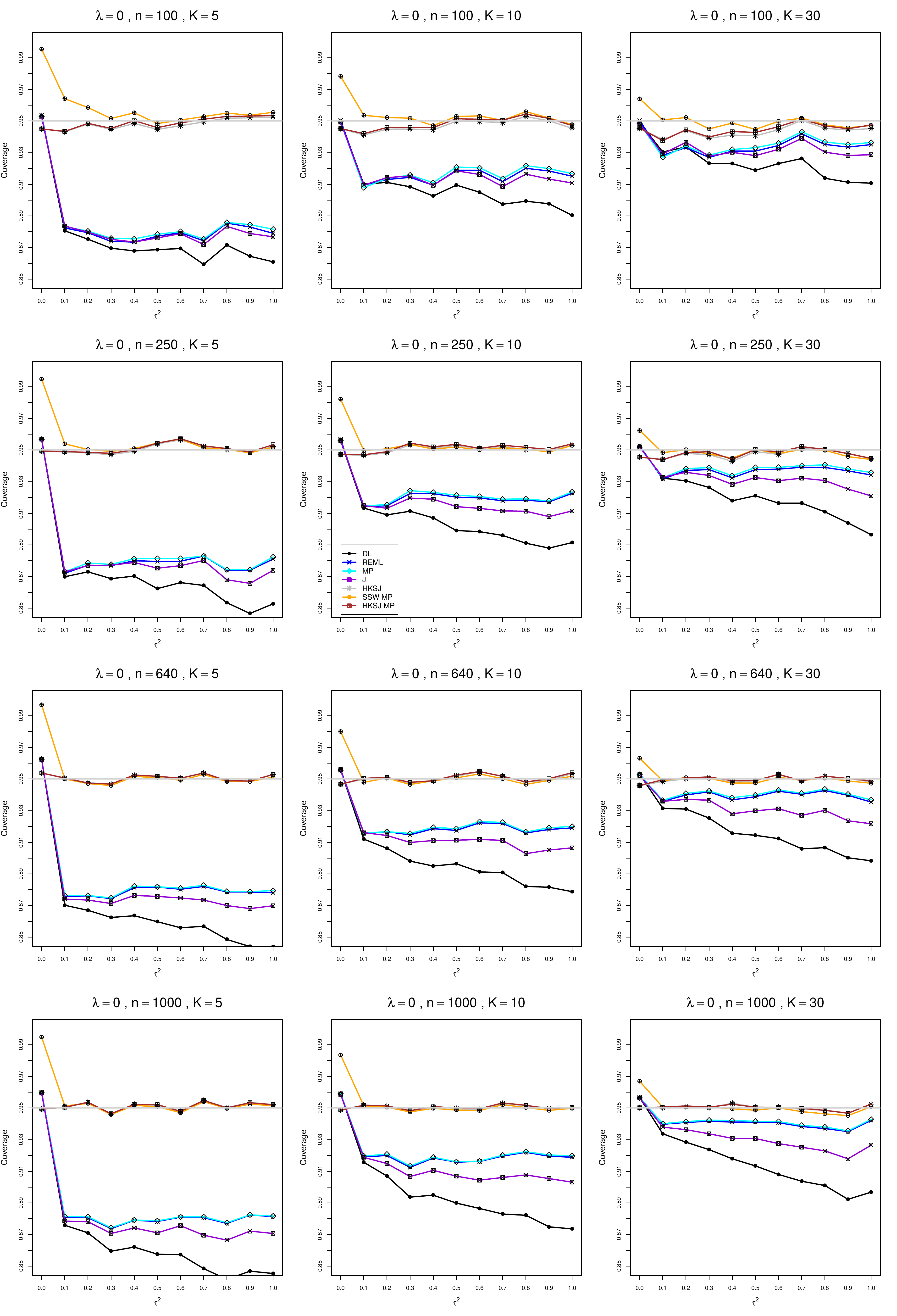}
	\caption{Coverage of 95\% confidence intervals for $\lambda$ when $\lambda=0$, $n = 100, \;250, \;640, \;10000$, and $K = 5, \;10, \;30$. Usual estimate of $\lambda_i$
		\label{CovThetaRoM0ln_largeN_small_K}}
\end{figure}

\begin{figure}[t]
	\includegraphics[scale=0.35]{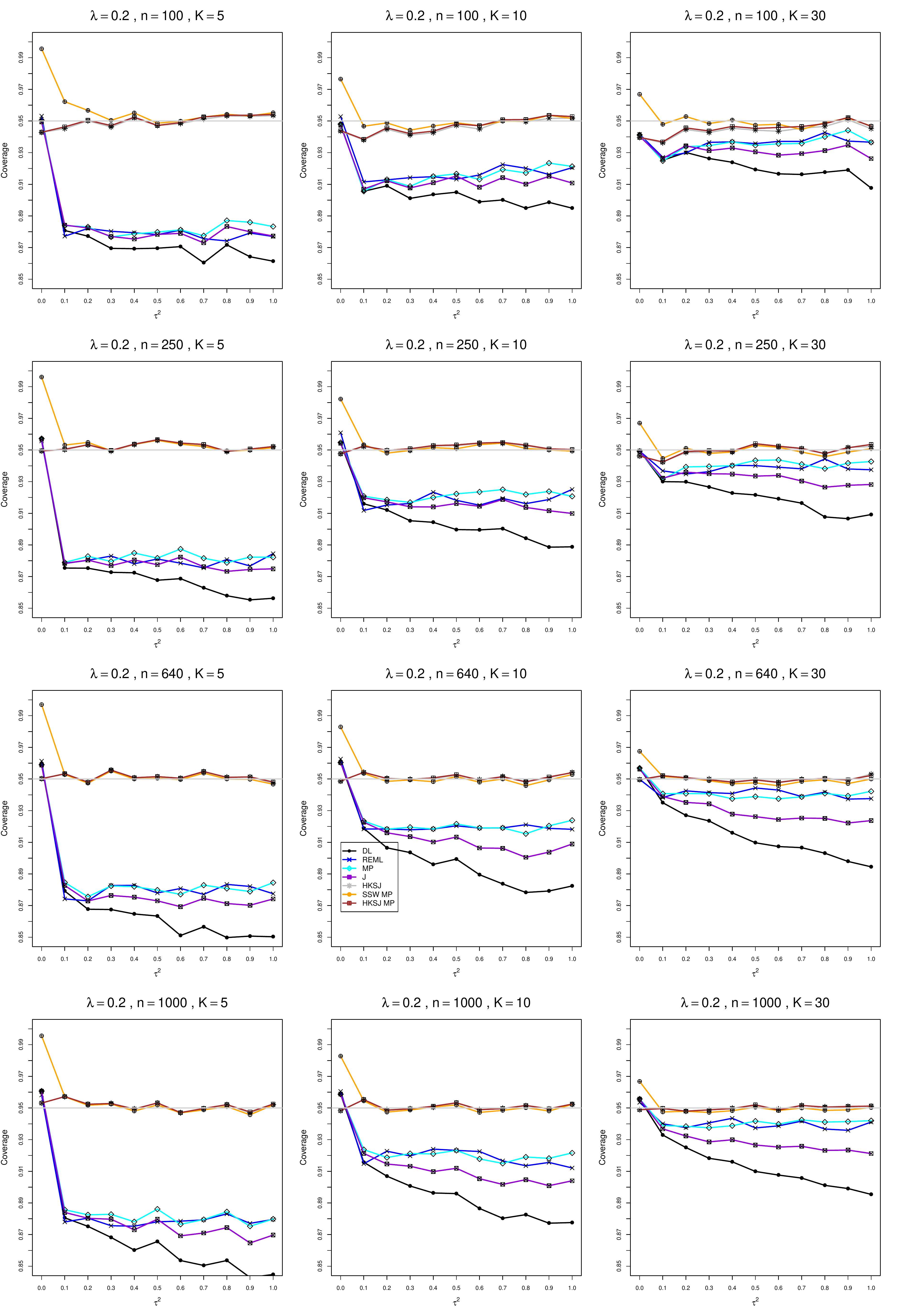}
	\caption{Coverage of 95\% confidence intervals for $\lambda$ when $\lambda=0.2$, $n = 100, \;250, \;640, \;10000$, and $K = 5, \;10, \;30$. Usual estimate of $\lambda_i$
		\label{CovThetaRoM02ln_largeN_small_K}}
\end{figure}

\begin{figure}[t]
	\includegraphics[scale=0.35]{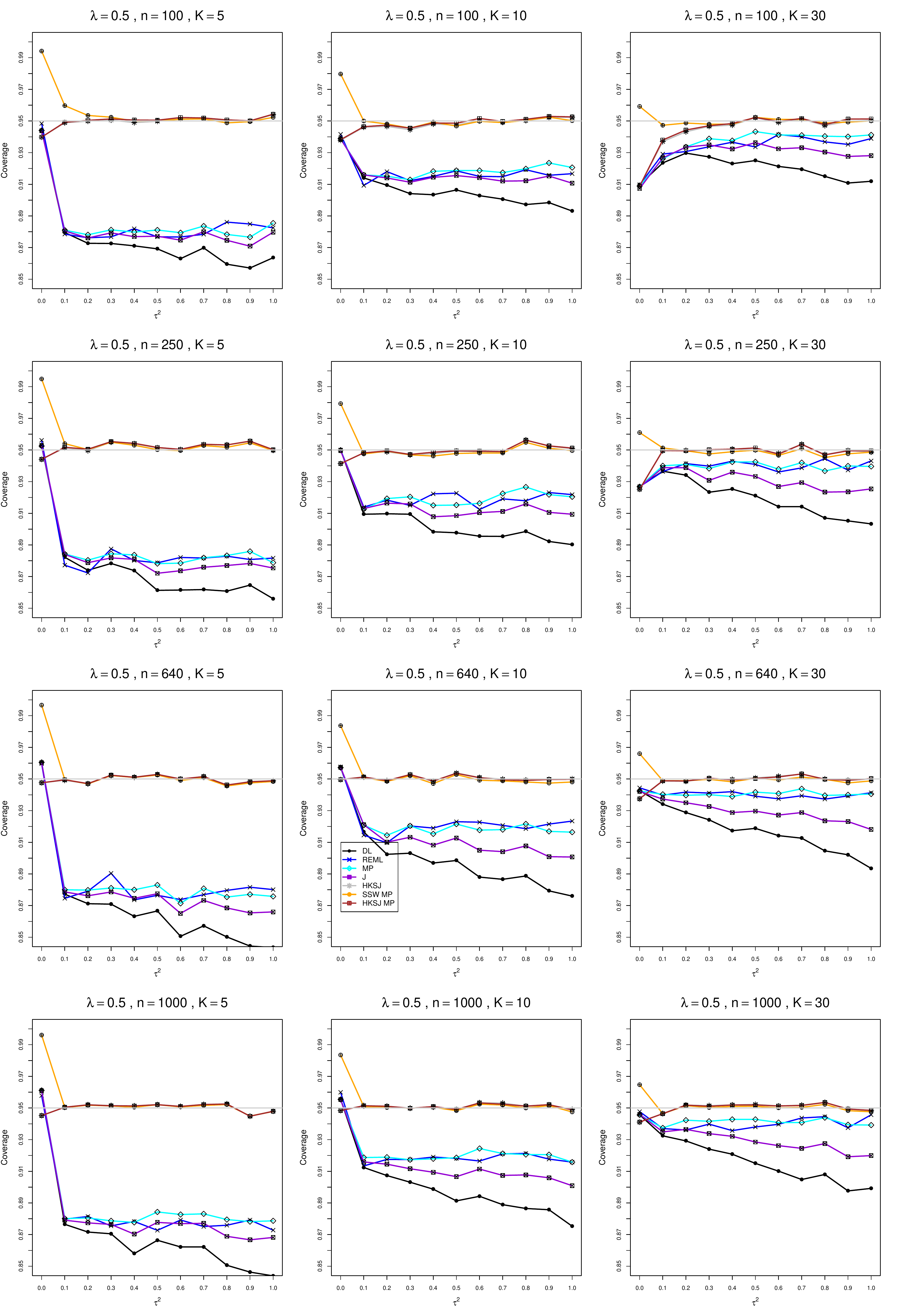}
	\caption{Coverage of 95\% confidence intervals for $\lambda$ when $\lambda=0.5$, $n = 100, \;250, \;640, \;10000$, and $K = 5, \;10, \;30$. Usual estimate of $\lambda_i$
		\label{CovThetaRoM05ln_largeN_small_K}}
\end{figure}

\begin{figure}[t]
	\includegraphics[scale=0.35]{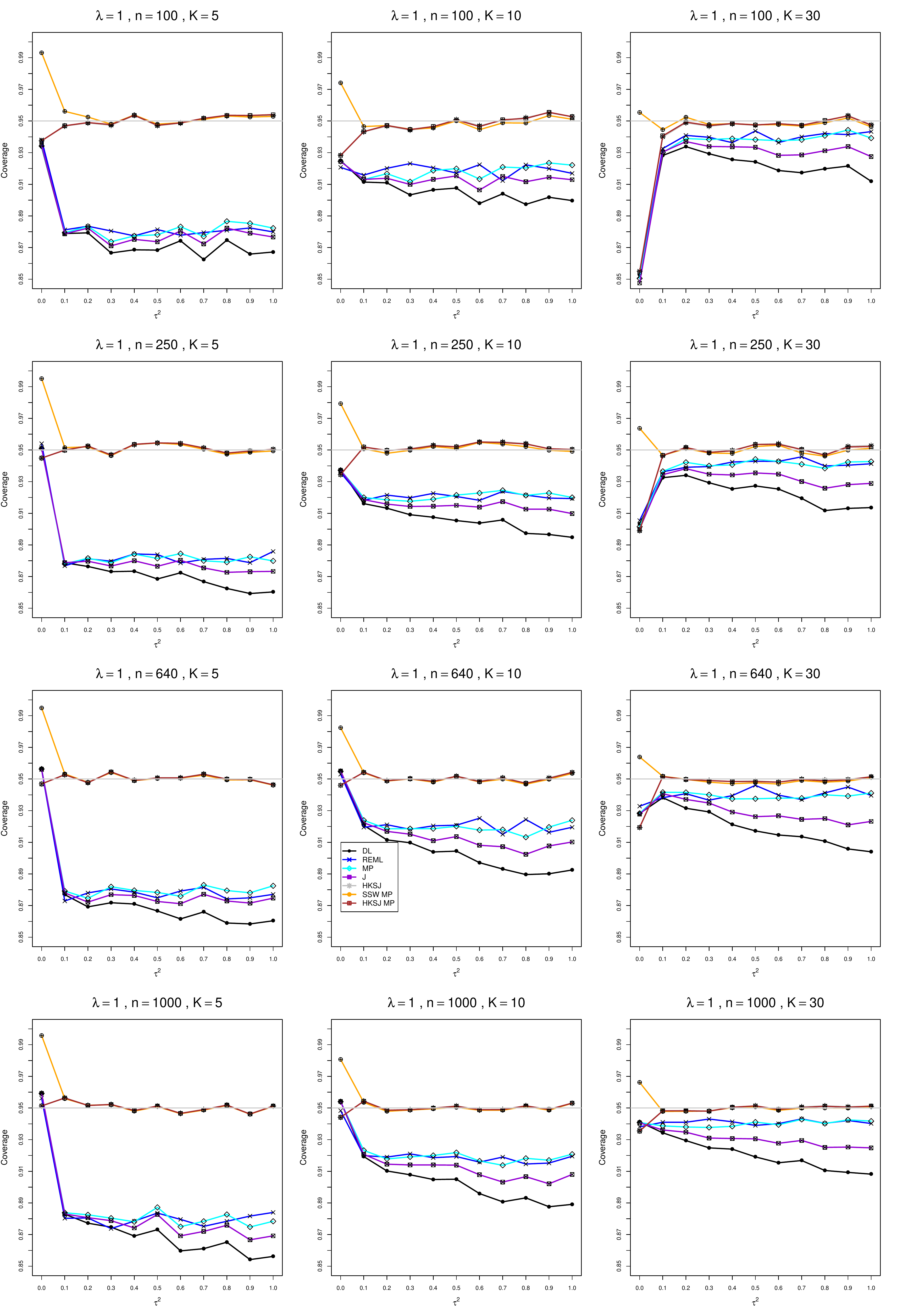}
	\caption{Coverage of 95\% confidence intervals for $\lambda$ when $\lambda=1$, $n = 100, \;250, \;640, \;10000$, and $K = 5, \;10, \;30$. Usual estimate of $\lambda_i$
		\label{CovThetaRoM1ln_largeN_small_K}}
\end{figure}

\begin{figure}[t]
	\includegraphics[scale=0.35]{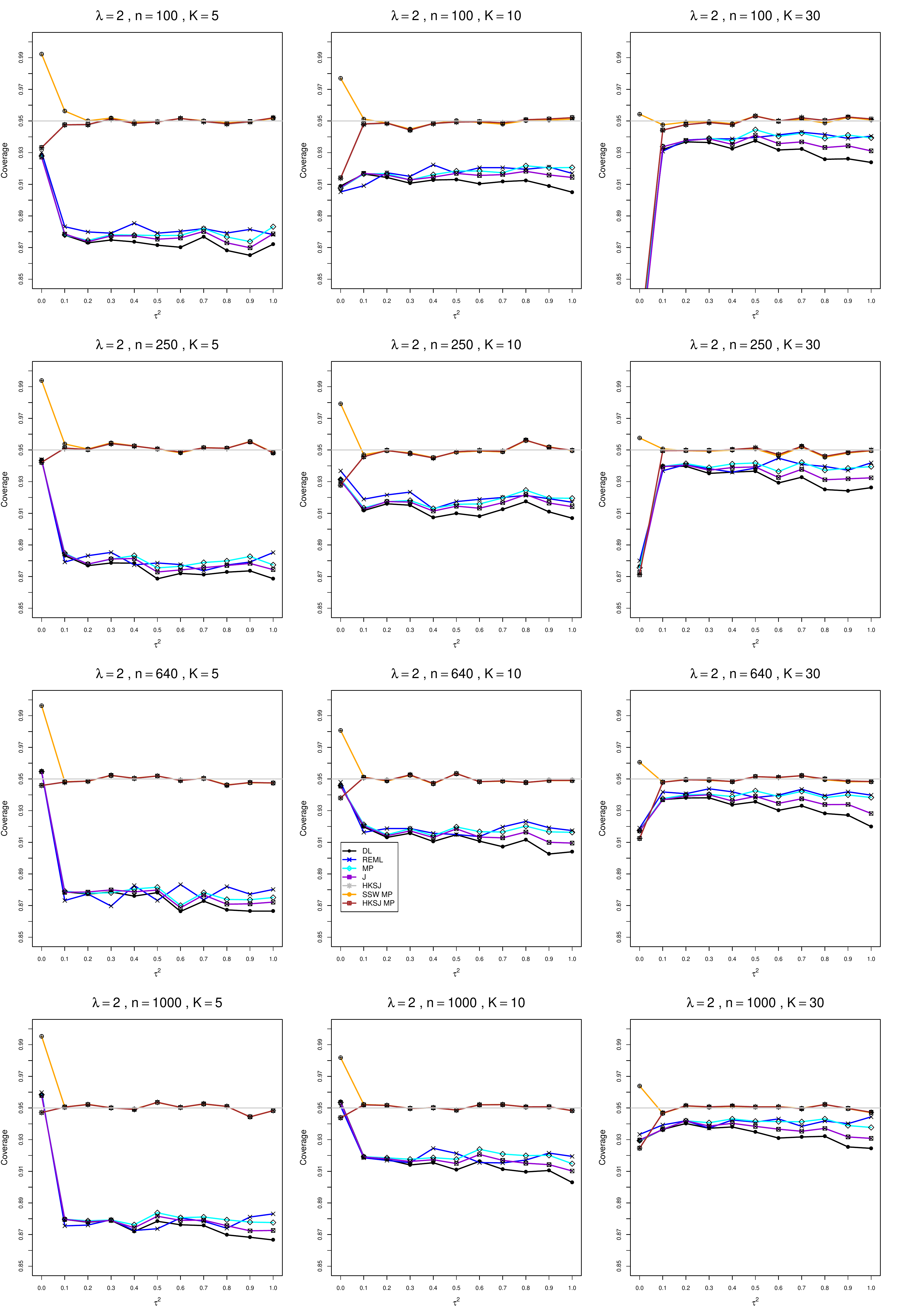}
	\caption{Coverage of 95\% confidence intervals for $\lambda$ when $\lambda=2$, $n = 100, \;250, \;640, \;10000$, and $K = 5, \;10, \;30$. Usual estimate of $\lambda_i$
		\label{CovThetaRoM2ln_largeN_small_K}}
\end{figure}

\clearpage
\section*{D2. Lognormal model, bias-corrected estimator of $\lambda_i$, $n= 100, 250, 640, 1000$, $K=5,10,30$}
\subsection*{D2.1 Bias of point estimators of $\lambda$}
Each figure corresponds to a value of $\lambda \;(= 0, 0.2, 0.5, 1, 2)$, a set of values of $n$ (= 100, 250, 640, 1000), and a set of values of $K$ (= 5, 10, 30).\\
Each panel corresponds to a value of $n$ and a value of $K$ and has $\tau^2 = 0.0(0.1)1.0$ on the horizontal axis.\\
The point estimators of $\lambda$ are
\begin{itemize}
	\item DL (DerSimonian-Laird)
	\item REML (restricted maximum likelihood)
	\item MP (Mandel-Paule)
	\item J (Jackson)
	\item SSW (sample-size-weighted)
\end{itemize}

\clearpage
\setcounter{figure}{0}
\renewcommand{\thefigure}{D2.1.\arabic{figure}}
\begin{figure}[t]
	\includegraphics[scale=0.33]{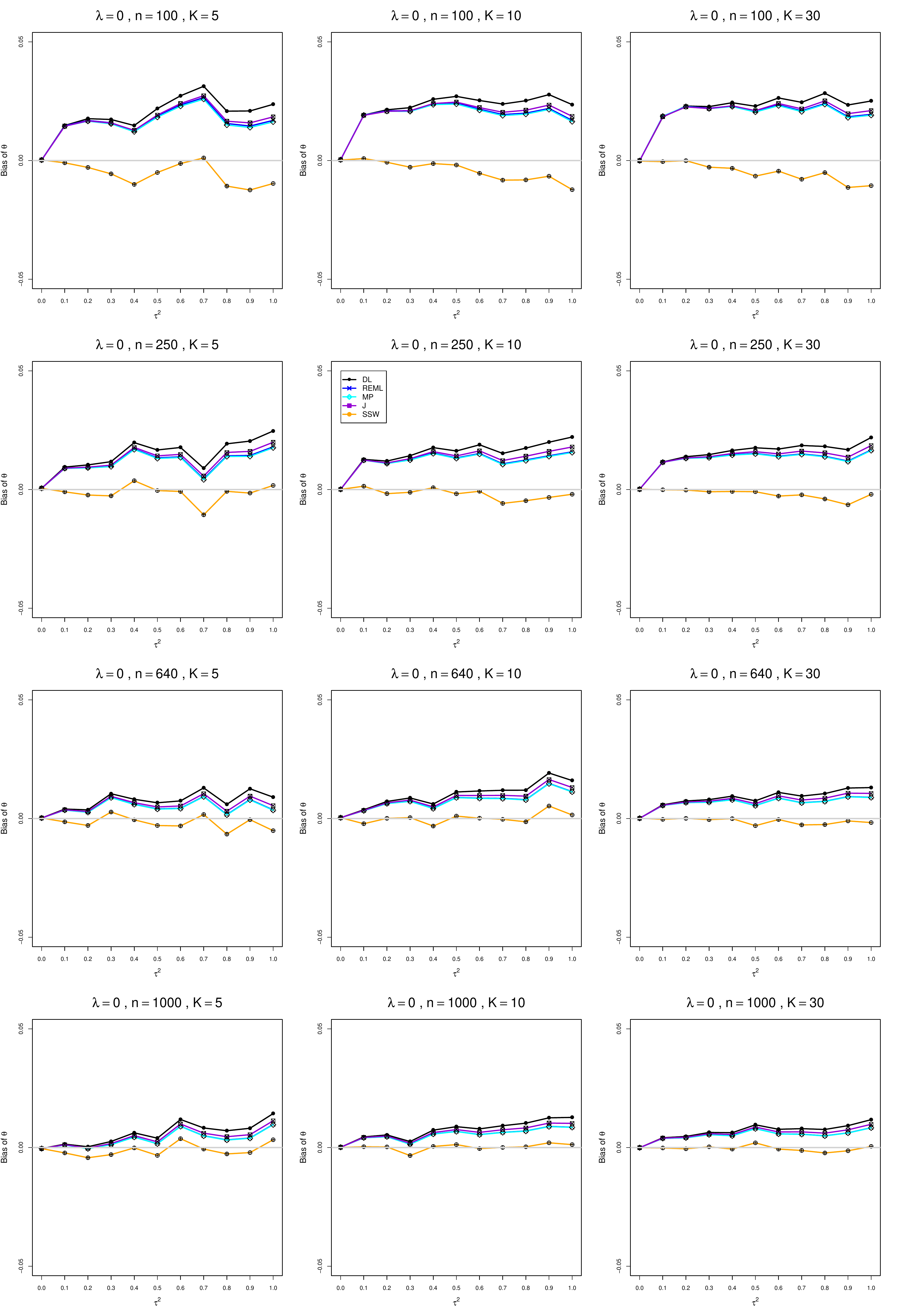}
	\caption{Bias of estimators of $\lambda$ for $\lambda=0$, $n = 100, \;250, \;640, \;1000$, and $K = 5, \;10, \;30$. Bias-corrected estimate of $\lambda_i$
		\label{BiasThetaRoM0lnCor_largeN_small_K}}
\end{figure}

\begin{figure}[t]
	\includegraphics[scale=0.33]{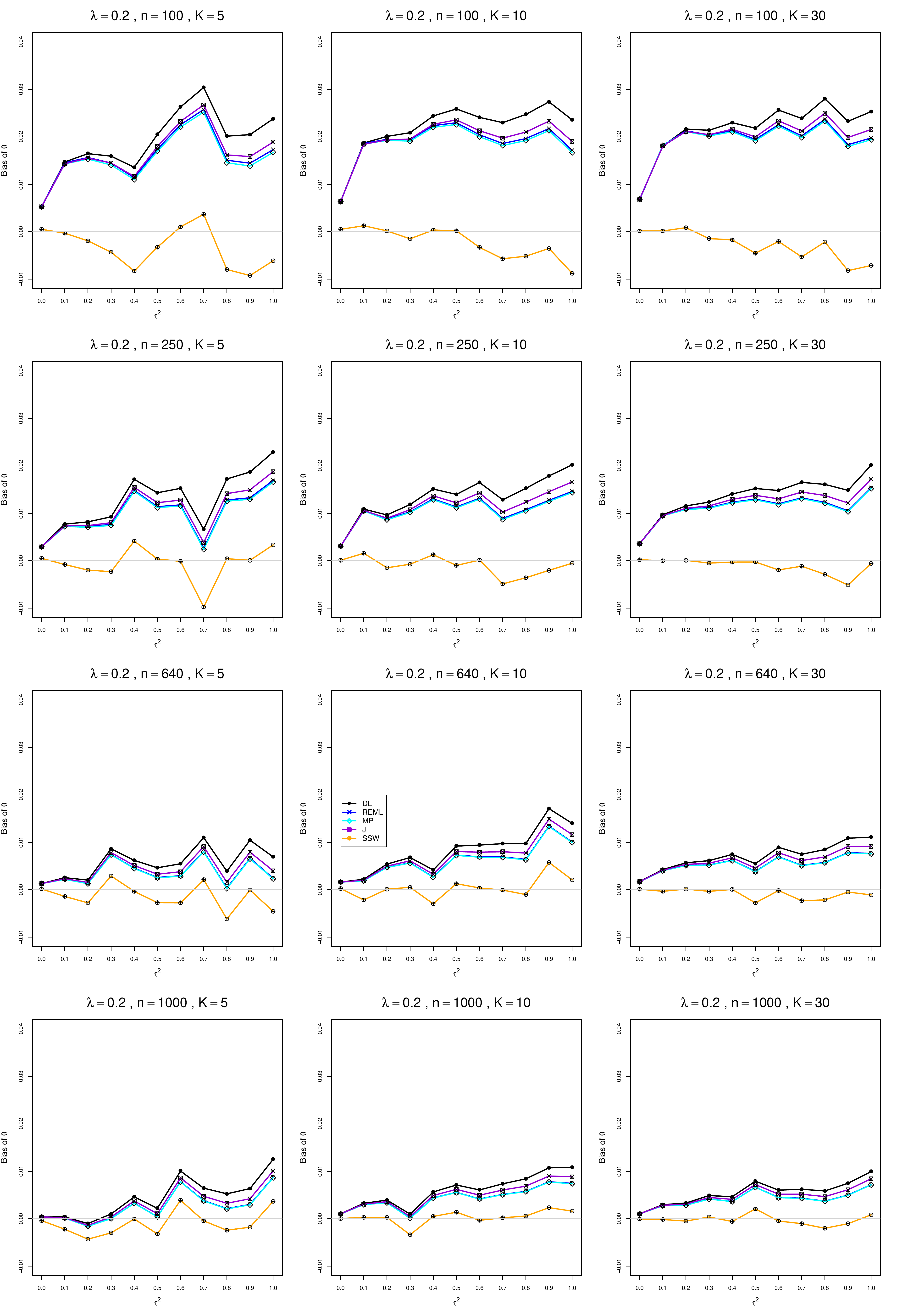}
	\caption{Bias of estimators of $\lambda$ for $\lambda=0.2$, $n = 100, \;250, \;640, \;1000$, and $K = 5, \;10, \;30$. Bias-corrected estimate of $\lambda_i$
		\label{BiasThetaRoM02lnCor_largeN_small_K}}
\end{figure}

\begin{figure}[t]
	\includegraphics[scale=0.33]{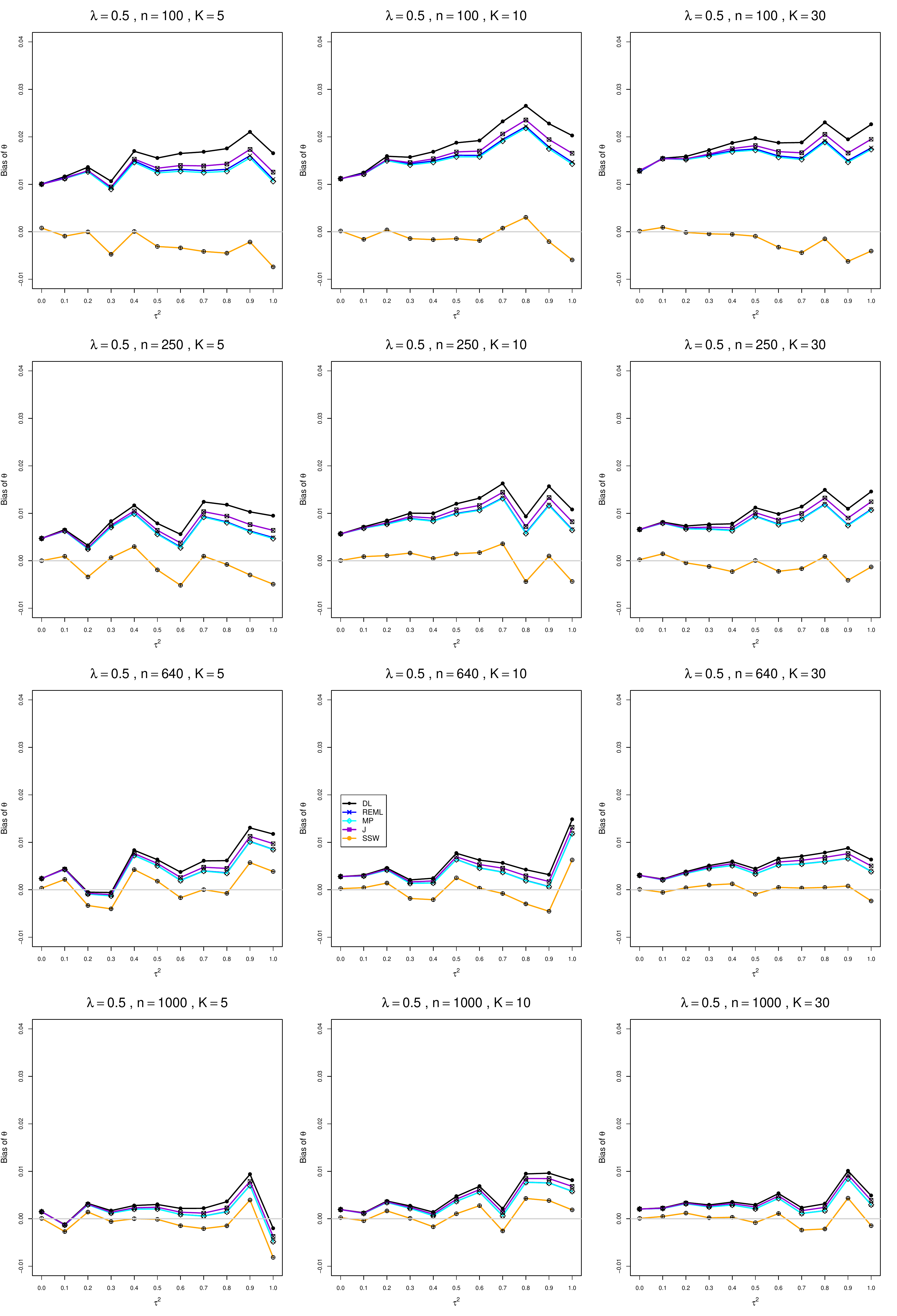}
	\caption{Bias of estimators of $\lambda$ for $\lambda=0.5$, $n = 100, \;250, \;640, \;1000$, and $K = 5, \;10, \;30$. Bias-corrected estimate of $\lambda_i$
		\label{BiasThetaRoM05lnCor__largeN_small_K}}
\end{figure}

\begin{figure}[t]
	\includegraphics[scale=0.33]{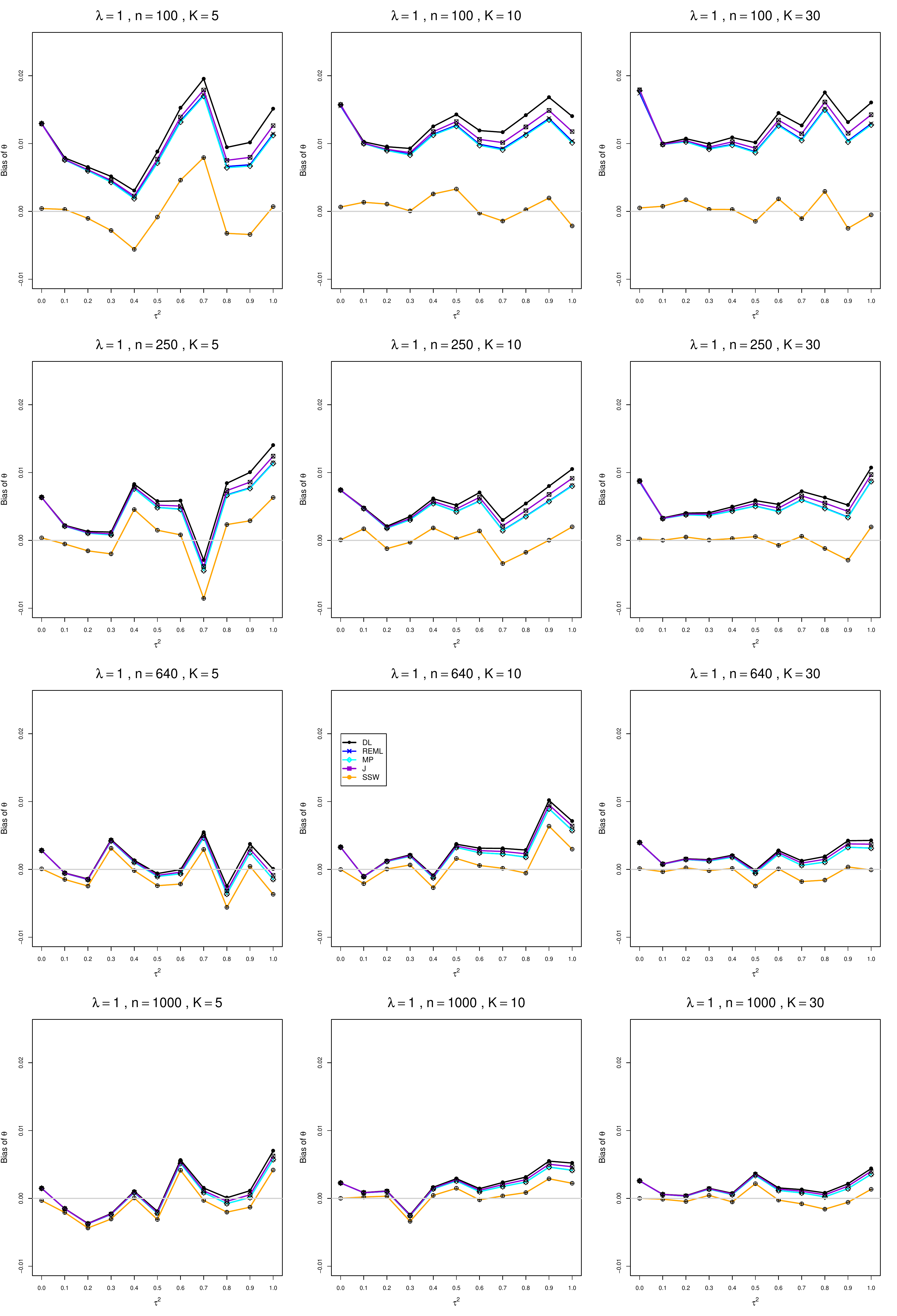}
	\caption{Bias of estimators of $\lambda$ for $\lambda=1$, $n = 100, \;250, \;640, \;1000$, and $K = 5, \;10, \;30$. Bias-corrected estimate of $\lambda_i$
		\label{BiasThetaRoM1lnCor__largeN_small_K}}
\end{figure}

\begin{figure}[t]
	\includegraphics[scale=0.33]{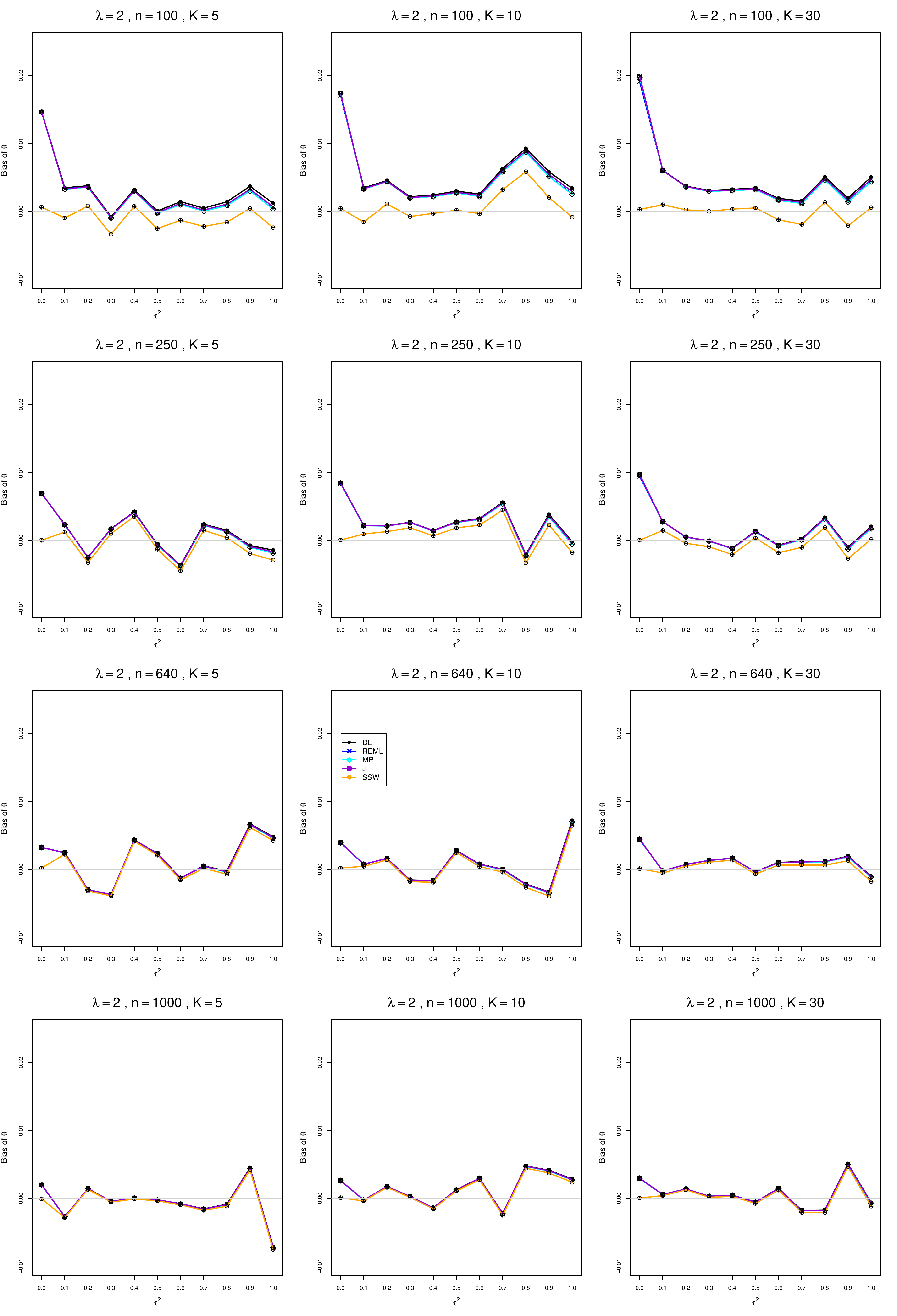}
	\caption{Bias of estimators of $\lambda$ for $\lambda=2$, $n = 100, \;250, \;640, \;1000$, and $K = 5, \;10, \;30$. Bias-corrected estimate of $\lambda_i$
		\label{BiasThetaRoM2lnCor__largeN_small_K}}
\end{figure}

\clearpage
\subsection*{D2.2 Coverage of interval estimators of $\lambda$}
Each figure corresponds to a value of $\lambda \;(= 0, 0.2, 0.5, 1, 2)$, a set of values of $n$ (= 100, 250, 640, 1000), and a set of values of $K$ (= 5, 10, 30).\\
Each panel corresponds to a value of $n$ and a value of $K$ and has $\tau^2 = 0.0(0.1)1.0$ on the horizontal axis.\\
The interval estimators of $\lambda$ are the companions to the inverse-variance-weighted point estimators
\begin{itemize}
	\item DL (DerSimonian-Laird)
	\item REML (restricted maximum likelihood)
	\item MP (Mandel-Paule)
	\item J (Jackson)
\end{itemize}
and
\begin{itemize}
	\item HKSJ (Hartung-Knapp-Sidik-Jonkman)
	\item HKSJ MP (HKSJ with MP estimator of $\tau^2$)
	\item SSW MP (SSW as center and half-width equal to critical value from $t_{K-1}$ times estimated standard deviation of SSW with $\hat{\tau}^2$ = $\hat{\tau}^2_{MP}$)
\end{itemize}

\clearpage
\setcounter{figure}{0}
\renewcommand{\thefigure}{D2.2.\arabic{figure}}
\begin{figure}[t]
	\includegraphics[scale=0.35]{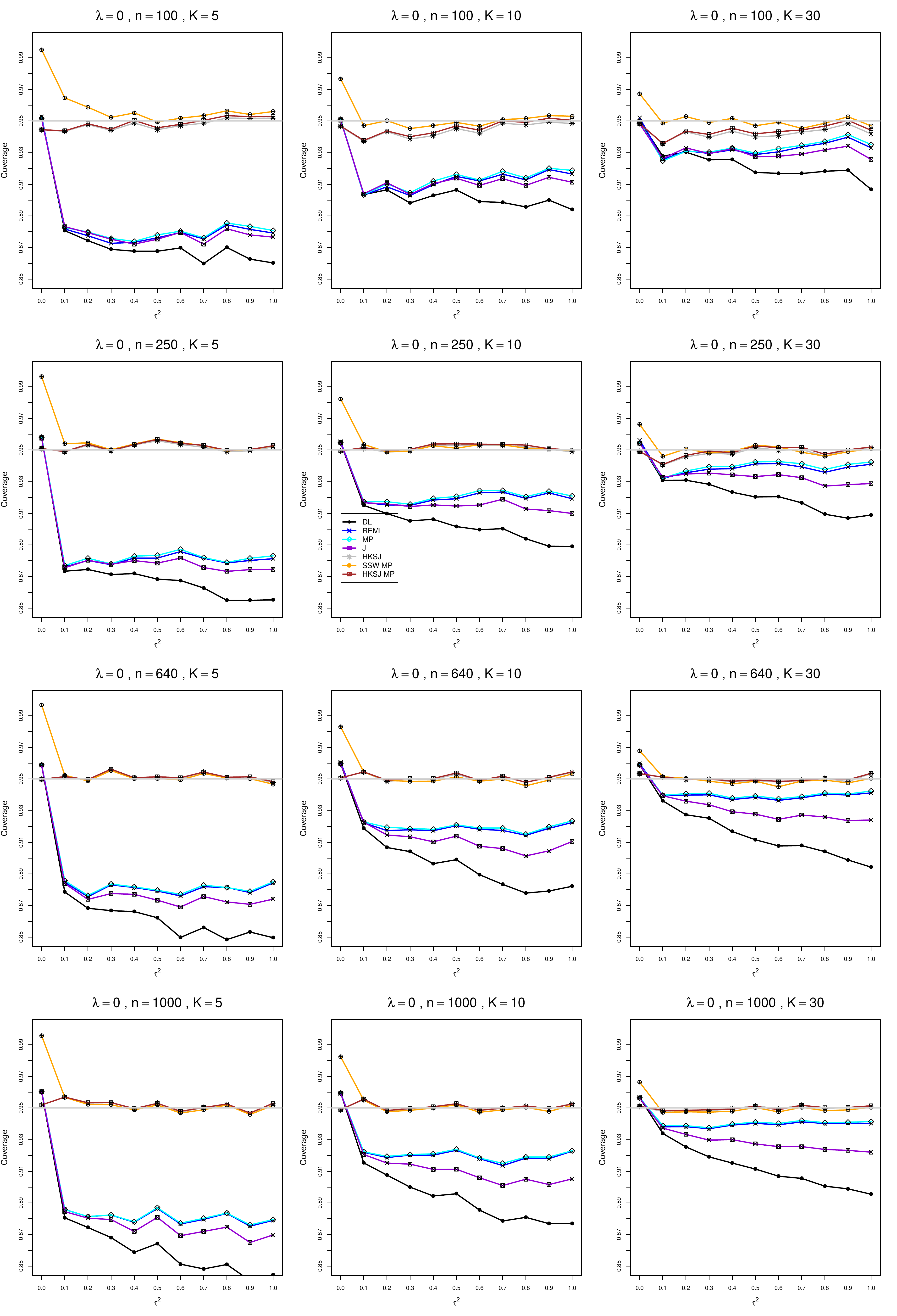}
	\caption{Coverage of 95\% confidence intervals for $\lambda$ when $\lambda=0$, $n = 100, \;250, \;640, \;1000$, and $K = 5, \;10, \;30$. Bias-corrected estimate of $\lambda_i$ 		
		\label{CovThetaRoM0lnCor_largeN_small_K}}
\end{figure}
\begin{figure}[t]
	\includegraphics[scale=0.35]{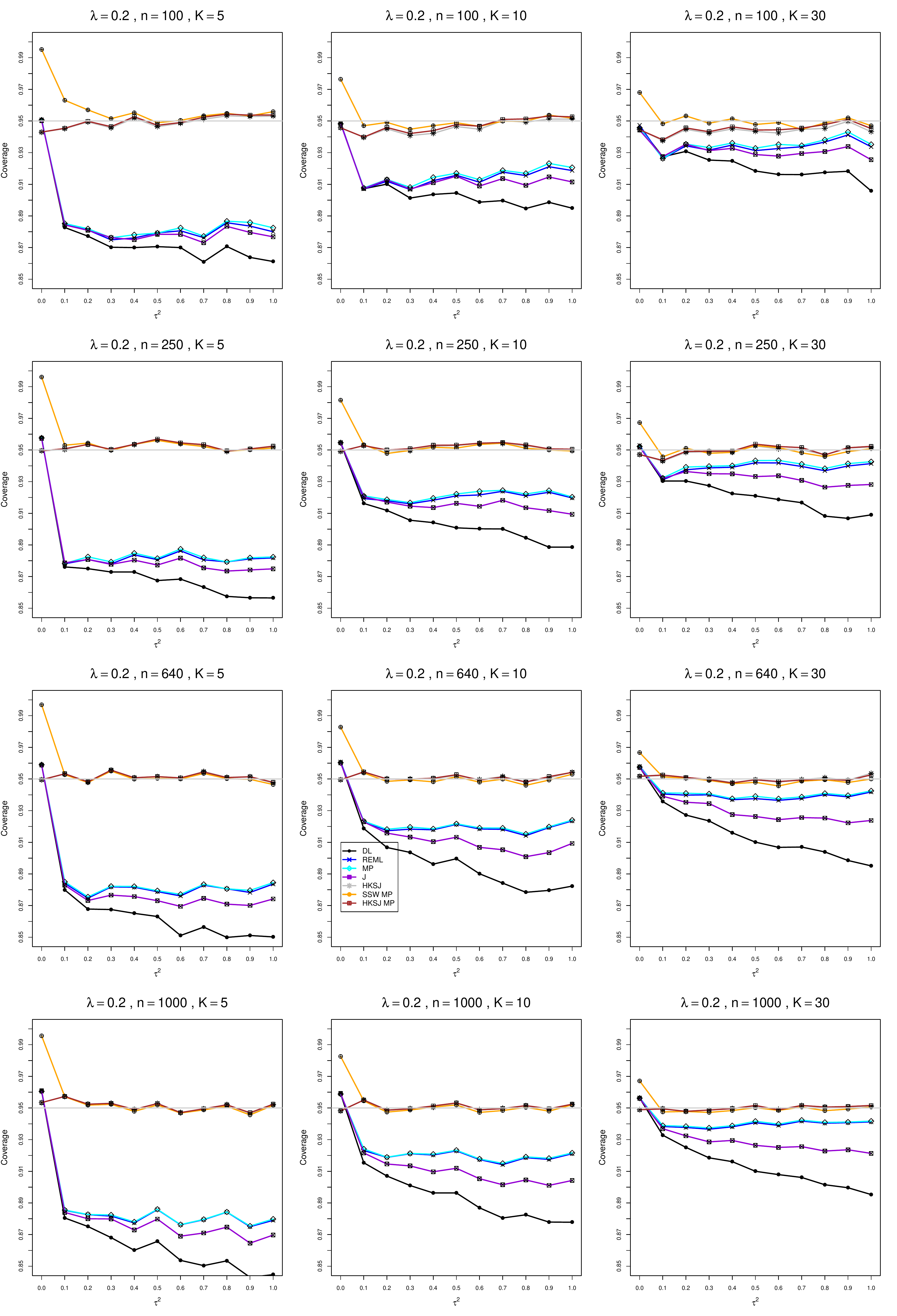}
	\caption{Coverage of 95\% confidence intervals for $\lambda$ when $\lambda=0.2$, $n = 100, \;250, \;640, \;1000$, and $K = 5, \;10, \;30$. Bias-corrected estimate of $\lambda_i$ 		
		\label{CovThetaRoM02lnCor_largeN_small_K}}
\end{figure}

\begin{figure}[t]
	\includegraphics[scale=0.35]{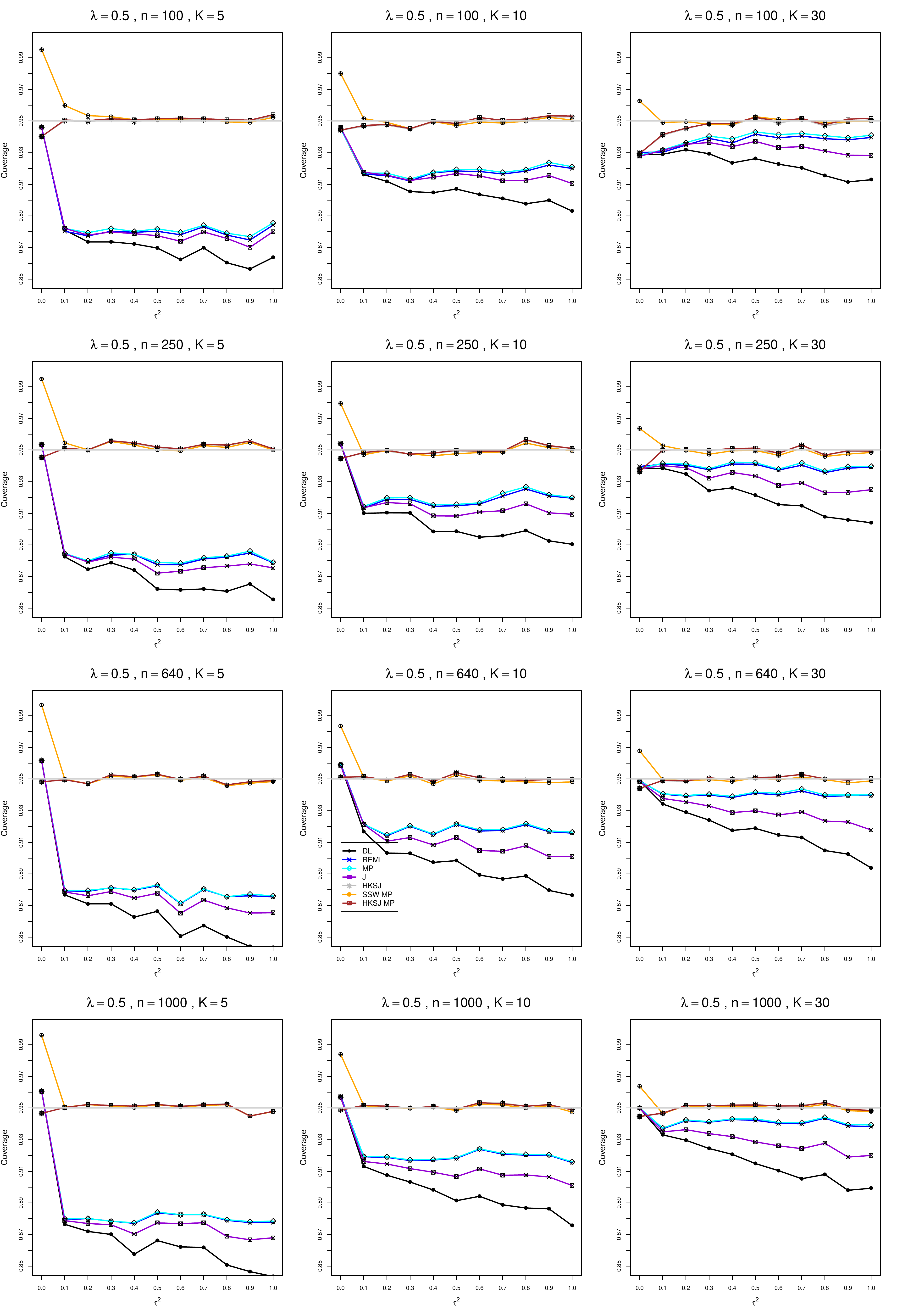}
	\caption{Coverage of 95\% confidence intervals for $\lambda$ when $\lambda=0.5$, $n = 100, \;250, \;640, \;1000$, and $K = 5, \;10, \;30$. Bias-corrected estimate of $\lambda_i$		
		\label{CovThetaRoM05lnCor_largeN_small_K}}
\end{figure}

\begin{figure}[t]
	\includegraphics[scale=0.35]{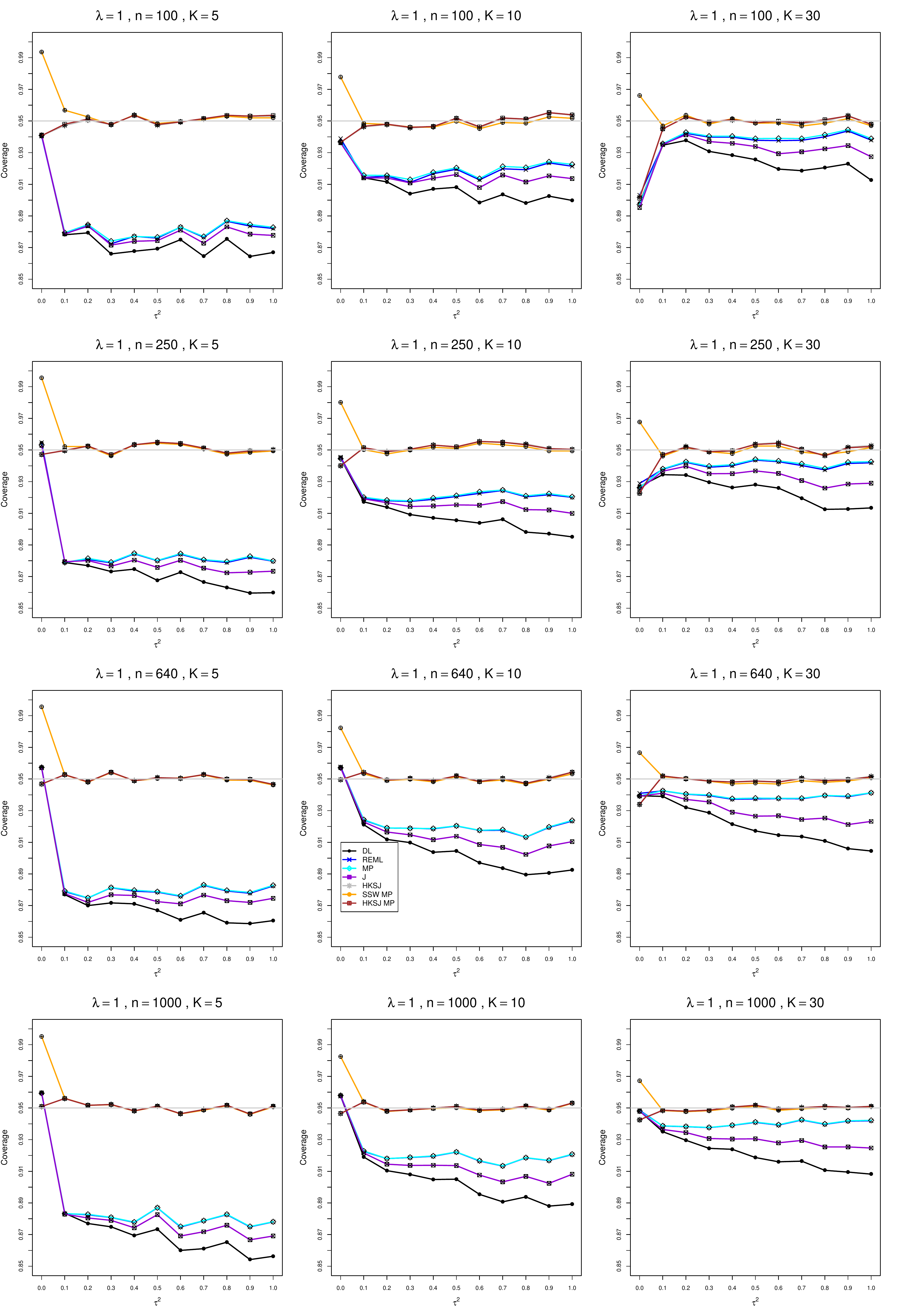}
	\caption{Coverage of 95\% confidence intervals for $\lambda$ when $\lambda=1$, $n = 100, \;250, \;640, \;1000$, and $K = 5, \;10, \;30$. Bias-corrected estimate of $\lambda_i$ 		
		\label{CovThetaRoM1lnCor_largeN_small_K}}
\end{figure}
\begin{figure}[t]
\includegraphics[scale=0.35]{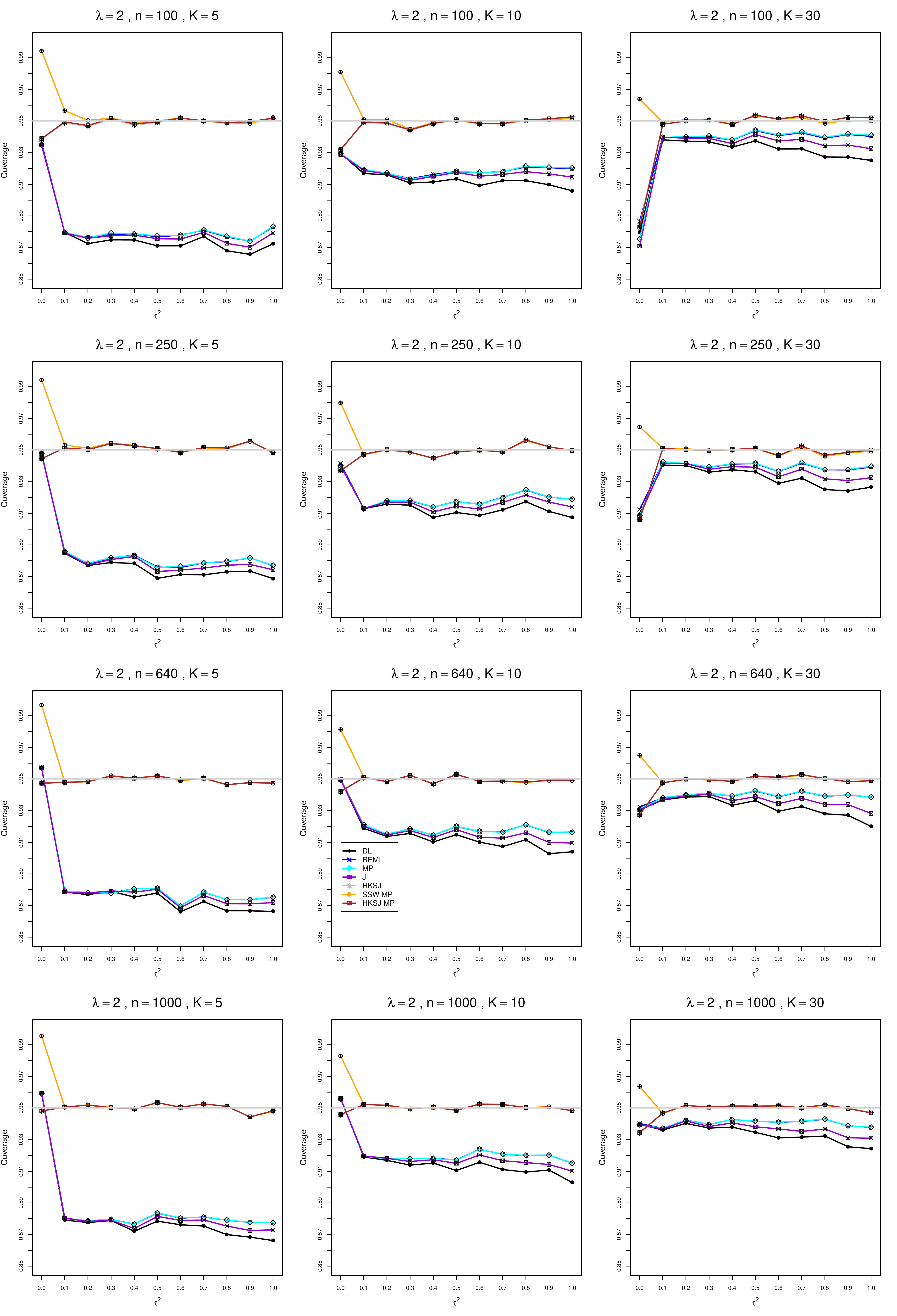}\caption{Coverage of 95\% confidence intervals for $\lambda$ when $\lambda=2$, $n = 100, \;250, \;640, \;1000$, and $K = 5, \;10, \;30$. Bias-corrected estimate of $\lambda_i$		\label{CovThetaRoM2lnCor_largeN_small_K}}
	\end{figure}

\clearpage
\section*{D3. Lognormal model, usual estimator of $\lambda_i$, $n= 100, 250, 640, 1000$, $K=50,100,125$}
\subsection*{D3.1 Bias of point estimators of $\lambda$}
Each figure corresponds to a value of $\lambda \;(= 0, 0.2, 0.5, 1, 2)$, a set of values of $n$ (= 100, 250, 640, 1000), and a set of values of $K$ (= 50, 100, 125).\\
	Each panel corresponds to a value of $n$ and a value of $K$ and has $\tau^2 = 0.0(0.1)1.0$ on the horizontal axis.\\
	The point estimators of $\lambda$ are
	\begin{itemize}
	\item DL (DerSimonian-Laird)
	\item REML (restricted maximum likelihood)
	\item MP (Mandel-Paule)
	\item J (Jackson)
	\item SSW (sample-size-weighted)
	\end{itemize}
	
	\clearpage
	\setcounter{figure}{0}
	\renewcommand{\thefigure}{D3.1.\arabic{figure}}
	
	\begin{figure}[t]
	\includegraphics[scale=0.33]{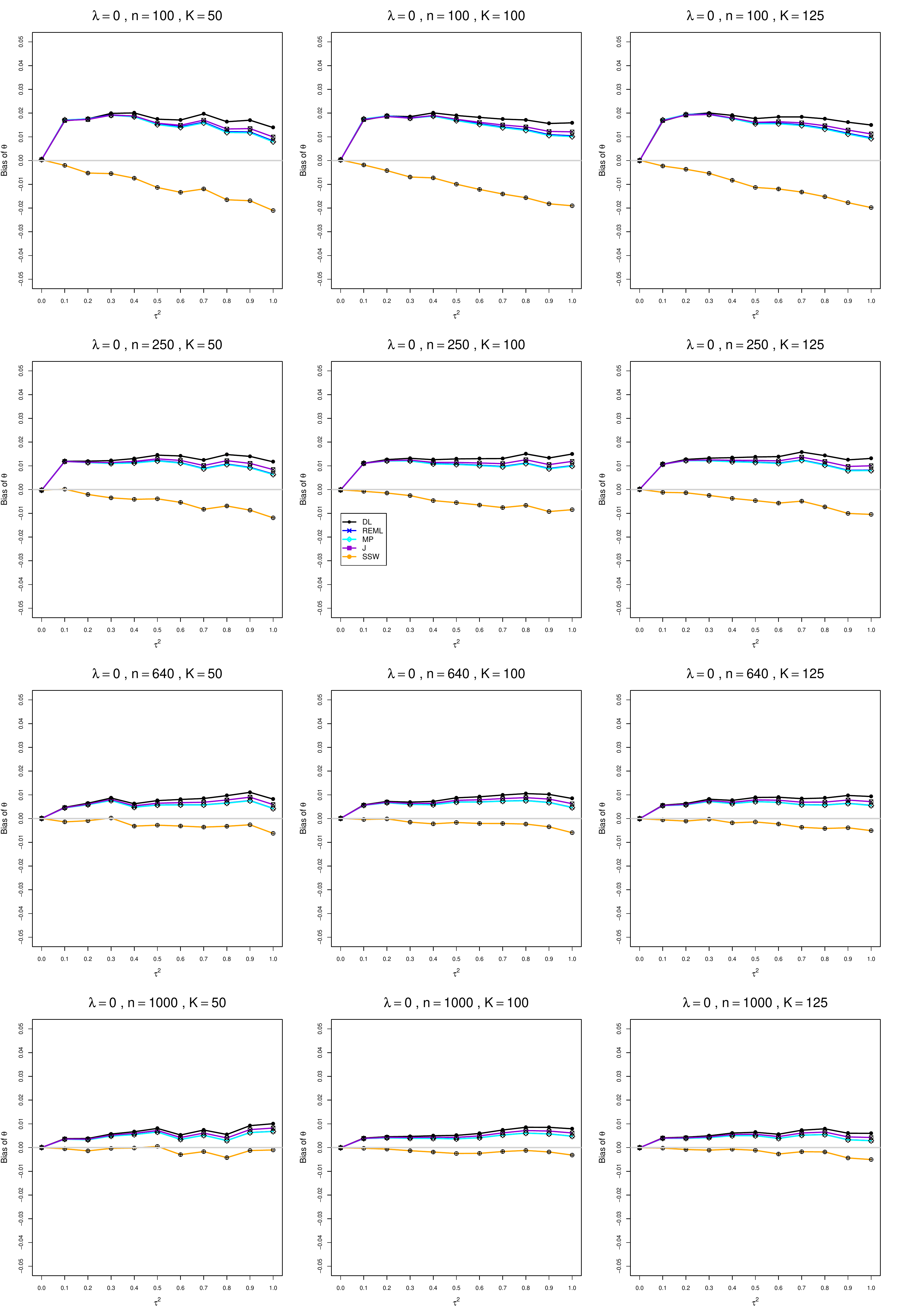}
	\caption{Bias of estimators of $\lambda$ for $\lambda=0$, $n = 100, \;250, \;640, \;1000$, and $K = 50, \;100, \;125$. Usual estimate of $\lambda_i$
		\label{BiasThetaRoM0ln_largeN_large_K}}
	\end{figure}
	
	\begin{figure}[t]
	\includegraphics[scale=0.33]{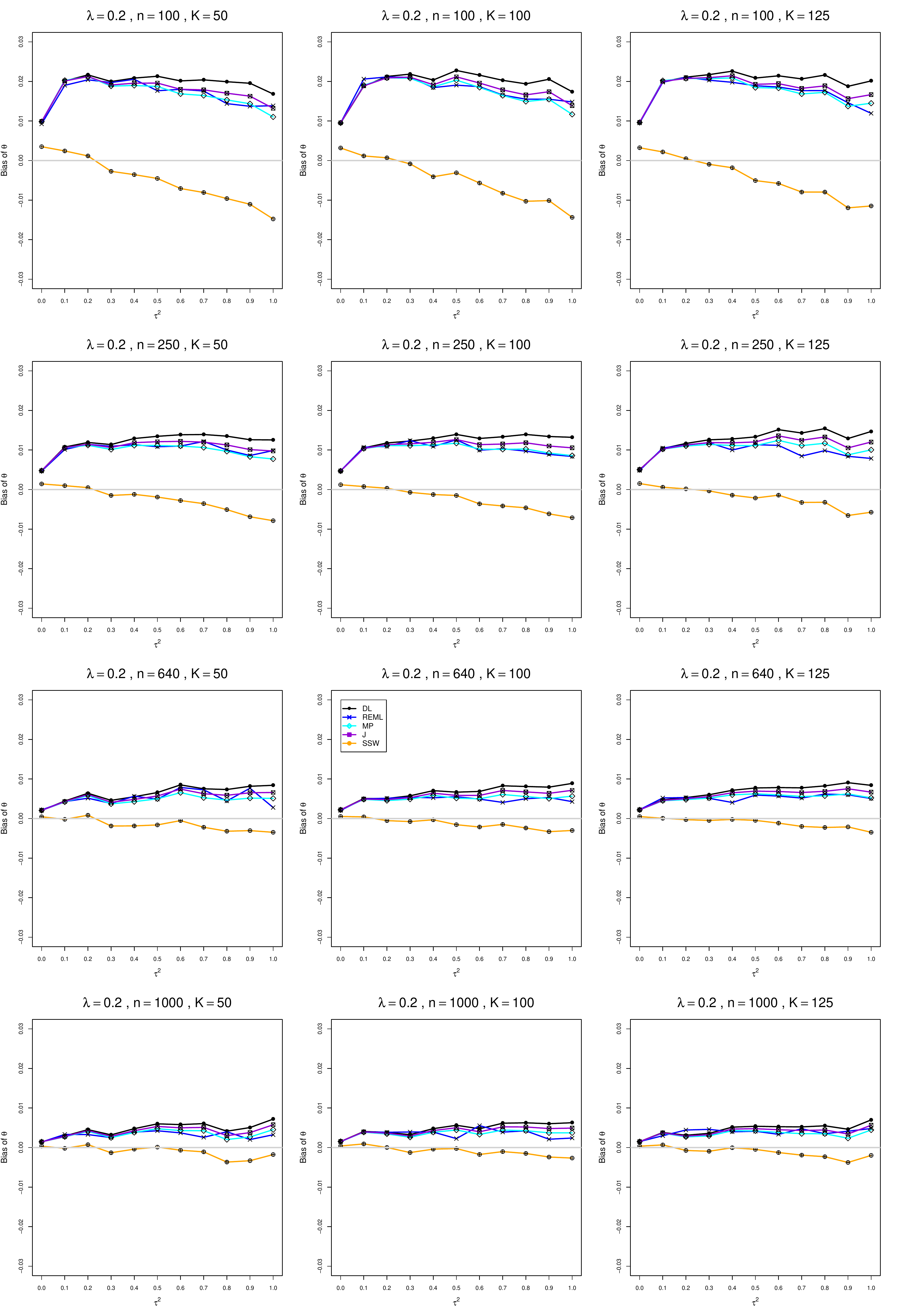}
	\caption{Bias of estimators of $\lambda$ for $\lambda=0.2$, $n = 100, \;250, \;640, \;1000$, and $K = 50, \;100, \;125$. Usual estimate of $\lambda_i$
		\label{BiasThetaRoM02ln_largeN_large_K}}
	\end{figure}
	
	\begin{figure}[t]
	\includegraphics[scale=0.33]{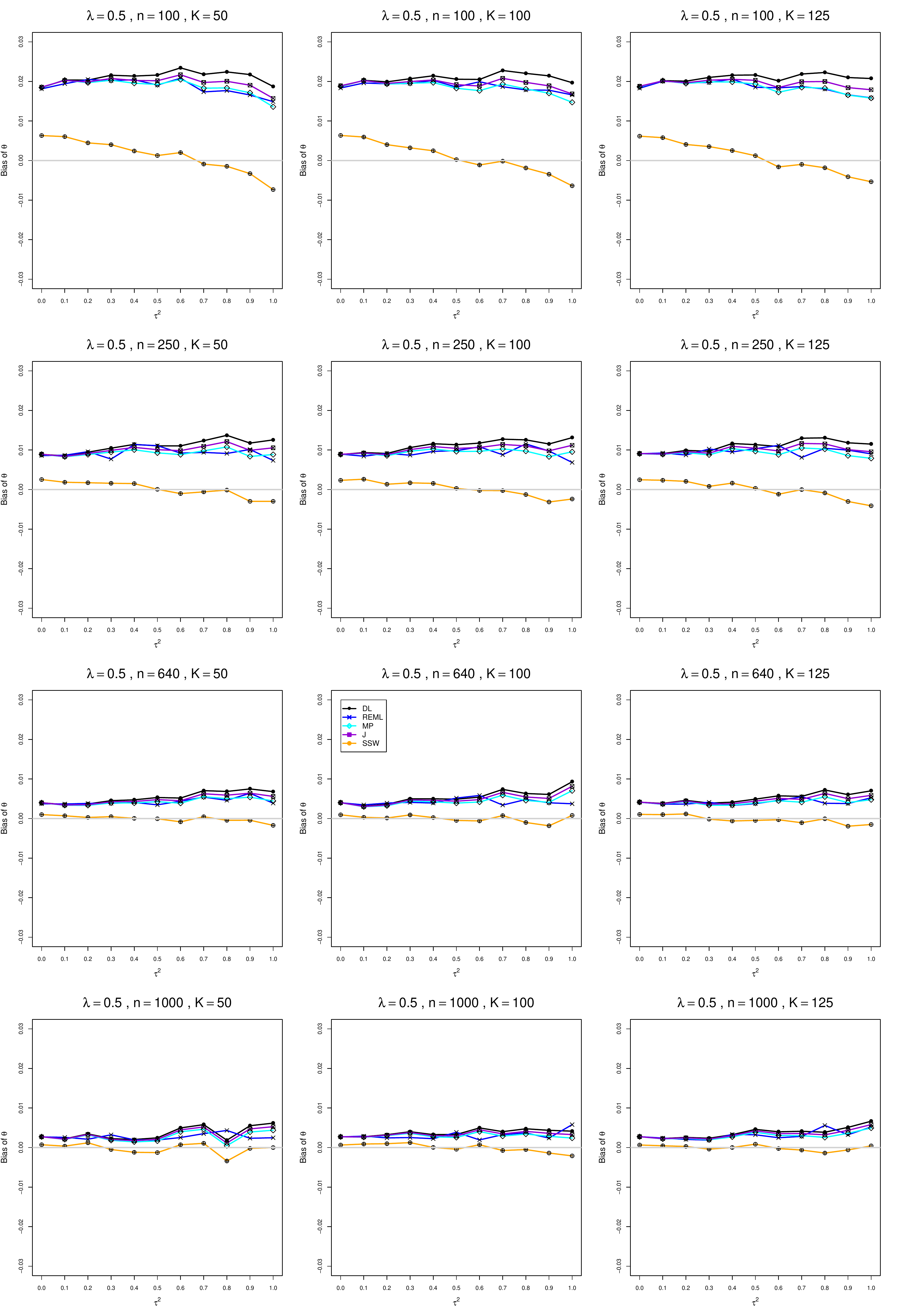}
	\caption{Bias of estimators of $\lambda$ for $\lambda=0.5$, $n = 100, \;250, \;640, \;1000$, and $K = 50, \;100, \;125$. Usual estimate of $\lambda_i$
		\label{BiasThetaRoM05ln_largeN_large_K}}
	\end{figure}
	
	\begin{figure}[t]
	\includegraphics[scale=0.33]{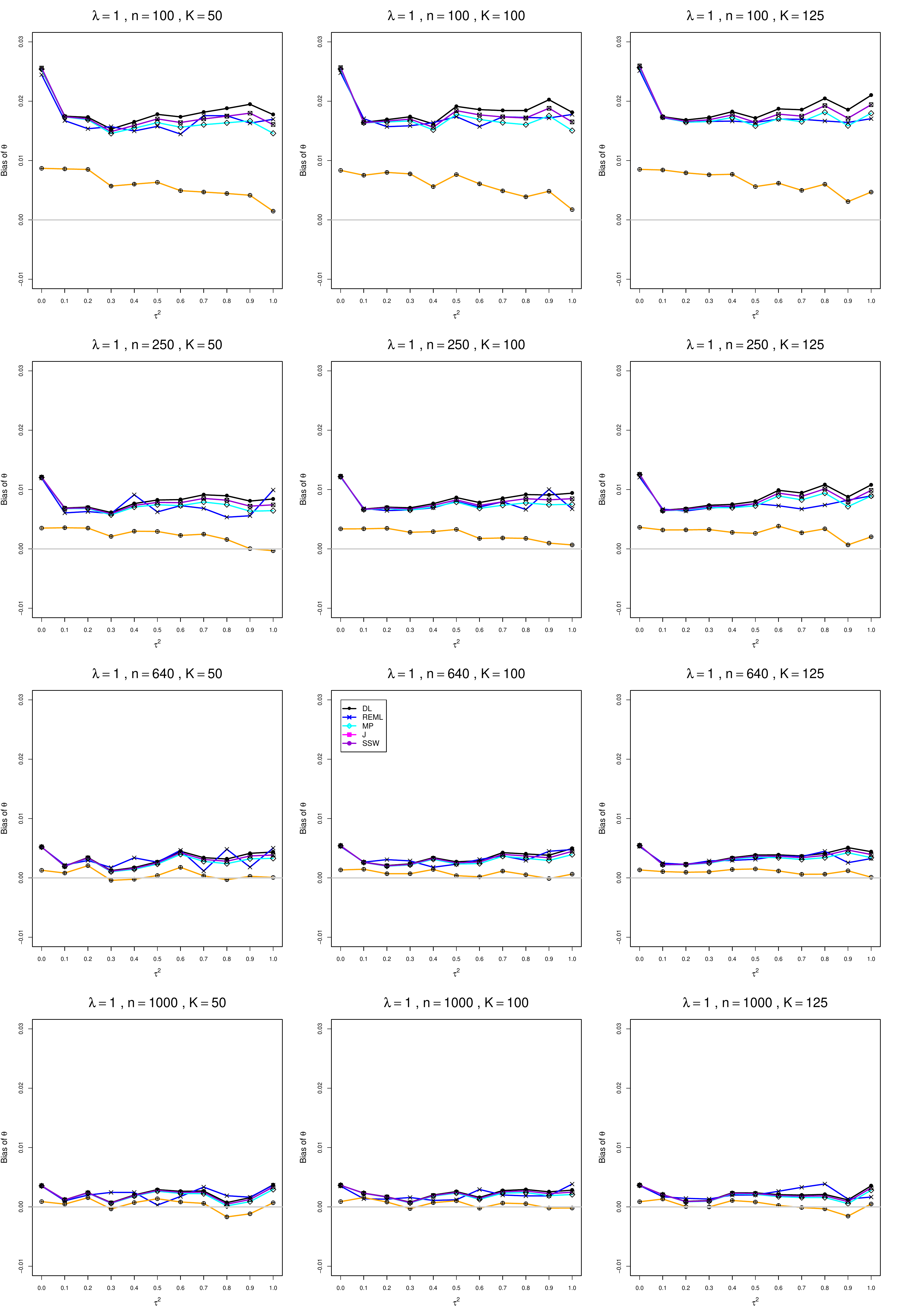}
	\caption{Bias of estimators of $\lambda$ for $\lambda=1$, $n = 100, \;250, \;640, \;1000$, and $K = 50, \;100, \;125$. Usual estimate of $\lambda_i$
		\label{BiasThetaRoM1ln_largeN_large_K}}
	\end{figure}
	
	\begin{figure}[t]
	\includegraphics[scale=0.33]{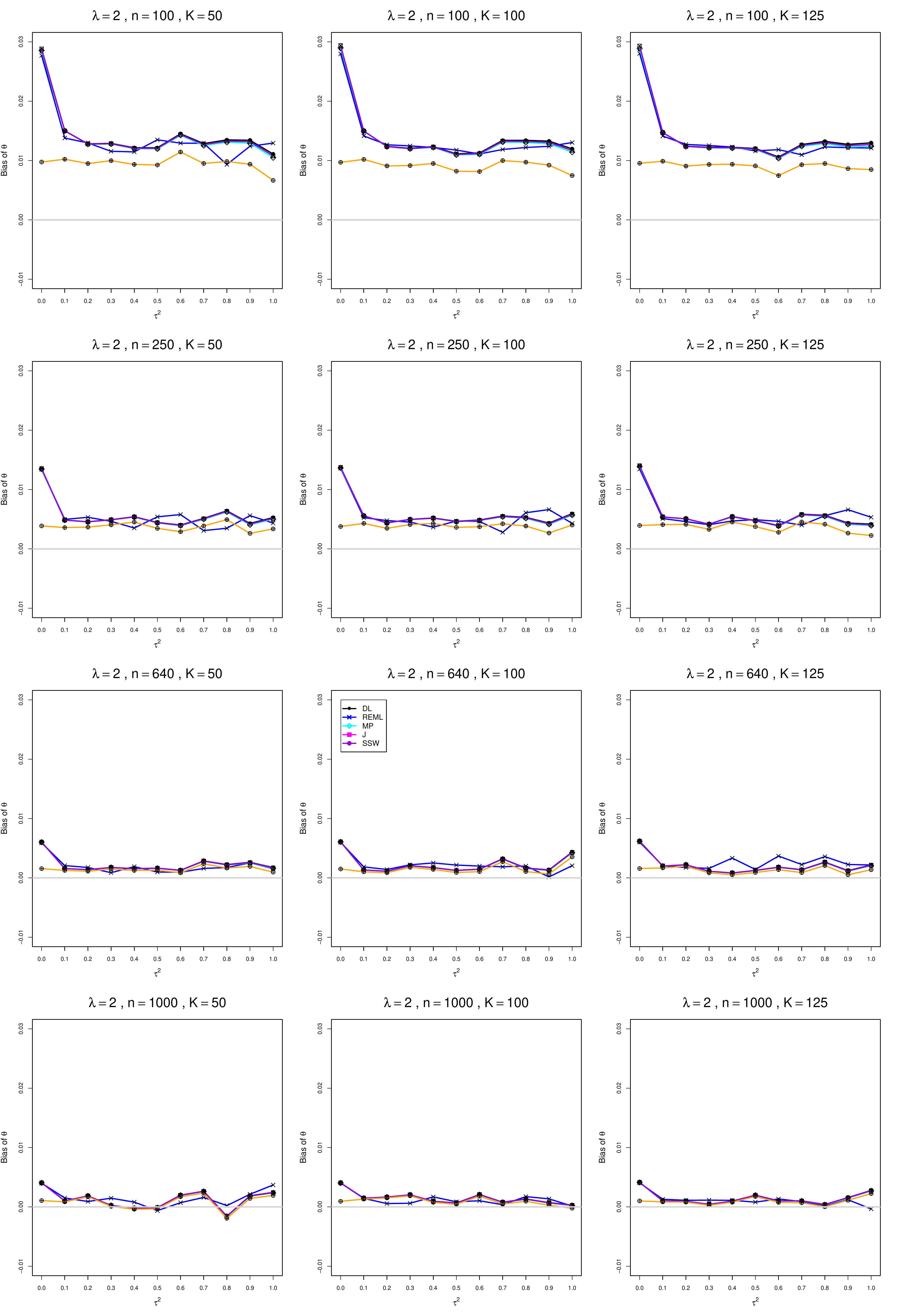}
	\caption{Bias of estimators of $\lambda$ for $\lambda=2$, $n = 100, \;250, \;640, \;1000$, and $K = 50, \;100, \;125$. Usual estimate of $\lambda_i$
		\label{BiasThetaRoM2ln_largeN_large_K}}
	\end{figure}

	\clearpage
	\subsection*{D3.2 Coverage of interval estimators of $\lambda$}
	Each figure corresponds to a value of $\lambda \;(= 0, 0.2, 0.5, 1, 2)$, a set of values of $n$ (= 100, 250, 640, 1000), and a set of values of $K$ (= 50, 100, 125).\\
	Each panel corresponds to a value of $n$ and a value of $K$ and has $\tau^2 = 0.0(0.1)1.0$ on the horizontal axis.\\
	The interval estimators of $\lambda$ are the companions to the inverse-variance-weighted point estimators
	\begin{itemize}
	\item DL (DerSimonian-Laird)
	\item REML (restricted maximum likelihood)
	\item MP (Mandel-Paule)
	\item J (Jackson)
	\end{itemize}
	and
	\begin{itemize}
	\item HKSJ (Hartung-Knapp-Sidik-Jonkman)
	\item HKSJ MP (HKSJ with MP estimator of $\tau^2$)
	\item SSW MP (SSW as center and half-width equal to critical value from $t_{K-1}$ times estimated standard deviation of SSW with $\hat{\tau}^2$ = $\hat{\tau}^2_{MP}$)
	\end{itemize}

	\clearpage
	\setcounter{figure}{0}
	\renewcommand{\thefigure}{D3.2.\arabic{figure}}
	\begin{figure}[t]
	\includegraphics[scale=0.35]{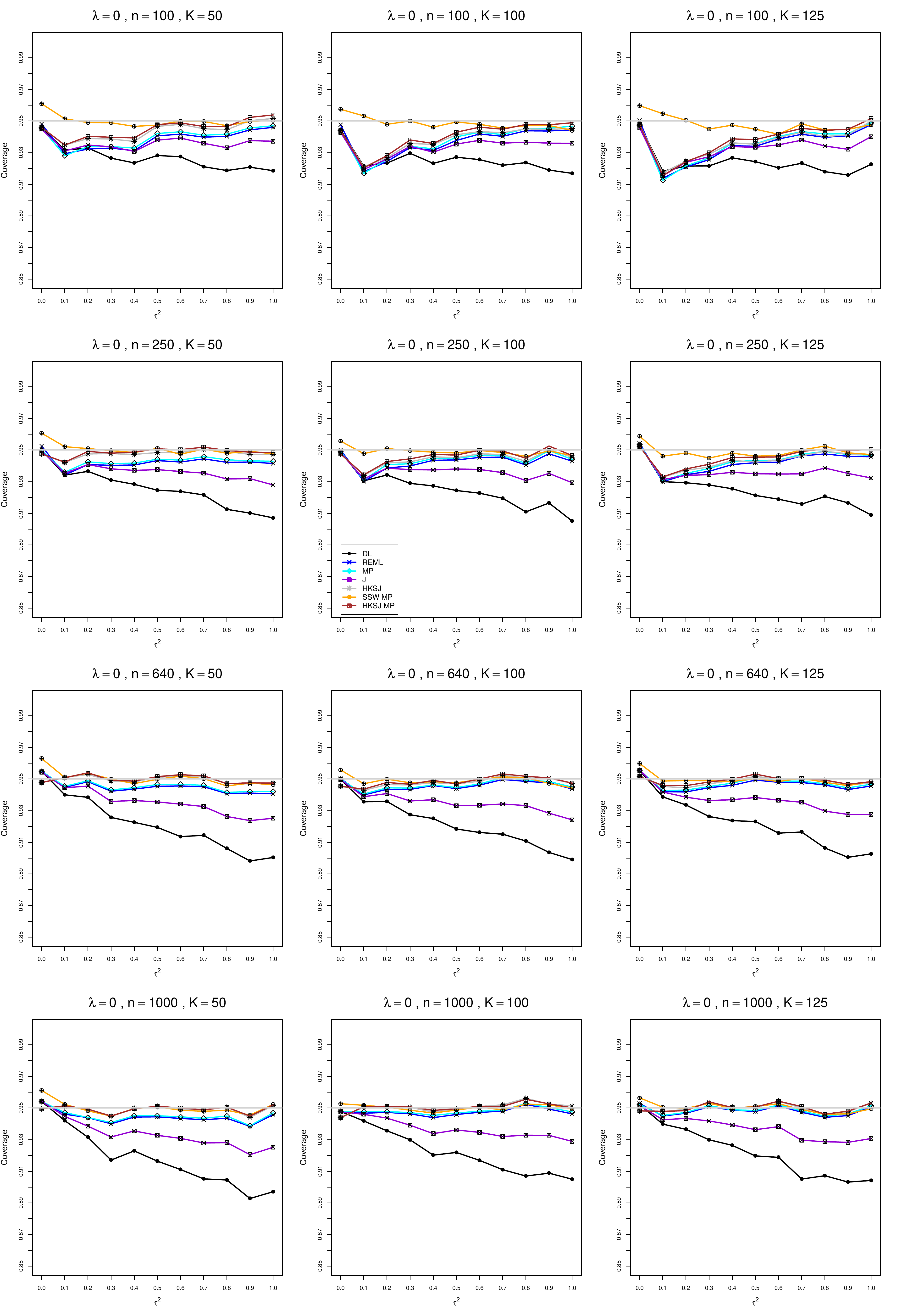}
	\caption{Coverage of 95\% confidence intervals for $\lambda$ when $\lambda=0$, $n = 100, \;250, \;640, \;1000$, and $K = 50, \;100, \;125$. Usual estimate of $\lambda_i$
		\label{CovThetaRoM0ln_largeN_large_K}}
	\end{figure}

	\begin{figure}[t]
	\includegraphics[scale=0.35]{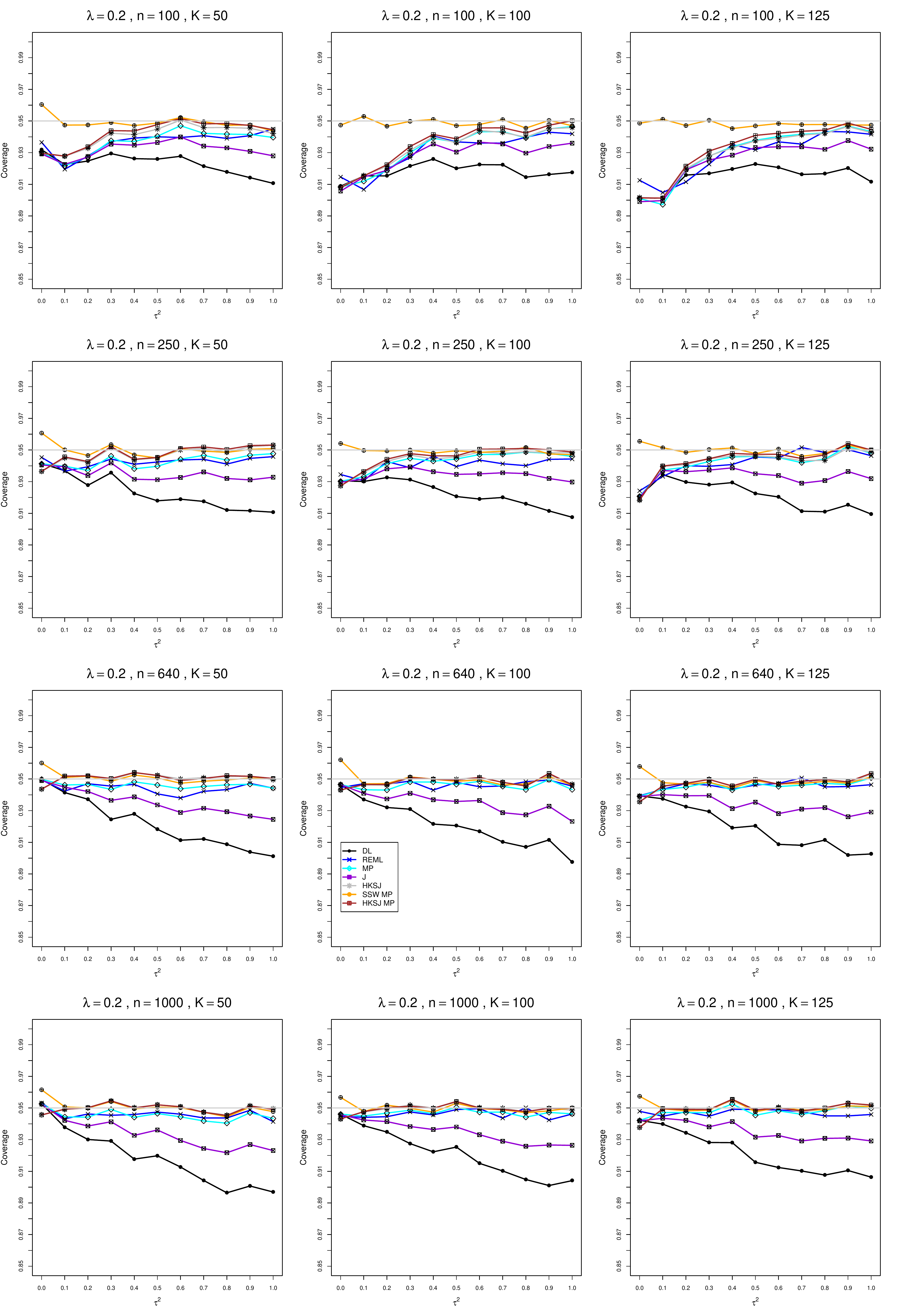}
	\caption{Coverage of 95\% confidence intervals for $\lambda$ when $\lambda=0.2$, $n = 100, \;250, \;640, \;1000$, and $K = 50, \;100, \;125$. Usual estimate of $\lambda_i$.
		\label{CovThetaRoM02ln_largeN_large_K}}
	\end{figure}

	\begin{figure}[t]
	\includegraphics[scale=0.35]{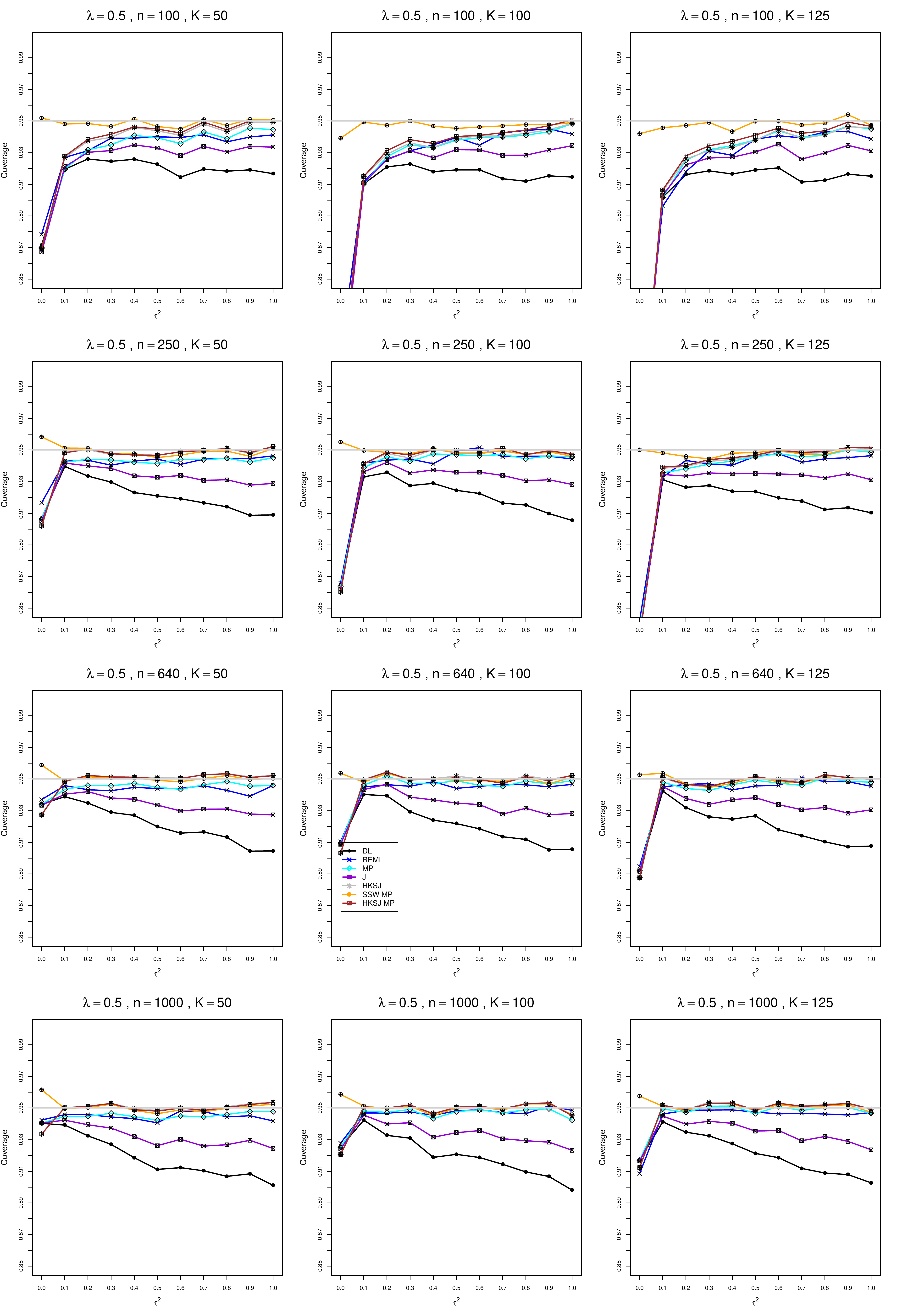}
	\caption{Coverage of 95\% confidence intervals for $\lambda$ when $\lambda=0.5$, $n = 100, \;250, \;640, \;1000$, and $K = 50, \;100, \;125$. Usual estimate of $\lambda_i$
		\label{CovThetaRoM05ln_largeN_large_K}}
	\end{figure}

	\begin{figure}[t]
	\includegraphics[scale=0.35]{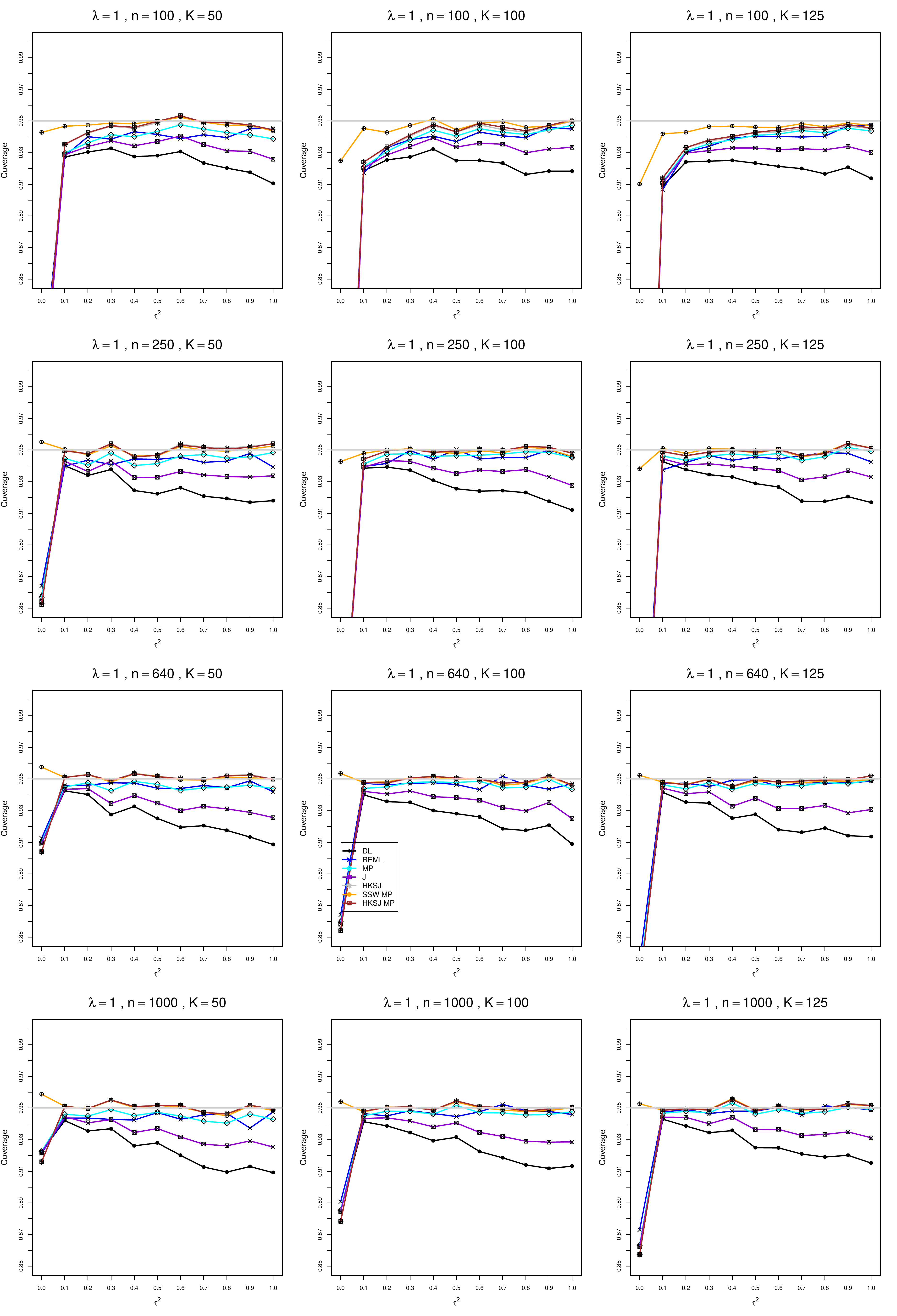}
	\caption{Coverage of 95\% confidence intervals for $\lambda$ when $\lambda=1$, $n = 100, \;250, \;640, \;1000$, and $K = 50, \;100, \;125$. Usual estimate of $\lambda_i$
		\label{CovThetaRoM1ln_largeN_large_K}}
	\end{figure}
	
	\begin{figure}[t]
	\includegraphics[scale=0.35]{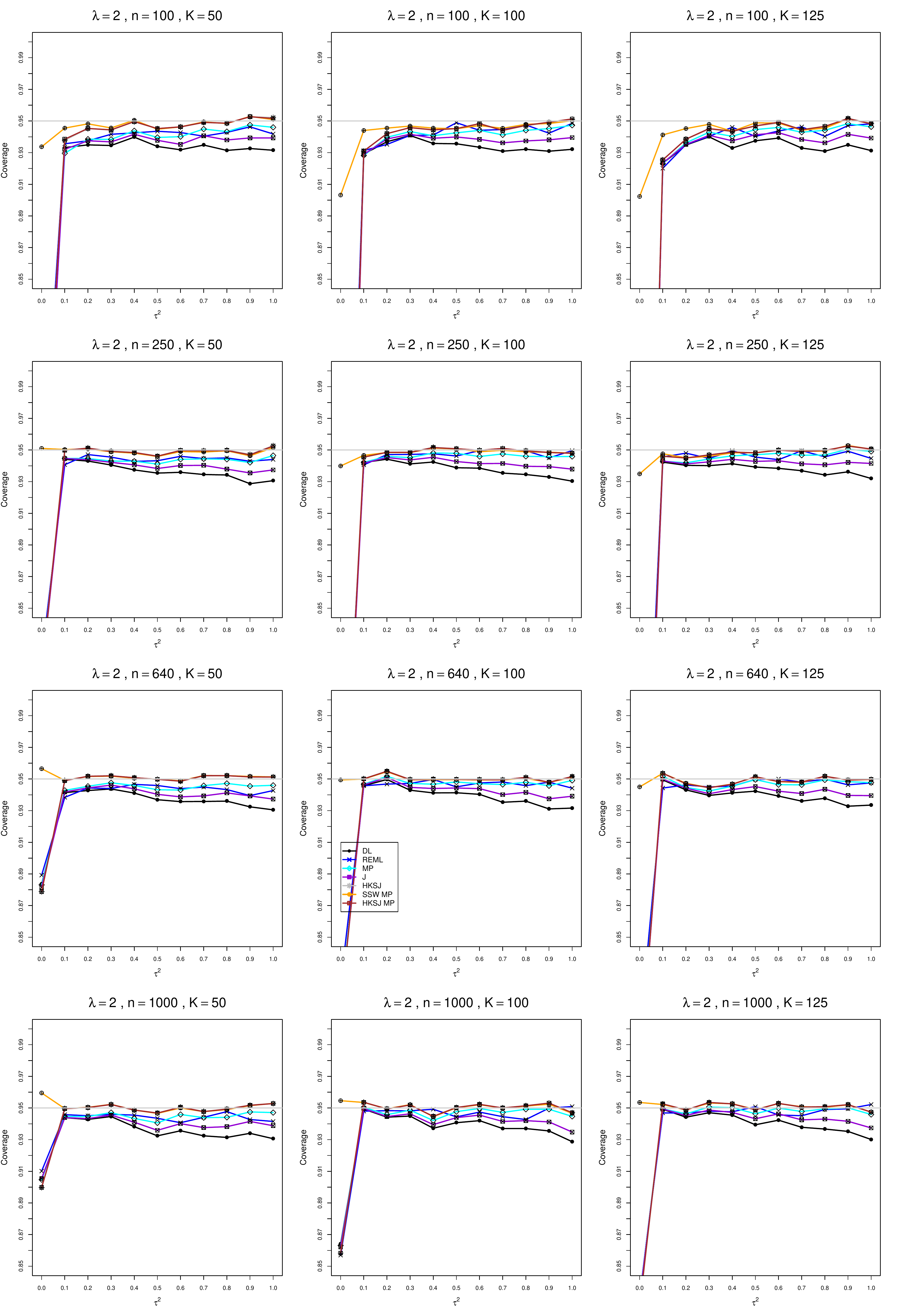}
	\caption{Coverage of 95\% confidence intervals for $\lambda$ when $\lambda=2$, $n = 100, \;250, \;640, \;1000$, and $K = 50, \;100, \;125$. Usual estimate of $\lambda_i$
		\label{CovThetaRoM2ln_largeN_large_K}}
	\end{figure}

	\clearpage
	\section*{D4. Lognormal model, bias-corrected estimator of $\lambda_i$, $n= 100, 250, 640, 1000$, $K=50,100,125$}
	\subsection*{D4.1 Bias of point estimators of $\lambda$}
	Each figure corresponds to a value of $\lambda \;(= 0, 0.2, 0.5, 1, 2)$, a set of values of $n$ (= 100, 250, 640, 1000), and a set of values of $K$ (= 50, 100, 125).\\
	Each panel corresponds to a value of $n$ and a value of $K$ and has $\tau^2 = 0.0(0.1)1.0$ on the horizontal axis.\\
	The point estimators of $\lambda$ are
	\begin{itemize}
	\item DL (DerSimonian-Laird)
	\item REML (restricted maximum likelihood)
	\item MP (Mandel-Paule)
	\item J (Jackson)
	\item SSW (sample-size-weighted)
	\end{itemize}

	\clearpage
	\setcounter{figure}{0}
	\renewcommand{\thefigure}{D4.1.\arabic{figure}}
	\begin{figure}[t]
	\includegraphics[scale=0.33]{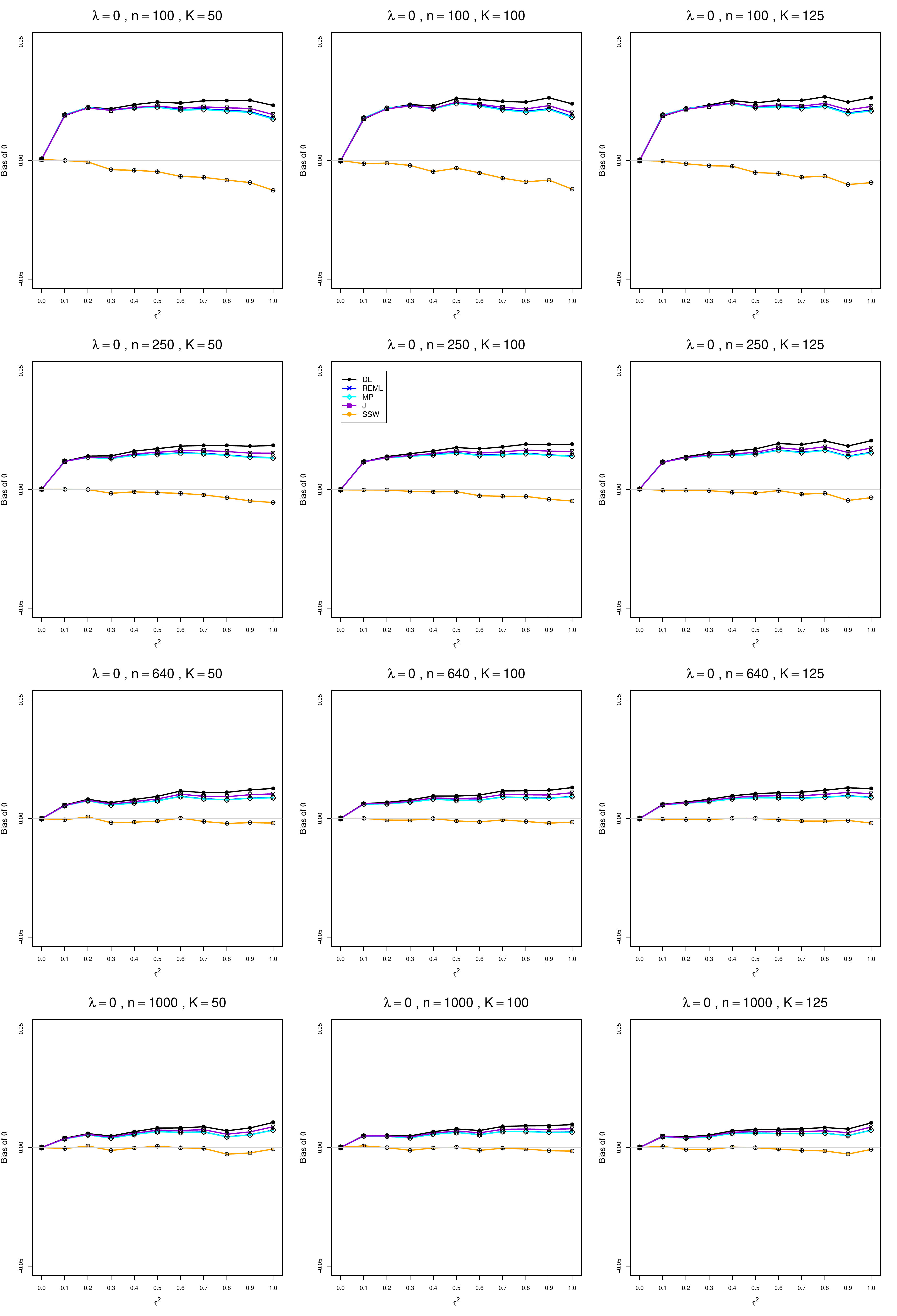}
	\caption{Bias of estimators of $\lambda$ for $\lambda=0$, $n = 100, \;250, \;640, \;1000$, and $K = 50, \;100, \;125$. Bias-corrected estimate of $\lambda_i$
		\label{BiasThetaRoM0lnCor_largeN_large_K}}
	\end{figure}
	
	\begin{figure}[t]
	\includegraphics[scale=0.33]{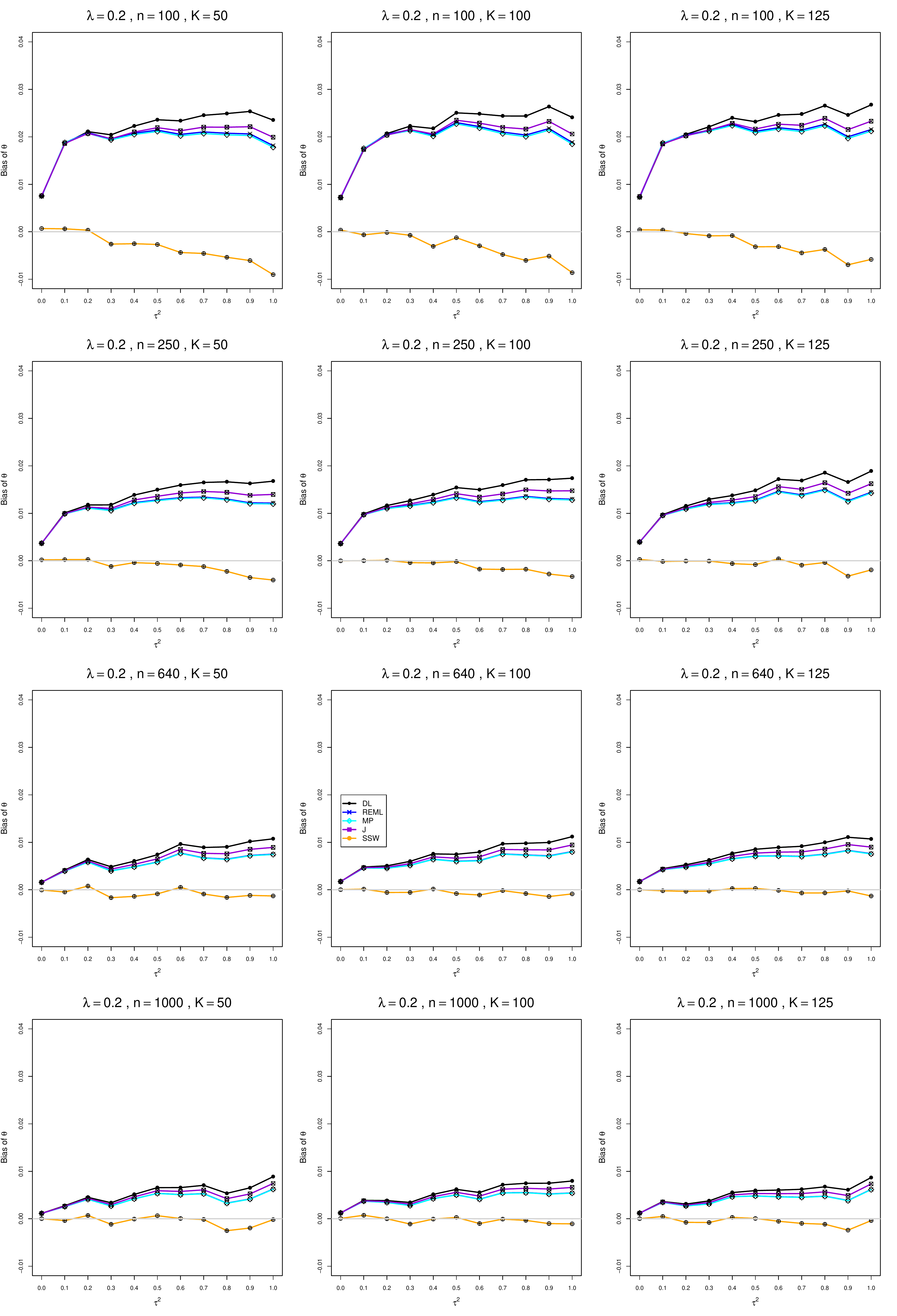}
	\caption{Bias of estimators of $\lambda$ for $\lambda=0.2$, $n = 100, \;250, \;640, \;1000$, and $K = 50, \;100, \;125$. Bias-corrected estimate of $\lambda_i$
		\label{BiasThetaRoM02lnCor_largeN_large_K}}
	\end{figure}
	
	\begin{figure}[t]
	\includegraphics[scale=0.33]{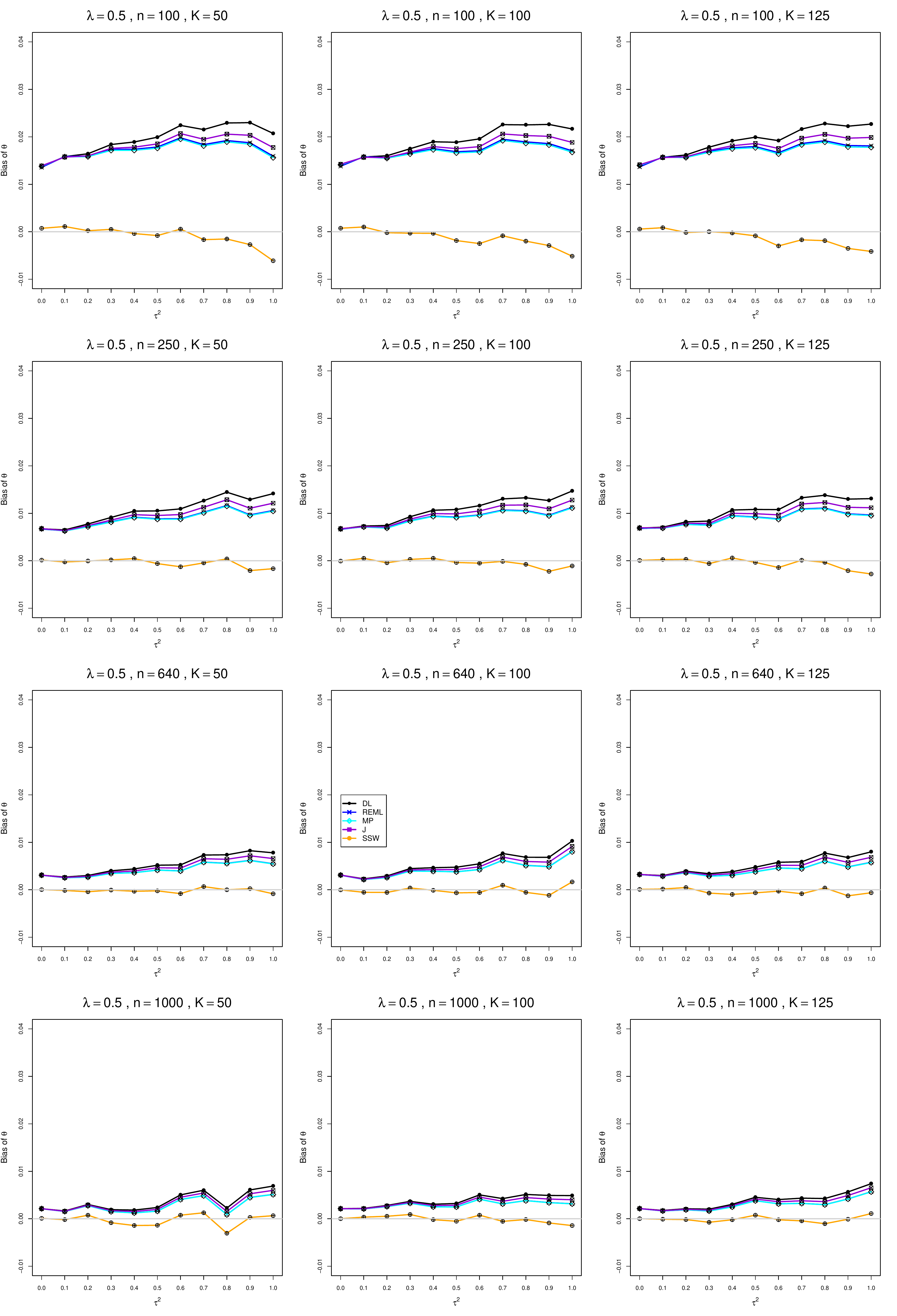}
	\caption{Bias of estimators of $\lambda$ for $\lambda=0.5$, $n = 100, \;250, \;640, \;1000$, and $K = 50, \;100, \;125$. Bias-corrected estimate of $\lambda_i$
		\label{BiasThetaRoM05lnCor_largeN_large_K}}
	\end{figure}
	
	\begin{figure}[t]
	\includegraphics[scale=0.33]{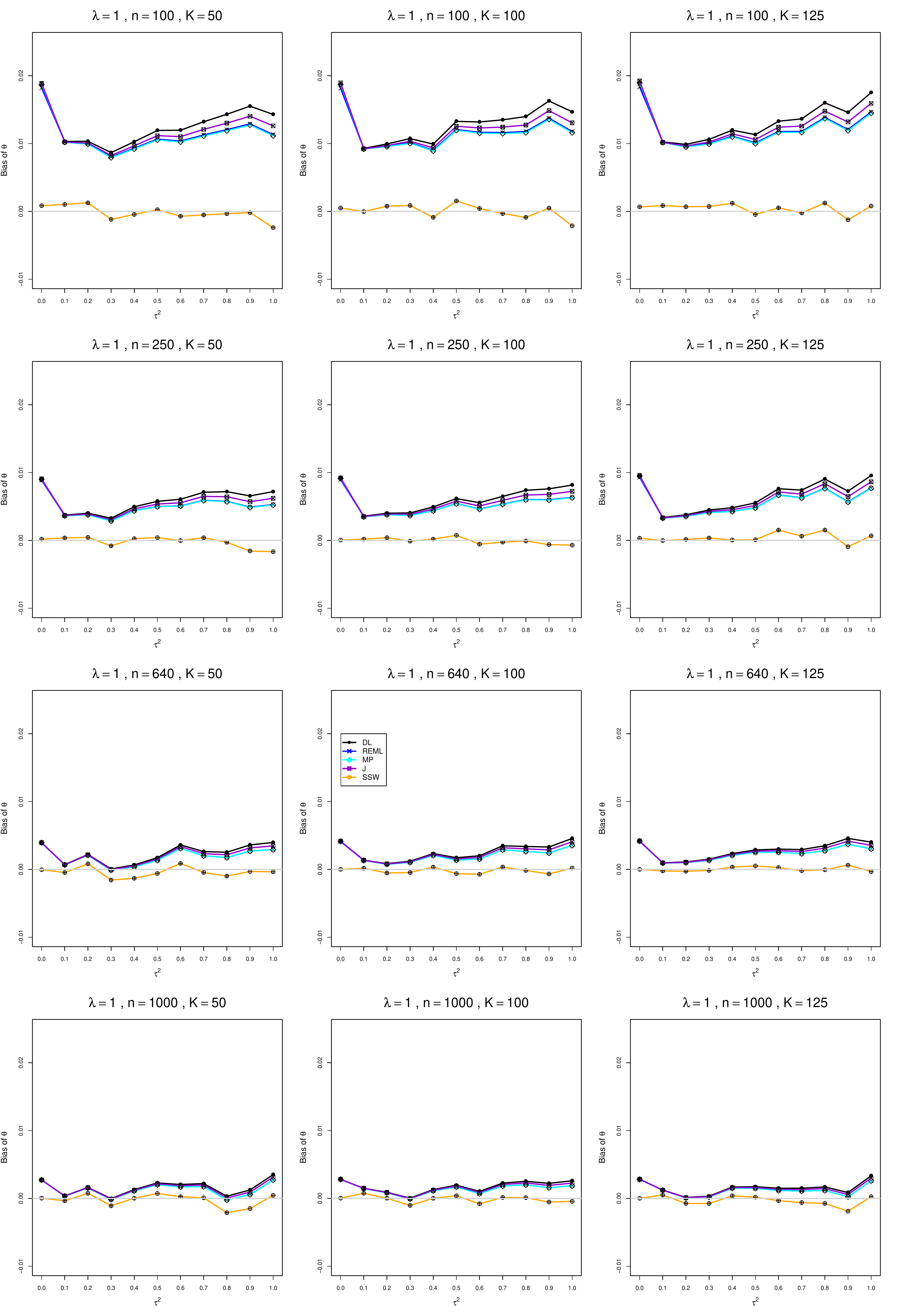}
	\caption{Bias of estimators of $\lambda$ for $\lambda=1$, $n = 100, \;250, \;640, \;1000$, and $K = 50, \;100, \;125$. Bias-corrected estimate of $\lambda_i$
		\label{BiasThetaRoM1lnCor_largeN_large_K}}
	\end{figure}
	
	\begin{figure}[t]
	\includegraphics[scale=0.33]{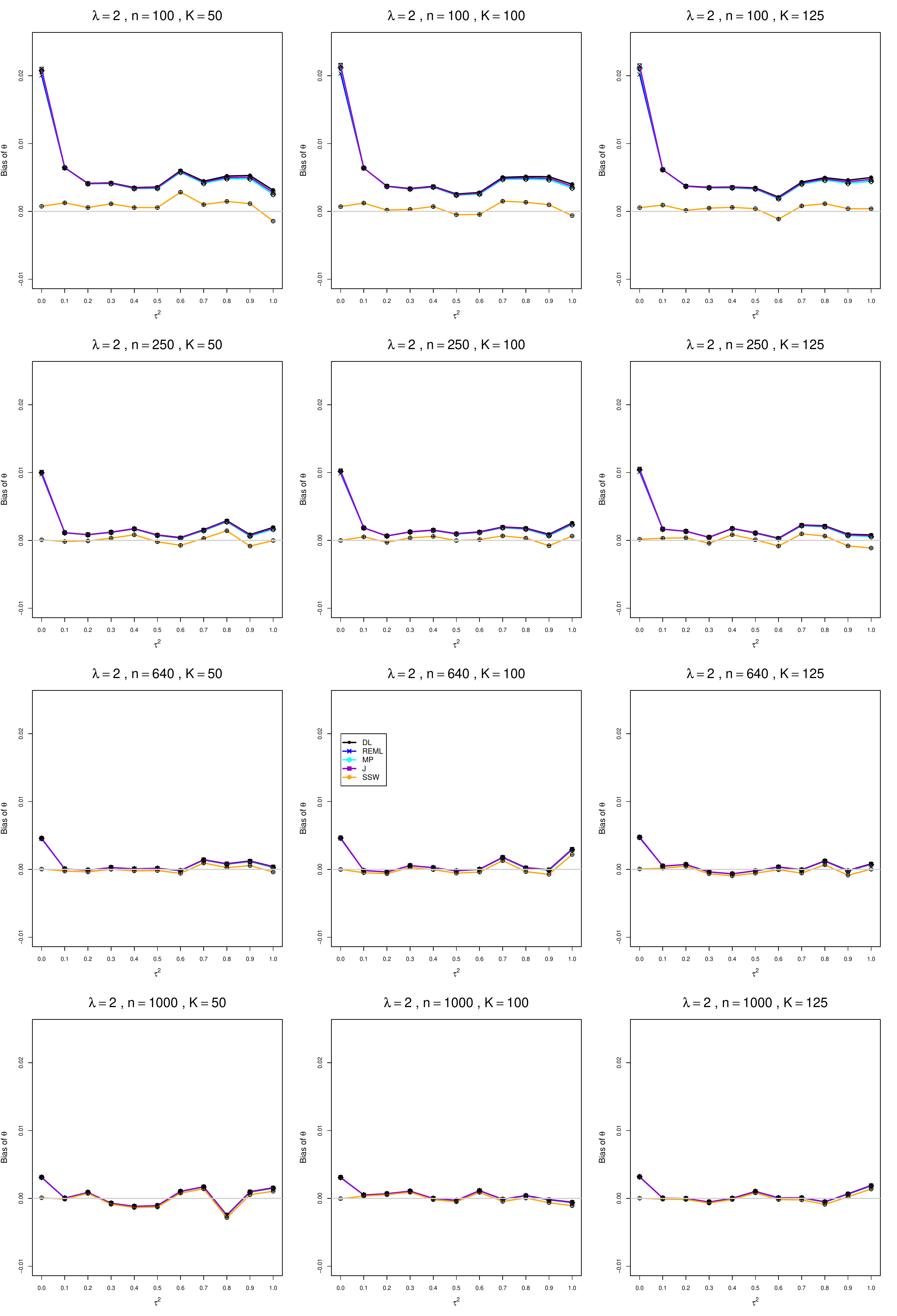}
	\caption{Bias of estimators of $\lambda$ for $\lambda=2$, $n = 100, \;250, \;640, \;1000$, and $K = 50, \;100, \;125$. Bias-corrected estimate of $\lambda_i$
		\label{BiasThetaRoM2lnCor_largeN_large_K}}
	\end{figure}
	
	\clearpage
	\subsection*{D4.2 Coverage of interval estimators of $\lambda$}
	Each figure corresponds to a value of $\lambda \;(= 0, 0.2, 0.5, 1, 2)$, a set of values of $n$ (= 100, 250, 640, 1000), and a set of values of $K$ (= 50, 100, 125).\\
	Each panel corresponds to a value of $n$ and a value of $K$ and has $\tau^2 = 0.0(0.1)1.0$ on the horizontal axis.\\
	The interval estimators of $\lambda$ are the companions to the inverse-variance-weighted point estimators
	\begin{itemize}
	\item DL (DerSimonian-Laird)
	\item REML (restricted maximum likelihood)
	\item MP (Mandel-Paule)
	\item J (Jackson)
	\end{itemize}
	and
	\begin{itemize}
	\item HKSJ (Hartung-Knapp-Sidik-Jonkman)
	\item HKSJ MP (HKSJ with MP estimator of $\tau^2$)
	\item SSW MP (SSW as center and half-width equal to critical value from $t_{K-1}$ times estimated standard deviation of SSW with $\hat{\tau}^2$ = $\hat{\tau}^2_{MP}$)
	\end{itemize}

	\clearpage
	\setcounter{figure}{0}
	\renewcommand{\thefigure}{D4.2.\arabic{figure}}
	\begin{figure}[t]
	\includegraphics[scale=0.35]{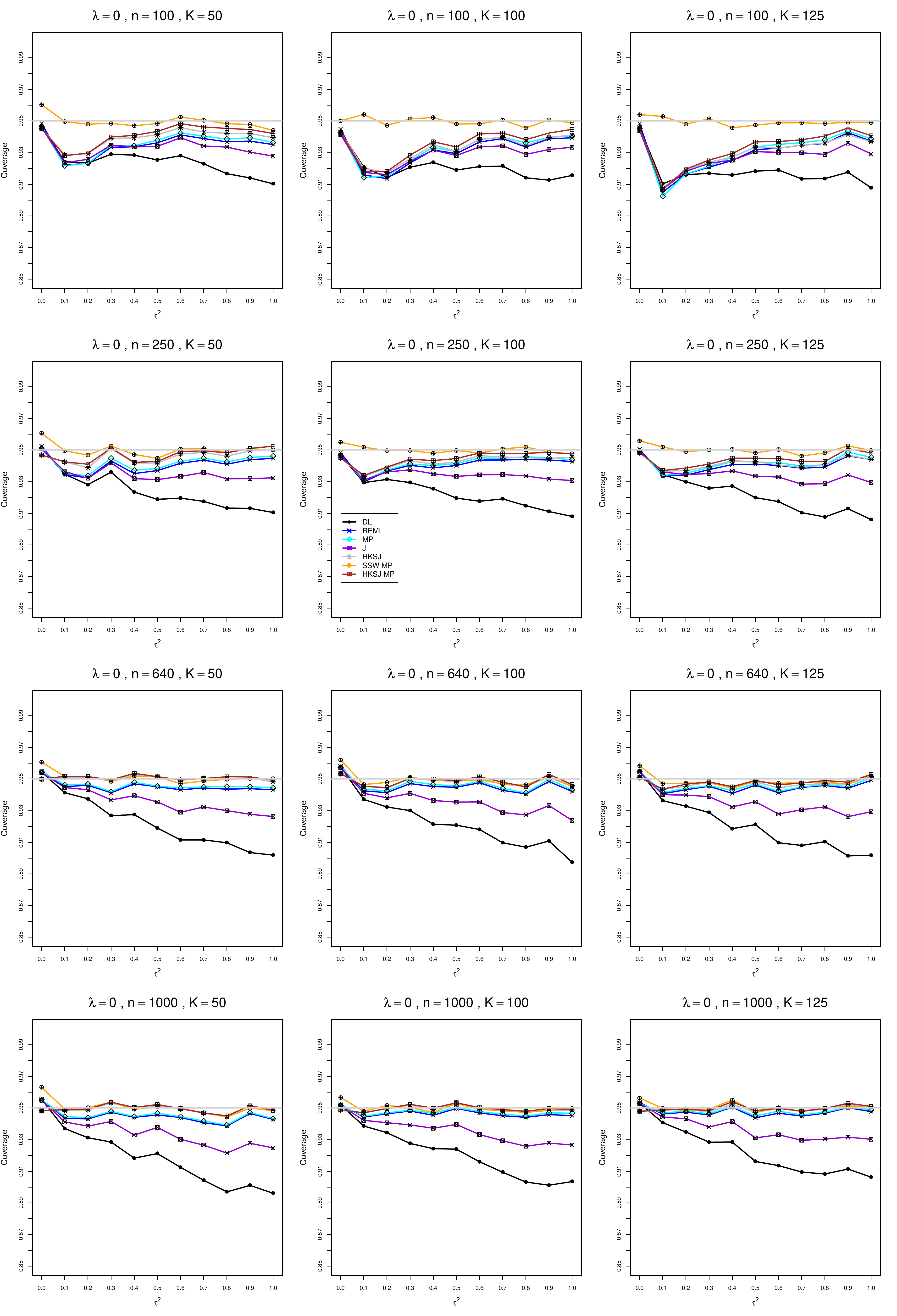}
	\caption{Coverage of 95\% confidence intervals for $\lambda$ when $\lambda=0$, $n = 100, \;250, \;640, \;1000$, and $K = 50, \;100, \;125$. Bias-corrected estimate of $\lambda_i$ 		\label{CovThetaRoM0lnCor_largeN_large_K}}
	\end{figure}
	\begin{figure}[t]
	\includegraphics[scale=0.35]{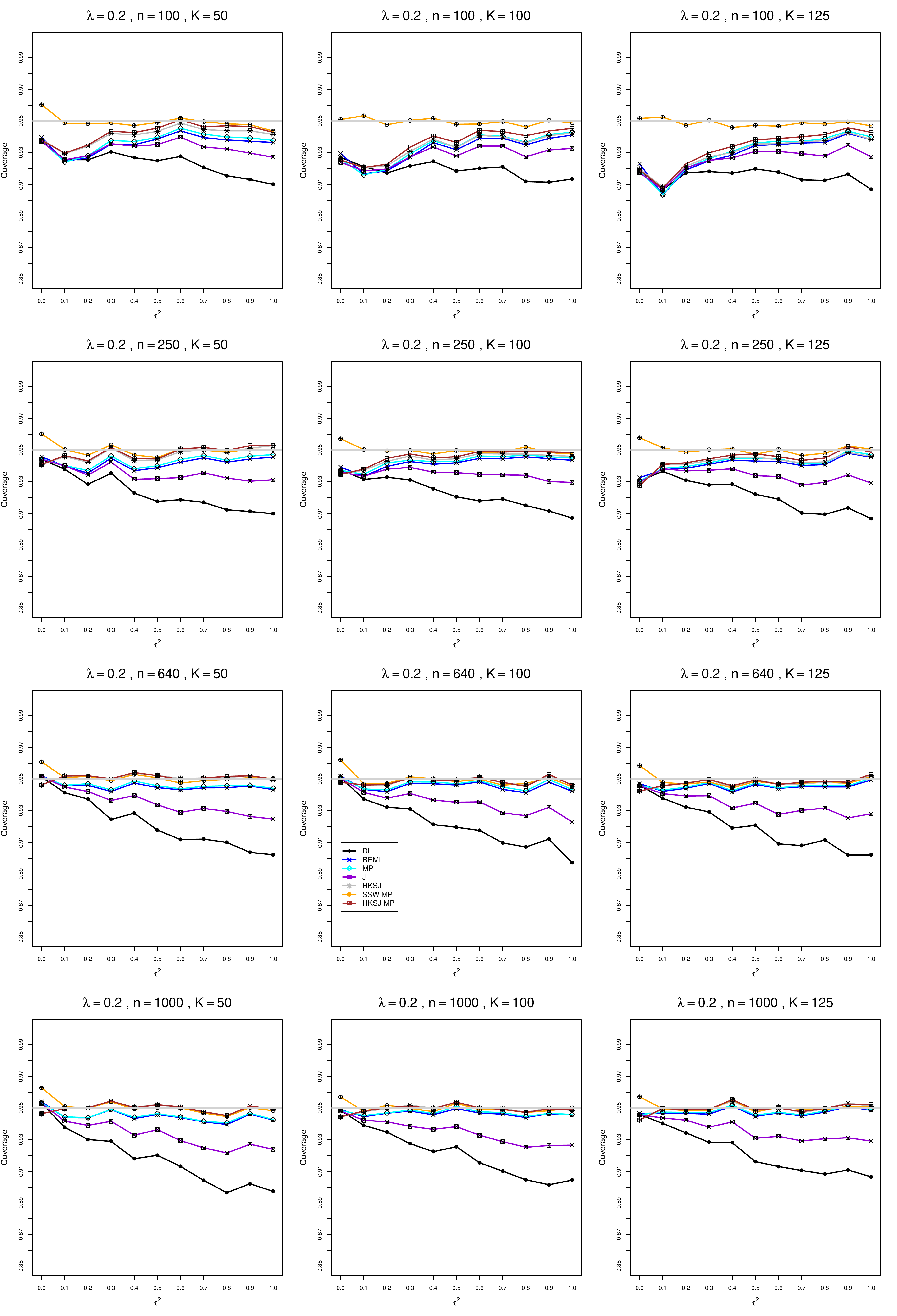}
	\caption{Coverage of 95\% confidence intervals for $\lambda$ when $\lambda=0.2$, $n = 100, \;250, \;640, \;1000$, and $K = 50, \;100, \;125$. Bias-corrected estimate of $\lambda_i$ 		\label{CovThetaRoM02lnCor_largeN_large_K}}
	\end{figure}
	
	\begin{figure}[t]
	\includegraphics[scale=0.35]{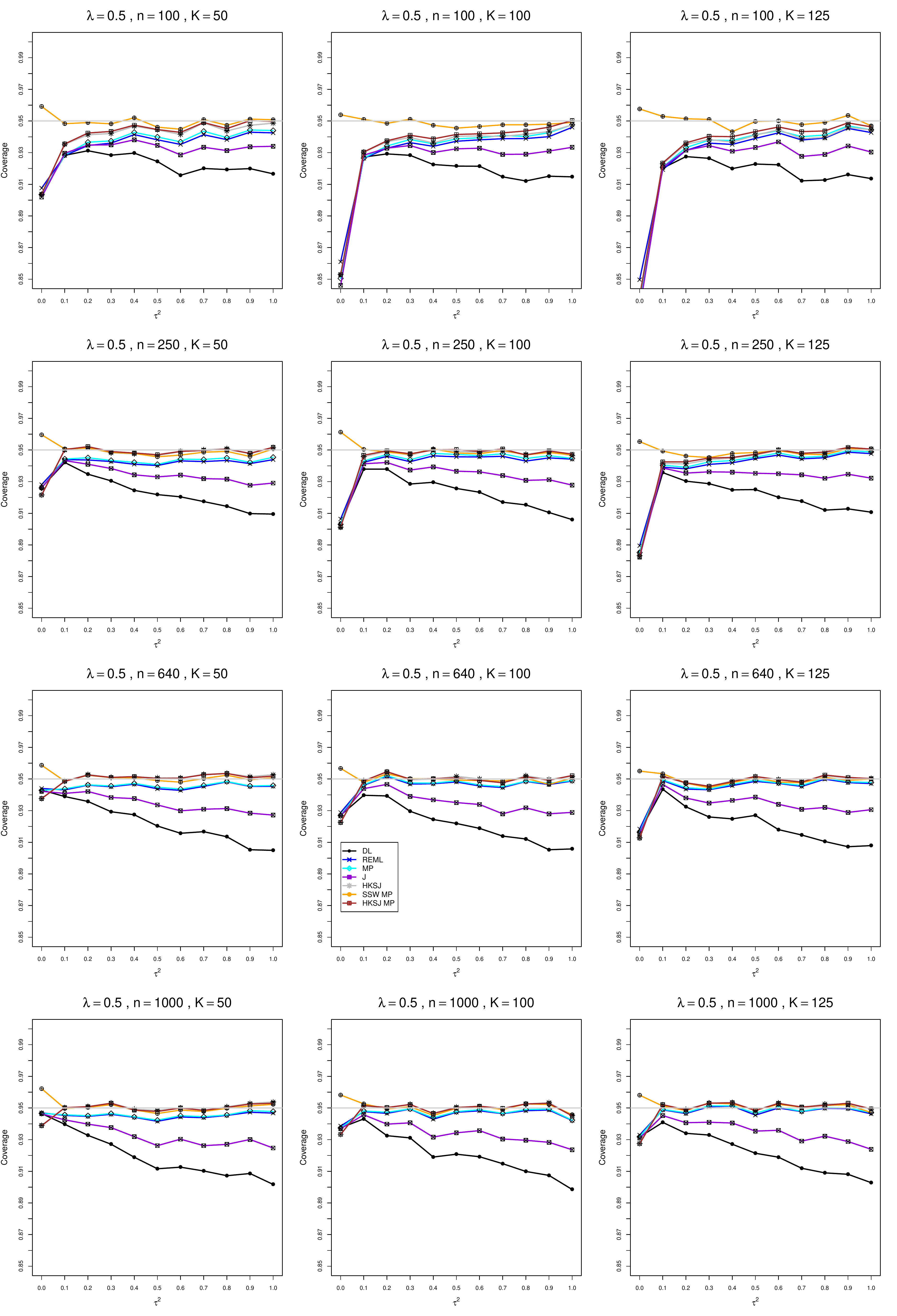}
	\caption{Coverage of 95\% confidence intervals for $\lambda$ when $\lambda=0.5$, $n = 100, \;250, \;640, \;1000$, and $K = 50, \;100, \;125$. Bias-corrected estimate of $\lambda_i$ 		\label{CovThetaRoM05lnCor_largeN_large_K}}
	\end{figure}
	
	\begin{figure}[t]
	\includegraphics[scale=0.35]{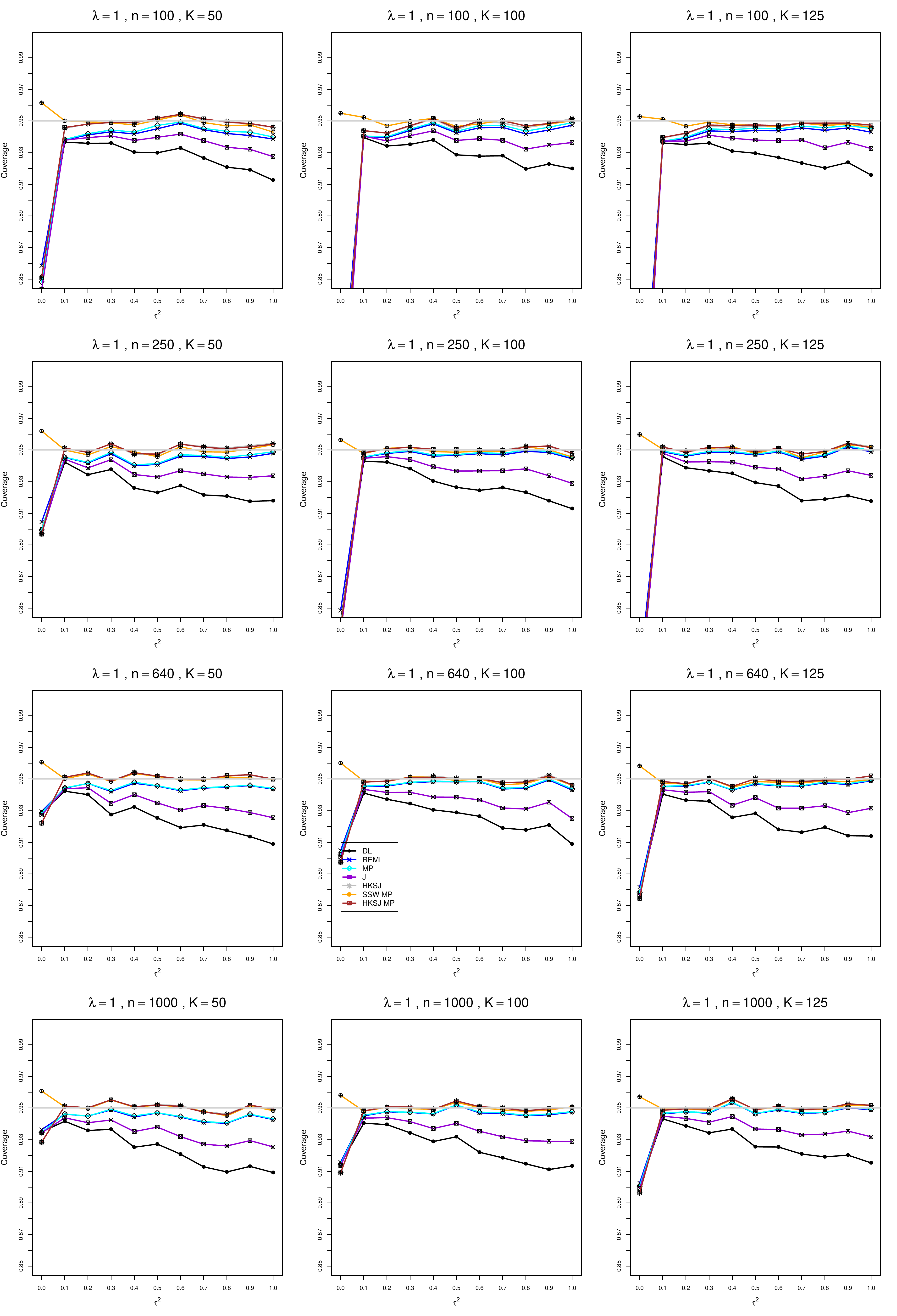}
	\caption{Coverage of 95\% confidence intervals for $\lambda$ when $\lambda=1$, $n = 100, \;250, \;640, \;1000$, and $K = 50, \;100, \;125$. Bias-corrected estimate of $\lambda_i$ 		\label{CovThetaRoM1lnCor_largeN_large_K}}
	\end{figure}
	\begin{figure}[t]
	\includegraphics[scale=0.35]{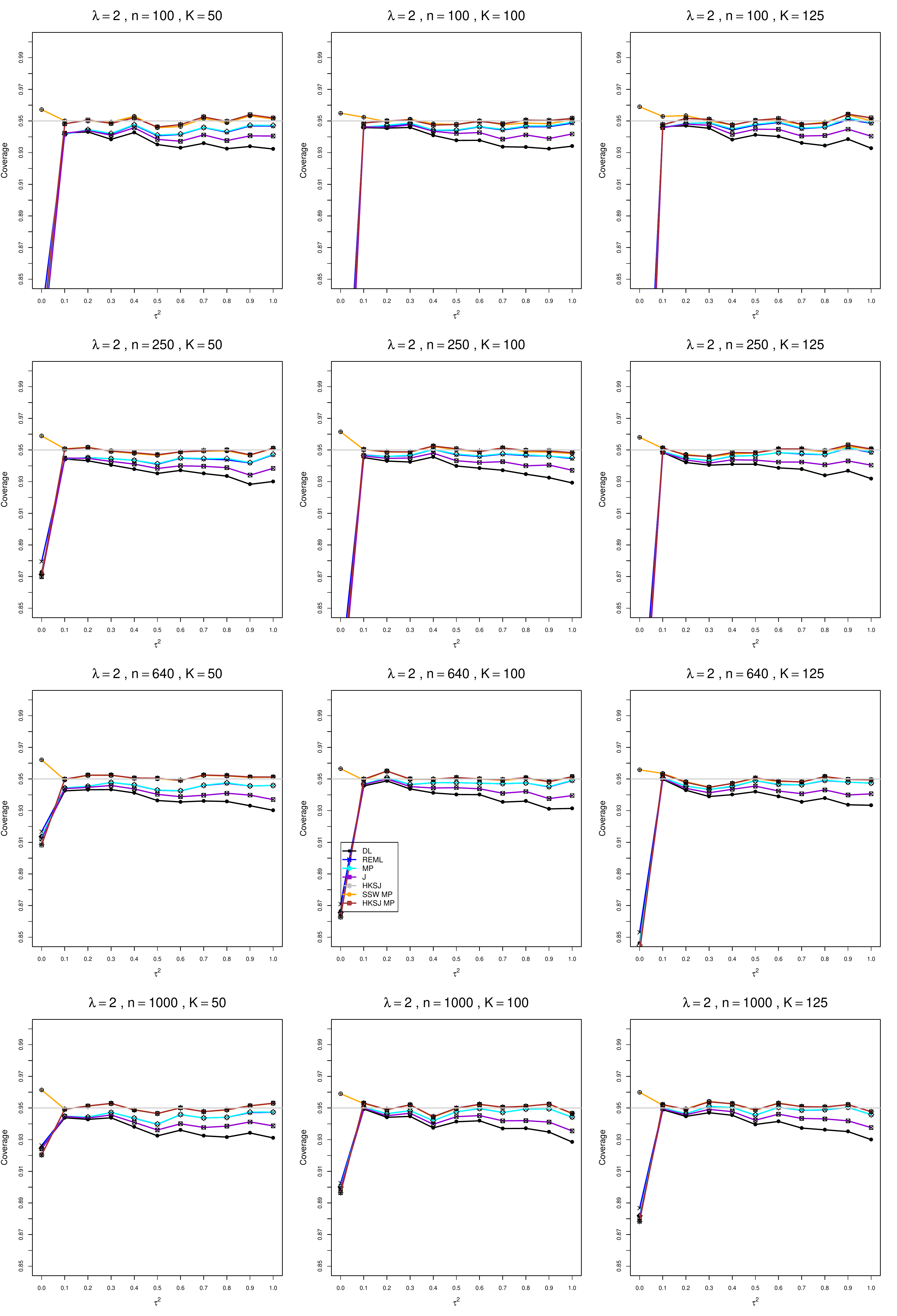}
	\caption{Coverage of 95\% confidence intervals for $\lambda$ when $\lambda=2$, $n = 100, \;250, \;640, \;1000$, and $K = 50, \;100, \;125$. Bias-corrected estimate of $\lambda_i$ 		\label{CovThetaRoM2lnCor_largeN_large_K}}
	\end{figure}

\end{document}